\documentclass[11pt,a4paper,twoside,final]{book}

\usepackage[utf8]{inputenc}  % nastavuje použité kódování, uživatelé unix zamění cp1250 za latin2
\usepackage[english]{babel}
\usepackage{natbib}
\usepackage{wrapfig}
\usepackage{array}
\usepackage{times}
\usepackage{url}
\usepackage[font=small]{caption}
\usepackage{mathptmx}
\usepackage{longtable,lipsum}
\usepackage{tocloft,calc}
\usepackage{url}
\usepackage[titletoc,toc,page]{appendix}
\usepackage{multicol}
\usepackage[usenames,dvipsnames]{xcolor}
\usepackage{graphicx} % nezbytné pro standardní vkládání obrázků do dokumentu
\usepackage{amssymb}
\usepackage{amsmath}
\usepackage{amsthm}
\usepackage{txfonts}
\usepackage{color}
\definecolor{DarkBlue}{rgb}{0.0,0.1,0.4}
\definecolor{Red}{rgb}{0.6,0.2,0.1}
\definecolor{DarkRed}{rgb}{0.5,0.0,0.5}
\definecolor{Blue}{rgb}{0.8,0.1,0.8}
\definecolor{Zelinkava}{rgb}{0.2,0.5,0.2}
\definecolor{Pink}{rgb}{0.0,0.7,0.7}
\definecolor{White}{rgb}{1.0,1.0,1.0}
\usepackage{hyperref}
\usepackage{textcomp}
\usepackage[fit]{truncate}
\hypersetup{
    colorlinks=true,
    linkcolor=Blue,
    citecolor=Blue,
    filecolor=Blue,     
    urlcolor=Blue,
}
\usepackage[Bc,Barevne]{sci.muni.thesis}

\NazevUstavu{\'{U}stav teoretick\'{e} fyziky a astrofyziky}{Department of
Theoretical Physics and Astrophysics}

\RokOdevzdaniPrace{2015}

\AkademickyRok{2015/2016}

\Autor{Petr Kurf\"urst}{Petr Kurf\"{u}rst}

\NazevPrace{Models of hot star decretion disks}{Modely odt\'{e}kaj\'{i}c\'{i}ch disk\r{u} horkých hvězd}
{Models of hot star decretion disks}

\VedouciPraceSTituly{Prof. Mgr. Ji\v{r}\'{i} Krti\v{c}ka, Ph.D.} %% Neni potreba pro rigorozni prace

\StudijniProgram{Fyzika}{Physics}

\StudijniObor{Teoretick\'{a} fyzika a astrofyzika}{Theoretical Physics and
Astrophysics}

\PocetStran{20\,$+$\,167}

\KlicovaSlova{hv\v{e}zdy: ztr\'{a}ta hmoty -- hv\v{e}zdy: v\'{y}voj -- hv\v{e}zdy: rotace -- hydrodynamika}
{stars: mass-loss -- stars: evolution -- stars: rotation -- hydrodynamics}

\Abstrakty
{\noindent Hmotné hvězdy mohou během svého vývoje dosáhnout stadia kritické (nebo velmi rychlé, téměř kritické) 
rotace, kdy další nárůst rotační rychlosti již z kinematického hlediska není ``povolen''.
Výrony hmoty a odtok momentu hybnosti z rovníkové oblasti takto rychle rotující hvězdy mohou vést ke zrodu a následné existenci odtékajícího 
(dekrečního) disku v oblasti okolo hvězdy. Dostatečně výkonný mechanismus, který umožní takový odtok hmoty a 
momentu hybnosti, je zajišťován anomální viskozitou materiálu. Vnější nadzvukové oblasti disku mohou zabíhat až do~mimořádně velkých vzdáleností od 
centrální hvězdy, přesné radiální rozměry disků nicméně neznáme, převážně díky nejistotám v radiálních průbězích teploty a viskozity.

Studujeme velmi podrobně vývoj a chování hydrodynamických veličin v odtékajících discích, tedy profily hustoty, radiální a azimutální rychlosti a také profil tempa ztráty momentu hybnosti,
až po~extrémně vzdálené oblasti. Zkoumáme také závislost těchto charakteristik na průběhu teploty a viskozity.
Studujeme rovněž radiální průběh magnetorotační nestability, kterou lze považovat za zásadní zdroj anomální viskozity v odtékajícím disku, 
do určité úrovně studujeme také vzájemně provázané rozložení hustoty a teploty ve dvourozměrném radiálně-vertikálním modelu . 

Předběžné stacionární modely disků jsme počítali pomocí Newtonovy-Raphsonovy metody. Pro účely časově závislého modelování jsme vyvinuli vlastní 
dvourozměrný hydrodynamický a magnetohydrodynamický numerický kód, založený na explicitním Eulerovském schématu a meto\-dě konečných diferencí, s tzv.~oddělenou sítí,
zahrnující všechny členy Navierovy-Stokesovy stři\-hové viskozity.
Pro výpočet radiálního průběhu diskové magnetorotační nestability jsme zvolili částečně analytický přístup, kdy na základě 
numerického časově závislého hydrodynamického modelu analyticky studujeme stabilitu odtékajícího disku, vnořeného do magnetického pole cent\-rální hvězdy. 

Vzdálenost zvukového bodu a maximální míra tempa ztráty momentu hybnosti silně závisejí na průběhu teploty a
jsou téměř nezávislé na průběhu viskozity. Rotační rychlost, stejně jako míra tempa ztráty momentu hybnosti, jeví
ve velkých vzdálenost rychlý pokles, radiální průběh tohoto poklesu silně závisí na zvoleném průběhu teploty a viskozity. 
Celkové množství hmoty a momentu hybnosti, obsažené v disku, vzrůstá s klesajícím průběhem teploty a viskozity. Doba vývoje disku také
výrazně roste s radiálně klesající teplotou a viskozitou.
Modely s centrální hvězdou nedosahující kritické rotační rychlosti vykazují pulzace diskové hustoty a radiální rychlosti v~blízkosti hvězdy,
modely s extrémně malým (``nulovým'') počátečním profilem hustoty mohou vést k charakteristickým hustotním poklesům (``jámám''), připomínajícím rázové vlny hvězdných větrů, 
v oblasti dotyku vněj\-šího okraje disku a mezihvězdné látky. 
Magnetorotační nestabilita se rozvíjí v oblasti blíže ke~hvěz\-dě, pokud hodnota plazmatického parametru je dostatečně vysoká. Ve větších vzdále\-nostech, 
poblíž oblasti kde se disková rotační rychlost zhruba vyrovnává rychlosti zvuku, tato nestabilita mizí. Navzdory tomu, že se jedná o zásadní zdroj viskozity,
která pohání vytváření a vývoj disku do~vzdálenějších oblastí, časově závislé modely ukazují rozšíření disku do nekonečna, 
především pak díky vysokým radiálním rychlostem látky ve vzdálených nadzvukových oblastech.

Časově závislé jednorozměrné modely v zásadě potvrzují výsledky našich předchozích stacionárních modelů a stejně tak i uvedených analytických předpokladů. 
Nefyzikálnímu propadu rotační rychlosti disku i míry tempa ztráty momentu hybnosti ve velkých vzdálenostech (který se objevuje v některých modelech) 
se lze systematicky vyhnout pomocí modelů s radiálně klesající teplotou a viskozitou.
Radiální průběh magnetorotační nestability je neměnný přibližně až do~oblasti zvukového bodu, v tomto smyslu můžeme vzdálenost zvukového bodu zhruba považovat za 
vnější okraj disku. Protože radiální průběh tempa ztráty momentu hybnosti dosahuje svého ma\-xima přibližně ve vzdálenosti zvukového bodu 
a nadále se zde stává plochým, můžeme při~výpočtu míry tempa ztráty momentu hybnosti a míry tempa ztráty hmoty disku
považovat poloměr zvuko\-vého bodu za efektivní vnější poloměr disku. V souvislosti s~dosaženou úrovní výsledků dvou\-rozměrných
modelů popisujeme také používané možnosti variantního uspořádání výpočetní sítě v~rámci~válcového souřadného systému a 
diskutujeme o jejich výhodách, případně jejich schůdnosti v~zájmu dosažení
optimálního kompromisu mezi přesností a využitelností více\-rozměrného modelu a jeho enormní výpočetní náročností.}
{\noindent Massive stars can during their evolution reach the phase of critical (or very rapid, near-critical) 
rotation when further increase in rotation rate is
no longer kinematically allowed. The mass ejection and angular momentum outward transport from such rapidly rotating star's equatorial surface may lead to formation and 
supports further existence of a circumstellar outflowing 
(stellar decretion)
disk. The efficient mechanism for the outward transport of the mass and angular momentum is provided by the anomalous viscosity. The outer supersonic regions of the
disks can extend up to a significantly large distance from the parent star, the exact radial extension is however basically unknown, partly 
due to the uncertainties in radial variations of temperature and viscosity.

We study in detail the behavior of hydrodynamic quantities, i.e., the evolution of density, radial and azimuthal velocity, and angular momentum loss rate in stellar decretion disks
out to extremely distant regions. We investigate the dependence of these physical characteristics on the distribution of temperature and
viscosity. We also study the magnetorotational instability, which we regard to be the source of anomalous viscosity in such outflowing disks 
and to some extent we provide the preliminary models of the two-dimensional radially-vertically correlated distribution of the disk density and temperature.

In the preliminary phase we calculated stationary models using the Newton-Raphson method. For time-dependent hydrodynamic modeling we
developed our own two-dimensional hydrodynamic and magnetohydrodynamic numerical code based on an explicit Eulerian finite difference scheme on 
staggered grid, including full Navier-Stokes shear viscosity.
We use semianalytic approach to investigate the radial profile of magnetorotational instability, where on the base of 
the numerical time-dependent hydrodynamic model we analytically
study the stability of outflowing disks submerged to the magnetic field of central star.

The sonic point distance and the maximum angular momentum loss rate strongly depend on the temperature distribution and
are almost independent of the profile of viscosity. The rotational velocity as well as the rate of the angular momentum loss 
at large radii rapidly drop according to assumed temperature and viscosity
distribution. The total amount of the disk mass and angular momentum increase with decreasing temperature and viscosity.
The disk evolution time significantly increases with radially decreasing temperature and viscosity. The models with subcritically rotating star indicate the 
pulsations in the disk density and radial velocity close to the star, while the models with extremely low (``zero'') initial density profile may lead to characteristic density bumps,
resembling the bow shocks in stellar winds,
in the contact region of the disk and interstellar medium. 
The magnetorotational instability develops in the region  close to the star if the plasma parameter is large enough. At larger radii the
instability disappears close to the radius where the disk orbital velocity is roughly equal to the sound speed. Despite being the main source of
viscosity that provides the driving mechanism for the disk outward development, the time-dependent simulations with vanishing viscosity show the
dissemination of the disk to the infinity, mainly due to a very high gas radial velocity in the distant supersonic regions.

The time-dependent one-dimensional models basically confirm the preliminary results from our stationary models as well
as the assumptions introduced within the analytical approximations. The unphysical drop of the disk rotational velocity and the angular momentum loss rate at large radii (which is present
in some models) can be systematically avoided in the models with radially decreasing temperature and viscosity.
The radial profile of the magnetorotational instability holds up approximately to the disk sonic point region, in that sense we can roughly regard the disk sonic 
point radius as an outer disk edge. Since the radial profile of the angular momentum loss rate reaches its maximum approximately at the sonic point distance and flattens there, 
we can quantify the disk angular momentum loss rate as well as the disk mass loss rate 
considering the sonic point radius being the effective disk outer radius. Within 
the performed level of results of two-dimensional models we introduce various possible coordinate geometries and discuss their 
advantages and their feasibility in order to achieve the optimum balance between the accuracy and efficiency of the self-consistent multidimensional calculations
and their enormous computational cost.}

\TextPodekovani
{\noindent Among many persons whom I want to express my gratitude for support and help during my 
PhD studies era, the most important one is my scientific supervisor prof.~Mgr.~Ji\v r\'i Krti\v cka,~Ph.D. 
His advice and suggestions as well as his universal organizational support and assistance were
all the time absolutely invaluable and indispensable. He also helped me to open the doors to quite useful and amazing 
experiences such as multiple study visits at the Potsdam University, presentations on various international astrophysical conferences, and other important scientific events. 

I have to thank also to other teachers and advisers, above all to the excellent consultants prof.~Dr.~Achim Feldmeier
from Institut f\"ur Physik und Astronomie, Universit\"at Potsdam, and doc.~RNDr.~Ji\v r\'i Kub\'at,~CSc. 
from Department of Stellar Physics, Astronomical Institute of The Czech Academy of Sciences.

I am also very grateful to all my friends and colleagues, particularly to Mgr.~Milan Prvák for significant help with computational system problems, 
to Mgr.~Brankica \v Surlan,~Ph.D. and Mgr.~Kl\' ara \v Sejnov\'a for cooperative efforts in solving various problems, and many, many others.

Special gratitude deserves my wife Jana for her sustained long-term support and patience as well as all other members of my family. Thank you all.\\

\noindent The access to computing and storage facilities owned
by parties and projects contributing to the National Grid Infrastructure
MetaCentrum, provided under the program “Projects of Large Infrastructure
for Research, Development, and Innovations” (LM2010005) is appreciated. 
}

\TextProhlaseni
{I declare that this thesis is my own work with literature that is cited.}

\DatumProhlaseni{1st July 2015}

\makeindex

\begin{document}

\VytvorPovinneStrany

\AbstraktyNaDvouStranach

\PodekovaniAProhlaseni

\pdfbookmark{Contents}{Contents}
\VytvorObsah
\cleardoublepage

\HlavickaUvod
%\pdfbookmark{Introduction}{Introduction}
\addcontentsline{toc}{chapter}{Introduction}
{\chapter*{Introduction}
%{textPrace/Introduction}
%{\chapter{Introduction}
\thispagestyle{plain}
\label{intro}
{The existence of stars with outflowing (decretion) disks is known for a long time. 
Typical representatives of this class of stars are Be stars (Sect.~\ref{Bephen}), a B-spectral type non-supergiant stars with prominent hydrogen (Balmer) 
emission lines in their spectrum. 
In the year 1866, the cleric, theologian and astronomer Padre Pietro Angelo Secchi,
who was the first to classify stars according to their spectra, published an observation of the star $\gamma$ Cas (classified as the star of type $\text{B0.5 IV}$), 
noting that \citep{Suchy,Rivi2013a}
the spectral line $\text{H}\beta$ shows \textit{``une particularit\'e curieuse \ldots une ligne lumineuse tr$\grave{e}$s-belle 
et bien plus brillante que tout le reste du spectre''} (which means that the line is particularly unusual \ldots a very nice luminous line 
which is brighter than the rest of the spectrum) in the report that was written by the ``Vaticano-Italian'' astronomer in French language to a German Journal 
(this may be also a noteworthy aspect of the epochal scientific event, especially in that era of emerging nationalisms). 
This was however the first star recognized as a Be star and one of the first stars with
spectral emission lines ever observed. A wide variety of types of stars (not only Be stars) in various stages of their evolution 
are currently recognized as the parent objects of circumstellar disks (or disk-like density enhancements): 
B[e] stars, luminous blue variables (LBVs), post-AGB stars, etc. (see Sect.~\ref{stelassoutfldisks} for the review).

Why do we associate these spectral emission lines with circumstellar disks? The link between these phenomena lies actually in several factors (cf.~Sect.~\ref{stelassoutfldisks}).
The natural explanation of the \textit{double-peaked} Balmer emission lines is that they originate in a circumstellar
material orbiting the star, while the infrared radiation excess reveals a free-free emission in an ionized circumstellar gas. The modern observational methods
like polarimetry and optical interferometry led more recently to the direct resolution of the disks and even of their more detailed structure \citep[see,~e.g.,][among others]{kvilda}. 

The material of the circumstellar disk is certainly ejected
from the star’s atmosphere, the detailed mechanism that creates and maintains such
a disk however still remains unclear \citep[see][]{Porter,Rivi2013a}.
The first theories and models of the stars with decretion disks, especially those of Be stars \citep[e.g.,][]{struve,marlboro}, 
suggested the direct centrifugal ejection of matter from equatorial surface of stars with sufficiently high (critical) rotational velocities (Sect.~\ref{gasejcl}).
However, observational studies have shown that many of e.g.,
Be stars rotate subcritically \citep[see,~e.g.,][for a review]{Rivi2013a}, mostly at about 70-80 \% of their critical velocity \citep{portyr}.
Even if we take into account the effects of gravitational darkening which can lead to a systematic
underestimate of the rotational velocity \citep[see, e.g.,][]{Towny04},
some additional supporting mechanisms (such as nonradial pulsations, binary companions, etc.) for an efficient transport of stellar angular momentum into the disk 
have to be suggested \citep[e.g.,][see also the review in Sect.~\ref{diskfome}]{Osaki,Lee,Bjorkman}.
The generally accepted (and likely the most realistic) scenario, explaining the disk formation and evolution, is the viscous decretion disk \citep[e.g.,][see Sect.~\ref{viskodisks}]{Lee}.
Moreover, current observational evidence strongly supports the idea that such disks are Keplerian (at least to a certain distance from parent star), 
in vertical hydrostatic equilibrium and nearly isothermal \citep{Karfiol08}. 
Most of the models and analyzes properly investigate
merely the inner, observationally significant, parts of the disk, whereas the disk structure and its behavior close to the sonic point or even in 
the distant supersonic regions (up to some possible outer disk edge) has not yet been
quite intensely studied. We expect that the decretion disks which are almost exactly Keplerian near a star
become in the very distant (sonic and supersonic) regions angular-momentum conserving \citep[in this point see, e.g.,][]{okazaki,Krticka,Kurf1,Kurf2,kurfek}.

The hydrodynamics (gas dynamics) of the disk gaseous medium behavior is thus determined by 
correlated effects of viscosity and thermal structure. As a main source of the disk anomalous turbulent viscosity, responsible for the
diffusive outward transport of mass and angular momentum, we regard the magnetorotational instabilities developed in a sufficiently weakly magnetized gas 
\citep[][see also Sect.~\ref{magnetousci}]{Balbus,Krticka_2014}.
The thermal structure of the disk however seems to be very complex phenomenon, affected mainly by the irradiation anisotropically emerging from rotationally oblate star 
(which determines the 
disk thermal equilibrium) as well as by the internal viscous heating effects which are dependent of the mass loss rate of the star-disk system, 
i.e.,~of the amount and density of the gas concentrated particularly in (the equatorial layers of) the inner disk region (see Sect.~\ref{templajznik}).

In our work we give the basic overview of the various either ``historic'' or currently accepted scenarios of the origin of stellar outflowing disks and disk-like formations connected with 
various types of stellar objects (Chap.~\ref{Steloutfl}), in Chap.~\ref{hydrobase} we review the physics (hydrodynamics) of the gas flows, including the basic hydrodynamic equations 
and the detailed physics of stresses. In Chap.~\ref{kombhgu} we apply the gas dynamics laws for the astrophysical case of the circumstellar disks, while Chap.~\ref{sdfge}
outlines the basic magnetohydrodynamics and its application for the magnetorotational instabilities and their development in weakly magnetized disks submerged in the external stellar
dipolar field. 
Chap.~\ref{numerovizci} describes the numerical background of our development of hydrodynamic and magnetohydrodynamic computational codes and demonstrates the basic results of their testing on 
the characteristic physical and essential astrophysical problems. Chap.~\ref{largemodel} shows the results of one-dimensional models of hydrodynamic structure of 
the very large circumstellar decretion disks 
of the critically as well as subcritically rotating stars with various initial conditions, we also describe the semianalytic solution of the disk radial profile of magnetorotational 
instabilities. Chap.~\ref{twodajmmodel} opens the problem of the current stage of our self-consistent two-dimensional models of the disk radial-vertical hydrodynamic and thermal structure 
where the geometry of the rotationally oblate central star is involved. In Chap.~\ref{sumarix} we briefly summarize the results. Appendix~\ref{appendix1}-Appendix~\ref{briefmention} 
gives the mathematical supplements of the described physics and of the demonstrated results. 

Since the basic principles (including the numerical structure and principles of the developed computational codes) of the decretion disk physics are essentially similar 
to the principles of the physics 
of accretion disks (as well as the protostellar and protoplanetary disks, etc.), we systematically use the main conclusions \citep{Pringle,Frank} 
of the accretion disks theory as a basis to the description of decretion disks. 
The physics of the stellar decretion disks represents thus the important astrophysical discipline that widely contributes to our understanding of the physics of 
stellar structure and evolution as well as of interaction between various types of astrophysical objects (star-disk, disk-interstellar medium, disk-stellar or planetary companion),
its results may also represent a promising interaction with other astrophysical and physical research fields.
\cleardoublepage}}

\renewcommand{\chaptermark}[1]{\markboth{\thechapter. #1}{}}
\renewcommand{\sectionmark}[1]{\markright{\thesection. #1}{}}
\pagenumbering{arabic}

\HlavickaKapitoly
\chapter{Stellar outflowing disks}
\label{Steloutfl}
\section{Disk formation mechanisms overview}\label{diskfome}
The following short review of the disk formation mechanisms
may be in some aspects regarded as rather ``historical'', because e.g., the wind-compressed disk theory or the magnetically
compressed disk theory
were found likely not to provide an explanation of the formation of the dense Keplerian outflowing disks \citep{ovoce1,dula1,Hilly1}.
They however brought much deeper and complex insight into the disk physics and, though it is not fully clear,
e.g., the magnetically directed stellar winds can play a significant role in forming the aspherical circumstellar environment 
of sgB[e] and post-AGB stars (see Sects.~\ref{macom}, \ref{Beforbphen}, \ref{postaks})
as well as of early OB stars \citep[see,~e.g.,][]{udulak14,nahac}.
We consider it therefore highly reasonable to give here the basic description of all these approaches,  
together with an indication of their limitations and possible contradictions.
The basic common requirement for all these theories was to provide the natural explanation for gas ejection from
the stellar photosphere resulting into generation of a dense enough region in the stellar equatorial 
plane and maintaining the kinematic supply of mass and angular momentum to the disk.

The analysis of stellar internal physical mechanism that may be responsible for bringing the angular momentum to the stellar 
equatorial surface region does not provide unambiguous answers \citep{Meynet1,maeder}. The main contributors to this process
are supposed to be the stellar meridional circulation %(see Fig.~\ref{figone}) 
which due to the large horizontal thermal imbalance 
in rotationally oblate stars \citep{vincek} may play a key role in the outwardly directed transport of angular momentum in 
equatorial plane, as well as a convection and stellar internal magnetic fields due to the enforcement of the stellar solid body-like rotation.
Horizontal turbulence is also the common phenomenon in fast rotating stars \citep{maeder} that produce shear instabilities 
between layers of different rotational velocities which may further facilitate the outward transport of angular momentum within 
stellar interiors.
\subsection{Direct centrifugal gas ejection from stellar surface}\label{gasejcl}
One of the first, although rather intuitive, explanations of the possible disk forming physical mechanism, given by
\citet{struve}, suggested the direct centrifugal matter ejection from the star, which rotates equally or close
to the critical velocity $V_{\text{crit}}=\sqrt{GM_{\star}/R_{\text{eq}}}=\sqrt{2GM_{\star}/3R_{\text{pole}}}$ (with $M_{\star}$ being the mass
of the star, $R_{\text{eq}}$ is the equatorial and $R_{\text{pole}}$ is the polar radius of a critically rotating 
star) into a rotating circumstellar disk. But since most of conventionally accepted estimations of the average Be star rotation velocity 
(based on observed line-broadenings) typically give roughly 70-80\% of the critical one (see Sect.~\ref{Bephen}), 
several processes that may help to provide additional boost of material into circumstellar orbit has been identified \citep{ovoce2}:
flow of material (stellar wind), radiatively driven by scattering of stellar photons and by line absorption appears 
as an obvious candidate in hot stars with $T_{\text{eff}}\gtrapprox 10^4\,\text{K}$. Indeed, the models of radiatively driven orbital
mass ejection have been proposed, suggesting additional radiative force in the regions of enhanced brightness (bright spots) on stellar surface
\citep[e.g.,][]{ovocekran}. 
However, this ``Radiatively Driven Orbital Mass Ejection'' (RDOME) scenario needs 
rather specialized modifications of the stellar surface conditions, 
for example the assumption of the prograde
radiative force that is expected in the region ahead of the bright spot and rather extreme brightness variations
as well as rather arbitrary
fine-tuning of the spot emission or line-force cutoff. Since
radiative driving is routinely capable of driving a stellar wind to speeds well in
excess of orbital launch speeds, it seems possible that the prograde force ahead of
the bright spot might impart material there with sufficient angular momentum
to achieve circumstellar orbit. For these reasons we may regard this scenario as questionable, further work is needed to determine
whether the brightness fluctuations, sometimes observed in Be stars, have a time-scale and magnitude that might be consistent with
this RDOME scenario \citep{ovoce2}.

Another candidate mechanism for the orbital mass ejection are stellar photospheric pulsations. The analyses of circumstellar 
spectral lines of Be star $\mu$ Centauri \citep{Rivi98,Rivi99} indicate a possible connection 
between nonradial pulsation modes and the material transfer from star to disk. In this point we may assume several effects induced by pulsations: 
although the direct driving force of the
pulsations is likely not able to lift material on the orbital trajectories \citep{ovoce6}, the effect of lowered gravity may 
locally support the action of radiation-driving forces to produce the orbital gas ejection \citep[and the reference therein]{Porter}.
Pulsations may also lead to instantaneous locally sufficiently large accumulation of angular momentum \citep[e.g.,][]{Osaki}, producing photospheric 
regions with supercritical velocities which facilitates matter to be thrown off the star. The pulsationally driven 
orbital mass ejection model suggests azimuthal density and velocity perturbations at star's equatorial
surface induced by nonradial pulsations 
with propagation phase being prograde (in the sense of stellar rotation), whose propagation velocities are very close to 
critical velocity at given equatorial radius.
The need of these special requirements may however also contradict the viability of this scenario
\citep[see, e.g.,][]{ovoce2}.  Various pulsation modes and velocities and their possible connection to upward
angular momentum transport are studied, e.g., in \citet{Towny5,roza,nemazal}. 

The models of thermally or magnetically driven local explosive outbursts in equatorial photospheric 
regions have been
proposed by, e.g.,~\citet[see also \citealt{Krohan}]{Kroll}. Although the simulations show in this case the natural formation of 
an equatorial disk \citep{Kroll}, such events, implying local temperature of the order $10^6\,\text{K}$, have not been (at least not in general) observed.
\subsection{Wind-compressed disks}\label{wicom}
The model proposed by \citet{Bjorkman} represents an important contribution to the disk theory and to the physics of stellar winds in case of stellar rotation. 
The work was inspired by the motion of the fluid in gravitational field and follows 
the one-dimensional models of the stellar winds from rotating stars of e.g., \citet{friba,lampaul}, etc. 
It was based on simple suggestion of purely radially directed radiative force that drives the stellar wind 
leaving the photosphere. In non-rotating case of hot star's wind the material flow is thus clearly confined to the radial streamlines. If 
the stellar rotation rate is sufficiently high, the streamlines of wind particles become in this model curved in the equatorward direction. 
The shock induced by the interception of opposingly directed flows of the wind from both hemispheres 
thus consequently produces the dense region confined to the stellar equatorial plane (disk). This is in brief the basic scenario of the
wind-compressed disk (WCD) model.
\begin{figure}[t!]
\begin{center}
\includegraphics[width=14.75cm]{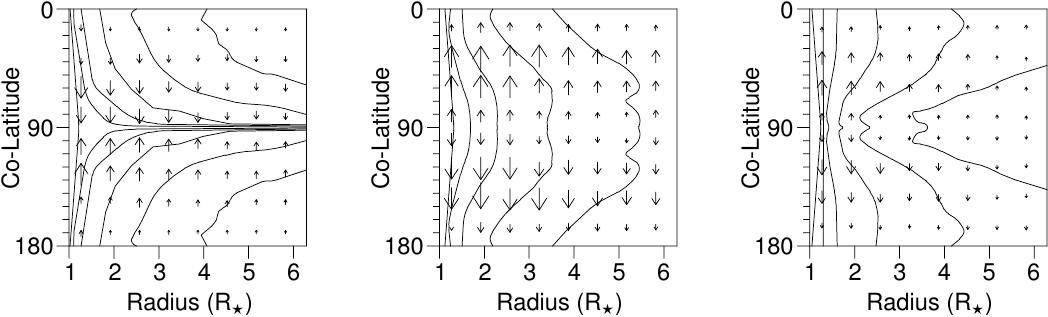}\\
\vspace{0.1cm}
\small{Fig.~\ref{owo2}a}~~~~~~~~~~~~~~~~~~~~~~~~~~~~~~~~~~~~~~~~~~~~\small{Fig.~\ref{owo2}b}
~~~~~~~~~~~~~~~~~~~~~~~~~~~~~~~~~~~~~~~~~~~~\small{Fig.~\ref{owo2}c}
\caption{The resulting graphs of several simulations performed by \citet{ovoce94} and \citet{ovoce1} to test the wind-compressed disk
paradigm. As a central star was in all the cases selected a B2-type star with equatorial rotation velocity about 350 km s$^{-1}$. 
Continuous lines are the density contours plotted in radial versus latitudinal plane.
Superposed arrows represent the magnitude and direction of the latitudinal velocity component $\varv_{\theta}$ of the wind: 
(a) the model assumes only radial forces and uniformly 
bright stellar surface, (b) the model of the corresponding wind density structure for the same star now including the rotationally 
induced dynamical forces but still with uniformly bright stellar surface,
(c) the model includes both nonradial dynamical forces and
the effects of gravity-darkened stellar surface. Only the case (a) leads to the disk formation. Adapted from \citet{ovorot}.}
\label{owo2}
\end{center}
\end{figure}

This paradigm was initially mostly confirmed by dynamical simulations given by \citet{ovoce94}. The resulting density contours,
calculated in this model, are performed in Fig.~\ref{owo2}a. Their detailed analysis of the velocity field found the strong inflow of the central part of the 
inner disk (unlike the fixed-outflow model suggested in \citet{Bjorkman}), representing thus a reaccretion of the wind material! However, subsequent work \citep{Krovoce} 
took the nonradial component of the radiative force into account. Since the vector of radiative force 
(and flux) is antiparallel to the vector of effective gravity $\vec{g}_{\text{eff}}$, given by the superposition of
gravitational and centrifugal acceleration \citep[see, e.g.,][]{maeder}, it is therefore (assuming roughly the Roche approximation where
the star rotates as a rigid body
and most of the mass is concentrated in stellar centre) normal to 
equipotential surfaces within the stellar body, and thus also to rotationally more or less oblate stellar surface. 
Inclusion of this poleward component effectively inhibits 
the effects of equatorial wind compression (see Fig.~\ref{owo2}b). Later \citet{ovoce1}, \citet{petronius}, etc., showed that inclusion of the effects of 
gravitational darkening even reverses the predictions of the equatorial wind compression, giving rise to the polar wind enhancement 
(induced by the so-called $\vec{g}_{\text{eff}}$ effect, see Fig.~\ref{owo2}c, see also Fig.~\ref{figtwo}). This generates a prolate density structure of the wind
with denser and faster outflow in the direction of the rotational poles and 
slower and thinner wind near the equatorial plane \citep{petronius}.
The WCD model also fails in explaining the observed disk IR excess \citep{Porter} as well as 
some line-profile features 
\citep{Rivi99,Hanus95} and does not seem to be compatible with observed disk kinematic structure 
(see, e.g.,~\citet{ovoce2} for details). The effects of radiatively driven wind outflow in case of stellar rotation are described in a simplified form in 
Appendix~\ref{vonZeipel}. 

Despite its currently obvious contradictions the wind-compressed disk model 
was however quite instructive step within the evolution of dynamical models. It was one of the first that
took into account not only a simple equatorial expulsion of matter
but rather a requirement of large enough angular momentum of the material particles to reach 
Keplerian orbits and thus to create and maintain the circumstellar disk.
\subsection{Effects of ionization and the bistability}\label{bistab}
\begin{figure}[t]
\begin{center}
\includegraphics[width=6.75cm]{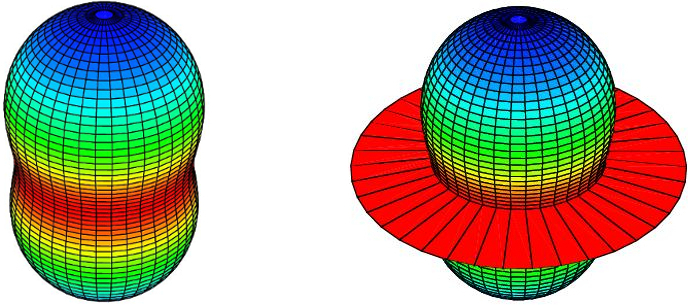}
\caption{The simple illustrations of stellar wind enhancement in case of rapidly rotating star. Left figure: 
the ``peanut shape'' symbolizes the 
strong polar wind induced by the $\vec{g}_{\text{eff}}$ effect in a star with average $T_{\text{eff}}=30\,000\,\text{K}$. 
Right figure: the equatorial disk is produced as a result of $\kappa$ effect in a cooler
equatorial region where sufficiently high opacity is determined by a latitudinal change in ionization structure in a star 
with average $T_{\text{eff}}=25\,000\,\text{K}$ (for details see
Sects.~\ref{wicom}, \ref{bistab}, \ref{vonZeipel}). Adapted from \citet{Madejak}.}
\label{figtwo}
\end{center}
\end{figure}

The effects of line-driven radiative acceleration on mass loss rate and terminal velocity 
of the stellar wind outflow are strongly correlated with the ionization structure of the
bottom layers of the flow, i.e., the star's regions where the wind basically originates \citep[e.g.,][]{CAK1,Lamercas}. 
If the ionization structure and therefore the opacity at the base of the wind undergoes a sufficient change (for example in dependence on
the latitude due to the stellar equatorial darkening) it may provoke
a so-called bistability jump in the wind behavior \citep[see also, e.g., \citealt{maeder,Hilly1}]{lampaul}.
\citet{lampa} defined the bistability as an abrupt discontinuity in hot stars stellar wind characteristics that occurs near 
effective temperatures about $21\,000\,\text{K}$. They expected that the ratio of the 
wind terminal and escape velocity ($\varv_{\infty}/\varv_{\text{esc}}$) drops due to this effect by a factor of nearly two (they 
assume that in the region of the bistability jump the mass-loss rate and the terminal velocity are anticorrelated due to 
roughly constant quantity $\dot{M}\varv_{\infty}$). 
\citet{vink2001} found bistability jump around $T_{\text{eff}}\approx 25\,000\,\text{K}$,
where mainly Fe IV recombines to Fe III with the latter ion being
a more efficient line driver than the first. They conclude that this may produce an
increase in $\dot{M}$ of about a factor of five and subsequently $\varv_{\infty}$ drops by a factor of two,
when going from high to low temperatures. Additional bistability jumps may occur at 
higher temperatures where CNO elements may provide the dominant line driving, 
especially for low metallicity stars \citep{vink2001}. 
From the detailed study of large sample of early to mid B supergiants~\citet{kroutil}~found 
that there is a gradual decline in the ratio $\varv_{\infty}/\varv_{\text{esc}}$ from $\sim3.4$ for stars with $T_{\text{eff}}$ above $24\,000\,\text{K}$ and
$\sim2.5$ for stars with $T_{\text{eff}}$ in the range $20\,000$-$24\,000\,\text{K}$ to $\sim1.9$ for stars with with $T_{\text{eff}}$ below $20\,000\,\text{K}$.
They also conclude that no significant increase in mass loss rate below $T_{\text{eff}}=24\,000\,\text{K}$ is observationally identified.
The enhanced mass ejection provoked by higher opacity in 
metallic lines in cooler equatorial region is the so-called $\kappa$ effect 
(cf.~Appendix~\ref{vonZeipel} for details, see also Fig.~\ref{figtwo}).

Although this mechanism is supposed to be responsible for the disk formation in some types of B[e] stars \citep{CassLBV,polyp},
the calculations however show that the bistability induced equatorial wind flow is quite weak to produce 
equatorial vs.~polar density ratio larger than about 10 \citep[see, e.g.,][and the reference therein]{Hilly1}, which is too low to
explain models of B[e] disks (cf.~Sect.~\ref{Beforbphen}) in general. In addition, \citet{curda} proposed the model where the standard solution (denoted there as
the fast solution) of the so-called modified CAK wind model 
\citep[hereafter m-CAK model,][see also Sect.~\ref{lajncakis}]{friba,papuk} 
vanishes in case of $V_{\text{rot}}\approx 0.7\text{-}0.8\,V_{\text{crit}}$. There may thus exist much
denser and slower solution than this standard (fast) m-CAK solution (denoted as the slow solution in \citet{kurial}). 
The slow solution may produce the equatorial vs.~polar density contrast 
of the order of $10^2\text{-}10^4$ near the stellar
surface ($r\lessapprox 2R_{\star}$), such disk may extend up to approximately one hundred stellar radii. 
This solution however involves large uncertainties, e.g., in rotation
speeds of B[e] supergiants,
they also do not take into account the nonradial forces resulting from the
change of shape of the star or the gravity darkening (see Sect.~\ref{cepelin}, see also \citet{kurial} for details).
Although the bistability jump may
lead to some density enhanced equatorial wind structure, 
no two-dimensional dynamical models thus confirm the idea of such bistability induced dense equatorial disks (the bistability mechanism and
its context was discussed,~e.g.,~in \citet{ovorot}).
\subsection{Magnetically compressed disks}\label{macom}
The idea of the wind compressed disk (WCD) model has been further 
expanded by inclusion of the effects of magnetic fields. The strength of magnetic field in B-type stars, required to significantly influence the behavior and trajectories of stellar wind, is 
of the order of hundreds of gauss (see \citet{Porter} review, see also remarks in Sect.~\ref{Bephen}). Although there is no clear observational evidence 
of sufficiently strong large-scale magnetic fields in classical Be stars, there is however strong
evidence of relatively stable and strong large-scale fields in related types of hot stars that may also form some types of disk-like
equatorial density enhancements (see Sect.~\ref{stelassoutfldisks} for the overview). The recent MiMeS (``Magnetism in Massive Stars'') analysis of the  
sample of approximately 100 Be stars yields no detection of magnetic fields. In this point \citet{Wade14}
conclude that classical Be stars and B stars with strong, organized magnetic fields
appear to be mutually exclusive populations. This is supported by their very different
distributions of the $\varv\,\text{sin}\,i$ projections of rotational velocities, 
which peak at nearly $250\,\text{km\,s}^{-1}$ for Be stars (with extension up to approximately 
$400\,\text{km\,s}^{-1}$), while the peak of the $\varv\,\text{sin}\,i$ distribution is very low ($\leq 50\,\text{km\,s}^{-1}$) for known
magnetic B stars \citep[see also \citealt{Wade14}]{small13}. 
Strong fields (up to the order of $10$ kilogauss) were revealed
for example in B0-B2 type helium-strong stars \citep[e.g.,~$\sigma$ Ori E,][]{landbok}, while relatively weaker fields have been found in
some high mass B-type stars \citep[which may however interfere with Be phenomenon, e.g., the B1 \!IIIe star $\beta$ Cephei, see][]{jindra},
in slowly pulsating (SPB-type) B stars \citep[typically $\zeta$ Cassiopeiae, e.g.,][]{nezmar} or in O-type stars \citep[e.g.,~$\theta^1$ Ori C,][]{donbaba}.
The latter served as the test star for the model that preceded the magnetically compressed disks and was originally developed by 
\citet{baba1, baba2} for chemically peculiar so-called Ap stars 
(that possess magnetic fields stronger than normal A or B-type stars). Stellar magnetic fields are generally characterized as dipole fields with the magnetic axes
arbitrarily tilted relative to the axes of rotation \citep{Towny2}.

One of the first magnetically compressed disk models was developed by \citet{kasa02}. They concluded that in regions where
magnetic energy density dominates over the 
wind matter kinetic energy density, the flow particles are forced to move along the magnetic field lines. 
Since the latter may in equatorial region form closed loops, the wind matter 
flow is thus confined from both magnetic hemispheres to the magnetic equatorial plane, where, similarly to the wind compressed disk paradigm, it may produce a shock
with consequent formation of the dense disk. Magnetohydrodynamic model was proposed by \citet{dula1}, who regard
the ratio 
of the magnetic energy density to the kinetic energy density of the matter (denoted here as $\eta$ or $\eta_*$ for the particular case of the dipole configuration)
as the key parameter, which dynamically affects the wind behavior. 
They have suggested that in case of the dipole field, when this $\eta_*$
ratio is significantly small ($\eta_*<0.1$), the dynamic effect of the magnetic field is negligible. On the other hand in case of magnetic field 
dominance ($\eta_*>1$) this field becomes the key factor in wind dynamics, ``forcing the wind to follow the magnetic field lines and thus giving rise to 
a thin outflowing disk in the magnetic equatorial plane'' \citep{dula1}. Their numerical simulations of the wind flow, subjected to dipolar magnetic field, 
basically confirm the simplified analytic models in case of non-rotating star, resulting however in dynamically 
ambiguous disk behavior when the inner part of the disk 
appears to reaccrete back on star producing the zone of stagnation of radial flow. The rigidly rotating
magnetosphere (RRM) model of \citet{Towny2} further develops the principles of the magnetically compressed disk 
theory and numerically corroborates the idea of the formation of warped equatorial 
thin disk-like distribution, ``whose mean surface normal lies
between the tilted axis of magnetic dipole field and the rotation axis of the star'' (see Appendix~\ref{eqconfirot} and \ref{rindzak}). 
The predictions of this model were supported by observations of the distribution of circumstellar emission 
in strong magnetic B stars \citep[see, e.g.,][]{Towny55,Karfiol13,Grunt13}.

The two-dimensional magnetohydrodynamic dynamical models that involve the stellar rotation \citep{ovocedul} however show that instead of producing 
the ``magnetically torqued disk'', postulated in non-rotating analyses, the disk material behavior tends either to the inward movement in the inner disk 
regions or is trying to escape the disk in the outer parts. Although in case of very strong fields they suggest 
the centrifugally supported, ``magnetically confined rigid disk'', in which the field not only forces the matter to rotate as a rigid body, but may also 
prevent the centrifugally induced breakout of the outflowing material. They conclude that such magnetically torqued rigid disks are basically ill-suited 
to explain the rapidly rotating Be stars dense equatorial disks emission, on the other hand there may be a good chance for the model to explain the properties
of circumstellar environment of magnetically strong Ap and Bp stars \citep[see also Appendix~\ref{rindzak}]{ovocedul}. 
There are also very few
O-type stars (which are presumed to be the progenitors of B[e] supergiants) in which relevantly 
strong magnetic field have been detected (one of them is $\theta^1$ Ori C, see, e.g., \citet{donbaba}). \citet{Hilly1} thus points out that there is even an uncertainty 
about the role of the magnetic fields in these sgB[e] stars (see Sect.~\ref{Beforbphen}).
\begin{figure}[t]
\begin{center}
\includegraphics[width=12cm]{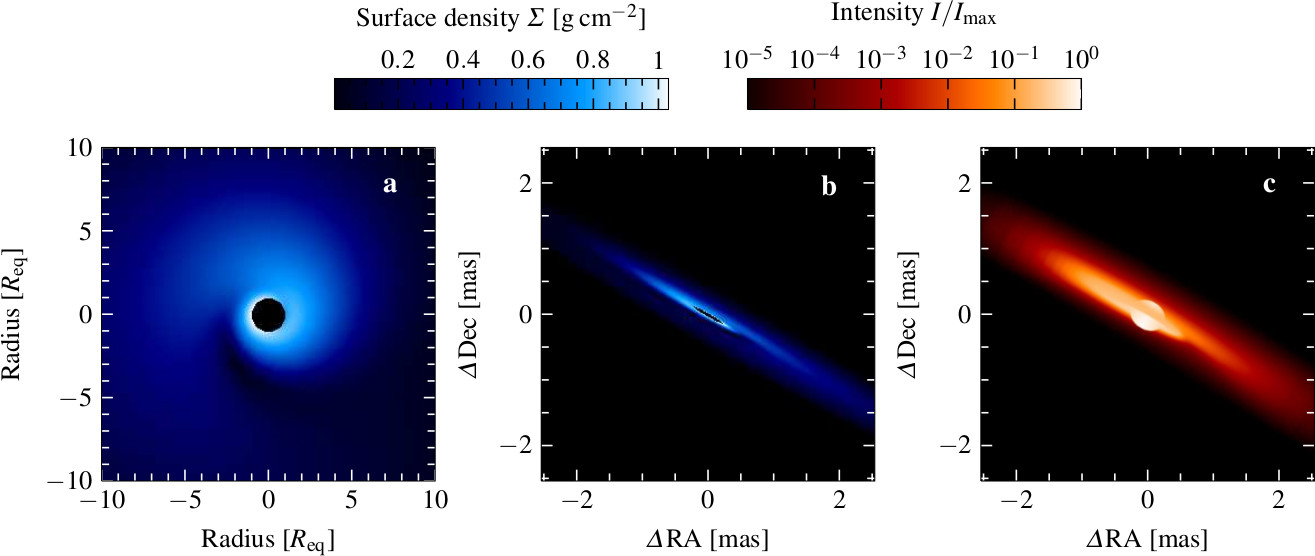}
\caption{Models for Be star $\zeta$ Tauri: Pattern of density perturbation of the global oscillation model 
seen from above the disk (left panel) and projected onto the real sky plane (middle panel). Right panel shows the intensity 
of modeled continuum at 2.16 $\mu$m wavelength. Adapted from \citet{Rivi2013a}, corresponding models were calculated by \citet{Karfiol09}.}
\label{figfour}
\end{center}
\end{figure}
\subsection{Viscous disk}\label{viskodisks}
The viscous decretion disk is currently supposed to be the most viable scenario leading to the formation
of dense equatorial circumstellar disks (such as, e.g.,~Be disks, see Sect.~\ref{stelassoutfldisks}). Once the matter and angular momentum
leave the star and enter the inner disk boundary, they diffuse outward through the disk
under the action of (presumably) turbulent magnetohydrodynamic viscosity \citep{Porter}. 
The clear explanation of how the disk is supplied by the matter and angular momentum however still
remains the basic uncertainty, some possible mechanisms are outlined in Sect.~\ref{gasejcl}.
The physics of viscous decretion disk models is quite similar to that of viscous accretion 
disks \citep[e.g.,][]{Shakura,Pringle,Frank}. Except the opposite signs of inflow vs.~outflow 
it differs in its basic form only in boundary conditions: in accretion disks it is assumed that there is a
torque free radius at or somewhere very near to the inner boundary. The accretion on central object is thus
allowed. The angular velocity of inflowing matter must there slow down and equalize the angular velocity of central object which is mostly
significantly smaller than the critical (Keplerian) value, $\Omega_{\star}<\Omega_K(R_{\star})$. 
On the other hand, in decretion disks we assume the Keplerian (or near-Keplerian) stellar equatorial rotational velocity at the inner boundary, while 
we expect the viscous torque free outer disk edge (or some region that can be considered as the outer disk radius).

Steady-state viscous decretion disks, i.e.,~the disks where we assume the constant rate of mass and angular momentum supply and a constant $\alpha$ viscosity parameter
in time and space \citep[among many others, see also Sects.~\ref{eqraddisk}, \ref{thindisk}]{Shakura} 
have been studied theoretically and observationally by many authors \citep[etc.]{Lee,Bjorky97,okazaki,Krticka,kurfek,Krticka_2014}. The common conclusion is that 
we can basically distinguish the inner disk region (see Sects.~\ref{Bephen} and \ref{timemod}), where the rotational velocity is roughly Keplerian and 
the almost negligible radial outflow linearly grows, and the outer disk region, where the radial outflow velocity exceeds the rotational velocity
which is no longer Keplerian and becomes angular momentum conserving, $V_{\phi}\sim R^{-1}$. This different behavior essentially results 
from the ratio between gravity and pressure terms in the 
momentum equation where the gravity term dominates in the inner region while it is very small in the distant regions where
predominates the radial pressure gradient term. This implies the matter rotating on nearly Keplerian orbits with 
the power law surface density decrease, $\Sigma\sim R^{-2}$ \citep[see also Sect.~\ref{thindisk}]{okazaki}, in the inner disk and 
the large radial outflow together with more steeply decreasing angular momentum conserving rotational velocity in the outer part of the disk \citep{okazaki,Krticka,kurfek}. 
The transitional area between these regions roughly corresponds to area of the sonic point location, and the outer radial outflow velocity 
is therefore supersonic. As a consequence, above this transitional area the specific angular momentum loss rate no longer 
increases with radius \citep{Krticka,kurfek}.
This behavior becomes even more complicated in binaries where the disks may be truncated at some radius,
the angular momentum is then transferred to the binary system at this so-called truncation radius (though this expression may be misleading, since the disks
do not cease to exist past that radius \citep[see also \citealt{Rivi2013a} for review]{okacnegr}.

Since the sonic point radius usually extends to the distance of hundreds stellar radii, the observational evidence of the disk features in this region is possible only 
in radio wavelengths and a full observational analysis of this area must still be carried out. There are however objects 
that particularly fit the testing of the steady-state viscous decretion disks models. This is the case of, e.g.,~the binary star $\zeta$ Tau, due 
to the long and well documented stable period with constant disk properties \citep{stefik}. Fig.~\ref{figfour} illustrates the model
of \citet{Karfiol09} where, using the theoretical predictions, the spectral energy distribution of $\zeta$ Tau 
(among other attributes, such as the HI spectral line profile) was successfully calculated in range 
from visible to far infrared wavelength region.

Steady-state viscous decretion disk model is however merely an idealization, since decretion disks never experience long-term stability. In fact they are 
either in phase of growth or decay \citep{okac07}. \citet{Houba12} pointed out the relation between the disk behavior and the ratio of two timescales: 
the \textit{disk mass injection} timescale that depends on number and length of star's surface events that may feed the disk with mass and angular momentum
(see Sect.~\ref{gasejcl}), and the \textit{disk mass redistribution} timescale, i.e.,~the time needed for distribution of the injected material through
the entire volume of the disk (which however clearly depends on the volume considered). If the first timescale is much longer
than the latter, the disk grows until the mass and angular momentum supply is turned off. Then follows the disk dissipation phase, 
characterized by a dual behavior of the inner and outer part of the disk, when the latter part further decretes while the inner 
part reaccretes inwards. The case when the two timescales are approximately equal implies in this model
the periodic disk density behavior in space and time, which may in detail strongly depend on various parameters 
(e.g., in case of large value of $\alpha$ viscosity parameter the disk grows much faster, etc.). \citet{leo13} suggests mechanisms for the disk
formation around rapidly rotating Be stars: the angular momentum supply is provided by the low frequency global oscillations excited by 
the opacity bump mechanism in case of SPB-type stars (see Sect.~\ref{macom}), stochastically 
excited by convective motions in the stellar convective core in case of early Be stars \citep[see also][]{leo14}, or it is excited by tidal interactions in case of the binaries.

Recent study of \citet{Karfiol12} may provide significant corroboration of some aspects of viscous decretion disk scenario. 
The results confirm the agreement of the theoretical models with the observed dissipation curve 
of time-variable dissipating disk of the Be star $\omega$ Canis Maioris during the period of the years 2003-2008. By fitting this dissipation curve
they estimate the disk viscosity parameter $\alpha\approx 1.0$ (the description of the viscosity is given in 
Sects.~\ref{stresstens} and \ref{eqraddisk}) and the disk mass injection rate $\dot{M}$ being approximately $3.5\times 10^{-8}\,\text{M}_{\odot}\,\text{yr}^{-1}$.
The authors conclude that the high value of $\alpha$ parameter results ``from turbulent viscosity induced by disk instability whose growth 
is limited by shock dissipation'' and the value of $\dot{M}$ exceeds the expected average value of 
spherically symmetric stellar wind 
mass loss rate of B stars at least by an order of magnitude \citep{puls1}.

Many other detailed disk features are well explained by the viscous decretion disk model, 
for example the global disk oscillations \citep[see also Sect.~\ref{Bephen}]{okac91}.
Since most of the models and calculations presented within the thesis are based on the viscous disk scenario, other details of the theory will
be yet described in respect to various particular problems. Most results, regarding the viscous decretion disk theory, are also summarized in the 
recent review of Be stars phenomenon given by \citet{Rivi2013a}.

\section{Stellar types associated with outflowing disks}\label{stelassoutfldisks}
In this section we give a brief overview of the main stellar types that may be basically associated with outflowing disks in general.
Not all these types of stars form the disks that can be classified as the dense equatorial Keplerian outflowing ones. 
We refer also to stellar types
with merely the disk-like equatorial stellar wind density enhancement, which may be formed due to, e.g., magnetic compression of the wind 
or binarity 
(some classes of B[e] type stars, LBVs, post-AGB stars) or, on the other hand, to stellar types with accretion inflows (young HAeB[e] stars).
\subsection{Be phenomenon}\label{Bephen}
\begin{figure}[t]
\begin{center}
\includegraphics[width=12.5cm]{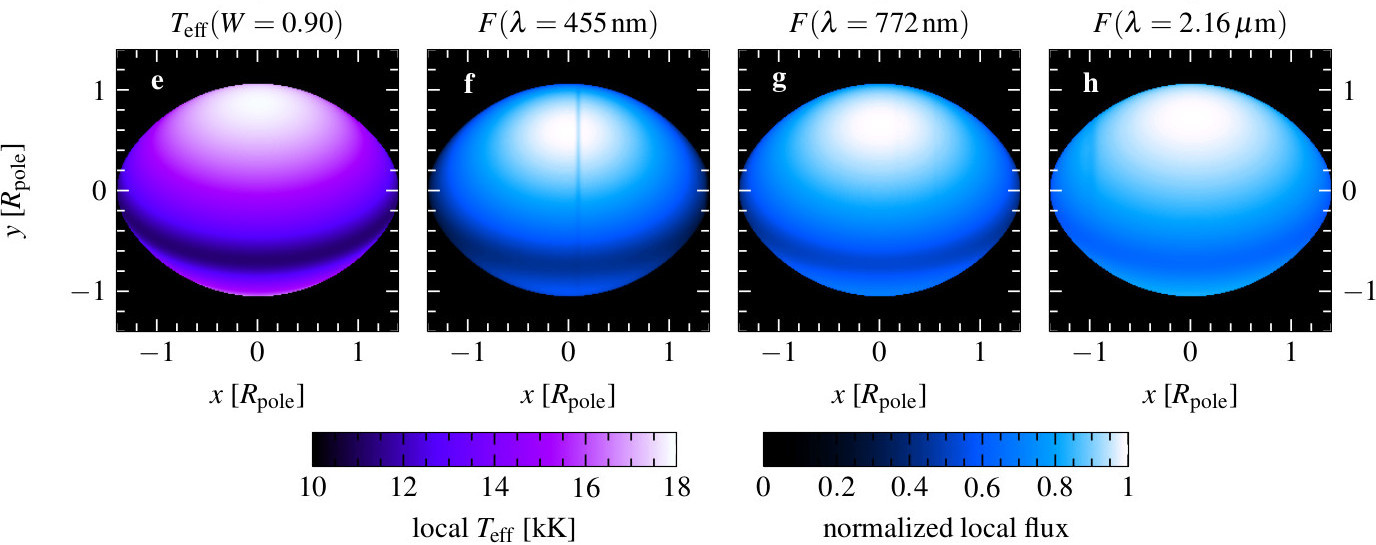}
\caption{Surface appearance of a gravity darkened star rotating at the ratio $W=0.90$, where the factor $W=V_{\text{rot,eq}}/V_{\text{orb,eq}}$
with $V_{\text{orb,eq}}=\sqrt{GM_{\star}/R_{\text{eq}}}$ (this may not yet be the critical velocity defined as 
$V_{\text{crit}}=\sqrt{2GM_{\star}/3R_{\text{pole}}}$). Adapted from \citet{Rivi2013a}. From left to right the panels show the 
temperature and the continuum flux distribution at given wavelengths. The images were computed with use of physical parameters 
found for $\alpha$ Eridani by \citet{Dominik3}.}
\label{figthree}
\end{center}
\end{figure}

Be stars are rapidly rotating non-supergiant stars that form gaseous circumstellar disks \citep{Rivi2013a}.
The first star recognized as a Be star was $\gamma$ Cassiopeiae, observed by Italian astronomer 
Angelo Secchi in the year 1866. It was one of the first stars with detected spectral emission lines \citep[see][]{Suchy}.
Since then, the list of identified Be stars has considerably expanded. For example, one of the most famous Be stars is
$\alpha$ Eridani (Achernar), whose equatorial rotation
rate was interferometrically confirmed to be 95\% of the critical one \citep{Dominik2}. It is also,
due to its relative proximity, the hottest star, for which the detailed photospheric information (including the rotationally induced oblateness) is available
\citep[see also Fig.~\ref{figthree}]{Dominik1,faja}. Moreover, Achernar is presently building a new disk and offers thus an excellent opportunity to 
observe this process from relatively close-up \citep{Rivi2013b,faja}. The viscous decretion disk model \citep[among others]{Lee} can actually best 
explain all studied properties of Be stars' disks. This model has already
been successfully applied to systems that show stable continuum emission: e.g. $\zeta$ Tauri \citep{Karfiol09},
$\chi$ Ophiuchi \citep{Tycka1} and $\beta$ Canis Minoris \citep{koloman} 
as well as to systems that exhibit a more variable photometric activity \citep[e.g., 28 CMa,][]{Karfiol09}.   
Recent observations support the fact that so far
all studied Be star disks rotate (at least in their inner well observable regions) in Keplerian mode \citep{Houba1}.

Although the Be phenomenon can be observed in some stars of late spectral type O and early type A, this designation
almost exclusively refers to stars of spectral type B \citep[see also \citealt{Rivi2013a} for a review]{Porter}.
The generally accepted definition of Be stars was given by \citet{Kolja}, stating that Be star is ``a 
non-supergiant star of spectral type B, whose spectrum has, or had at some time, one or more Balmer spectral lines in emission.''
This definition however covers wide range of stellar types, hence the term ``classical'' Be stars is
used in order to exclude the types
such as Herbig AeBe stars, Algol systems, $\sigma$ Ori E, etc., for which the definition introduced by Collins 
fits as well. More recently the term Be stars has become widely regarded in the sense of the ``classical'' Be stars \citep{Porter}.

Within this specification, Be stars are on average the fastest rotators among all other (nondegenerate) types of stars, their equatorial rotation 
rate is closest to its critical limit \citep{Rivi2013a}. Whether the rotation of Be stars is (at least for significant fraction of them) exactly critical, 
or mostly remains more or less subcritical (see Sect. \ref{diskfome}), is still an open question  
that is furthermore intensively discussed. Determination of the observable projected rotational velocity $\varv\,\text{sin}\,i$
%(in the following text we strictly distinguish between distance $r$ and velocity $\varv$, referring to the spherical geometry, while $R$ and $V$ refer to all other geometries)
is in case of rapid rotators strongly affected by the stellar gravity darkening, 
which makes the characteristics of the fast~rotating equatorial area barely observable \citep[see, e.g.,][cf.~also Fig.~\ref{figthree}]{Towny04}.
Examining the interferometric measurements of this effect, e.g.,~\citet{krasavec} states that 
``actual oblateness values are always in excess of the simple prediction from $\varv\,\text{sin}\,i$''. Moreover, 
the rotation rate can be yet biased by additional line emission or by line absorption in large circumstellar disks, whose line profiles 
are narrower, or by 
photospheric absorption of arbitrary companion in undetected binaries, which is most likely a slower rotator than the Be star itself \citep{Rivi2013a}.

Another common attribute of all Be stars, regardless of their spectral subtype, is their multiperiodic variability 
(in addition to other variabilities on basically all timescales) driven by pulsations. 
The variability of only the early subtypes
is large enough to be observable from the ground, the late types nevertheless pulsate as well, but with smaller amplitudes
\citep[e.g.,][]{sako}. The Be stars
variability can be in detail sorted into numerous types, according to various mechanisms that excite the pulsation modes
(for example the $\kappa$ effect, convection, etc.). On the other hand, no magnetic field has been reliably detected in any Be star, 
the MiMeS project (cf.~Sect.~\ref{macom})
did not find in sample of about 100 observed Be stars 
neither the organized large-scale surface magnetic fields (stronger than approximately 250 G), 
nor even the localized small scale fields, e.g., magnetic loops (see \citet{Rivi2013a} for detailed review of classical Be stars, 
see also \citet{Wade14} for most recent results of the MiMeS survey).
Another unknown within our understanding of internal structure and evolution of rotating hot stars is the mechanism that brings 
these stars so close to the critical rotation rate. The evolution of stellar internal angular momentum distribution
as a result of the effects of convective motion, meridional circulation, etc., was discussed,~e.g.,~by 
\citet{Heger2000,Madr2000}.

The viscous outflowing (decretion) disks formation mechanism described in Sect.~\ref{diskfome} 
fully applies in case of the Be stars' disks \citep[see also,~e.g.,][]{Porter,Hilly1} which are
regarded as the typical examples of the dense equatorial stellar decretion disks
with the Keplerian rotation velocity 
%\begin{align}
$V_{\phi}=\sqrt{{GM_{\star}}/{R}}$
%\end{align}
($M_{\star}$ denotes the mass of central star), 
at least within observationally significant inner parts of the disk. The radial flow in these inner regions is practically negligible,
while in the outer parts of the disks (regarding the distance of hundreds of stellar radii or more) the fast radial 
outflow significantly exceeds the disk rotation rate which at that distance becomes very low.  

The viscosity
plays a key role in the outward transport of matter and angular momentum 
\citep[][among others]{Lee} and 
therefore governs the processes of the disk creation and its further feeding and evolution
(the basic disk kinematic relations are introduced in Sects.~\ref{eqraddisk} and \ref{thindisk}, while 
the basic theoretical background for the viscosity physics is given in Sect.~\ref{stresstens}).
In general, the disks are supposed to be rotationally supported and geometrically very thin in the region close to the star.
Assuming an isothermal gas, the disks are in vertical hydrostatic equilibrium with
the Gaussian density and pressure profiles (see Sect.~\ref{thindisk} for details). The vertical disk thickness (and even the disk
vertical opening angle)
grows with radial distance, we may therefore talk about the flaring disk. The vertical disk scale height obeys in this case the relation
\citep[e.g.,][cf.~Eq.~\eqref{Shakurakos} in Sect.~\ref{eqraddisk}]{Bjorky97}
\begin{align}
H(R)=\frac{aR^{3/2}}{V_{\text{eq}}R^{1/2}_{\text{eq}}},
\end{align}
where $a$ denotes the isothermal speed of sound, $V_{\text{eq}}$ and $R_{\text{eq}}$ are the stellar equatorial velocity and 
stellar equatorial radius, respectively.
The disk scale height is therefore proportional to the ratio of the sound speed and the equatorial rotational velocity of the central star and 
flares with the $3/2$ power of the radial distance. Numerous analyses of the disk 
equatorial plane radial density profile \citep[e.g.,][]{Karfiol08,Tycka1,Sigi1} assume the power law 
relation
\begin{align}
\rho_0=\rho\big(R_{\text{eq}}\big)\left(\frac{R_{\text{eq}}}{R}\right)^d,
\end{align}
where $\rho\big(R_{\text{eq}}\big)$ is the disk midplane density at the disk inner radius (the disk base density), i.e., 
at closest point to the stellar equatorial surface. The power of radial density slope $d$ is usually 
found to be in range between $3$-$4$, the inner boundary disk midplane density is estimated to lie between
$10^{-9}$ to almost $10^{-6}\,\text{kg}\,\text{m}^{-3}$ \citep[see, e.g.,][]{granada}.
The detailed models of the disk temperature structure \citep[e.g.,][]{Millar1,Millar2,Karfiol08} show that the radial temperature profile
in the optically thick region close to the star falls-off with
$T(R)\sim R^{-0.75}$
while in the optically thin outer regions the disk is found to be nearly isothermal with
very gradual temperature decline of about $1000~\text{K}$ in range from $10$ to $50$ stellar radii.
\begin{figure}[t]
\begin{center}
\includegraphics[width=10cm]{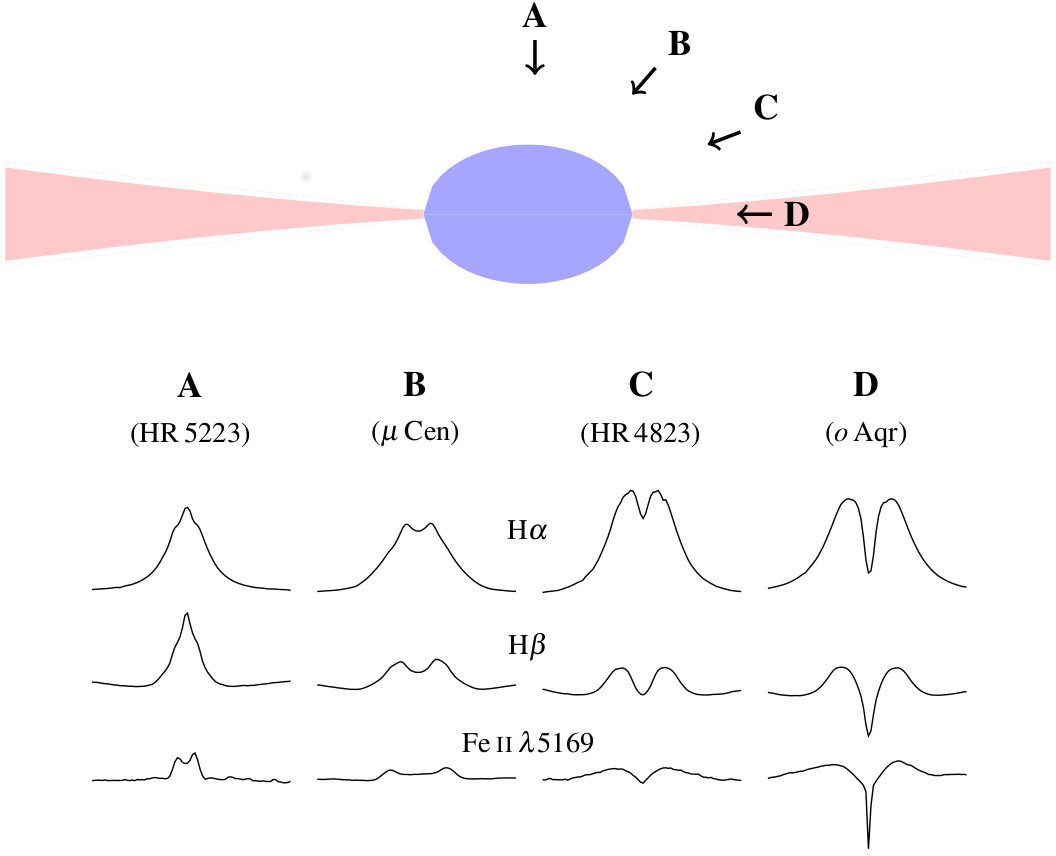}
\caption{Examples of schematic spectral profiles of various Be stars. The sequence of spectra is 
ordered from pole-on direction to the equatorial plane view (directions A-D). Be star $\omicron$ Aquarii 
(D spectrum on the right) is a shell star \citep{Hanus1}. The upper part shows the equator-on scheme of 
critically rotating Be star with flared equatorial disk. The figure is adapted from \citet{Rivi2013a}.}
\label{figfive}
\end{center}
\end{figure}

The basic observable feature of Be stars, i.e., the emission lines,
mostly dominated by the lines of HI, HeI, FeII, sometimes also SiII, MgII \citep{Porter}, 
are produced in the dense equatorial circumstellar 
material that orbits the star. Their typical double peaked appearance with central absorption as the result of the rotating
nonspherically shaped hot gas was first observed by \citet{struve}, who described it as \textit{``a nebular ring which revolves around
the star and gives rise to emission lines''}.
The separation of the two emission peaks shows strong correlation with the observed line width. For example the so-called shell stars 
that were defined in \citet{Hanus1} as ``Be stars with central absorption that reaches below the stellar undisturbed flux'',
have the highest measured photospheric line widths among Be stars and their emission peaks
when present, they also have the largest separation \citep[see also Fig.~\ref{figfive}]{Porter}. 
Most Be stars have both peaks
equally high, there is however a significant minority (\citet{Hanus1} gives the value of about 13\%) with
cyclically variable so-called violet-to-red ratio $V/R$. The periods of these cyclic
variations last from years to decades. Stable Be stars with $V=R$ may change to the $V/R$ variability and back.
Theoretical explanation of these $V/R$ variations, proposed by \citet{okac91}, suggested that these variations 
are caused by global disk oscillations with the azimuthal wavenumber $m=1$, i.e.,~by the one-armed 
density waves. The model calculated by \citet{papa1} showed that inclusion of the quadrupole 
gravitational potential of the rotationally oblate star induces the prograde precession motion of 
these waves. The period time of the precession ranges from years to decades and therefore highly 
exceeds the period of the disk rotational motion. Furthermore, the increase of the precession period 
with the distance from the central star creates the spiral structure in the disk (see the 
left panel in Fig.~\ref{figfour}). \citet{okac97} examined the scenario of one-armed oscillations and he confirmed the 
results of \citet{papa1} in case of late-type Be stars. 
The oscillation theories of \citet{okac91,papa1,okac97} and \citet{oliva} 
were observationally tested by spectrointerferometric measurements of the Be star $\zeta$ Tauri by \citet{Karfiol09}, who 
calculated the model of $\zeta$ Tauri's disk (see Fig.~\ref{figfour}) using these data and using the global disk oscillation model of \citet{okac91} and \citet{papa1} 
to describe the perturbations in density.

Another important factor that deeply influences the physics of Be stars disks is binarity. The estimations
of the fraction of binary systems (or multiple systems in general) among Be stars do not exceed 30-35\%, this is 
however typical also for non-Be stars \citep{Porter}. In ``closer''
systems (in case of circular orbits we may regard them as the systems with orbital periods of about a year or less)
may the companion, due to tidal interactions, help the matter to leave the equatorial region of Be star 
and feed the disk, as well as the process of the spin-up of Be star can be strongly affected 
by the physics of interactions in binary systems \citep[e.g.,][]{Harma02}. In case of highly eccentric
orbits with much longer orbital periods the systems may be strongly affected by tidal interactions during
periastron. Famous and intensively studied case of such system is for example the Be star $\delta$ Scorpii
\citep[see, e.g.,][]{deco,Miro1,Karfiol06}.
There is also evidence of warping of the disk in binaries, 
however, the binarity is not the required mechanism for inducing warping \citep{Porter}.
Moreover, the binarity may cause the disk outer truncation, which we suppose to occur at 
the radius where tidal torque, exerted by the companion, balances the disk viscous torque \citep{okac02}.
Be/X-ray binaries provide the evidence of compact companions in Be star binary systems. 
In fact, neutron stars were mostly observationally confirmed as companions in Be/X-ray binaries so far, black holes are expected to be rare \citep[see][however, one was recently revealed 
by \citet{kanarek} as a companion of the Be star HD 215227]{podsada}. Surprisingly even 
white dwarfs have not yet been fully confirmed, $\gamma$ Cas is a strong candidate \citep[e.g.,][]{smid} as well as another candidate, the Be (or Oe) star + white dwarf system XMMU
J010147.5-715550, located in the Small Magellanic Cloud, was discovered by \citet{sturm}. Be/X-ray binaries are the 
largest subgroup from the group of high-mass X-ray binaries. The disk properties in Be/X-ray binaries are almost identical
to those of single Be stars, however, the intensive accretion of the disk material onto the compact object  
produces the X-ray emission. Most recent detailed review of the problematics of Be/X-ray binaries
was given by \citet{Reig}.

This brief characterization of Be stars is far from being exhaustive. Many other observationally and theoretically
corroborated features of this field can yet be discussed (e.g.,~the interpretation of polarization effects, mechanisms of
disk growth and decay, etc.), 
as well as there are many other open questions: e.g.,~the rotation rate 
(are most of Be stars rotate critically or near-critically or is it just the 70-80\% of the critical value), the mechanism 
of mass and angular momentum transfer that creates and feeds the disk, and what is the role of the binarity in this point, etc.
For the more recent detailed
overview of the problematics of Be stars that covers practically all relevant published results, see, e.g.,~\citet{Rivi2013a}.
\subsection{B[e] phenomenon}\label{Beforbphen}
The term ``B[e] stars'' was first introduced by \citet{Conti} and it was based on a detailed investigation 
of the defining B[e] star HD 45677 made by \citet{Swing1}.  The B[e] stars are defined as the B-type stars
which show forbidden emission lines in their optical spectrum 
(the notation ``[e]'' refers to forbidden lines) indicating a large amount of very rarefied circumstellar 
gas. The forbidden spectral lines and the dust-type infrared excess
make them distinct from the classical Be stars, indicating a different
population with likely a different mechanism responsible for the filling of the 
circumstellar environment with gas \citep[which may nevertheless be disk-shaped,][]{Rivi2013a}.
The characteristic criteria for B[e] stars (B[e] phenomenon) are reviewed in \citet
[see also \citealt{Allenswings}, \citealt{Zicki1}]{Lamers1}:
\begin{enumerate}
\item Strong Balmer emission lines: their presence, as well as the presence of other
permitted low ionization lines, indicates a very large emission rate
(i.e., the volume integral of squared electron density $n_{\text{e}}^2$) of singly ionized gas in the 
atmospheric layers above the stellar continuum forming region.
\item Low excitation permitted emission lines of (predominantly)
low ionization metals in the optical spectrum, e.g.,~Fe II: the presence of these lines indicates
the temperature of the emitting gas about $10^4$ K. The gas is most likely ionized 
by the radiation (impinging irradiation) from the central B star.
\item The presence of forbidden emission lines [Fe II] and [O I] in the optical
spectrum implies a large amount of very rarefied circumstellar gas that is likely extended
to a significant distance from the star.
\item A strong near or mid-infrared excess emitted by hot circumstellar
dust: it indicates the dust region with a temperature varying in range from 500 to 1000 K. The dust equilibrium
temperature decreases with the distance as $T_{\text{dust}}\approx T_{\text{eff}}(2r/R_{\star})^{-0.4}$ 
\citep[see also \citealt{Lamers1}]{Lamercas}, where $T_{\text{eff}}$ is the stellar effective temperature,
$r$ is the (spherical) radial distance and $R_{\star}$ is the radius of the star. The distance of the dust region must be 
at least about $r\geq 500\,R_{\star}$ in case of, e.g.,~a star with $T_{\text{eff}}=20\,000\,\text{K}$.
\end{enumerate}
In addition there are introduced other more detailed possible criteria (e.g.,~the detection of He II or O III lines emission), but these criteria are 
not regarded as defining characteristics. We also classify the stars of the type FS CMa 
\citep[][see also, e.g.,~\citealt{Kricek}]{Miro2} as the separate group of B[e] stars, which are supposed to be surrounded by the aspherical gaseous ring and 
by the ring of hot circumstellar dust whose properties are still not fully explained. 

Despite the more or less homogenous local physical conditions in the circumstellar emitting regions, 
the group of stars classified as B[e] stars is quite 
heterogenous and contains objects of large difference in mass and in
evolutionary phase. The nature and the geometry of the physical mechanism that produces the characteristic forbidden 
emission lines can be quite different and ranges, e.g.,~from the outflowing disk to the presence of a hot companion (in this 
point it shows basic similarity to classical Be stars). \citet{Lamers1} therefore proposed the 
term ``stars with the B[e] phenomenon'' and in their improved classification they divided the group of B[e] stars
into five categories with the following basic criteria. The name of each criterion refers to its evolutionary phase:
\begin{itemize}
\item[(a)] sgB[e]: supergiants which show the B[e] phenomenon. This group of stars
is formed by the supergiants of spectral type B with relative luminosity 
$\text{log}(L_{\star}/L_{\odot})\gtrsim 4.0$. Their spectra show above all strong H$\alpha$ emission line,
many emission lines of He I, Fe II and the forbidden emission [O I]. 
The mass loss of this type of stars can be indicated from the Balmer lines that show for example P Cygni profiles or 
double peaked emission profiles with the central absorption. Since the sgB[e] supergiants in the Galaxy are 
located in great distances and near the galactic plane, their distance and luminosity is much more uncertain than those of 
Magellanic Clouds, which create a quite homogenous group of this category of objects (for a review see \citet{Zicki1}).
The origin of the B[e] phenomenon in evolved massive stars like sgB[e]s is still an open question. There is usually detected a disk
or ring of high-density material in the surroundings of these objects, containing gas and dust \citep{Michaela13}. However, the disk formation mechanism 
is still not clear. So far we know about at least two sgB[e] candidates for rapid rotators with the rotation velocity reaching a substantial part of the critical velocity: 
the stars LHA 115-S 23 \citep{Michaela08} and LHA 115-S 65 \citep{Zicki2, Michaela10}, both in Small Magellanic Cloud.
Another famous example of this type of objects is the Galactic eccentric binary system GG Car with a circumbinary ring,
where two possible scenarios for its origin are discussed: either the nonconservative Roche lobe overflow or the
classical Be star phase underwent by the primary component during the
previous evolution of the system \citep{Michaela13}.
Particular (less luminous) subgroup of B[e] stars can also be identified with rapid rotators on a blue loop in
HR diagram \citep{Heger}. Their disks might be formed by the spin-up mechanism induced by stellar contraction (the complete 
process may be nevertheless far more complicated, for the details see \citet{Heger}) during the transition phase
from red supergiant star toward the blue region on HR diagram.
\item[(b)] HAeB[e] or pre-main sequence stars that show the B[e] phenomenon and are related to the Herbig Ae/Be stars
\citep{Caj}: the group of very young stars in the star forming 
regions that show the spectroscopic evidence of circumstellar matter inflow rather than outflow and show significant variability \citep{Grinin},
caused by irregularities in the circumstellar dust distribution. 
Only a few stars of this category are known \citep{Caj}, since they are younger than 2.5 Myr. Their luminosities are 
$\text{log}(L_{\star}/L_{\odot})\lesssim 4.5$, the Balmer lines are often double peaked, the separation of the peaks ranges from
60 to 300 km s$^{-1}$ \citep{Fernandez, Rejpa}. 
The spectrum of, e.g., the star V~380~Ori
shows strong permitted and forbidden Fe II emission lines and strong H$\alpha$ emission. 
The inverse P Cygni profile of H$\beta$ line gives evidence of matter infall \citep{Lamers1}.
Various scenarios, where all of them assume Keplerian accretion disk, are suggested to explain the emission 
\citep[e.g.,][see also \citealt{Bem}]{Rosak}.
\item[(c)] cPNB[e]: compact planetary nebula stars that show the B[e] phenomenon. They form a group
of low mass stars that are evolving into planetary nebula \citep{Ciatti}, 
their luminosities are $\text{log}(L_{\star}/L_{\odot})\lesssim 4.0$. The stellar spectra indicate possible
nebularity and show forbidden lines of higher ionization states, e.g., [O III], [S III], [Ne III] 
\citep{Allenswings} as well as strong infrared excess. Many of these objects (e.g.,~the star HD 51585 = OY Gem) show significant excess of
the infrared radiation that indicates the existence of a circumstellar dust shell \citep{Archipa}.
\item[(d)] SymB[e]: symbiotic stars that show the B[e] phenomenon. A category of interacting 
binary stars with a hot compact object and a cool giant, surrounded by a nebula. The hot component can be 
detected from recombination lines (e.g., He II), the cool star is manifested by the spectral molecular bands of TiO 
or by the infrared spectral features. SymB[e] stars show permitted and forbidden metallic emission lines of 
low excitation and, since they are irregular (photometric and spectroscopic) variables, near maximum of luminosity they show 
Balmer emission lines \citep{Ciatti}. Symbiotic stars, including their spectral characteristics, are listed e.g. in \citet{Allensymbi}. 
\item[(e)] unclB[e]: unclassified stars that show the B[e] phenomenon. The stars do not fit clearly to any of the previous classes.
Galactic stars HD 50138 and HD 87643 may be the examples where the first of them shows some features of HAeB[e] star and of Be star 
with a presence of short-term variability \citep{Fernando}. The luminosity of the star HD 50138 indicates a kind of main sequence star.
The latter is likely the pre-main sequence star but unlike HAeB[e] star it shows no evidence of infall indicating a different kind of 
circumstellar disk (possibly outflowing) \citep{Lamers1}.
\end{itemize}
\noindent There is thus an ample evidence that the stars showing B[e] phenomenon are surrounded by the circumstellar gas and dust disk or ring.
The dust formation can be facilitated due to the large density of the gaseous disk material and therefore due to its ability
to shield the impinging stellar radiation \citep{Lamers1}.
In case of B[e] supergiants (and possibly some other types) the disk is outflowing due to their rapid rotation. But
whether all B[e] supergiants are indeed rapid rotators is not known \citep{Michaela13}.
The disk formation mechanism does not in this case differ from the mechanism that is assumed to form the outflowing disks in general 
(see Sect.~\ref{Bephen}). An overview of various mechanisms, proposed to explain the formation of the dense 
equatorial disk in case of B[e] phenomenon, is given, e.g., in \citet{Hilly1}.
\subsection{Luminous blue variables (LBVs)}\label{Lumbluvars}
There is a good evidence that also evolved stars of the luminous blue variable (LBV) type have large envelopes with equatorially enhanced density
distribution \citep{schulda,Davy}. This cannot be probably regarded as the outflowing disk in the sense of dense equatorial material in Keplerian orbits.
The geometry of the denser equatorial region can however lead to shielding that allows the formation of dust and molecules (molecular emission
of CO is seen in the LBV Star HR Car) relatively close to the star \citep{CassLBV}. LBVs 
are surrounded by aspherical expanding nebulae that can in general be bipolar, elliptical or irregular. A bipolarity in the wind 
geometry as well as the equatorially enhanced wind may occur
if the stellar rotation rate is close to critical. The star's radiative flux is therefore latitudinally dependent due to stellar oblateness and
gravity-darkening (see Sect.~\ref{diskfome}) and the star is therefore close to bistability temperature jump \citep[e.g.,][]{Davy}.
In case of famous Homunculus Nebula around LBV star $\eta$ Carinae, \citet{Smith} assume bipolar lobes and an 
equatorial disk that is simultaneously produced by a rotating surface explosion. They proposed the semianalytic model  
in which rotating hot stars can produce bipolar and equatorial mass ejections. Although motivated by $\eta$ Carinae, this model
can be applicable to the cases where mass ejection due to fast rotation is expected, 
including other luminous blue variable stars, B[e] stars, the nebula around SN 1987A, or possibly even bipolar supernova explosions.

\subsection{Post-AGB stars}\label{postaks}
Post-AGB stars are low and intermediate 
stars (with initial mass $\leq 8$-$9\,M_{\odot}$) which underwent the phase of large 
mass-loss at the end of their evolution on the asymptotic giant branch (AGB).
During that phase the whole stellar envelope was almost completely expelled.
They are evolving on a fast evolutionary track with constant luminosity 
when the central star crosses the HR-diagram from a
cool asymptotic giant branch to the region, where the temperature is high enough for the ionization of the expanding gas. Given the short evolutionary 
timescale of about $10^4$ years, not many post-AGB stars are known \citep[e.g.,][]{vanWicki1,Ruyter}.

The spin-up mechanism, proposed by \citet{Heger} for the transition phase \textit{red supegiant} $\rightarrow$ \textit{blue loops}
(see Sect.~\ref{Beforbphen}) can 
work in the post-AGB phase just as well (and similarly in the transition \textit{red supergiant} $\rightarrow$ \textit{Wolf-Rayet star}).
Whether the post-AGB stars can reach critical rotation due to this spin-up or not is however not clear \citep{Heger}.
In any case, this spin-up mechanism may be relevant for bipolar outflows from central
stars of proto-planetary nebulae and from stars in the
transition phase from the red supergiant stage to the Wolf-Rayet stage.

\citet{Matt} proposed a model where the disk is formed from the winds of a single asymptotic giant branch star. The model assumes the dipole magnetic field
on the surface of the star, magnetic forces can redirect the isotropically accelerated stellar wind plasma towards the equatorial plane, 
forming a disk.
MHD simulations, performed within that model, demonstrated a dense equatorial disk
produced in case of a dipole field strength of only a few gauss
on the surface of the star. A disk formed by that model could be dynamically important for the shaping of
planetary nebulae.
One can also mention, e.g.,~\citet{Ruyter}, who presented the study of a sample of 51 post-AGB binary 
objects, based on broad-band spectral energy distribution
characteristics, indicating the circumstellar Keplerian rotating dust disks. Since the sample contained significant fraction
of known post-AGB stars, \citet{Ruyter} concluded that binarity is the widespread phenomenon among AGB and post-AGB stars.
They assumed that the disks were evolved during a phase of strong binary interaction when the primary star was much larger than the secondary.
They also estimated that the disks are small in radial direction. 
They however conclude that the structure of such disks, their formation, stability and evolution are still not very well understood. 

\subsection{Population III stars}\label{poptri}
The stars with zero initial metallic
abundance - named Population III stars (or sometimes First Stars in Universe)
- are expected to have different structure and evolution from those in which heavier elements are present.
Evolutionary models of stars with zero initial metallicity, covering a large range of
initial masses ($0.7\,M_{\odot}\leq M\leq 100\,M_{\odot}$), including evolutionary tracks and isochrones, 
were calculated, e.g.,~by \citet{marigo}. The link between stars with very low metallicity and the
problem of the evolution of the surface velocity was studied by \citet{ekstrom}. Since these stars lose negligible 
amount of mass and angular momentum by radiatively driven stellar winds (whose acceleration is dominantly produced in metallic spectral lines 
which however lack in Population III stars),
they are much more prone to reach the critical rotational limit during their evolution \citep[see also][]{Madr2001,Meynet2}.
We suppose that at this critical rotational limit (as in Be stars) the stellar matter is launched into an outflowing
equatorial disk. Significant mass fraction of these extremely massive stars is therefore
expected to be carried away by the disks (up to $10^{-5}\,M_{\odot}\,\text{yr}^{-1}$, see, e.g.,~\citet{ekstrom}).
In that sense the disks of the Population III stars might play an
important role in kinematical and chemical evolution of early universe. 

Mass and angular momentum loss of low and zero metallicity stars was studied by \citet{Krticka09,Krticka}, who examined the potential
role of viscous coupling in outward transport of angular momentum in stellar decretion disks. They calculated stationary models
of the profiles of hydrodynamic quantities (density, radial and azimuthal velocity, angular momentum loss rate) in Population III star's disks. Stationary as well as
time-dependent models of the hydrodynamic quantities of Population III star's disks \citep{kurfek}, calculated using our time-dependent hydrodynamic code 
(see Sect.~\ref{timemod}), are also 
presented within this thesis in Sects.~\ref{statcalc} and \ref{timemod}.

\HlavickaKapitoly
\chapter{Basic hydrodynamics}
\label{hydrobase}
\section{Boltzmann kinetic equation} 
The general form of the Boltzmann kinetic equation for a particle of the type $\alpha$, used in
gas (plasma) kinetic theory \citep[see, e.g.,][for details]{Bittyk}, is
\begin{align}\label{boki1}
\frac{\partial f_\alpha}{\partial t}+\vec{\varv}\cdot\vec{\nabla}f_\alpha+
\vec{a}_{\text{ext}}\cdot\vec{\nabla}_{\varv}\,f_\alpha=\left(\frac{\delta f_\alpha}{\delta t}\right)_{{\text{coll}}}.
\end{align}
The particle distribution function $f_\alpha(\vec{r},\vec{\varv},t)$ is defined as the density of the particles of the type 
$\alpha$ in phase space in the form
\begin{align}\label{boki2}
f_\alpha({\vec{r}},\vec{\varv},t)\,\text{d}^3r\,\text{d}^3\varv=\text{d}^6N_\alpha(\vec{r},\vec{\varv},t).
\end{align}
The quantity $\text{d}^6N_\alpha(\vec{r},\vec{\varv},t)$ denotes the number of the particles of the type $\alpha$ inside the phase space volume 
$\text{d}^3r\,\text{d}^3\varv$ with coordinates
$(\vec{r},\,\vec{\varv})$ at instant time $t$. In Eq.~\eqref{boki1} the quantity $\vec{a}_{\text{ext}}$ means the acceleration induced by external force,
$\vec{\nabla}_{\varv}$ is the so-called velocity gradient $\partial/\partial\vec{\varv}$ 
and the collision term on the right hand side refers to time rate of change
of particle distribution function due to particle collisions.
The average value of the arbitrary physical property $\chi({\vec{r}},\vec{\varv},t)$ for the particles of type $\alpha$ is given by 
\citep[see][for the detailed explanation of the derivation of the following relations in this section]{Bittyk}
\begin{align}\label{boki3a}
\langle\chi(\vec{r},\vec{\varv},t)\rangle_\alpha=\frac{1}{n_\alpha(\vec{r},t)}\int\chi(\vec{r},\vec{\varv},t)\,f_\alpha(\vec{r},\vec{\varv},t)\,\text{d}^3\varv,
\end{align}
where the quantity $n_\alpha(\vec{r},t)$ denotes the number density of the particles of type $\alpha$ in configuration space with coordinate
$\vec{r}$ at instant time $t$, defined as integral of the particle distribution function $f_\alpha(\vec{r},\vec{\varv},t)$ over the whole velocity space,
\begin{align}\label{concentrix}
n_\alpha(\vec{r},t)=\int f_\alpha\!\left({\vec{r}},\vec{\varv},t\right)\,\text{d}^3\varv.
\end{align}

We now multiply the Boltzmann kinetic equation for a particle of the type $\alpha$
by an arbitrary physical quantity $\chi(\vec\varv)$, which is independent of time and space and is, in general, a function of the particle velocity.
Integrating it over the whole velocity space, we obtain 
\begin{align}\label{boki3}
\int\chi\frac{\partial f_\alpha}{\partial t}\,\text{d}^3\varv+\int\chi\,\vec{\varv}\cdot\vec{\nabla}f_\alpha\,\text{d}^3\varv+
\int\chi\,\vec{a}_{\text{ext}}\cdot\vec{\nabla}_\varv\,f_\alpha\,\text{d}^3\varv=
\int\chi\left(\frac{\delta f_\alpha}{\delta t}\right)_{{\text{coll}}}\text{d}^3\varv.
\end{align}
Since $\vec{r},\,\vec{\varv}$ and $t$ are independent variables, the spatial gradients of the velocity dependent quantities in expanded
Eq.~\eqref{boki3}, i.e.,~$\vec\nabla\cdot\vec{\varv}$ and
$\vec\nabla\chi(\vec\varv)$, are zero. The force component $F_i$ is generally independent of the corresponding component of velocity $\varv_i$,
the velocity gradient of the acceleration induced by external forces, $\vec{\nabla}_\varv\cdot\vec{a}_{\text{ext}}$,
therefore vanishes (this restriction however 
does not apply in case of, e.g.,~the magnetic force where ${F_{L,\,i}=q_\alpha\,\epsilon_{ijk}\,\varv_jB_k}$).
The expansion of Eq.~\eqref{boki3} gives the equation for the average value $\langle\chi\rangle_\alpha$ of the velocity dependent quantity 
(or other quantities
in the brackets $\langle\,\,\rangle_\alpha$) with respect to velocity space for the particles of type $\alpha$. 
Equation \eqref{boki3} for a particle $\alpha$ thus takes the form
\begin{align}\label{bol}
\frac{\partial}{\partial t}\left(n_\alpha\langle\chi\rangle_\alpha\right)+\vec{\nabla}\cdot\left(n_\alpha\langle\chi\vec{\varv}\rangle_\alpha\right)-
n_\alpha\langle\vec{a}_{\text{ext}}\cdot\vec{\nabla}_\varv\,\chi\rangle_\alpha=
\left[\frac{\delta}{\delta t}\left(n_\alpha\langle\chi\rangle_\alpha\right)\right]_{{\text{coll}}},
\end{align}
where the collision term on the right-hand side of Eq.~\eqref{bol} denotes the 
time rate of change of the quantity $\chi$ for the particles of the type $\alpha$ due to collisions.
Following Eq.~\eqref{concentrix} we define the mass density for the $\alpha$-type particles as $\rho_\alpha=n_\alpha m_\alpha$ (where $m_\alpha$ denotes
the mass of particle of the type $\alpha$).

The velocity $\vec{\varv}$ of the particle of the type $\alpha$ is the vector sum 
\begin{align}\label{boki4}
\vec{\varv}=\vec{V}_\alpha+\vec{C}_\alpha
\end{align}
of the average velocity $\vec{V}_\alpha(\vec{r},t)$, which is defined as the macroscopic drift 
(or flow) velocity of the $\alpha$-type particles near the position $\vec{r}$ at instant time $t$, 
\begin{align}\label{velocitrix}
\vec{V}_\alpha(\vec{r},t)=\frac{1}{n_\alpha(\vec{r},t)}\int\vec{\varv}\,\,f_\alpha\!\left({\vec{r}},\vec{\varv},t\right)\,\text{d}^3\varv,
\end{align}
and the random or peculiar velocity of thermal motion $\vec{C}_\alpha$, defined as the velocity of the $\alpha$-type particles, relative 
to the average velocity $\vec{V}_\alpha(\vec{r},t)$.
The average value of the drift velocity (as a macroscopic collective property) is $\langle\vec{V}_\alpha\rangle=\vec{V}_\alpha$,
the average random thermal velocity has to be zero, $\langle\vec{C}_\alpha\rangle=\vec{0}$.
The average global $\alpha$-type particle velocity is thus equal to the particle drift (or flow)
velocity, 
\begin{align}\label{boki5}
\langle\vec{\varv}\rangle_\alpha=\langle\vec{V}_\alpha\rangle=\vec{V}_\alpha.
\end{align}
From the kinetic theory of gases also follows that the mass and momentum density of the matter must obey the relations
\begin{align}\label{boki8}
\rho=\sum\limits_\alpha\rho_{\alpha},\,\,\,\,\,\,\,\,\,\,\,\,\rho\vec{V}=\sum\limits_\alpha\rho_{\alpha}\vec{V}_\alpha,
\end{align}
where $\rho$ and $\vec{V}$ are the density and the flow velocity of the whole medium, respectively.
We introduce further the diffusion velocity $\vec{w}_\alpha$ defined as the vector subtraction of mean velocity $\vec{V}_\alpha$ of particle $\alpha$ 
and the global mean velocity $\vec{V}$ of the medium,
\begin{align}\label{boki6}
\vec{w}_\alpha=\vec{V}_\alpha-\vec{V}.
\end{align} 
The diffusion velocity can be regarded as the mean velocity of the particle of the type $\alpha$ in a reference frame that is co-moving with the 
global average velocity $\vec{V}$ of the whole medium (where the global momentum density is regarded simply as the sum 
of particular momentum densities of the species $\alpha$). 
Since the diffusion velocity $\vec{w}_\alpha$ is clearly the macroscopic quantity, there applies
$\langle\vec{w}_\alpha\rangle=\vec{w}_\alpha$.

We define also the global random velocity $\vec{C}_{\alpha 0}$ for particles of the type $\alpha$ relative to the global mean velocity of the 
fluid,
which, employing Eqs.~\eqref{boki4} and \eqref{boki6}, gives
\begin{align}\label{boki7}
\vec{C}_{\alpha 0}=\vec{\varv}-\vec{V},\,\,\,\,\,\,\,\text{that is}\,\,\,\,\,\,\,\vec{C}_{\alpha 0}=\vec{C}_\alpha+\vec{w}_\alpha.
\end{align}
Kinetic stress tensor $T_{\alpha ij}$ for the particle $\alpha$ (where $i,j$ are spatial components) 
and the global kinetic stress tensor $T_{ij}$ are defined as
\begin{align}\label{boki9}
T_{\alpha ij}=-\rho_{\alpha}\langle C_{\alpha i}C_{\alpha j}\rangle,\,\,\,\,\,\,\,\,\,\,\,\,T_{ij}=
-\sum\limits_\alpha\rho_{\alpha}\langle C_{\alpha 0i}C_{\alpha 0j}\rangle.
\end{align}
Scalar pressure $P_{\alpha}$ for the particle $\alpha$ and the global scalar pressure for the whole fluid $P$ are defined as 
one third of a trace of corresponding stress tensor (with opposite sign),
\begin{align}\label{boki10}
P_{\alpha}=\frac{1}{3}\,\rho_{\alpha}\langle C_{\alpha}^2\rangle,\,\,\,\,\,\,\,\,\,\,\,\,P=
\frac{1}{3}\sum\limits_\alpha\rho_{\alpha}\langle C_{\alpha 0}^2\rangle.
\end{align}
The vector of the flux of thermal energy $\vec{q}_{\alpha}$ for the particle $\alpha$ and the global vector of the 
flux of thermal energy $\vec{q}$ become
\begin{align}\label{boki11}
\vec{q}_{\alpha}=\frac{1}{2}\,\rho_{\alpha}\langle C_{\alpha}^2\vec{C}_{\alpha}\rangle,\,\,\,\,\,\,\,\,\,\,\,\,\vec{q}=
\frac{1}{2}\sum\limits_\alpha\rho_{\alpha}\langle C_{\alpha 0}^2\vec{C}_{\alpha 0}\rangle.
\end{align}
The physics of the stress in a Newtonian fluid, i.e.,~in the fluid where the shearing stress is linearly related to the rate of shearing strain
in the medium,
is shown in Sect.~\ref{stresstens}.
The explicit form of the stress tensor for a Newtonian fluid, 
written for various coordinate systems, is given in Appendix~\ref{stressappend}.
\section{Mass conservation (continuity) equation}
\label{Massconserve}
The following notation will be strictly kept hereinafter: the capital letters $R,\,V$ for radial distance and velocity (including the 
corresponding vectors and their components)
in Cartesian and cylindrical coordinate systems (as well as in the coordinate system introduced in Sect.~\ref{flarecoords}), while the small letters 
$r,\,\varv$ are used for the same quantities exclusively within the spherical coordinate system.

If we substitute in Eq.~\eqref{bol} the mass $m_\alpha$ of the particle $\alpha$ for the arbitrary physical property $\chi$, 
we obtain the mass conservation (continuity) equation (0-th moment of Boltzmann kinetic equation that corresponds to the mass conservation law) for the particle $\alpha$, 
\begin{align}\label{conticollide}
\frac{\partial\rho_\alpha}{\partial t}+\nabla\cdot\left(\rho_\alpha\vec{V}_\alpha\right)=S_\alpha.
\end{align}
The collision term $S_\alpha$ on the right-hand side of Eq.~\eqref{conticollide} takes the general form (cf. Eq.~\eqref{boki3}) 
\begin{align}\label{collisionix}
S_\alpha=m_\alpha\int\left(\frac{\delta f_\alpha}{\delta t}\right)_{{\text{coll}}}\text{d}^3\varv=\left(\frac{\delta\,\rho_\alpha}{\delta t}\right)_{{\text{coll}}}.
\end{align}
The term $S_\alpha$ refers to the rate of production or destruction (their transformation to another particle type) 
of particles $\alpha$ due to particle interactions, i.e., due to, e.g., 
ionization, recombination (collisional or radiative - see the equations of radiative transfer and statistical equilibrium in, e.g.,~\citet{Mihalas1}), charge transfer, etc. In case 
of non-isolated physical system with external interactions it may also refer to source (or sink) of mass, it is therefore sometimes called source term.
The explicit expression of the collision term $S_\alpha$ may be in general very complex. It involves the 
inelastic collisions that lead to production or loss of a particle type. For example in case of electrons, the most important interactions are ionization and 
recombination (neglecting,~e.g.,~the electron capture ionization, etc.). Denoting $k_i$ and $k_r$ the collision rate coefficients for ionization and recombination, respectively
(which are theoretically derived as well as experimentally verified and are proportional to the number of electrons produced per unit time in case of ionization and 
to the product of the electron and specific ion type number densities in case of recombination), 
we can express such simplified collision term for electrons $S_e$ in the explicit form \citep{Bittyk}
\begin{align}\label{electrocollide}
S_e=m_e\left(k_in_e-k_rn_e^2+\,\ldots\,\right).
\end{align}
Inclusion of the radiative rate coefficients leads to the equation of statistical equilibrium \citep[e.g.,][]{Mihalas1}, which is currently beyond the scope of this study
(this is also the case of the equation of motion in Sect.~\ref{momentumconserve} and the energy equation in Sect.~\ref{energyconserve}).
If we sum Eq.~\eqref{conticollide} over all particle types $\alpha$,
the collision term $S_{\!\!\alpha}$ vanishes as a consequence of 
the total mass conservation in the closed system and we obtain the continuity equation for the medium as a whole,
\begin{align}\label{masscylinder001}
\frac{\partial\rho}{\partial t}+\vec{\nabla}\cdot\left(\rho\vec{V}\right)=0.
\end{align}

Alternative approach for the derivation of the continuity equation
is based on consideration that there has to be an
equilibrium between the number of particles 
leaving some volume $\Omega$ through its closed surface ${A}$ (particle flux) and the time rate of particle 
number density decrease inside this volume. 
Assuming the certain mass $m_{\alpha}$ of the particle of type $\alpha$, this gives
\begin{align}\label{fluidysol}
\sum_\alpha\int_An_{\alpha}m_{\alpha}\vec{V}_{\alpha}\cdot\text{d}\vec{A}=
-\sum_\alpha\frac{\partial}{\partial t}\int_{\Omega}n_{\alpha}m_{\alpha}\text{d}\Omega,
\end{align}
where the quantity $n_{\alpha}m_{\alpha}$ is the mass density $\rho_{\alpha}$ of the particle $\alpha$.
Since the result of Eq.~\eqref{fluidysol} must be valid for any arbitrary volume $\Omega$ \citep[see, e.g.,][among others]{Landau6,Bittyk}, 
applying the Gauss's theorem and summing over all particles $\alpha$, we obtain the expression \eqref{masscylinder001}.

The explicit form of the continuity equation in cylindrical and spherical coordinates is expanded in Appendix~\ref{contappend}.
In axisymmetric case is the cylindrical expression \eqref{conticylap}
reduced to the axisymmetric cylindrical continuity equation \eqref{masscylinder1}.
The spherically symmetric continuity equation (see Eq.~\eqref{contispherelap}) becomes the form of Eq.~\eqref{masscylinder002}.

\section{Equation of motion}
\label{momentumconserve}
Substituting the momentum of the particle  $m_\alpha\vec{\varv}$ for the arbitrary
physical quantity $\chi(\vec{\varv})$ in Eq.~\eqref{bol}, 
we obtain the equation of motion (momentum equation) for the particle $\alpha$ 
(1st moment of Boltzmann kinetic equation that corresponds to the momentum conservation law),
\begin{align}\label{momentumcollide}
\frac{\partial}{\partial t}\left(\rho_\alpha\vec{V}_\alpha\right)+\vec\nabla\cdot\left[\rho_\alpha\left(\vec{V}_\alpha\otimes\vec{V}_\alpha
+\left\langle\vec{C}_\alpha\otimes\vec{C}_\alpha\right\rangle\right)\right]-n_\alpha\vec{F}_\alpha=\vec{A}_\alpha,
\end{align} 
where $\vec{V}_\alpha\otimes\vec{V}_\alpha$ and $\vec{C}_\alpha\otimes\vec{C}_\alpha$ are the tensor (dyadic) products of vectors.
The third left-hand side term $\vec\nabla\cdot\left(\rho_\alpha\left\langle\vec{C}_\alpha\otimes\vec{C}_\alpha\right\rangle\right)$ 
expresses the divergence of the stress tensor $-\nabla_j T_{\alpha}^{ij}$ (cf.~Eq.~\eqref{boki9}), 
the fourth term $-n_\alpha\vec{F}_\alpha$ is the sum of external forces (multiplied by particle 
number density) acting on the particle $\alpha$, i.e.,~the gravity 
(Eqs.~\eqref{gravcylexpli} and \eqref{spheregravka} in Appendix \ref{appendix1}), radiative force (see Eq.~\eqref{CAK8} in Sect.~\ref{lajncakis}), etc. 
The collision term $\vec{A}_\alpha$ on the right-hand side refers to the interchange of momentum due to the collisions,
creation and destruction of particles (cf.~the collision term $S_\alpha$ in Eq.~\eqref{conticollide}).
Its explicit expression is usually given as the linear approximation for a small difference in velocities \citep{Bittyk},
\begin{align}\label{momcolisix}
\vec{A}_\alpha=-\rho_\alpha\sum_\beta\nu_{\alpha\beta}\left(\vec{V}_\alpha-\vec{V}_\beta\right),
\end{align}
where we assume that the force exerted on the particles of type $\alpha$ by colliding particles of the type $\beta$ is
proportional to the difference of mean velocities $\vec{V}_\alpha,\,\vec{V}_\beta$ of the particles. The constant of proportionality $\nu_{\alpha\beta}$
is called the collision frequency for transfer of momentum.

Subtracting the continuity equation \eqref{conticollide}, multiplied by $V_\alpha$ from Eq.~\eqref{momentumcollide}, and noting that
$\partial\vec{V}_\alpha/\partial t+\vec{V}_\alpha\cdot\vec{\nabla}\vec{V}_\alpha=\text{d}\vec{V}_\alpha/\text{d}t=\vec{a}_\alpha$
(acceleration of the particle $\alpha$), 
we obtain the momentum equation in the compact form
\begin{align}\label{momos1}
\rho_\alpha{a}_\alpha^i=\nabla_j T_{\alpha}^{ij}+{F}^i_\alpha+{A}^i_\alpha,
\end{align} 
where the indices $i,j$ are the spatial components. 
If we sum Eq.~\eqref{momos1} over all particle types $\alpha$, the collision term ${A}^i_\alpha$ in the right-hand side vanishes due to 
the conservation of the total momentum of fluid particles in the closed system. 
Eq.~\eqref{momos1} thus becomes the momentum equation for the whole fluid that can be written as
\begin{align}\label{moment}
\rho{a}^i=\nabla_j T^{ij}+{F}^i.
\end{align} 
The expanded acceleration term on the right-hand side of Eq.~\eqref{moment}, written in indexed component form, is
\begin{align}\label{moment1}
\rho{a}^i=\rho\frac{\text{d}V_i}{\text{d}t}=\rho\left(\frac{\partial V_i}{\partial t}+V^j\nabla_j V^i\right),
\end{align}
while the same equation, written in the vector form, becomes
\begin{align}\label{moment2}
\rho\vec{a}=\rho\left[\frac{\partial\vec{V}}{\partial t}+\rho\left(\vec{V}\cdot\vec{\nabla}\right)\vec{V}\right].
\end{align}
Using vector calculus identities, we can express the advection term in Eq.~\eqref{moment1} in a different way,
\begin{align}\label{moment3}
\left(\vec{V}\cdot\vec{\nabla}\right)\vec{V}=\frac{1}{2}\vec\nabla V^2-\vec{V}\times\left(\vec\nabla\times\vec{V}\right),
\end{align}
where the two terms on the right-hand side refer to a separated laminar flow and to a rotational (turbulent) motion, respectively. 
We express the detailed form of the stress tensor, including the derivation of its components, 
in Sect.~\ref{stresstens} and in Appendix~\ref{stressappend}.
The explicit expression of the momentum equation in cylindrical and in spherical polar coordinates is given in 
Sect.~\ref{momappend}.

Alternative approach for deriving the momentum equation is based on Newton's second law \citep[e.g.,][]{Landau6},
whose most basic form $m\vec{a}=\vec{F}$ can be written as the following sum of forces exerted on the particle $\alpha$,
\begin{align}\label{newtmoment1}
\sum_\alpha\int_{\Omega}\frac{\text{d}\left(\rho_\alpha\vec{V}_\alpha\right)}{\text{d}t}\text{d}\Omega=
\sum_\alpha\left(\int_{\Omega}\vec{F}_\alpha^\Omega\text{d}\Omega+\int_{A}\vec{F}_\alpha^A\text{d}A\right).
\end{align}
We follow here the notation used in Eq.~\eqref{fluidysol}, $\vec{F}_\alpha^\Omega$ is a density of volume forces that act
throughout the volume $\Omega$ and $\vec{F}_\alpha^A$ is a density of surface forces acting on the surface $A$.
Applying the divergence theorem on the surface integral (the second term on the right-hand side),
where the surface force is the pressure (negative stress) force
represented by the divergence of the stress tensor ${T}_\alpha$ from Eq.~\eqref{boki9}, the equation \eqref{newtmoment1} obviously leads to the 
same expression~\eqref{moment}.

\section{Stress tensor}
\label{stresstens}
\begin{figure}[t]
\begin{center}
\includegraphics[width=8cm]{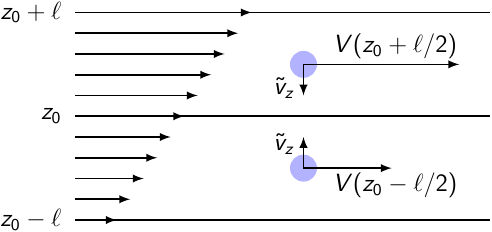}
\caption{Schema of the planar shear stress with vertically increasing magnitude of velocity $V$ of horizontal layers. 
The reference horizontal plane is denoted $z_0$ while $\ell$ is the mean free path of 
a particle or a fluid parcel (that is regarded as a subject of the shear stress force) 
and $\tilde{\varv}_z$ is the magnitude of $z$-component of a mean velocity of random motion 
of particles or fluid parcels. Adapted from \citet{Frank}.}
\label{planestrih}
\end{center}
\end{figure}

\noindent The stress tensor for a Newtonian fluid (cf.~Eq.~\eqref{boki9} of the kinetic stress tensor) may be written in general form as
\begin{align}\label{stress}
T_{ij}=-P\,\delta_{ij}+\sigma_{ij},
\end{align}
where $P$ is the scalar pressure and $\sigma_{ij}$ is the nondiagonal symmetric viscous stress tensor. 
Since the tensor components must have physical dimension of the pressure, 
where the $i$-th component of the force $F_i=\text{d}\Pi_i/\text{d}t$ is tangential to a surface area $A_j$, we may write
\begin{align}\label{stresswritten}
\sigma_{ij}=\frac{1}{A_j}\frac{\text{d}{\Pi_i}}{\text{d}t},\,\,\,\,\text{where}\,\,\,\,
\frac{\text{d}\Pi_i}{\text{d}t}=\rho\Omega\frac{\text{d}V_i}{\text{d}t}=
\rho A_j\ell\frac{\text{d}V_i}{\text{d}t}.
\end{align}
The quantity $\Pi_i$ denotes the $i$-th component of momentum of fluid particles in case of particle (molecular) viscosity,
or the ``particles'' of material, e.g., the gas eddies, in case of macroscopic turbulent viscosity.
In Eq.~\eqref{stresswritten} the quantity $\Omega$ denotes a fluid volume, $\Omega=A_j\ell$, 
where the distance $\ell$ expresses the mean free path of the particles or ``particles''. We denote the mean random velocity of 
those particles as $\tilde{\varv}$.
Considering one-dimensional planar shear viscous stress, illustrated in Fig.~\ref{planestrih}, where we denote the horizontal and vertical direction 
as $x$ and $z$, respectively,
we obtain
\begin{align}\label{planarvisco}
\sigma_{xz}=\rho\ell\frac{\text{d}V_x}{\text{d}z}\frac{\text{d}z}{\text{d}t}=
\rho\ell\,\tilde{\varv}_z\,\frac{\text{d}V_x}{\text{d}z}=\eta\frac{\text{d}V_x}{\text{d}z}. 
\end{align}
The factor of proportionality in Eq.~\eqref{planarvisco} is called the 
coefficient of dynamical viscosity $\eta$. Its physical meaning is $\rho\ell\,\tilde{\varv}_z$ or $f\rho\ell\,\tilde{\varv}$, where the numerical factor $f$ 
is approximated as
$1/3$, which refers to an average fraction of fluid particles moving in $z$ direction (its exact value depends on the type of particle interactions \citep{maeder}).
In a general case, assuming large deformations, one can write the expression for 
the viscous stress tensor (using the Einstein summation convention) in the form \citep{Landau1,Mihalas2,Frank}
\begin{align}\label{viscous}
\sigma_{ij}=\eta\left(\nabla_iV_j+\nabla_jV_i+\nabla_iV_k\nabla_jV_k\right)+\lambda\,\nabla_kV^k\delta_{ij}= 
2\eta\,E_{ij}+\lambda\,\nabla_kV^k\,\delta_{ij}, 
\end{align}
where $\eta$ is the (introduced) dynamic or shear viscosity coefficient, $\lambda$ is the dilatational or second viscosity coefficient (the formalism of
covariant derivatives in curvilinear coordinate systems is described in Sect.~\ref{diffcylinder} and in Appendix~\ref{stressappend}).

In Eq.~\eqref{viscous} we employ the Green-Lagrangian strain tensor, its covariant form is
\begin{align}\label{glstrain}
E_{ij}=\frac{1}{2}\left(\nabla_iV_j+\nabla_jV_i+\nabla_iV_k\nabla_jV_k\right).
\end{align}
\noindent In case of small deformations the nonlinear term in the Green-Lagrangian strain tensor is omitted and
we obtain the symmetric Cauchy strain tensor, 
\begin{align}\label{castrain}
E_{ij}=\frac{1}{2}\left(\nabla_iV_j+\nabla_jV_i\right).
\end{align}
\noindent The viscous stress tensor can be written in the form
\begin{align}\label{viskoza}
\sigma_{ij}=\eta\left(\nabla_iV_j+\nabla_jV_i+\nabla_iV_k\nabla_jV_k-\frac{2}{3}\nabla_kV^k\,\delta_{ij}\right)+\zeta\,\nabla_kV^k\delta_{ij},
\end{align}
where $\zeta=\lambda+\frac{2}{3}\eta$ is the coefficient of bulk viscosity.
\noindent Using the shear and bulk viscosity coefficients, the stress tensor can be written as
\begin{align}\label{stresik1}
T_{ij}=-P\,\delta_{ij}+\eta\left(\nabla_iV_j+\nabla_jV_i+\nabla_iV_k\nabla_jV_k\right)+
\left(\zeta-\frac{2}{3}\eta\right)\,\nabla_kV^k\delta_{ij},
\end{align}
or, using the strain tensor notation,
\begin{align}\label{stresik2}
T_{ij}=-P\,\delta_{ij}+2\eta\,E_{ij}+\left(\zeta-\frac{2}{3}\eta\right)\,\nabla_kV^k\delta_{ij}. 
\end{align}
The full explicit form of Eq.~\eqref{stresik2} in various coordinates, involving the Cauchy strain tensor \eqref{castrain}, is given in 
Appendix~\ref{stressappend}.

Another important quantities, which describe the properties of viscosity, are the ratio of the dynamic viscosity $\eta$ 
to the fluid density $\rho$ designated as the kinematic viscosity $\nu$,
and the characteristic timescale $t_{\text{visc}}$ of the viscous effects (viscous timescale),% whose order results from the equation of motion,
\begin{align}\label{kincl}
\nu=\frac{\eta}{\rho}=\frac{1}{3}\ell\,\tilde{\varv},\,\,\,\,\,\,\,\,\,\,\,\,\,\,\,\,\,\,\,\,\,\,\,\,\,\,\,t_{\text{visc}}\sim\frac{L^2}{\nu}, 
\end{align}
where $L$ denotes a typical length scale within the system.
In fluid mechanics is the influence of viscosity usually expressed using a dimensionless Reynolds number $\mathcal{R}e$, given as \citep[e.g.,][]{maeder}
\begin{align}\label{rejnok}
\mathcal{R}e=\frac{\rho VL}{\eta}=\frac{VL}{\nu},
\end{align}
where $V$ is a typical velocity within the system. The Reynolds number thus expresses the ratio of inertial forces to viscous forces. The extent of viscous effects 
is scaled by a critical value $\mathcal{R}e_{\text{crit}}$, which however varies in dependence on the fluid type and the geometry of the system (e.g., in 
stellar interiors it may take the value $\mathcal{R}e_{\text{crit}}\approx 10$ \citep[see,][]{sleha1}).

\section{Equation of energy}
\label{energyconserve}
Substituting the kinetic energy of the particle $\frac{1}{2}m_\alpha{\varv}^2$ for the arbitrary
physical property $\chi(\vec{\varv})$ in Eq.~\eqref{bol}, we obtain the energy equation for the $\alpha$-type particle 
(2.~moment of Boltzmann kinetic equation that corresponds to the energy conservation law),
\begin{align}\label{energos1}
\frac{\partial}{\partial t}\left(\frac{1}{2}\,\rho_\alpha{{V}}_\alpha^2\right)+\frac{\partial}{\partial t}\left(\frac{1}{2}\,\rho_\alpha\left\langle{C}_\alpha^2\right\rangle\right)
+\nabla\cdot\left(\frac{1}{2}\,\rho_\alpha{V}_\alpha^2\vec{V}_\alpha\right)
+\nabla\cdot\left(\frac{1}{2}\,\rho_\alpha\left\langle{C}_\alpha^2\right\rangle\vec{V}_\alpha\right)+\nonumber\\
+\,\nabla\cdot\left(\rho_\alpha\left\langle\vec{C}_\alpha\otimes\vec{C}_\alpha\right\rangle\vec{V}_\alpha\right)
+\nabla\cdot\left(\frac{1}{2}\,\rho_\alpha\left\langle{C}_\alpha^2\vec{C}_\alpha\right\rangle\right)-n_\alpha\left(\vec{F}\cdot\vec{V}\right)_\alpha={M}_\alpha.
\end{align} 
The term $n_\alpha\left(\vec{F}\cdot\vec{V}\right)_\alpha$ is the flux of the sum of external forces acting on the particle $\alpha$
(multiplied by particle number density,
cf.~Eq.~\eqref{momentumcollide}). 
The collision term ${M}_\alpha$ on the right-hand side represents the rate of energy change due to collisions,
creation and destruction of particles (cf.~the collision terms $S_\alpha$ in Eq.~\eqref{conticollide} and $\vec A_\alpha$ in Eq.~\eqref{momentumcollide}),
\begin{align}
M_\alpha=\frac{1}{2}m_\alpha\int\varv^2\left(\frac{\delta f_\alpha}{\delta t}\right)_{{\text{coll}}}\text{d}^3\varv=
\left[\frac{\delta\left(\frac{1}{2}\rho_\alpha\left\langle\varv^2\right\rangle_\alpha\right)}{\delta t}\right]_{{\text{coll}}}.
\end{align}
If we sum Eq.~\eqref{energos1} over all particle types $\alpha$ (see Eqs.~\eqref{boki8}-\eqref{boki11}), we obtain
the energy equation for the whole medium,
\begin{align}\label{energos2}
\frac{\partial}{\partial t}\left(\frac{1}{2}\,\rho{V}^2+\frac{3P}{2}\right)
+\nabla_j\left(\frac{1}{2}\,\rho{V}^2{V}^j+\frac{3P}{2}{V}^j-V_iT^{ij}+q^j\right)-V_i{F}^i=0,
\end{align}
where, due to the conservation of the total energy in the closed system,  
the collision term ${M}_\alpha$ (taking into account only particle collisions while omitting the radiative rates, cf. Sect.~\ref{Massconserve}) vanishes when summed over all particle species. 
The terms $\displaystyle\frac{1}{2}\,\rho_\alpha\left\langle{V}_\alpha^2\right\rangle$ and 
$\displaystyle\frac{1}{2}\,\rho_\alpha\left\langle{C}_\alpha^2\right\rangle$ in Eq.~\eqref{energos1} 
represent the kinetic and internal (thermal) energy of the particle type $\alpha$, while the corresponding expression 
$\displaystyle\frac{1}{2}\sum_{\alpha}\rho_\alpha\left\langle{C}_{\alpha 0}^2\right\rangle$ (cf.~Eq.~\eqref{boki10}) 
expresses the internal energy $\rho\epsilon$ of the whole fluid. Following Eq.~\eqref{boki10}, we set
\begin{align}\label{energos3}
\frac{3P}{2}=\rho\epsilon,
\end{align}
where $\epsilon$ denotes the specific internal energy (internal energy per unit mass). Equation \eqref{energos2} can therefore be written as
\begin{align}\label{energy}
\frac{\partial}{\partial t}\left(\rho\epsilon+\frac{1}{2}\,\rho{V}^2\right)+\nabla_j\left[\rho{V}^j\left(\epsilon+\frac{1}{2}{V}^2\right)
-V_iT^{ij}+q^j\right]=V_i{F}^i.
\end{align}
Multiplying the equation of motion \eqref{moment} by the velocity $V_i$ and subtracting the continuity equation, we
obtain the equation of the mechanic (kinetic) energy for the whole medium,
\begin{align}\label{energos4}
\frac{\partial}{\partial t}\left(\frac{1}{2}\,\rho{V}^2\right)+\nabla_j\left(\rho{V}^j\frac{1}{2}{V}^2\right)=V_i\nabla_jT^{ij}+V_i{F}^i.
\end{align}
Subtraction of Eq.~\eqref{energos4} from the total energy equation \eqref{energy} gives the equation of internal (thermal) energy in the form
\begin{align}\label{energos5}
\frac{\partial\left(\rho\epsilon\right)}{\partial t}+\nabla_j\left(\rho\epsilon\,{V}^j\right)=T^{ij}\nabla_jV_i-\nabla_j\,q^j.
\end{align}
Subtracting equation of continuity \eqref{masscylinder001} from Eq.~\eqref{energos5} and following the identity $\text{d}/\text{d}t=\partial/\partial t+V^j\,\nabla_j$, 
we can write Eq.~\eqref{energos5} as 
\begin{align}\label{internalenergy}
\rho\frac{\text{d}\epsilon}{\text{d}t}=T^{ij}\nabla_jV_i-\nabla_j\,q^j.
\end{align}\\
\noindent Following expression~\eqref{stresik2} for the stress tensor, we may write the first term on the right-hand side of Eq.~\eqref{internalenergy} in the form
\begin{align}\label{stresstensordetail}
T^{ij}\nabla_jV_i=\left[-P\delta^{ij}+2\eta E^{ij}+\left(\zeta-\frac{2}{3}\eta\right)\nabla_kV^k\delta_{ij}\right]\nabla_jV_i.
\end{align}
Equation~\eqref{stresstensordetail} can be explicitly written in the form
\begin{align}\label{stresstensordetailek}
T^{ij}\nabla_jV_i=-P\,\nabla_iV^i+2\eta E^{ij}\nabla_jV_i+\left(\zeta-\frac{2}{3}\eta\right)\left(\nabla_iV^i\right)^2.
\end{align}
We provide the explicitly derived detailed form of this term in various coordinate systems in Sect.~\ref{enappend} where, 
using the Cauchy strain tensor formalism, Eq.~\eqref{stresstensordetailek} becomes 
\begin{align}\label{innererga}
T^{ij}\nabla_jV_i=-P\,\nabla_iV^i+2\eta E_{ij}E^{ij}+\left(\zeta-\frac{2}{3}\eta\right)\left(\nabla_iV^i\right)^2
\end{align}
(see also Eq.~\eqref{picapendour} in Appendix~\ref{enappend}).
The first term on the right-hand side is the reversible work done by the expanding matter, while the 
second and third term on the right-hand side represent the dissipation function, i.e.,~the energy of the viscous dissipation in the gas 
\citep[see, e.g.,][]{Mihalas2}. The dissipation function is conventionally denoted as $\Phi$, we call it $\Psi$ in order not to be confused 
with gravitational potential. We write
\begin{align}\label{Phi}
\Psi=2\eta E_{ij}E^{ij}+\left(\zeta-\frac{2}{3}\eta\right)\left(\vec\nabla\cdot\vec{V}\right)^2.
\end{align}
For the dissipation function always applies $\Psi\geq 0$ (see, e.g.,~\citet{Mihalas2} for explicit proof).
The equation of the internal (thermal) energy \eqref{internalenergy} in the vector notation thus becomes
\begin{align}\label{internalenergy1}
\rho\,\frac{\text{d}\epsilon}{\text{d}t}=-P\,\vec{\nabla}\cdot\vec{V}+\Psi-\vec\nabla\cdot\vec{q}.
\end{align}
The reversible work done by pressure forces (the first term on the right-hand side of Eq.~\eqref{internalenergy1}) vanishes in case of 
noncompressible fluid ($\vec{\nabla}\cdot\vec{V}=0$). The other terms are contributions to the heat energy - 
the second term on the right-hand side of Eq.~\eqref{internalenergy1} is the (already introduced) irreversible 
energy of dissipation while the third term can be regarded as 
a reversible contribution from heat conduction and from other heat energy sources (e.g., radiation, chemical reactions, etc.). It thus corresponds in general to 
the divergence of $\vec{q}$ in Eq.~\eqref{boki11}.

The energy equation can be alternatively derived from the first law of thermodynamics \citep[see, e.g.,][]{Landau6}, 
involving the continuity equation \eqref{masscylinder001} and the equation of motion \eqref{moment}.
Partial time derivative of the energy (where all the quantities are per unit volume) gives
\begin{align}\label{firstthermo}
\frac{\partial}{\partial t}\left(\rho\epsilon+\frac{\rho V^2}{2}\right)=\rho\frac{\partial\epsilon}{\partial t}-\left(\epsilon+\frac{V^2}{2}\right)
\vec\nabla\cdot\rho\vec{V}+\rho\vec{V}\frac{\partial\vec{V}}{\partial t}.
\end{align}
Noting that $\partial\epsilon=\text{d}\epsilon-\vec{V}\cdot\vec\nabla\epsilon$, the first term on the right-hand side of Eq.~\eqref{firstthermo},
using the first law of thermodynamics can be written as
\begin{align}\label{int}
\rho\,\frac{\partial\epsilon}{\partial t}=\rho T\frac{\text{d}s}{\text{d}t}-P\,\vec\nabla\cdot\vec{V}-\rho\vec{V}\cdot\vec\nabla\epsilon,
\end{align}
where $s$ is the specific entropy (entropy per unit mass). 
We can formally express the third term on the right-hand side of Eq.~\eqref{firstthermo} 
from equation of motion \eqref{moment}. Using Eq.~\eqref{moment1} and the vector calculus identity~\eqref{moment3}, 
we may write
\begin{align}\label{mech}
\rho\vec{V}\,\frac{\partial\vec{V}}{\partial t}=-\rho\vec{V}\cdot\vec\nabla\frac{1}{2}V^2-
{\rho\vec{V}\cdot\left(\vec\nabla\times\vec{V}\right)\times\vec{V}}
-\vec{V}\cdot\vec\nabla\cdot\mathcal{P}+\vec{V}\cdot\vec{F},
\end{align}
where the second term on the right-hand side vanishes 
(since the vector identity $\vec{A}\cdot\left(\vec{B}\times\vec{A}\right)=0$ for any arbitrary vectors $\vec{A}$ and $\vec{B}$) 
and where $\mathcal{P}$ in the third term on the right-hand side is the 
pressure tensor $\mathcal{P}_{ij}=-T_{ij}$. In that sense the pressure tensor represents the same physics as the stress tensor, however, with the opposite sign.
Using Eqs.~\eqref{firstthermo}, \eqref{int} and \eqref{mech}, we can write the equation of the total energy \eqref{energos2} in the form 
\begin{align}\label{firstthermo1}
\frac{\partial}{\partial t}\left(\rho\epsilon+\frac{\rho V^2}{2}\right)+\vec\nabla\cdot\left[\rho\vec{V}\left(\epsilon+\frac{V^2}{2}\right)+
P\vec{V}\right]=\vec{V}\cdot\vec\nabla\cdot\sigma_P+\rho T\frac{\text{d}s}{\text{d}t}+\vec{V}\cdot\vec{F},
\end{align}
where the quantity $\sigma_P$ denotes the nondiagonal part of the pressure tensor, i.e.,~the tensor of the viscous pressure (which represents the 
same physics as the negatively signed viscous stress tensor \eqref{viskoza}).
\noindent Equation \eqref{int} can be written in more compact form,
\begin{align}\label{cunce}
\rho\,\frac{\text{d}\epsilon}{\text{d}t}=\rho T\frac{\text{d}s}{\text{d}t}-P\,\vec\nabla\cdot\vec{V},
\end{align} 
and by comparing Eqs.~\eqref{cunce} and \eqref{internalenergy} we directly obtain
\begin{align}
\rho T\frac{\text{d}s}{\text{d}t}=\Psi-\vec\nabla\cdot\vec{q}.
\end{align}
From Eq.~\eqref{Phi} follows the relation $\Psi=\sigma_P\cdot\vec\nabla\vec{V}$ (cf.~Eqs.~\eqref{stresstensordetail}-\eqref{innererga}
where the diagonal scalar pressure term is omitted). 
Including this into Eq.~\eqref{firstthermo}, we obtain the expression
\begin{align}\label{firstthermo2}
\frac{\partial}{\partial t}\left(\rho\epsilon+\frac{\rho V^2}{2}\right)+\vec\nabla\cdot\left[\rho\vec{V}\left(\epsilon+\frac{V^2}{2}\right)+
\mathcal{P}\vec{V}+\vec{q}\,\right]=\vec{V}\cdot\vec{F},
\end{align}
which is clearly identical to the total energy expression given in Eq.~\eqref{energy}.

Another frequently used expression of the energy equation involves the Fourier's law of heat conduction, 
which states that the heat flux in any material is proportional to its internal temperature gradient  
and that the heat flows from hotter to cooler regions \citep{Mihalas2}. The term $-\,\vec\nabla\cdot\vec{q}$ for the heat energy flux
can be expanded as 
\begin{align}-\,\vec\nabla\cdot\vec{q}=\vec\nabla\cdot\left(K\,\vec\nabla{T}\right)+q_R,\end{align}
where the constant of proportionality $K$ is the material conductivity and 
the term $q_R$ refers to the heat sources other than conduction \citep{Hirsch}, i.e.,~to radiation, chemical reactions, etc.
The structure of the term $q_R$
can be quite complex and it depends on the detailed physics of internal and external heat sources.

\section{Equation of state}
\label{statak}
The system of hydrodynamic equations has to be closed with the equation of state that refers to the (thermodynamic) properties of the state 
variables in the fluid. Following the notation introduced in Sect.~\ref{energyconserve}, we can write the first law of thermodynamics (expressed per unit mass) in the form
\citep[e.g.,][]{Landau6}
\begin{align}\label{statix1}
\text{d}q=\text{d}\epsilon-\frac{P}{\rho^2}\,\text{d}\rho,
\end{align}
which, in case of adiabatic conditions, is obviously equal to zero. The adiabatic transformation of a perfect (ideal) gas is given by the relation $P/\rho^\gamma=\text{const.}$, 
where the adiabatic exponent $\gamma$ is defined as the ratio of specific heats at constant pressure $P$ and constant volume $V$, $\gamma=c_P/c_V=
(\text{d}\,\text{ln}\,P/\text{d}\,\text{ln}\,\rho)_{\text{ad}}$. Since the specific heats $c_P,\,c_V$ are defined as
$c_P=(\text{d}q/\text{d}T)_{P}$ and $c_V=(\text{d}q/\text{d}T)_{V}$, Eq.~\eqref{statix1} gives
\begin{align}\label{statix1a}
\text{d}h=c_P\,\text{d}T,\quad\text{where}\quad h=\epsilon+P/\rho\,\,(\text{enthalpy})\quad\quad\quad\text{and}\quad\quad\quad\text{d}\epsilon=c_V\,\text{d}T.
\end{align}
For adiabatic
transformations in a fluid described by general equation of state (in case of nonideal gas) is the basic relation similar, $P/\rho^{\Gamma_1}=\text{const.}$, where 
$\Gamma_1$ is the general adiabatic exponent (see, e.g., \citet{maeder} for details).

The law of ideal gas relates the pressure $P$, volume $V$ and temperature $T$ by the equation $PV/T=\text{const}$.
Assuming $V$ is the volume occupied by one particle, $V=\mu\,m_u/\rho$, where the quantity $\mu$ is the mean molecular weight (see Eq.~\eqref{statix3})
and $m_u$ is the atomic mass unit (i.e., 1/12 of the mass of the neutral $^{12}\text{C}$ atom), this relation becomes
\begin{align}\label{statix2}
P=\frac{k}{\mu\,m_u}\,\rho T, 
\end{align}
where $k$ is the Boltzmann constant and the term $(\mu\,m_u)$ is the average mass of the particles (electrons, ions, atoms or molecules) in the gas. 
The mean molecular weight $\mu$ is defined in a medium with various elements $j$ \citep[see, e.g.,][]{maeder} as 
\begin{align}\label{statix3}
\frac{1}{\mu}=\sum_j\frac{X_j}{A_j}\left(1+n_{\text{e},j}\right), 
\end{align}
where $X_j$ is the mass fraction of element $j$ with atomic mass $A_j$ and $n_{\text{e},j}$ is the number of free electrons per 1 atom (ion) of element $j$.
The number density (concentration) $n_\alpha$ of particles of the type $\alpha$, i.e.,~the number of particles $\alpha$ per unit volume is $\rho_\alpha/(\mu_\alpha\,m_u)$
where $\mu_\alpha$ is the mean molecular weight.
Equation \eqref{statix2} in this case gives $P_\alpha=n_\alpha kT$ and consequently $P=n\,kT$ when summed over all the particle types.

Integrating the equation for internal energy per unit mass (the last expression in Eq.~\eqref{statix1a}), we obtain $\epsilon=c_V\,T$.
For a perfect gas one has $c_P-c_V=\mathcal{R}=k/(\mu m_u)$, where $\mathcal{R}$ is the so-called specific gas constant.
The state equation \eqref{statix2} thus becomes
\begin{align}\label{statix4}
P=\left(\gamma-1\right)\,\rho\epsilon. 
\end{align}
This basically corresponds to Eq.~\eqref{energos3}, where we implicitly assume the fluid (plasma) of particles without 
rotational or vibrational degrees of freedom (atoms, ions, etc.),
for which the adiabatic exponent $\gamma=5/3$.

We can also relate the pressure, density and temperature via the speed of sound $a$ that is in general given by 
$a^2=\partial P/\partial\rho$.
In adiabatic case we obtain the relation between pressure and density and the relation between adiabatic speed of sound and temperature,
respectively, in the following form \citep[e.g.,][]{Landau6}
\begin{align}\label{statix5}
P=\frac{a^2_{\text{ad}}\,\rho}{\gamma},\quad\quad\quad a^2_{\text{ad}}=\frac{\gamma}{\mu\,m_u}\,kT. 
\end{align}
In isothermal case the index of polytrope is $n=1$ (in adiabatic case $n=\gamma$). Following Eqs.~\eqref{statix2} and \eqref{statix5} (where in this case $T=\text{const.}$),
we obtain the same relations in the form
\begin{align}\label{statix6}
P=a^2\rho,\quad\quad\quad a^2=\frac{kT}{\mu\,m_u}. 
\end{align}
The speed of sound plays a key role in gas dynamics, the flow characteristics are basically determined by whether the flow velocity is subsonic 
or supersonic \citep[see, e.g.,][]{zelinar}. In astrophysics is the temperature of the gas very often determined by thermal balance between heat sources and radiative cooling,
the relaxation timescale is in this case quite short compared to the timescale of the sound wave travel \citep[e.g.,][]{supak}. Hence, the isothermal speed of sound may be applied 
if the cooling time is much shorter than the propagation speed of the sound waves. It is therefore convenient to use the isothermal sound speed in situations 
where the temperature of the gas (plasma) is independent of the density, i.e.,~if the gas temperature is determined by external processes 
(e.g.,~by irradiation of external sources, etc.).

\HlavickaKapitoly
{\chapter{Hydrodynamic structure of circumstellar disks}
\label{kombhgu}
\section{Basic equations of radial disk structure}
\label{eqraddisk} 
\noindent In this chapter we derive the basic equations of viscous disk (see Sect.~\ref{viskodisks} for its general description).
The natural geometrical reference frame is the cylindrical coordinate system ($R,\,\phi,\,z$), where the radial coordinate 
we denote as $R$ (unlike the spherical coordinate frame, where the coordinate axes are denoted $r,\,\theta,\,\phi$, see also 
Appendix \ref{appendix1}).
The mathematical description introduced in this chapter follows in main points the principles given in \citet{Pringle,Frank}.
It provides formally rather alternative approach for the derivation of the disk physics, compared to the basic hydrodynamic formalism provided
in Chap.~\eqref{hydrobase}. It is however highly instructive for understanding some key issues, referred in following chapters.
      
We introduce the vertically integrated density (designated alternatively as disk column or surface density) of the disk, defined as
\begin{align}\label{sigmik}
\Sigma=\int\limits_{-\infty}^{\infty}\rho\,\text{d}z\,.                                                       
\end{align}
The mass $\Delta M$ of radial disk segment lying between fixed radii $R$ and $R+\Delta R$ where $\Delta R\ll R$, with vertically integrated density $\Sigma$, is
\begin{align}
\Delta M=2\pi R\,\Delta R\Sigma.                                                      
\end{align} 
We can express the rate of change of the mass of the radial disk segment (radial mass flux) in the form
\begin{align}\label{massflux}
\frac{\partial}{\partial t}\left(2\pi R\,\Delta R\Sigma\right)&=V_R(R,t)\,2\pi R\,\Sigma(R,t)\nonumber\\
&-V_R(R+\Delta R,t)\,2\pi\left(R+\Delta R\right)\,\Sigma(R+\Delta R,t).                                                       
\end{align}
Expanding the right-hand side of Eq.~\eqref{massflux} to first order in $R$ (in the limit $\Delta R\rightarrow 0$), we obtain
\begin{align}
\frac{\partial}{\partial t}\left(2\pi R\,\Delta R\Sigma\right)=-2\pi\Delta R\frac{\partial}{\partial R}\left(R\Sigma V_R\right),                                         
\end{align}
which consequently takes the form of the radial part of mass conservation equation in cylindrical 
coordinates (see Eq.~\eqref{masscylinder1}),
\begin{align}\label{massfluxtot}
R\frac{\partial\Sigma}{\partial t}+\frac{\partial}{\partial R}\left(R\Sigma V_R\right)=0\,.                                         
\end{align}

Analogously we obtain the angular momentum conservation equation for the same radial disk segment: 
the angular momentum $\Delta J$ contained between the fixed radii $R$ and $R+\Delta R$ where $\Delta R\ll R$, with the vertically integrated density $\Sigma$, is
\begin{align}
\Delta J=2\pi R\,\Delta R\Sigma R^2\Omega,                                                      
\end{align} 
where $\Omega(R,t)$ is the angular velocity of the segment.
The expression of the rate of change of angular momentum $\text{d}(\Delta J)/\text{d}t$ 
of the radial disk segment (radial angular momentum flux) is similar to Eq.~\eqref{massflux}.
We must however include the transport due to the net effects of the viscous torque $\mathcal{G}(R,t)$ \citep{Pringle,Frank}, 
which represent the difference between viscous torque
exerted on radii $R$ and $R+\Delta R$,

\begin{align}\label{angularmomentumflux}
\frac{\partial}{\partial t}\left(2\pi R\,\Delta R\Sigma R^2\Omega\right)&=V_R(R,t)\,2\pi R\,\Sigma(R,t)\,R^2\Omega(R,t)\nonumber\\
&-V_R(R+\Delta R,t)\,2\pi\left(R+\Delta R\right)\,\Sigma(R+\Delta R,t)
\left(R+\Delta R\right)^2\,\Omega(R+\Delta R,t)+\frac{\partial\mathcal{G}}{\partial R}\Delta R.
\end{align}
Expansion of the right-hand side of Eq.~\eqref{angularmomentumflux} to first order in $R$ in the limit $\Delta R\rightarrow 0$ gives
\begin{align}\label{angularmomentumfluxa}
\frac{\partial}{\partial t}\left(2\pi R\,\Delta R\Sigma R^2\Omega\right)
=-2\pi\Delta R\frac{\partial}{\partial R}\left(R\Sigma V_RR^2\Omega\right)+\frac{\partial\mathcal{G}}{\partial R}\Delta R,                                                      
\end{align}
which consequently takes the form of the radial part of the angular momentum conservation equation in cylindrical 
coordinates (see Eq.~\eqref{masscylinder1}),
\begin{align}\label{angularmomentumflux1}
R\frac{\partial}{\partial t}\left(\Sigma R^2\Omega\right)+\frac{\partial}{\partial R}\left(R\Sigma V_RR^2\Omega\right)
=\frac{1}{2\pi}\frac{\partial\mathcal{G}}{\partial R}.                                                       
\end{align} 

We neglect the particle (atomic, molecular, etc.) viscosity (where $\ell$ and $\tilde{\varv}$ in 
Eq.~\eqref{stresswritten} are the mean free path and mean
thermal velocity of the particles), which in fact is not able to 
produce the large scale viscous effects such as angular momentum transport or viscous dissipation
that may influence the hydrodynamics of the disk \citep[e.g.,][]{Shakura,Pringle,Frank}.
We involve the macroscopic (turbulent) viscosity where the fluid ``particles'' that produce the viscous effects
are considered to be the shearing or interpenetrating parcels of the fluid (gas eddies).
In this case the mean free path $\ell$ in Eq.~\eqref{stresswritten} is a characteristic spatial scale of turbulent
motion and $\tilde{\varv}$ is typical average velocity of the turbulent cells (eddies) of the gas. 

We assume that the turbulent motion of the gas eddies is collisionless (see Fig.~\ref{frankiecylinder}) within a distance $\ell$, between the planes $z=0$ and $z=H$ of
cylindrical polar coordinate system ($R,\phi,z$).
The fluid blobs, 
denoted as A and B, cross the radius $R=\text{const.}$ and exchange their angular momenta within the region $R-\ell/2,\,R+\ell/2$.
\begin{figure}[t]
\begin{center}
\includegraphics[width=9.6cm]{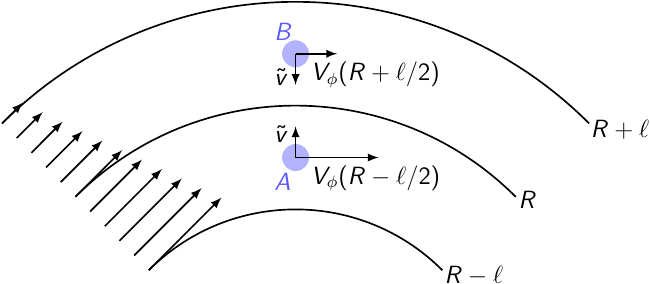}
\caption{Schema of the cylindrical shear stress with radially decreasing magnitude of the velocity $V_\phi$ of cylindrical segments. 
The reference radius is denoted as $R$, while $\ell$ is the mean free path of 
fluid parcels (which are considered as a subject of the viscous stress) 
and $\tilde{\varv}$ is the magnitude of $R$-component of a mean velocity of random motion 
of the fluid parcels (gas eddies). The fluid blobs, denoted A and B, can cross the radial line $R$ on the distance $\ell$ and 
contribute to the exchange of angular momentum within the region $R-\ell/2,\,R+\ell/2$, where $\ell\ll R$. The depicted gas flow
is considered to lie between vertical planes $z=z_0$ and $z=z_0+\ell$ (cf.~Fig.~\ref{planestrih}).
The viscous torque exerted by the outer disk ring on the inner ring at the cylindrical interface $R+\ell$ (since $\ell\ll R$, we may 
write $R+\text{d}R$) is $\mathcal{G}(R+\text{d}R)$ and the viscous torque exerted at the boundary $R$ is $\mathcal{G}(R)$.
Adapted from \citet{Frank}.}
\label{frankiecylinder}
\end{center}
\end{figure}
The mass crosses the surface $R=\text{const.}$ at equal rates in both directions, of the
order $H\rho\tilde{\varv}$ per unit arc length, where $\rho(R)$ is the mass density. 
We can thus express the net outward component of the angular momentum flux per unit arc length 
(where we account for the exchanged fluid parcels, i.e., the parcel with higher rotation velocity $V_{\phi}(R-\ell/2)$ is displaced along the mean free path $\ell$ to the 
new position $R+\ell/2$ and vice versa)
in the form 
\begin{align}
\rho\tilde{\varv}\,H(R+\ell/2)\,V_{\phi}(R-\ell/2)\,-\,\rho\tilde{\varv}\,H(R-
\ell/2)\,V_{\phi}(R+\ell/2).
\label{angus}
\end{align}

Since the typical size of 
largest turbulent eddies as well as the length of their mean free path cannot exceed the disk vertical thickness \citep{Pringle,Frank}, 
we regard the length scale $H$ as the disk vertical scale height. 
Stationary expansion of the expression given in Eq.~\eqref{angus} to first order in $R$ (assuming $\ell\ll R$) gives
\begin{align}
-\rho\tilde{\varv}\,\ell H R^2\frac{\partial\,\Omega(R)}{\partial R}.
\label{anousek}
\end{align}
The net outward angular momentum flux \eqref{anousek} defines the torque exerted by the inner ring on the outer ring.
Setting $\rho H=\Sigma$ and $\tilde{\varv}\ell=\nu$ (kinematic viscosity, see Eq.~\eqref{kincl}),
we obtain the expression for the viscous torque (per circumference of the disk ring) exerted by the outer ring on the inner ring,
\begin{align}\label{viscoustorque}
\mathcal{G}(R)=2\pi R\nu\Sigma R^2\frac{\partial\,\Omega(R)}{\partial R}.
\end{align}

We can explicitly expand the net effect of the viscous torque acting on the inner and outer boundary of a radial disk segment (i.e.,~between $R$ and $R+\text{d}R$) 
as
\begin{align}\label{this}
\mathcal{G}(R+\text{d}R)-\mathcal{G}(R)=\frac{\partial\,\mathcal{G}(R)}{\partial R}\,\text{d}R.
\end{align}
Multiplying \eqref{this} by angular velocity gives the rate of work of the viscous torque (i.e.,~the power per disk ring) that can be expanded 
into the form
\begin{align}
\Omega\,\frac{\partial\,\mathcal{G}}{\partial R}\,\text{d}R=\left[\frac{\partial}{\partial R}(\mathcal{G}\Omega)-
\mathcal{G}\,\frac{\partial\Omega}{\partial R}\right]\,\text{d}R.
\label{torqpower}
\end{align}
The first term on the right-hand side of Eq.~\eqref{torqpower}, 
\begin{align}
\frac{\partial}{\partial R}(\mathcal{G}\Omega)\,\text{d}R,  
\end{align}
represents the rate of ``radial convection'' of rotational energy through the gas induced by torques \citep{Frank}, while the 
second term on the right-hand side of Eq.~\eqref{torqpower},
\begin{align}
-\,\mathcal{G}\,\frac{\partial\Omega}{\partial R}\,\text{d}R,
\end{align}
gives the rate of loss of mechanical energy within one disk segment of radial width $\text{d}R$ (per circumference of the disk ring)
that has to be transformed into heat energy, i.e.,~the (negatively signed) rate of production of viscous dissipation energy.
Adopting the notation of \citet{Lee},
we express this dissipation energy production rate in terms of viscous heat (or thermal) flux $F_{\text{vis}}(R)$ 
(since the physical dimension of both the quantities is $\text{Wm}^{-2}$) per unit area of the disk surface (taking into account two sides 
of the disk) as 
\begin{align}\label{viscounitpower1}
F_{\text{vis}}(R)=\frac{1}{4\pi R}\,\mathcal{G}\,\frac{\partial\Omega}{\partial R}.
\end{align}
Substituting \eqref{viscoustorque} into \eqref{viscounitpower1} we obtain the explicit expression 
for the dissipation energy production rate (viscous heat flux) $F_{\text{vis}}(R)$ in the form
\begin{align}\label{viscounitpower1a}
F_{\text{vis}}(R)=\frac{1}{2}\nu\Sigma\left(R\frac{\partial\Omega}{\partial R}\right)^2=
\frac{1}{2}\nu\Sigma\left[R\frac{\partial}{\partial R}\left(\frac{V_\phi}{R}\right)\right]^2.
\end{align}
If we consider the rotational velocity of the gas as exactly Keplerian, i.e.,~$V_\phi=V_K$, 
the viscous heat flux $F_{\text{vis}}(R)$ per unit area of the (one-sided) disk surface is
\begin{align}\label{viscounitpower2}
F_{\text{vis}}(R)=\frac{9}{8}\,\nu\Sigma\,\frac{GM}{R^3}.
\end{align} 

The turbulences remain however the major uncertainty within the fluid dynamics and we cannot exactly quantify their effect
on the length scale $\ell$, on the mean turbulent velocity $\tilde{\varv}$, etc.
For this reason we relate the kinematic viscosity $\nu$ to the sound speed using the dimensionless $\alpha-$prescription 
(cf. Sect.~\ref{viskodisks}), first 
introduced in the fundamental paper \citet{Shakura}. This prescription parameterizes the ratio of the turbulent and sound velocities, 
$\alpha=\tilde{\varv}/a$. Using relations \eqref{kincl} and regarding
the mean free path of turbulent gas eddies $\ell$ as the disk scale height $H$ (cf.~Eq.~\eqref{angus}), we write
\begin{align}\label{Shakurakos}
\nu=\alpha aH,
\end{align}
where we assume $\alpha<1$ \citep[e.g.,][among others]{Shakura,Penna}.
In this point it is important to note that Eq.~\eqref{Shakurakos} is merely a parameterization: all uncertainties 
about the viscosity mechanism have been incorporated rather than in kinematic viscosity $\nu$ into this dimensionless number
$\alpha<1$. There is however no relevant reason
to expect that the $\alpha$-parameter is constant throughout the whole disk structure 
\citep[despite many authors use this assumption for the calculations of the disk structure close to the star, see, e.g.,][for details]{Penna}; 
even the assumption $\alpha<1$
may be violated in small regions of the disk where some physical input (e.g.,~the stellar pulsations, see Sect.~\ref{gasejcl})
continually feeds the supersonic turbulences \citep{Frank}. The $\alpha$-prescription however enables a semi-empirical approach to
the viscosity problem. Many authors \citep[][etc., see also Sect.~\ref{viskodisks}]{Penna,Karfiol12} have tried to estimate the value of 
$\alpha$-parameter by a comparison of theory and observations. 
Moreover, the $\alpha$-prescription can be identically formulated for viscosity that is induced by winding-up of 
chaotic instabilities in the disk magnetic fields \citep[e.g.,][]{Balbus,Krticka_2014}; the estimate
$\alpha<1$ seems to be valid with high probability also in this case (see Sects.~\ref{largemodelix} and \ref{magnetousci} below).

\section{Vertical disk structure}
\label{vertikalekdisk} 
Assuming the vertical hydrostatic equilibrium, 
\begin{align}\label{hydroequix}
\frac{\text{d}P}{\text{d}z}=\rho g_z, 
\end{align}
and the vertically constant speed of sound (see Sect.~\ref{statak}) in vertically isothermal disk
(the vertical component $g_z$ of gravitational
acceleration is given in Eq.~\eqref{gravcylexpli}),
we obtain the vertical density profile in the form
\begin{align} \label{fullvertigo}
\rho(R,z)=\rho(R,0)\,\text{e}^{{\textstyle{-\frac{GM_{\star}}{a^2R}+\frac{GM_{\star}}{a^2\sqrt{R^2+z^2}}}}},
\end{align}
where $M_{\!\star}$ is mass of the central star and $\rho(R,0)$ is the density in the equatorial plane (disk midplane), where $z=0$.
Equation~\eqref{rotframe2} in Appendix \ref{appendix1} gives for the centripetal acceleration ${a}_{\text{c}}$ of a particle that is at rest in a steadily rotating frame 
(where $\vec{V}^\prime=0$ and $\text{d}\vec{\Omega}/\text{d}t=0$), the simple relation
${a}_{\text{c}}=-\Omega^2{R}$. 
Regarding ${a}_{\text{c}}$ being induced by external gravitational acceleration, 
${a}_{\text{c}}=-\partial{\Phi}/\partial R$, where $\Phi(R,z$) is the gravitational potential
of the central star (see also Appendix~\ref{momappend}),
\begin{align} \label{cylindricalgravpot}
\Phi(R,z)=-\frac{GM_{\star}}{(R^2+z^2)^{1/2}},
\end{align}
we obtain the general relation $\Omega^2=R^{-1}\partial{\Phi}/\partial R$,
which applies for any gravitational potential $\Phi$ \citep[e.g.,][]{binney}.
In the cylindrical reference frame takes the 
radial derivative of the gravitational potential induced by the central star 
(since the gravitational potential of the outflowing disk, i.e.,~the self-gravity, can be neglected,
cf.~subroutine \textit{selfgrav} in Appendix~\ref{dvojdimhydrous}) 
the form of the radial component of gravitational acceleration $g_R$ (Eq.~\eqref{gravcylexpli}).
The squared angular velocity therefore is 
\begin{align} \label{unphysical}
\Omega^2(R,z)=\frac{GM_{\star}}{(R^2+z^2)^{3/2}}.
\end{align}
In the very proximity of the central star such disk kinematics is however clearly unphysical
outside the regions where $|z|\ll R$. Equation~\eqref{unphysical}, expressing $\Omega(R,z)$,
together with the relations \eqref{gravcylexpli} for radial and vertical cylindrical 
components of gravitational acceleration need to be involved in two-dimensional numerical schema (see, e.g.,~Sect.~\ref{hydrostatous})
as the initial function for the angular (or azimuthal) velocity.

If we assume that the sound speed varies along the $z$-direction, $a=a(R,z)$, and following the condition of the physical relevance
in the inner disk region, $|z|\ll R$, we
expand Eq.~\eqref{fullvertigo} to first order in $z/R$. We obtain 
\begin{align} \label{slightvertigo}
\rho(R,z)=\rho(R,0)\,\text{e}^{{\textstyle{-\frac{GM_{\star}}{a^2(R,z)\,R^3}\frac{z^2}{2}}}}=
\rho(R,0)\,\text{e}^{{\textstyle{-\frac{\Omega^2(R,0)}{a^2(R,z)}\frac{z^2}{2}}}}.
\end{align}
From Eq.~\eqref{slightvertigo} clearly follows that the vertical density profile 
approximately exhibits the Gaussian curve scaled by the dispersion $a^2/\Omega^2$.
We may therefore regard the standard deviation $a/\Omega$ 
as the vertical scale height $H$. Using this scaling identity, Eq.~\eqref{Shakurakos} takes the form
\begin{align}\label{shakuravelophi}
\nu=\frac{\alpha a^2(R,z)}{\Omega(R,0)}=\frac{\alpha a^2(R,z)\,R}{V_\phi(R,0)},
\end{align}
which thus applies everywhere in the disk, assuming the vertical hydrostatic equilibrium condition.
This leads to the vertical scaling of the thin disk. Since we require that the local Kepler velocity 
should be highly supersonic ($V_\phi(R,0)\gg a(R,0)$) and the disk scale height (cf. Eq.~\eqref{slightvertigo})
\begin{align}\label{kepscaleheight}
H(R)=\frac{a(R,0)}{\Omega(R,0)}=\frac{a(R,0)R}{V_\phi(R,0)}, 
\end{align}
we may write $a(R,0)/V_\phi(R,0)=H(R)/R\ll 1$ within the inner part of the disk. The ``thin disk'' assumption then implies the relation $H\ll R$ \citep{Frank}.

Following Eq.~\eqref{shakuravelophi}, we may approximate the vertical disk structure in hydrostatic and thermal equilibrium
\citep[cf.][]{Lee}, which may be valid in optically thick equatorial disk area that is not too far from the central star, in the following way: 
Eq.~\eqref{viscounitpower1a}, using Eqs.~\eqref{sigmik} and \eqref{shakuravelophi}, can be written as
\begin{align}\label{kocour} 
F_{\text{vis}}(R,z)=\frac{1}{2}\frac{\alpha a^2(R,z)}{\Omega(R,0)}\left[R\frac{\partial\Omega(R,0)}{\partial R}\right]^2
\int\limits_{0}^{z}\rho(R,z^\prime)\,\text{d}z^\prime,
\end{align}
where we integrate from $0$ to $z$ (while integrating from $-\infty$ to $\infty$ we obtain $F_{\text{vis}}(R)$)
and where the condition for the viscous heat vertical flux $F_{\text{vis}}(R,0)=0$ must obviously apply. Neglecting vertical 
variations of angular velocity (which is reasonable for $z\ll R$) and considering only vertical variations of 
pressure, we write $P(R,z)=a^2(R,z)\,\rho(R,z)$. From Eq.~\eqref{kocour} we thus obtain the simple expression 
for vertical slope of the viscous heat flux in the form
\begin{align}\label{kocourek} 
\frac{\text{d}F_{\text{vis}}(R,z)}{\text{d}z}=\frac{1}{2}\frac{\alpha P(R,z)}{\Omega(R,0)}\left[R\frac{\partial\Omega(R,0)}{\partial R}\right]^2.
\end{align}
From the vertical hydrostatic equilibrium condition (Eq.~\eqref{hydroequix}, see also Eq.~\eqref{gravcylexpli}) comes the identity for the pressure vertical dependence,
\begin{equation}\label{presour}
P(R,z)=-\int\limits_{0}^{z}\rho(R,z^\prime)\frac{GM_{\star}z'}{(R^2+{z^\prime}^2)^{3/2}}\,\text{d}z^\prime.
\end{equation}
Involving \eqref{unphysical} in \eqref{presour}, we may write the equation of the vertical pressure gradient \eqref{hydroequix} in the simple form,
\begin{equation}\label{presgradour}
\frac{\text{d}P(R,z)}{\text{d}z}=-\rho\left(R,z\right)\,\Omega^2(R,z)\,z.
\end{equation}
The vertical gradient of the disk optical depth $\tau$ we may express as 
\begin{equation}\label{presgradour2}
\frac{\text{d}\tau(R,z)}{\text{d}z}=-\kappa\rho\left(R,z\right),
\end{equation}
where $\kappa$ is the opacity of the disk material \citep[e.g.,][]{Mihalas1}. Assuming local thermodynamic equilibrium (LTE) 
in optically thick regions (cf.~Sect.~\ref{templajznik}) and neglecting the external radiative flux $\mathcal{F}_{\text{irr}}$ 
(see Sect.~\ref{smakec}), the vertical gradient of the temperature \citep[cf.][]{Lee} 
caused by the dissipative heating is
\begin{equation}\label{radacek} 
\frac{\text{d}T(R,z)}{\text{d}z}=-\frac{3\,F_{\text{vis}}(R,z)}{16\sigma T^3(R,z)}\frac{\text{d}\tau(R,z)}{\text{d}z},
\end{equation}
if the condition $\nabla_{\text{rad}}<\nabla_{\text{ad}}$ \citep{Lee,maeder} is fulfilled ($\sigma$ is 
the Stefan-Boltzmann constant). The physical meaning of these gradients is defined in the following way:
$\nabla_{\text{rad}}$ (radiative gradient) is the thermal gradient which is necessary for carrying the viscous heat flux $F_{\text{vis}}$
only by radiation and it is therefore directly given by $\text{d}T/\text{d}z$ (Eq.~\eqref{radacek}), 
while the adiabatic gradient $\nabla_{\text{ad}}$ is $(\partial\,\text{ln}\,T/\partial\,\text{ln}\,P)_{\text{ad}}$
and using Eq.~\eqref{statix1} it is $(\gamma-1)/\gamma$
for an ideal gas. Since the adiabatic gradient for monoatomic perfect gas $\nabla_{\text{ad}}=2/5$, it is quite unlikely that the convection develops in the optically 
thick disk region (see Sect.~\ref{templajznik} for details of the models).

We can however analytically study the connection between the vertical disk density and thermal structure by
assuming the vertical thermal dependence of the sound speed.
Following the considerations of \citet{Karfiol08}, we 
integrate the left-hand side of the vertical hydrostatic equilibrium equation \eqref{hydroequix} in the form
\begin{align}\label{verttempisoundbroadix}
\frac{1}{\rho a^2}\frac{\partial (\rho a^2)}{\partial z}=\frac{1}{a^2}\frac{\partial}{\partial z}\left[\frac{GM_{\star}}{(R^2+z^2)^{1/2}}\right],
\end{align}
in the limits from the disk midplane $(R,0)$ to some arbitrary disk vertical distance $(R,z)$. We 
thus obtain the logarithm of the ratio of these quantities in this vertical distance and in the disk midplane,
\begin{align}\label{Trholeftix}
\int\limits_{0}^z\frac{1}{\rho a^2}\frac{\partial (\rho a^2)}{\partial z^\prime}\,\text{d}z^\prime=
\text{ln}\left[\frac{a^2(R,z)\,\rho(R,z)}{a^2(R,0)\,\rho(R,0)}\right]=\text{ln}\left[\frac{T(R,z)\,\rho(R,z)}{T(R,0)\,\rho(R,0)}\right],
\end{align}
where all the constants, given in Eq.~\eqref{statix6}, cancel.
Integration of the right-hand side of Eq.~\eqref{verttempisoundbroadix}, assuming the constraint $z\ll R$, 
can be written within the limits that range between the disk midplane
and the 
% squared 
arbitrary vertical distance $z$ as
% \begin{align}\label{Trhorightix}
% \int\limits_{0}^{z^2}\frac{1}{a^2}\frac{\partial}{\partial z^2}\left[\frac{GM_{\star}}{(R^2+z^2)^{1/2}}\right]\text{d}z^2=
% -\frac{1}{2}\int\limits_{0}^{z^2}\frac{GM_{\star}}{a^2\left(R^2+z^2\right)^{3/2}}\,\text{d}z^2.
% \end{align}
\begin{align}\label{Trhorightix}
\int\limits_{0}^{z}\frac{1}{a^2(R,z^\prime)}\frac{\partial}{\partial z^\prime}\left[\frac{GM_{\star}}{(R^2+z^{\prime\,2})^{1/2}}\right]\text{d}z^\prime=
-\int\limits_{0}^{z}\frac{GM_{\star}z^\prime}{a^2(R,z^\prime)\,R^3}\,\text{d}z^\prime=-\int\limits_{0}^{z}\frac{\Omega^2(R,0)}{a^2(R,z^\prime)}\,z^\prime\,\text{d}z^\prime.
\end{align}
Expanding the fraction in the latter expression by a squared sound speed in the disk equatorial plane $a^2(R,0)$, we can write Eq.~\eqref{Trhorightix} in terms of the disk 
scale height $H$ (see Eq.~\eqref{kepscaleheight}),
\begin{align}\label{Trhoscalix}
=-\frac{1}{H^2(R)}\int\limits_{0}^{z}\frac{a^2(R,0)}{a^2(R,z^\prime)}\,z^\prime\text{d}z^\prime=
-\frac{1}{H^2(R)}\int\limits_{0}^{z}\frac{T(R,0)}{T(R,z^\prime)}\,z^\prime\text{d}z^\prime.
\end{align}
Combining Eqs.~\eqref{Trholeftix} and \eqref{Trhoscalix}, we can write the expression for the density structure 
in case of non-isothermal speed of sound in the form
\begin{align}\label{Trhoscalefarix}
\rho(R,z)=\rho(R,0)\frac{T(R,0)}{T(R,z)}\,\text{exp}\,\Bigg[-\frac{1}{H^2(R)}\int\limits_{0}^{z}\frac{T(R,0)}{T(R,z^\prime)}\,z^\prime\text{d}z^\prime\Bigg].
\end{align}
From Eq.~\eqref{Trhoscalefarix}, with use of Eq.~\eqref{sigmik}, we obtain the expression for the integrated surface density $\Sigma(R)$ and consequently
the expression for the disk midplane density $\rho(R,0)$.
Substituting these quantities back into Eq.~\eqref{Trhoscalefarix}, we get the relation for the density $\rho(R,z)$, expressed 
in terms of the surface density $\Sigma(R)$ and of the vertical temperature structure.
To solve these equations self-consistently, we must calculate the detailed temperature structure as a function of position together with 
the density structure, given by hydrodynamic equations. Since such calculations are extremely difficult 
\citep[cf.~e.g.,][]{Karfiol08} and computationally prohibitively costly, we describe below another approximations 
that provide the feasibility of the calculations while not losing the desired physical relevance (see Sects.~\ref{timemod} and \ref{teplous}).

\section{Stationary thin disk approximation}\label{thindisk}

From the stationary ($\partial/\partial t=0$), vertically integrated ($\partial/\partial z=0$), axisymmetric ($\partial/\partial\phi=0$)
equation of continuity \eqref{massfluxtot}, written in the form
\begin{align}
R\Sigma V_R=\text{const.},
\end{align}
where we assume the radial velocity component in outflowing disks $V_R>0$, we obtain in axisymmetric (cylindrical) case the mass conservation 
equation in the form
\begin{align} \label{angularmomentumfluxix0}   
\dot M=2\pi R\Sigma V_R=\text{const.}
\end{align}
Integrating the stationary part of Eq.~\eqref{angularmomentumflux1}, we obtain
\begin{align}\label{stationaryangmomix}
2\pi R\Sigma V_R R^2\Omega=\mathcal{G}+C,
\end{align}
where the constant of integration $C$ can be expressed, e.g.,~by adopting the (often used) concept of an outer disk radius $R_{\text{out}}$
\citep[][etc.]{Pringle,Frank,Lee} 
where we can set the constraint (outer boundary condition)
\begin{align}\label{korix}
\mathcal{G}\left(R_{\text{out}}\right)\rightarrow 0.
\end{align}
The constant of integration $C$ in this case becomes
\begin{align}\label{korilix}
C=2\pi\left(R^3\Sigma V_R\,\Omega\right)_{\text{out}}=\dot{M}R^2_{\text{out}}\,\Omega\!\left(R_{\text{out}}\right),        
\end{align}
where, according to Eq.~\eqref{angularmomentumfluxix0}, we denote 
the outer boundary disk mass loss rate $\dot{M}\left(R_{\text{out}}\right)$ simply as $\dot{M}$.
Substituting Eq.~\eqref{korilix} together with the expression for the viscous torque from Eq.~\eqref{viscoustorque} into
equation for the stationary angular momentum \eqref{stationaryangmomix}, we obtain the relation
\begin{align}\label{nusigmix1}
\nu\Sigma(R)=\frac{\dot M}{2\pi R}\,\frac{\text{d}R}{\text{d}\Omega}\left[\Omega(R)-
\frac{R^2_{\text{out}}\,\Omega\!\left(R_{\text{out}}\right)}{R^2}\right].         
\end{align}
By substituting \eqref{nusigmix1} into Eq.~\eqref{viscounitpower1a}
of the viscous heat flux $F_{\text{vis}}(R)$ per unit area of the disk, we get
\begin{align}\label{viscounitpowerix3}
F_{\text{vis}}(R)=\frac{\dot M}{4\pi}R\,\frac{\text{d}\Omega(R)}{\text{d}R}\left[\Omega(R)-
\frac{R^2_{\text{out}}\,\Omega\!\left(R_{\text{out}}\right)}{R^2}\right],
\end{align} 
which in case of the Keplerian angular rotation velocity $\Omega_K(R)$ becomes
\begin{align}\label{viscounitpowerix4}
F_{\text{vis}}(R)=\frac{3}{8\pi}\dot M\Omega^2_K(R)\Bigg(\!\!\!\sqrt{\frac{R_{\text{out}}}{R}}-1\Bigg).
\end{align}
We now integrate Eq.~\eqref{viscounitpowerix3} for the viscous heat flux 
$F_{\text{vis}}(R)$ over the whole disk surface
to obtain the total disk luminosity produced by the viscous heat energy. In Sect.~\ref{Bephen} we introduced the stellar equatorial radius $R_{\text{eq}}$ 
of critically rotating stars as $3/2\,R_{\text{pole}}$ in standard Roche approximation \citep[e.g.,][among others]{maeder},
we can therefore write the equation for the total luminosity of the viscous dissipation energy 
of an outflowing stellar disk in the general form
\begin{align}\label{viscounitpowerix7}
L_{\text{vis}}=2\int\limits_{R_{\text{eq}}}^{R_{\text{out}}}F_{\text{vis}}(R)\,2\pi R\,\text{d}R,
\end{align}
where the factor $2$ refers to two sides of the disk.
Substituting Eq.~\eqref{viscounitpowerix3} into \eqref{viscounitpowerix7}, we get
\begin{align}\label{viscounitpowerix8}
L_{\text{vis}}=\dot{M}\int\limits_{R_{\text{eq}}}^{R_{\text{out}}}R^2\frac{\text{d}\Omega(R)}{\text{d}R}\left[\Omega(R)-
\frac{R^2_{\text{out}}\,\Omega\!\left(R_{\text{out}}\right)}{R^2}\right]\,\text{d}R.
\end{align}
Assuming again the Keplerian angular rotation velocity $\Omega_K$, the equation for the luminosity \eqref{viscounitpowerix8} takes the form
\begin{align}\label{viscounitpowerix9}
L_{\text{vis}}=\frac{GM_{\star}\dot{M}}{R_{\text{eq}}}\Bigg(\!\!\!\sqrt{\frac{R_{\text{out}}}{R_{\text{eq}}}}
-1\Bigg)\,\,-\,\,\frac{GM_{\star}\dot{M}}{2}\Bigg(\!\frac{1}{R_{\text{eq}}}-\frac{1}{R_{\text{out}}}\!\Bigg).
\end{align}
The second term on the right-hand side of Eq.~\eqref{viscounitpowerix9} represents 
one half of the potential energy of the disk matter needed to ascend the disk potential well from $R_{\text{eq}}$ to $R_{\text{out}}$, i.e.,~(one half of) 
the energy that is beyond supporting the material at 
exactly circular orbiting. Since the potential energy of the gas increases, this term is subtracted from the total disk luminosity according to the virial theorem. 
The first term on the right-hand side of Eq.~\eqref{viscounitpowerix9} thus corresponds to the work done by the viscous stress at the inner disk boundary, i.e., the work that is necessary 
just to support 
the steady disk extending from the $R=R_{\text{eq}}$ to $R_{\text{out}}$ \citep[cf.][]{Lee} (in other words to prevent the disk matter from falling inwards). 

In this point we can make a simple instructive comparison of Eq.~\eqref{viscounitpowerix4} with the similar equation for the 
viscous heat flux $F_{\!\text{accr,~vis}}(R)$ in case of an accretion disk, i.e.,~in case of inflowing instead of outflowing disk.
The physics of both the cases is practically identical (except of the change of the sign of radial velocity), the 
determining difference is there however in the boundary problem. Instead of the condition \eqref{korix}
we assume that the inwardly increasing Keplerian angular velocity of the accretion disk begins to decrease to the value of the 
angular velocity $\Omega_{\star}$ of a central object, which in realistic situation is 
smaller than the critical (Keplerian) value, $\Omega_{\star}<\Omega_K(R_{\star})$ (cf.~Sect.~\ref{viskodisks}). Hence, there exists a radius (in the very proximity of the 
central object radius $R_{\star}$, see \citet{Frank}), at which  
$\text{d}\Omega/\text{d}R=0$. We can therefore associate the inner accretion disk boundary $R_{\text{in}}\approx R_{\star}$ with the torque free radius, setting
\begin{align}\label{korinix}
\mathcal{G}\left(R_{\star}\right)\rightarrow 0.
\end{align}
Involving this 
condition and assuming Keplerian angular velocity, we obtain the accretion disk analogy of Eq.~\eqref{viscounitpowerix4} in the form
\citep[e.g.,][]{Frank}
\begin{align}\label{viscounitaccrpowerix4}
F_{\text{accr,~vis}}(R)=\frac{3}{8\pi}\dot M\Omega^2_K(R)\Bigg(1-\sqrt{\frac{R_{\star}}{R}}\,\Bigg).
\end{align}
We can calculate the luminosity produced by the viscous heat energy for Keplerian accretion disk (cf.~Eq.~\eqref{viscounitpowerix8}
for the outflowing disk) as (negative) integral of Eq.~\eqref{viscounitpowerix7} with substituted Eq.~\eqref{viscounitaccrpowerix4} 
within the range from infinity to the radius $R_\star$ of the central object. In this case we obtain
\begin{align}\label{viscounitaccrpowerix10}            
L_{\text{accr,~vis}}=\frac{GM_{\star}\dot{M}}{2R_\star}=\frac{1}{2}L_{\text{accr}},
\end{align}
where $L_{\text{accr}}$ is the total disk luminosity, i.e., where we suppose that the entire potential energy of 
the accreting matter is radiated away, $L_{\text{accr}}=\text{d}E_p/\text{d}t$. 
Assuming the constant mass of the central object,
$M_\star\approx\text{const.}$, the total accretion disk luminosity becomes $L_{\text{accr}}=GM_\star\dot{M}/R_\star$ \citep[e.g.,][]{Frank}.
The luminosity of the viscous heating energy in Keplerian accretion disk represents one half 
of the total accretion luminosity $L_{\text{accr}}$ (which is however the special case of the virial theorem).

Applying consistently the assumption that the disks 
are vertically very thin, $z\ll R$ (which is the case of astrophysical disks in observationally 
important inner regions, i.e.,~up to the distance of at least several 
stellar radii from 
the parent star, see Sects.~\ref{hydrostatous}) and neglecting the vertical variations of sound speed, $a=a(R)$ (denoted hence as $a$), Eq.~\eqref{slightvertigo} yet simplifies to the form
\begin{align} \label{thinvertigo}
\rho(R,z)=\rho(R,0)\,\text{e}^{{\textstyle{-\frac{GM_{\star}}{a^2R^3}\frac{z^2}{2}}}}.
\end{align}
Integrating Eq.~\eqref{thinvertigo} over the whole vertical range 
(in terms of Eq.~\eqref{sigmik}), we obtain the relation for the integrated disk density $\Sigma(R)$ in the 
elementary form
\begin{align} \label{Sigmasegva}
\Sigma(R)=\int\limits_{-\infty}^\infty\rho_0\,\text{e}^{{\textstyle{-\frac{GM_{\star}}{a^2R^3}\frac{z^2}{2}}}}\,\text{d}z=\sqrt{2\pi}\rho_0H,
\end{align}
where $\rho_0$ shortly denotes the disk midplane density $\rho(R,0)$.

Since in the thin disk approximation we work exclusively in the cylindrical radial direction while the external gravitational force 
of the central star
is naturally spherically symmetric, we need (analogously to integrated density) to integrate the gravity vertically. 
In this point we integrate the whole second right-hand side term $F^i$ in the momentum equation \eqref{moment} where we now 
consider that it represents 
only the cylindrical radial component of external gravitational force (the cylindrical radial component of 
gravitational acceleration represents the part of Eq.~\eqref{gravcylexpli}). We can write the desired vertical integration in the form
\begin{align}\label{gravvertixos}
\int\limits_{-\infty}^{\infty}\!F_R\,\text{d}z=-\int\limits_{-\infty}^{\infty}\!\rho g_R\,\text{d}z=
-\int\limits_{-\infty}^{\infty}\!\rho\frac{GM_{\star}R}{\left(R^2+z^2\right)^{3/2}}\,\text{d}z\approx
-\int\limits_{-\infty}^{\infty}\!\rho\frac{GM_{\star}}{R^2}\left(1-\frac{3}{2}\frac{z^2}{R^2}\right)\,\text{d}z
\end{align}
for $z\ll R$. Rewriting explicitly the last term of Eq.~\eqref{gravvertixos}, where we substitute 
the density $\rho(R,z)$ from Eq.~\eqref{thinvertigo} and the integrated density $\Sigma(R)$ from Eq.~\eqref{Sigmasegva}, we get
\begin{align}\label{gravvertixosix}
-\frac{GM_{\star}}{R^2}\int\limits_{-\infty}^{\infty}\!\rho\,\text{d}z+\frac{3}{2}\frac{GM_{\star}}{R^4}\!
\int\limits_{-\infty}^{\infty}\!\rho_0\,\text{e}^{{\textstyle{-\frac{GM_{\star}}{a^2R^3}\frac{z^2}{2}}}}\,z^2\,\text{d}z=
-\Sigma(R)\frac{GM_{\star}}{R^2}+\frac{3}{2}\frac{a^2\Sigma(R)}{R}.
\end{align}
The first term on the right-hand side is the radial gravity and the second term we may regard as the correction for the
decrease of the radial component of gravitational force $\rho g_R$ when moving away from the disk equatorial plane
\citep{Matsu,okazaki}. Since this term  
is of the same order as the pressure gradient term $\text{d}P/\text{d}R$, it must be included in the thin disk radial 
momentum equation (see Eq.~\eqref{radmomconserve} in Sect.~\ref{largemodelix}).

We can also make an analytical estimation of the disk radial velocity 
using Eqs.~\eqref{Shakurakos}, \eqref{angularmomentumfluxix0} and \eqref{nusigmix1} and assuming the 
Keplerian rotation. We can write in this case the relation for the radial velocity in the stationary
thin disk approximation,
\begin{align}\label{vradialkeplix}
V_R\approx\frac{3\alpha aH}{2R}\Bigg(\!\!\!\sqrt{\frac{R_{\text{out}}}{R}}-1\Bigg)^{-1}.         
\end{align}
Trying to analytically determine in this approximation the location of the critical (sonic) point $R_{\text{s}}$ (where $V_R=a$), 
we obtain from Eq.~\eqref{vradialkeplix}
the quadratic equation for $\!\sqrt{R_{\text{s}}}$ ,
\begin{align}
R_{\text{s}}-\sqrt{R_{\text{s}}\,R_{\text{out}}}+\frac{3}{2}\alpha H=0,        
\end{align}
whose explicit solution, omitting the lower root (tending in the next step to zero), is
\begin{align}
\sqrt{R_{\text{s}}}=\frac{1}{2}\,\Bigg(\!\!\!\sqrt{R_{\text{out}}}+\sqrt{R_{\text{out}}}\!\sqrt{1-6\alpha\frac{H}{R_{\text{out}}}}\,\Bigg).         
\end{align}
Since the term $6\alpha{H}/{R_{\text{out}}}\ll 1$ in the thin disk approximation, we may therefore regard 
the location of the sonic point in this approximation being at the distance
\begin{align}\label{rsrout}
R_{\text{s}}=R_{\text{out}},   
\end{align}
which implies that within the concept of the outer disk radius (see Eq.~\eqref{korix}) the disk should not extend to supersonic area.
Equation \eqref{vradialkeplix} also shows that in the region where $R\ll R_{\text{out}}$ we can write
$V_R\approx 3\alpha a^2R/\left(2\sqrt{GM_{\star}R_{\text{out}}}\right)$, 
that is, $V_R\sim \alpha a^2R$, while in the region where $R\to R_{\text{out}}$ we get the unphysical solution 
$V_R\to \text{infinity}$. The latter conclusion indicates that in the outer region the disk can no longer be Keplerian
(cf.~the analytical estimates of the radial thin disk structure in Sect.~\ref{radthin}).

Following the considerations of the disk nonisothermal vertical structure, introduced for a general case in Sect.~\ref{vertikalekdisk}, 
we can write a similar analysis also for a thin disk \citep[cf.][]{Karfiol06}.
From Eqs.~\eqref{Shakurakos} and \eqref{stresikvalecexplirphi}, with use of Eq.~\eqref{kincl} 
and using the definition of the vertical disk scale height from Sect.~\ref{vertikalekdisk},
we get in axisymmetric ($\partial/\partial\phi=0$) Keplerian case the relation for the $\phi R$ component of the stress tensor in the form 
\begin{align}\label{Trrf}
T_{\phi R}=\nu\rho\,R\frac{\partial\Omega_K}{\partial R}=-\frac{3}{2}\alpha a^2\rho.
\end{align}
The azimuthal component of the momentum equation \eqref{appendphimomcylinder1}, including equation \eqref{cylmomphi} and involving only the terms relevant 
for the stationary thin disk approximation (i.e.,~neglecting terms with explicit time derivative and derivatives in $\phi$ and $z$ direction),
can be written as
\begin{align}\label{anabjk1}
V_R\frac{\partial V_{\phi}}{\partial R}+\frac{V_RV_{\phi}}{R}=\frac{1}{\rho R^2}\frac{\partial}{\partial R}\big(R^2 T_{\phi R}\big),
\end{align}
Multiplying Eq.~\eqref{anabjk1} by $\rho R^2$ and integrating it over $\phi$ and $z$ (involving Eq.~\eqref{angularmomentumfluxix0}), we get
\begin{align}\label{anabjk2}
\dot{M}\frac{\partial}{\partial R}\big(RV_{\phi}\big)=\frac{\partial}{\partial R}\!\int\limits_{-\infty}^{\infty}\!R\,T_{\phi R}\,2\pi R\,\text{d}z.
\end{align}
Substituting Eq.~\eqref{Trrf} into Eq.~\eqref{anabjk2} and integrating it over $R$ (assuming constant $\alpha$ and Keplerian rotation velocity), we get
\begin{align}\label{anabjk3}
\dot{M}RV_K=-3\pi\alpha R^2\!\int\limits_{-\infty}^{\infty}\!a^2\rho\,\text{d}z+\text{const.}=
-3\pi\alpha R^2a_0^2\!\int\limits_{-\infty}^{\infty}\!\frac{a^2}{a_0^2}\rho\,\text{d}z+\text{const.},
\end{align}
where $V_K$ denotes the Keplerian azimuthal velocity ($V_K=R\,\Omega_K$) and
$a_0$ shortly denotes the speed of sound $a(R,0)$ in the disk equatorial plane. According to equation of state \eqref{statix6} we may write
the vertical thermal structure of the stationary thin disk in the form (where $T$ denotes the temperature and not the stress tensor)
\begin{align}\label{anabjk4}
\dot{M}RV_K=-3\pi\alpha R^2a_0^2\int\limits_{-\infty}^{\infty}\frac{T}{T_0}\rho\,\text{d}z+\text{const.},
\end{align}
where, similarly to $a_0$, we denote the disk midplane temperature $T(R,0)=T_0$. We may define the mass-weighted vertically averaged temperature
\citep{Karfiol06},
\begin{align}\label{anabjk5}
\langle T(R)\rangle=\frac{\int_{-\infty}^{\infty}T(R,z)\,\rho(R,z)\,\text{d}z}{\int_{-\infty}^{\infty}\rho(R,z)\,\text{d}z}=
\frac{\int_{-\infty}^{\infty}T(R,z)\,\rho(R,z)\,\text{d}z}{\Sigma(R)},
\end{align}
which leads to simplification of Eq.~\eqref{anabjk4} (where $a_0^2=kT_0/\mu m_u$, see Eq.~\eqref{statix6}) into the form
\begin{align}\label{anabjk6}
\dot{M}RV_K=-\frac{3\pi\alpha R^2k}{\mu m_u}\,\Sigma\,\langle T(R)\rangle+\text{const.}
\end{align}
We determine the constant of integration in Eq.~\eqref{anabjk6} from the boundary value $\Sigma(R_{\text{out}})\to 0$, which gives 
the condition $\dot{M}R_{\text{out}}V_{K}(R_{\text{out}})=\text{const}$.
The relation between the density and thermal disk structure can be analogously to Eq.~\eqref{Trhoscalefarix}, using the vertically 
integrated quantities $\Sigma$ and $\langle T(R)\rangle$, expressed as
\begin{align}\label{anabjk8}
\Sigma(R)=\frac{\dot{M}\!\!\sqrt{GM_\star}\,\mu m_u}{3\pi\alpha k\,R^{3/2}\,\langle T(R)\rangle}\left[\left(\frac{R_{\text{out}}}{R}\right)^{1/2}-1\right].
\end{align}
Equation \eqref{anabjk8} provides another useful analytical insight into the profiles of essential hydrodynamic quantities.
In case of the isothermal disk (constant $\langle T(R)\rangle$) we have $\Sigma\sim R^{-3/2}\big[(R_{\text{out}}/R)^{1/2}-1\big]$
and assuming also the inner region of large disk where 
$R/R_{\text{out}}\ll 1$, this relation becomes $\Sigma\sim R^{-2}$. Since the disk scale height $H=a/\Omega$,
Eq.~\eqref{kepscaleheight} implies $H\sim R^{3/2}$ in the Keplerian isothermal disk region (see Sect.~\ref{vertikalekdisk}) 
and the thin disk equatorial density $\rho_0\sim\Sigma/H$ (see Eq.~\eqref{Sigmasegva}). 
The radial dependence of the disk midplane density in the inner Keplerian disk region thus obeys
$\rho_0\sim R^{-7/2}$. Similarly the disk radial velocity $V_R=\dot{M}/(2\pi R\Sigma)$ leads to $V_R\sim R$
\citep[][cf.~also the analytical estimates of the radial thin disk structure in Sect.~\ref{radthin}]{okazaki,Krticka,kurfek}.

\section{Basic parameters of stellar irradiation}\label{kolobrik}
The disk ``surface'' is permanently impinged by the radiation from the central object (star), which mainly affects the
thermal structure of the circumstellar disks (see Sect.~\ref{smakec}).
In Sect.~\ref{cepelin} we describe the basic physics of the rapidly rotating oblate star's surface radiation,
while in Sect.~\ref{smakec} we focus on a description of the impinging irradiation at first in a simplified case of 
a spherically symmetric isotropically radiating central star (in more detail is the oblate star's irradiation described within Sect.~\ref{vikvous}).
The geometry of the problem is schematically illustrated in Fig.~\ref{irrad}.
\subsection{Stellar (equatorial) gravity darkening}
\label{cepelin}
Radiative equilibrium equation \eqref{CAK3} (resulting from the frequency integrated radiative transfer equation in the diffusion approximation), 
where the radiation pressure $P_{\text{rad}}=4\sigma T^4/(3c)$
($\sigma$ is the Stefan-Boltzmann constant, $c$ is the speed of light and $T$ is temperature), leads in uniformly rotating star (angular velocity $\Omega=\text{const.}$ in space and time) to the 
following rotationally and latitudinally dependent relation
\citep[e.g.,][]{maeder},
\begin{align}\label{fomis2}
\vec{F}_{\text{rad}}(\theta,\Omega)=-\chi\vec\nabla T(\theta,\Omega)\quad\text{with}\quad\chi=\frac{16\sigma T^3}{3\kappa\rho},
\end{align}
where $\theta$ denotes the stellar colatitude ($\theta=0$ at the pole of the star).
We also introduce the effective potential $\Psi(r,\theta)$ in a star with constant angular velocity $\Omega$ 
as the sum of a gravitational potential $\Phi(r)$ and a ``centrifugal'' potential $V(r,\theta)$,
where for given $\Omega$ the radial distance $r$ of the arbitrary point 
on the rotationally oblate stellar surface is however also a function of colatitude, $r=r(\theta)$ \citep[][cf.~also Eq.~\eqref{epicyclfreq}]{maeder}. 
Assuming the Roche model
(see also Sect.~\ref{wicom}), we write
\begin{align}\label{fomis1}
\Psi({r,\theta})=\Phi(r)+V(r,\theta)=-\frac{GM_\star}{r(\theta)}-\frac{1}{2}\Omega^2r^2(\theta)\,\text{sin}^2\,\theta.
\end{align}
The global hydrostatic equilibrium in case of a rigid body rotation (the so-called barotropic case where the density is a function of only pressure, $\rho=\rho(P)$), 
written in the form $\vec{\nabla}P/\rho=-\vec{\nabla}\Psi=\vec{g}_{\text{eff}}$,
implies the constant pressure on equipotentials (surfaces with constant $\Psi$), that is, $P=P(\Psi)$ and the equipotentials and isobaric surfaces coincide.
The vector of the effective gravitational acceleration  
$\vec{g}_{\text{eff}}$ 
is the vector sum of the gravitational and centrifugal acceleration (cf. Eq.~\eqref{geff})
and it may be also defined as the negative
gradient of the effective potential $\Psi$ \eqref{fomis1}.
This equilibrium also implies that $\rho=-\text{d}P/\text{d}\Psi$, the density is thus also a function of merely the effective potential,
$\rho=\rho(\Psi)$. Using the equation of state \eqref{statix2}, one also has $T=T(\Psi)$, 
and, since the opacity is also a function of $\rho$ and $T$, we write $\kappa=\kappa(\Psi)$ \citep{maeder}.
We may also expand the temperature gradient in the following way: 
$\vec\nabla{T}=(\text{d}T/\text{d}P)\,\vec\nabla{P}=(\text{d}T/\text{d}P)\,\rho\vec{g}_{\text{eff}}=-(\text{d}T/\text{d}P)\,\rho\vec\nabla{\Psi}$.
These relations show that the term $\chi$ as well as the whole expanded term $\rho\chi\,\text{d}T/\text{d}P$ are constant on equipotentials in the rigidly rotating star. 
From Eq.~\eqref{fomis2} we thus obtain, 
\begin{align}\label{fomis3}
\vec{F}_{\text{rad}}(\Omega,\theta)=\left(\rho\chi\frac{\text{d}T}{\text{d}P}\right)\vec\nabla\Psi(\Omega,\theta)\quad\text{which implies}\quad
L(\Omega)=\left(\rho\chi\frac{\text{d}T}{\text{d}P}\right)\int_S \vec\nabla\Psi(\Omega,\theta)\cdot\vec{n}\,\text{d}S.
\end{align}
In case of shellular rotation the centrifugal force cannot be derived from a potential and Eq.~\eqref{fomis1}
does not apply. In this non-conservative case the stars are said to be baroclinic, i.e.,~the density depends on both temperature and pressure, $\rho=\rho(P,T)$
(see, e.g., \citet{maeder} for details).

From Eq.~\eqref{fomis3} follows that the radiative flux is proportional to the gradient of the effective potential. 
We may therefore express the stellar luminosity on any equipotential surface as an integral of the flux over the
equipotential surface area $S$. Noting that the effective potential $\Psi$ (Eq.~\eqref{fomis1}) is the sum of the 
gravitational and ``centrifugal'' potential and employing the Gauss theorem, we obtain
\begin{align}\label{fomis4}
L(\Omega)=\left(\rho\chi\frac{\text{d}T}{\text{d}P}\right)\int_V\nabla^2\Psi\,\text{d}V=
\left(\rho\chi\frac{\text{d}T}{\text{d}P}\right)\int_V\left(4\pi G\rho-2\Omega^2\right)\,\text{d}V,
\end{align}
where we evaluate the Laplacian of the gravitational potential using the Poisson equation and the Laplacian of the ``centrifugal'' potential $\Omega^2R^2/2$ 
(where $R=r\,\text{sin}\,\theta$ is the cylindrical radial distance, 
see Eq.~\eqref{fomis1}) we evaluate using the radial part of Eq.~\eqref{laplacecyl}.
Considering the fact that the stellar surface is also the equipotential, we may write the radiative flux from rotationally oblate stellar surface (in 
dependence on the stellar angular velocity and colatitude), that is, the von Zeipel theorem \citep{vincek,maeder}, as
\begin{align}\label{omegacgamacflux}
\vec{F}_{\text{rad}}(\Omega,\theta)=-\frac{L_\star}{4\pi GM_\star\left(1-\frac{\Omega^2}{2\pi G\overline{\rho}_M}\right)}\vec{g}_{\text{eff}}(\Omega,\theta),
\end{align}
where $\overline{\rho}_M$ is the density averaged over the volume of the whole stellar body, $\overline{\rho}_M=M_\star/V$.
In case of critical rotation the term in bracket in the denominator is approximately $0.639$ \citep{maeder}.
The von Zeipel theorem in a differentially rotating stars with shellular rotation shows only minor differences with respect to Eq.~\eqref{omegacgamacflux}
\citep{Madr1999}.
The total stellar luminosity $L_\star$ we approximate as the luminosity for selected $T_{\text{eff}}$ of the oblate star with average stellar radius 
$\overline{r}$,
given as the radius at the colatitude $\text{sin}^2\theta=2/3$ \citep[which corresponds to the radius at the root $P_2(\text{cos}\,\theta)=0$ of the 
second Legendre polynomial, see][]{maeder}, where, in case of the critical rotation, $\overline{r}\approx 1.1503 R_\star$.
The vector of stellar radiative flux is thus parallel to the vector of effective gravity,
$\vec{F}_{\text{rad}}\sim\vec\nabla\Psi\sim\vec{g}_{\text{eff}}$ (cf.~Eq.~\eqref{omegacgamacflux}).
Applying the Stefan's law, $\vec{F}_{\text{rad}}=\sigma T_{\text{eff}}^4$, we obtain the stellar effective temperature $T_{\text{eff}}$ for the corresponding colatitude and rotation rate,
\begin{align}\label{teploukis}
T_{\text{eff}}(\Omega,\theta)=\left[\frac{L_\star}{4\pi\sigma GM_\star
\left(1-\frac{\Omega^2}{2\pi G\overline{\rho}_M}\right)}\right]^{1/4}\big[g_{\text{eff}}(\Omega,\theta)\big]^{1/4},
\end{align}
where $g_{\text{eff}}(\Omega,\theta)$ is the magnitude of the vector of local effective gravity \citep{maeder}.
The implications of the stellar equatorial gravity darkening for the stellar winds physics are in the main points described in Appendix \ref{vonZeipel}.

\subsection{Equations of the disk irradiation by a (spherically symmetric) star}\label{smakec}
\begin{figure}[t!]
\begin{center}
\includegraphics[width=13.5cm]{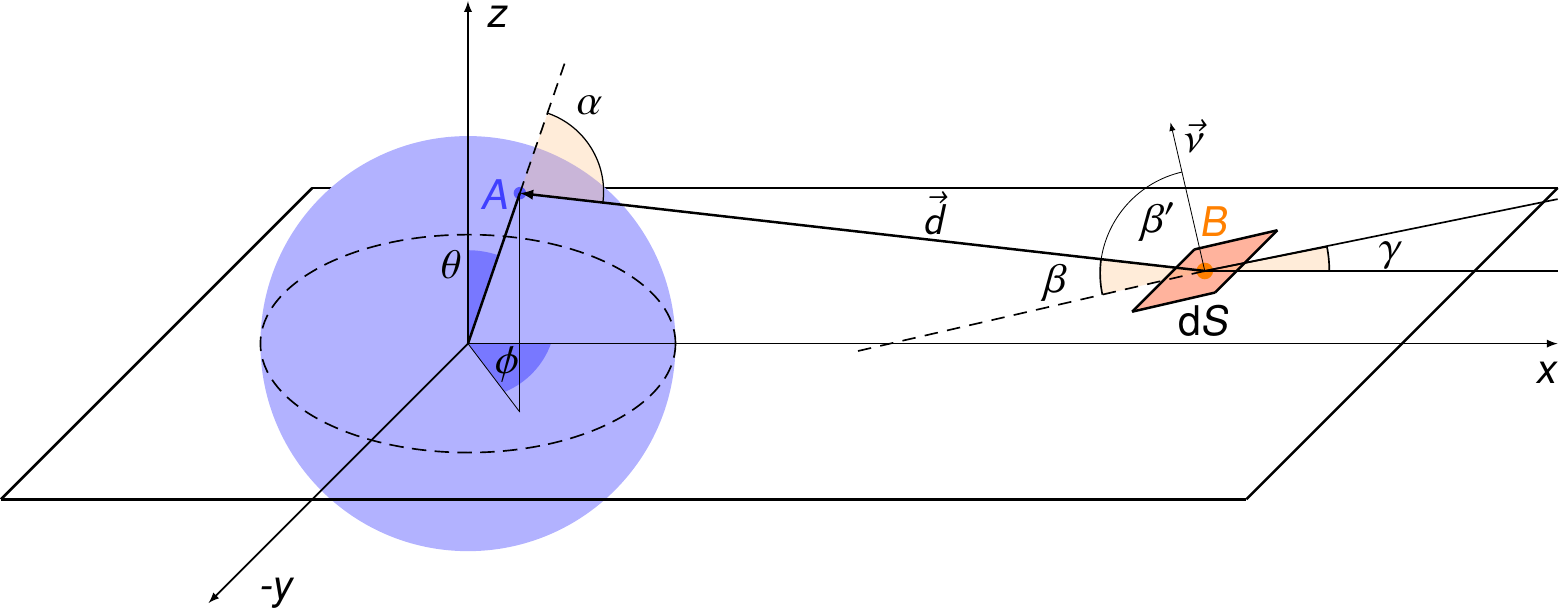}
\caption{Schema of the geometry of the irradiation of the disk layers by the central star. The disk ``surface'' element $\text{d}S$
(with the central point $B$)
is impinged by the stellar radiation that comes along the line-of-sight vector $\vec{d}$
from the stellar surface element $\text{d}S_{\!\!\star}$ that is characterized by its center point $A$. The angle $\alpha$ denotes the 
spatial deviation of the position vector $\vec{r}_A$ of the point $A$ and the vector $\vec{d}$
(the position vector of the point $B$ we denote $\vec{r}_B$). The angle $\beta$ is the spatial deviation of 
the inclination of the surface element $\text{d}S$ and the vector $\vec{d}$ (in fact it is the complementary angle to the deviation angle 
$\beta^\prime$ of the normal vector $\vec{\nu}$ of the surface element $\text{d}S$ and the line-of-sight vector $\vec{d}$). The angle $\gamma$ gives the
inclination of the surface element $\text{d}S$ in the disk $R$-$z$ plane. The basic idea is adapted from \citet{smak}.}
\label{irrad}
\end{center}
\end{figure}
Consider the radiating surface element $\text{d}S_{\!\!\star}$ of the star with the uniform radius $R_{\!\!\star}$, 
whose position vector $\vec{r}_A=\left(R_{\!\!\star}\,\text{sin}\,\theta\,\text{cos}\,\phi,\,R_{\!\!\star}\,\text{sin}\,\theta\,\text{sin}\,\phi,\,
R_{\!\!\star}\,\text{cos}\,\theta\right)$ in the Cartesian coordinate system, demonstrated in Fig.~\ref{irrad}. The radiation emerging from this element hits
the arbitrarily inclined (in the disk $R$-$z$ plane where the derivative $\text{d}z/\text{d}R$ may be arbitrary while there is no lateral inclination, $\text{d}z/\text{d}\phi=0$, 
see the schema in Fig.~\ref{irrad}) 
``surface'' element $\text{d}S$ of the disk, whose position vector $\vec{r}_B=\left(R,0,z\right)$ in the same Cartesian system,
where $R$ and $z$ are the cylindrical radial and vertical coordinates (the cylindrical $R$ coincides with Cartesian $x$
while the cylindrical and Cartesian $z$ coordinates are identical). The vertical
distance of the center of the disk ``surface'' element (point $B$) above the disk equatorial plane we thus denote $z_0$ 
and the projection of the position vector $\vec{r}_B$ into the disk equatorial plane
is $R_0$. We denote the inclination angle of the surface element $\text{d}S$ in the disk $R$-$z$ plane as $\gamma$, 
where $\text{tan}\,\gamma=(z-z_0)/(R-R_0)$. 
The line-of-sight vector $\vec{d}=\left(R-R_{\!\!\star}\,\text{sin}\,\theta\,\text{cos}\,\phi,\,-R_{\!\!\star}\,\text{sin}\,\theta\,\text{sin}\,\phi,\,
z-R_{\!\!\star}\,\text{cos}\,\theta\right)$, whose square magnitude $d^2$ (cf.~Sect.~\ref{vikvous}) is
$d^2=R^2+z^2+R_{\!\!\star}^2-2R_\star\left(R\,\text{sin}\,\theta\,\text{cos}\,\phi+z\,\text{cos}\,\theta\right)$,
connects the surface elements $\text{d}S$ and $\text{d}S_{\!\!\star}$ (the points $B$ and $A$). 
We denote the angles of vector $\vec{d}$ deviation from the position vector $\vec{r}_A$ and from the disk surface element $\text{d}S$
as $\alpha$ and $\beta$, respectively, where the $\beta$ angle is defined as the complementary angle to the deviation angle $\beta^\prime$ 
of the normal vector $\vec{\nu}$ of the surface element $\text{d}S$ and the line-of-sight vector $\vec{d}$ \citep{smak}.

The total radiative flux $\mathcal{F}$ from the star (we regard here the radiation integrated over all frequencies) 
that is intercepted by the surface element $\text{d}S$ of the disk (see the notation in Fig.~\ref{irrad}, cf.~\citet{Mihalas1}) is
\begin{align}\label{CenterFlux1}
\mathcal{F}_{\text{irr}}=\oint I(\text{cos}\,\alpha)\,\text{cos}\,\alpha\,\text{d}\omega,
\end{align}
where $I$ is the stellar radiation intensity and $\text{d}\omega$ is the solid angle subtended by the element $\text{d}S$ as seen from the center of the element 
$\text{d}S_{\!\!\star}$ (from the point $A$).
Following \citet{Mihalas1}, we may write $I\,\text{cos}\,\alpha\,\text{d}\omega=I\,\text{cos}\,\beta^\prime\,\text{d}\omega^\prime=I\,\text{sin}\,\beta\,\text{d}\omega^\prime$
where $\omega^\prime$ is the solid angle subtended by the element $\text{d}S_{\!\!\star}$, as seen from the center of the element 
$\text{d}S$ (from the point $B$), we therefore write $\text{d}\omega^\prime=\text{d}S_{\!\!\star}\,\text{cos}\,\alpha/d^2$.
In terms of the notation introduced above, the area of the surface element
of the star
$\text{d}S_{\!\!\star}=R_{\!\!\star}^2\,\text{sin}\,\theta\,\text{d}\theta\,\text{d}\phi$. Denoting here the uniform radiative flux of the star 
as ${F}_{\!\!\star}$, the specific intensity of stellar radiation is \citep[e.g.,][]{hejrak}
\begin{align}\label{icacak}
I=I_0\left(1-u+u\,\text{cos}\,\alpha\right),  
\end{align}
where we involve the most basic limb-darkening law, generally referred to as ‘‘linear limb darkening,’’ with $I_0=F_{\!\!\star}/\pi$ 
being the mean intensity of the
stellar surface and $u$ is the coefficient of the limb darkening that determines the shape of the limb-darkening profile.
The uniform radiative flux from the upper hemisphere of the star we may integrate as \citep[e.g.,][]{Mihalas1}
\begin{align}\label{frakacak}
{F}_{\!\!\star}=\int_{0}^{2\pi}\!\!\!\!\text{d}\phi\int_{\pi/2}^0\!\!\!\!I_0\left(1-u+u\,\text{cos}\,\alpha\right)\,\text{cos}\,\alpha\,\text{d}(\text{cos}\,\alpha),
\quad\text{giving}\quad {F}_{\!\!\star}=\pi I_0\left(1-\frac{u}{3}\right).
\end{align}
There are various (more or less complex) methods used for obtaining the $u$ coefficient \citep[e.g.,][]{klarov}. 
Since it is currently beyond the scope of this work, 
we do not introduce them (the values of $u$ which we adopt for the first approximation are given within the model description in Sect.~\ref{vikvous}).
Using the relations \eqref{CenterFlux1},\,\eqref{icacak} and \eqref{frakacak}, we can express the total stellar radiative flux 
that impinges the (arbitrarily inclined) disk ``surface'' unit area as
\begin{align}\label{CenterFlux2}
\mathcal{F}_{\text{irr}}=\frac{F_{\!\!\star}}{\pi}\!\!\iint_{\theta,\phi}\!\!
R_{\!\!\star}^2\text{sin}\,\theta\,\text{d}\theta\,\text{d}\phi
\,\frac{\left(1-u+u\,\text{cos}\,\alpha\right)\,\text{cos}\,\alpha\,\text{sin}\,\beta}{\left(1-u/3\right)\,d^2}.
\end{align}
The integration limits in $\phi$ and $\theta$ range within the angles
$\phi\in\langle-\pi/2,\,\pi/2\rangle,\,\theta\in\langle\theta_{\text{min}},\,\theta_{\text{max}}(\phi)\rangle$, where
the value of $\theta_{\text{max}}(\phi)$ is also the function of the inclination angle $\gamma$ of the disk surface element $\text{d}S$,
while the angle $\theta_{\text{min}}$ is a limiting angle subtended by a 
silhouette ring on stellar surface where the vectors $\vec{d}$ and $\vec{r}_A$ are perpendicular \citep{smak}.
In case of a thin disk, where $z\to 0$, we obtain $\text{sin}\,\theta_{\text{min}}=R_{\!\!\star}/R$ 
and the ($\phi$-independent) angle $\theta_{\text{max}}=\pi/2$ while at large 
distances, where $R\gg R_{\!\!\star}$, we thus obviously have $\theta_{\text{min}}\to 0$ .

We express the cosine of the angle $\alpha$ and the sine of the angle $\beta$ in Fig.~\ref{irrad}, respectively, as
\begin{align}\label{CenterFlux6}
\text{cos}\,\alpha=\frac{R_{\!\!\star}-R\,\text{sin}\,\theta\,\text{cos}\,\phi-z\,\text{cos}\,\theta}{d},\quad\quad
\text{sin}\,\beta=\frac{\left(R_{\!\!\star}\,\text{sin}\,\theta\,\text{cos}\,\phi-R\right)
\text{tan}\gamma+\left(z-R_{\!\!\star}\,\text{cos}\,\theta\right)}{d/\text{cos}\gamma}.
\end{align}
Using all the equations introduced above in this section, the full explicit form of Eq.~\eqref{CenterFlux2} of the total radiative flux 
from isotropically radiating spherically symmetric star that impinges the unit disk surface area may be rewritten as
\begin{gather}
\mathcal{F}_{\text{irr}}=\frac{F_{\!\!\star}}{\pi}\left(\frac{R_{\!\!\star}}{R}\right)^3\int_{-\frac{\pi}{2}}^{\frac{\pi}{2}}
\int_{\theta_{\text{min}}}^{\theta_{\text{max}}}
\!\!\!\frac{\left(1-u+u\,\text{cos}\,\alpha\right)\left(\frac{R_{\!\!\star}}{R}-\text{sin}\,\theta\,\text{cos}\,\phi-\frac{z}{R}\,\text{cos}\,\theta\right)}
{\left(1-u/3\right)\left[1+\left(\frac{z}{R}\right)^2+
\left(\frac{R_{\!\!\star}}{R}\right)^2-\frac{2R_{\!\!\star}}{R}\,\text{sin}\,\theta\,
\text{cos}\,\phi-\frac{2R_{\!\!\star}z}{R^2}\,\text{cos}\,\theta\right]^{2}}\nonumber\\
\times\left[\left(\text{sin}\,\theta\,\text{cos}\,\phi-\frac{R}{R_{\!\!\star}}\right)
\text{sin}\gamma+\left(\frac{z}{R_{\!\!\star}}-\text{cos}\,\theta\right)
\text{cos}\gamma\right]
\text{sin}\,\theta\,\text{d}\theta\,\text{d}\phi.
\label{pelousek}
\end{gather}

In the approximation of a thin disk (see Sect.~\ref{thindisk}) with $z_0\to 0$, $\gamma\to 0$,
the relation \eqref{pelousek} simplifies to
the form
\begin{align}\label{CenterFlux10}
\mathcal{F}_{\text{irr}}=\frac{2F_{\!\!\star}}{\pi}\left(\frac{R_{\!\!\star}}{R}\right)^3\int_{-\frac{\pi}{2}}^{\frac{\pi}{2}}\int_{\theta_{\text{min}}}^{\pi/2}
\frac{\left(1-u+u\,\text{cos}\,\alpha\right)\left(\text{sin}\,\theta\,\text{cos}\,\phi-\frac{R_{\!\!\star}}{R}\right)\,\text{sin}\,\theta\,\text{cos}\,\theta\,\text{d}\theta\,\text{d}\phi}
{\left(1-u/3\right)\left[1+\left(\frac{R_{\!\!\star}}{R}\right)^2-\frac{2R_{\!\!\star}}{R}\,\text{sin}\,\theta\,\text{cos}\,\phi\right]^{2}},
\end{align}
where the factor $2$ comes from two sides of the disk.
Assuming yet the very large distance approximation, $R_{\!\!\star}/R\approx 0$, after integration over $\theta$ and $\phi$ we obtain the 
approximate equation of the impinging radiative flux per unit disk surface in a 
quite simple and instructive form
\begin{align}\label{smaksimpl}
\mathcal{F}_{\text{irr}}\approx\frac{2}{3\pi}\frac{\left(1-u\right)}{\left(1-u/3\right)}F_{\!\!\star}\left(\frac{R_{\!\!\star}}{R}\right)^3.
\end{align}
The irradiative flux thus decreases with the third power of the radius, similarly to, e.g.,~a stellar magnetic dipole field radial decrease, described in Appendix~\ref{rindzak}.

If we compare the stellar irradiative flux $\mathcal{F}_{\text{irr}}$ given by Eq.~\eqref{smaksimpl} with the viscous heat flux $F_{\text{vis}}$ per unit area of the disk given by Eq.~\eqref{viscounitpowerix4},
for example at the distance $R/R_{\text{out}}=1/10$ (in the Keplerian region)
where for the stellar radiative flux applies $F_{\!\!\star}=L_{\!\!\star}/(4\pi R_{\!\!\star}^2)$ and neglecting the reduction of the irradiative flux by limb darkening, we obtain the ratio
\begin{align}\label{smakratio}
\frac{\mathcal{F}_{\text{irr}}}{{F}_{\text{vis}}}\approx\frac{4}{9\pi}\frac{L_{\!\!\star}R_{\!\!\star}}{GM_{\!\!\star}\dot M}\Bigg(\!\!\sqrt{\frac{R_{\text{out}}}{R}}-1\Bigg)^{-1}.
\end{align}
For the evaluation of Eq.~\eqref{smakratio} we substitute the mass, radius and luminosity of the order that is typical for stars of spectral type B: $M_{\!\!\star}=10M_{\odot},\, 
R_{\!\!\star}=4R_{\odot}$ and $\text{log}\,(L_{\!\!\star}/L_{\odot})=3,67$.
Using these parameters, we find the ratio $\mathcal{F}_{\text{irr}}/{F}_{\text{vis}}=1$ if the stellar mass loss rate $\dot{M}\approx 7.8\cdot 10^{24}\,\text{kg}\,\text{yr}^{-1}
\approx 4\cdot 10^{-6}\,M_{\odot}\,\text{yr}^{-1}$.
This implies that the disk viscous heating generated by the viscous flux ${F}_{\text{vis}}$ begins to dominate only if 
$\dot{M}\gtrsim 10^{-6}M_{\odot}\,\text{yr}^{-1}$ \citep[][see also the current results in Sect.~\ref{templajznik}]{Lee}.
\HlavickaKapitoly
\chapter{Influence of magnetic fields on the disk structure}\label{sdfge}
\section{Fundamental equations of ideal magnetohydrodynamics}
\label{basemaghydro}
The general expression for the electromagnetic Lorentz force \citep{Bittyk,Landau2} per unit volume is
\begin{eqnarray}\label{mhd0}
\vec{F}_L=\rho_\text{e}(\vec{E}+\vec{V}\times\vec{B}),
\end{eqnarray}
where $\rho_\text{e}$ is the electric charge density, 
$\vec{E}$ is the vector of the electric field intensity, $\vec{V}$ is the vector of the flow velocity of the matter (ionized gas) 
and $\vec{B}$ is the vector of the magnetic induction.
This equation we may further expand by using the vacuum form of the ``fourth'' Maxwell equation (Maxwell-Amp\`{e}re's law) 
\begin{eqnarray}\label{mhd1}
\vec{\nabla}\times\vec{B}=\mu_0\vec{J}+\mu_0\epsilon_0\frac{\partial\vec{E}}{\partial t},
\end{eqnarray}
where $\epsilon_0$ and $\mu_0$ are the vacuum electric permittivity and magnetic permeability, respectively, and $\vec{J}$ is the electric current density,
$\vec{J}=\rho\vec{V}$. 
Neglecting the term $\mu_0\epsilon_0\,\partial\vec{E}/\partial t$ we obtain the Amp\`{e}re's law 
\begin{eqnarray}\label{mhd1a}
\vec{\nabla}\times\vec{B}=\mu_0\vec{J},
\end{eqnarray}
which is the approximation normally used in MHD. 
Employing the Ohm's law, $J=\sigma E$, with $\sigma$ being the material-dependent conductivity, which for most of the fluids is typically greater than S/m (Siemens per meter), 
the dimensional analysis of Eq.~\eqref{mhd1} shows that $\epsilon_0 E/\tau\sim\sigma E$ in case of extremely small characteristic time $\tau$ for changes in the electric field
(i.e., that the term $\partial\vec{E}/\partial t$
cannot be neglected only in case of $\tau$ being of the order of $10^{-11}\,\text{s}$ or less). 

Now we use analogous dimensional analysis with the ``first'' Maxwell equation (Gauss's law) in the vacuum form $\vec{\nabla}\cdot\vec{E}=\rho/\epsilon_0\approx{E/\ell}$
(where $\ell$ is the characteristic length scale of the system) and with Eq.~\eqref{mhd1} in the described approximated form $\vec{\nabla}\times\vec{B}=\mu_0\vec{J}\approx{B/\ell}$.
Combining the two approximations and the ideal Ohm's law for a perfect conductor given by 
\begin{eqnarray}\label{mhd1b}
\vec{E}=-\vec{V}\times\vec{B}, 
\end{eqnarray}
whose dimensional expression we may write as $-VB$, we obtain (noting that $\epsilon_0\,\mu_0=c^{-2}$)
the approximate ratio $\rho E/(\vec{J}\times\vec{B})\approx(V/c)^2)\ll 1$ for the non-relativistic ideal MHD.
We may therefore neglect the electrostatic force term in Eq.~\eqref{mhd0}, 
writing the magnetic Lorentz force equation in the modified form
\begin{eqnarray}\label{mhd2}
\vec{F}_L=\vec{J}\times\vec{B}=\frac{1}{\mu_0}(\vec{\nabla}\times\vec{B})\times\vec{B}. 
\end{eqnarray}
Employing the vector identity for cross product of vector rotation,
Eq.~\eqref{mhd2} takes the form
\begin{eqnarray}\label{mhd4}
\vec{F}_L=\frac{1}{\mu_0}(\vec{B}\cdot\vec\nabla)\vec{B}-\frac{1}{2}\frac{\vec\nabla B^2}{\mu_0},
\end{eqnarray}
where the first term on the right-hand side expresses the advection of the magnetic field and the second term on the right-hand side expresses 
the gradient of the magnetic energy density \citep{Bittyk}.
Basic \textit{hydrodynamic} equations, including the Lorentz force and the induction equation \eqref{mhd8},
that is, the basic \textit{magnetohydrodynamic} (MHD) equations, can be written in the following way: the continuity equation 
\eqref{masscylinder1} remains unchanged,
\begin{eqnarray}\label{mhd5}
\frac{\partial\rho}{\partial t}+\vec\nabla\cdot\rho \vec{V}=\vec{0}, 
\end{eqnarray}
while the equation of motion \eqref{moment1} now takes the form
\begin{eqnarray}\label{mhd7}
\frac{\partial\vec{V}}{\partial t}+(\vec{V}\cdot\vec\nabla)\vec{V}+\frac{1}{\rho}\vec\nabla\cdot\mathcal{P}+
\frac{1}{2}\frac{\vec\nabla B^2}{\mu_0\rho}
-\frac{1}{\mu_0\rho}(\vec{B}\cdot\vec\nabla)\vec{B}+\vec\nabla\Phi=\vec{0},
\end{eqnarray}
where $\mathcal{P}$ is the pressure tensor and $\Phi$ is the gravitational potential. Including the Ohm's law for
\textit{ideally} conductive plasma
(where the electrical conductivity $\sigma\rightarrow\infty$) in the form of Eq.~\eqref{mhd1b},
from the ``{second}'' Maxwell equation (Faraday's law of induction)
\begin{eqnarray}\label{mhd7a}
\vec\nabla\times\vec{E}=-\frac{\partial\vec{B}}{\partial t} 
\end{eqnarray}
we consequently obtain its special form,
i.~e., the \textit{Maxwell-Faraday equation}. In the literature it is usually concisely referred to as \textit{induction equation} \citep{Bittyk},
\begin{eqnarray}\label{mhd8}
\frac{\partial\vec{B}}{\partial t}-\vec\nabla\times(\vec{V}\times\vec{B})=\vec{0}.
\end{eqnarray}
We neglect the diffusion term $\eta\vec{\nabla}^2\vec{B}$ in Eq.~\eqref{mhd8} where $\eta=1/(\mu_0\sigma)$ denotes the magnetic diffusivity 
\citep{Bittyk}.
The diffusion term plays a significant role merely in case of very low gas velocity or very small electric conductivity.

To derive the magnetohydrodynamic terms, which enter the energy equation, we expand the term $\vec{V}\cdot\vec{F}$ on the right-hand side of Eq.~\eqref{firstthermo1} 
where we regard the force $\vec{F}$ being the magnetic Lorentz force \eqref{mhd2}. 
Multiplying Eq.~\eqref{mhd2} by velocity, $\vec{F}_{L}\cdot\vec{V}=(\vec{J}\times\vec{B})\cdot\vec{V}=-(\vec{V}\times\vec{B})\cdot\vec{J}$, 
and using Eq.~\eqref{mhd1b}, we obtain $\vec{F}_{L}\cdot\vec{V}=\vec{E}\cdot\vec{J}=\vec{E}\cdot(\vec\nabla\times\vec{B})/\mu_0$,
where the last expression comes from the Amp\`{e}re's law \eqref{mhd1a}.
We expand the term $\vec{E}\cdot(\vec\nabla\times\vec{B})/\mu_0$ as $[\vec{B}\cdot(\vec\nabla\times\vec{E})-\vec\nabla\cdot(\vec{E}\times\vec{B})]/\mu_0$, while we 
consequently rewrite the first term 
using the Faraday's law of induction \eqref{mhd7a} and the second term using the Ohm's law for ideally conducting fluid 
(Eq.~\eqref{mhd1b}) into the form 
\begin{eqnarray}\label{mhd8a}
\vec{F}_{L}\cdot\vec{V}=\frac{1}{\mu_0}\left\{-\vec{B}\cdot\frac{\partial\vec{B}}{\partial t}+\vec\nabla\cdot\left[\left(\vec{V}\times\vec{B}\right)\times\vec{B}\right]\right\}. 
\end{eqnarray}
Employing the vector identity for the triple cross product,
we obtain the expression for the term $\vec{F}_{L}\cdot\vec{V}$ in the form
\begin{eqnarray}\label{mhd9}
\vec{F}_{L}\cdot\vec{V}=-\frac{1}{2}\frac{\partial}{\partial t}\frac{B^2}{\mu_0}+\frac{1}{\mu_0}\vec\nabla\cdot
\left[\left(\vec{B}\cdot\vec{V}\right)\vec{B}-B^2\vec{V}\right].
\end{eqnarray}
Inserting Eq.~\eqref{mhd9} into the energy equation~\eqref{firstthermo2}, we obtain the full (ideal) magnetohydrodynamic 
energy equation
in the form
\begin{eqnarray}\label{mhd10}
\frac{\partial E}{\partial t}+\vec\nabla\cdot\left[\left(E+\mathcal{P}\right)\vec{V}+\vec{q}\right]=\frac{1}{\mu_0}\vec\nabla\cdot
\left[\left(\vec{B}\cdot\vec{V}\right)\vec{B}-\frac{B^2\vec{V}}{2}\right]+\rho\vec{g}\cdot\vec{V},
\end{eqnarray}
where the first term in the square bracket on the right-hand side is the magnetic tension force that is trying to straighten the magnetic field lines,
the second term in the right-hand side's square bracket is the magnetic pressure flux and $\vec{g}$ denotes the vector of (external) gravitational acceleration. 
The explicit form of the total energy density $E$ in Eq.~\eqref{mhd10} is
\begin{align}\label{mhd11}
E=\rho\epsilon+\frac{\rho V^2}{2}+\frac{B^2}{2\mu_0},
\end{align}
consisting thus from the densities of internal, kinetic and magnetic energy, respectively.
\section{Shear instability in weakly magnetized disk (magnetorotational instability)}
\label{sivikis}
\subsection{Linear analysis of magnetorotational instability}\label{sheainst}
The problem of the hydrodynamic stability of a fluid, subjected to action of a magnetic field and rotation, 
has been studied for a long time \citep[e.g.,][among others]{chandrafer,chandrous,Fricke}, the analysis of \textit{magnetorotational} instability in circumstellar disks is however first 
quite systematically described in \citet{Balbus}. Because of the ubiquity of 
magnetic fields can the turbulences of magnetized disk matter
(where the gas is partially or fully ionized) act as a main source of anomalous viscosity. 
The disk is subjected to very strong shear instabilities caused by weak magnetic field
that can be far more destabilizing than a strong one (the strong field would rather
enforce the disk to rotate as a rigid body). Arbitrarily small magnetic field 
cannot therefore be neglected in linear analysis of the disk disturbances. 

The basic destabilizing mechanism act as follows: consider a differentially rotating disk that is perpendicularly threaded
by magnetic field, its field lines are therefore ``{vertically}'' oriented. 
The motion of the disk volume element (gas or plasma) that is displaced for example in the outward direction from its Keplerian orbit is
elastically controlled by the magnetic field. The field is trying to eliminate the effects caused by shear friction between radial disk 
segments by enforcement of rigid rotation while it simultaneously returns this element back to its original position (and thus eliminates stretching
of the field lines). 
The second effect is stabilizing, while the first effect acts as the source of the instability. Magnetic field is trying to 
force the gas element to rotate too fast for its new radial location, the excess of centrifugal force drives the element further 
outward.
At sufficiently long wavelengths (longer than a critical wavelength that corresponds 
to a critical wavenumber) is the returning force too weak and destabilization wins (see Sect.~\ref{limitanalyz}). 
The presence of the finite value of the vertical wavenumber of the magnetic field is essential, otherwise no axisymmetric instability occurs.
The magnetorotational instability generates thus the viscous couple, simply caused by an interpenetration of the gas volume elements with higher and lower angular momentum,
leading to turbulence.

The dispersion relation is described in \citet{Balbus} with the following assumptions: within the principal linear analysis of the strictly axisymmetric case is the radial component of the
magnetic field set to zero ($B_R=0$), while in the more advanced part of the description is analyzed also the general case with the nonzero radial component ($B_R\neq 0$).
The behavior of the fluid is restricted by the so-called \textit{Boussinesq approximation} 
(first described in \citet{busin}), which is considered to be valid for the noncompressive disturbances of interest.
This approximation assumes that the variations of density are so small that can be neglected (in the continuity equation as well as in the advection term in equation of motion 
we may substitute $\rho\to\rho_0=\text{const}.$). However, even weak density variations are important in buoyancy, and so we retain the density variations 
only in the right-hand side of the equation of motion and in the equation of state, while the pressure
perturbations are neglected in the equation of state \citep{Fricke}. 
In fluid flows driven by buoyancy (\textit{buoyancy-driven flows}) are the density perturbations connected merely with (reduced) gravity via 
the equation
\begin{align}\label{boussinesq1}
\delta g=\frac{\delta\rho}{\rho}g\quad\text{with}\quad\delta\rho=0\,\,\text{otherwise}, 
\end{align}
the fluid thus essentially acts as incompressible.
The Boussinesq approximation enables us to eliminate acoustic waves from consideration, since acoustic waves are the density perturbations in general. 
We consider constant angular velocity $\Omega$ on radial (cylindrical) segments with $V_R\approx 0,\,V_z\approx 0$ and
$\partial\Omega/\partial z=0$. Axisymmetric ($\partial/\partial\phi=0$) Eulerian space-time dependent perturbation $\delta\xi$ of an
arbitrary physical quantity $\xi$ 
with radial and vertical wavenumbers $k_R$ and $k_z$ and the angular frequency of the magnetorotational instability $\omega$ can be described as
\begin{align}\label{perturb}
\delta\xi=\xi_0\,\text{e}^{i(k_RR+k_zz-\omega t)}.
\end{align}
The above set of seven basic ideal magnetohydrodynamic equations, that is, one equation of continuity \eqref{mhd5}, three vector momentum equations 
\eqref{mhd7} and three vector Maxwell-Faraday (induction) equations \eqref{mhd8} we can thus expand to first order. Since the density in the continuity
equation does not vary, we can therefore set $\vec\nabla\cdot\vec{V}=\vec{0}$, and the equation takes the explicit form (with only
the largest terms retained)
\begin{align}\label{perturb1a}
\frac{1}{R}\frac{\partial}{\partial R}\left[R(V_R+\delta V_r)\right]+\frac{\partial}{\partial z}(V_z+\delta V_z)=0.
\end{align} 
We do not take into account the constant unperturbed radial and vertical components of velocity as well as we neglect any variations in radius $R$. 
After linearization of perturbations according to Eq.~\eqref{perturb}
we can write Eq.~\eqref{perturb1a} in quite simple form 
\begin{align}\label{mhd12}
k_R\delta V_R+k_z\delta V_z=0.
\end{align}
Explicit form of the radial component of momentum equation \eqref{mhd7}, taking into account Eq.~\eqref{trafounit8} (where the terms that contain the radial derivative of the 
radial magnetic field component $\partial B_R/\partial R$ mutually cancel), is
\begin{align}\label{mhd13}
\frac{\partial V_R}{\partial t}+V_R\frac{\partial V_R}{\partial R}+\frac{V_{\phi}}{R}\frac{\partial V_R}{\partial {\phi}}+
V_z\frac{\partial V_R}{\partial z}-\frac{V_{\phi}^2}{R}+\frac{1}{\rho}\frac{\partial P}{\partial R}+
\frac{1}{2\mu_0\rho}\frac{\partial}{\partial R}\left(B_{\phi}^2+B_z^2\right)-\nonumber\\
-\frac{1}{\mu_0\rho}\left(\frac{B_{\phi}}{R}\frac{\partial B_R}{\partial {\phi}}+
B_z\frac{\partial B_R}{\partial z}-\frac{B_{\phi}^2}{R}\right)+g_R=0.
\end{align}
According to the Boussinesq approximation \eqref{boussinesq1}, the radial component of the density perturbation 
(employing the hydrostatic equilibrium) can be written in the form
\begin{align}\label{boussinesq2}
\delta g_R=-\frac{\delta\rho}{\rho^2}\frac{\partial P}{\partial R}, 
\end{align}
and after linearization of perturbations according to Eq.~\eqref{perturb} with use of Eq.~\eqref{boussinesq2} one can write
\begin{align}\label{mhd14}
-\text{i}\omega\delta V_R-2\Omega\delta V_{\phi}+\frac{\text{i}k_R}{\rho}\delta P+\frac{\text{i}k_R}{\mu_0\rho}
\left(B_{\phi}\delta B_{\phi}+B_z\delta B_z\right)-\frac{\text{i}k_z}{\mu_0\rho}B_z\delta B_R
-\frac{\delta\rho}{\rho^2}\frac{\partial P}{\partial R}=0.
\end{align}
In quite similar way we analyze the vertical component of the momentum equation \eqref{mhd7}, employing Eq.~\eqref{trafounit10}. Its
explicit form (after mutual cancellation of the terms containing $\partial B_z/\partial z$) is
\begin{align}\label{mhd15}
\frac{\partial V_z}{\partial t}+V_R\frac{\partial V_z}{\partial R}+\frac{V_{\phi}}{R}\frac{\partial V_z}{\partial {\phi}}+
V_z\frac{\partial V_z}{\partial z}+\frac{1}{\rho}\frac{\partial P}{\partial z}+
\frac{1}{2\mu_0\rho}\frac{\partial}{\partial z}\left(B_R^2+B_{\phi}^2\right)-\nonumber\\
-\frac{1}{\mu_0\rho}\left(B_R\frac{\partial B_z}{\partial R}+\frac{B_{\phi}}{R}\frac{\partial B_z}{\partial {\phi}}\right)+
g^\prime_z=0.
\end{align}
The same linearization of perturbations, using the analog of Eq.~\eqref{boussinesq2}, leads to expression
\begin{align}\label{mhd16}
-\text{i}\omega\delta V_z+\frac{\text{i}k_z}{\rho}\delta P+\frac{\text{i}k_z}{\mu_0\rho}
B_{\phi}\delta B_{\phi}-\frac{\delta\rho}{\rho^2}\frac{\partial P}{\partial z}=0.
\end{align}
The explicit form of the azimuthal momentum component \eqref{mhd7},
omitting the negligible viscous terms (where we employ merely the scalar pressure term, while the terms containing the derivative $\partial B_\phi/\partial\phi$ mutually cancel), is
\begin{align}\label{mhd18}
\frac{\partial V_{\phi}}{\partial t}+V_R\frac{\partial V_{\phi}}{\partial R}+\frac{V_{\phi}}{R}\frac{\partial V_{\phi}}{\partial {\phi}}+
V_z\frac{\partial V_{\phi}}{\partial z}+\frac{V_RV_{\phi}}{R}+
\frac{1}{\rho}\frac{\partial P}{\partial\phi}
+\frac{1}{2\mu_0\rho}\frac{1}{R}\frac{\partial}{\partial\phi}\left(B_R^2+B_z^2\right)-\nonumber\\
-\frac{1}{\mu_0\rho}\left(B_R\frac{\partial B_{\phi}}{\partial R}+B_z\frac{\partial B_{\phi}}{\partial z}+\frac{B_RB_{\phi}}{R}\right)=0.
\end{align}
Linearization of perturbations in Eq.~\eqref{mhd18} according to Eq.~\eqref{perturb} leads
to the form
\begin{align}\label{mhd19}
-\text{i}\omega\delta V_{\phi}+\delta V_R\left(\frac{\partial V_{\phi}}{\partial R}+\frac{V_{\phi}}{R}\right)-\frac{\text{i}k_z}{\mu_0\rho}
B_z\delta B_{\phi}=0.
\end{align}
We can rewrite the term in bracket as $2\Omega+R\,\text{d}\Omega/\text{d}R=\kappa^2/(2\Omega)$, where the quantity $\kappa$ is the  
\textit{epicyclic frequency}, $\kappa^2=4\Omega^2+\text{d}\Omega^2/\text{d}\,\text{ln}\,R$ \citep{Balbus, binney}. 
In this point we note that we derive the general expression for the epicyclic frequency by considering 
a small radial 
displacement of an arbitrary particle (body or fluid parcel), orbiting at the circular trajectory with radius $R_0$ in a gravitational field whose gravitational potential is $\Phi$
(see Eq.~\eqref{fomis1}). 
The acceleration of the displacement (described by its only nonzero radial component),
$a_R=\ddot{R}-R\dot{\phi}^2$ (see Eq.~\eqref{trafounit7}). However, since $a_R$ must be also the negative radial derivative of a gravitational potential,
$a_R=-\partial\Phi/\partial R$, one thus obtains 
$\ddot{R}=-\partial\Phi/\partial R+j^{\,2}/R^3$, where the quantity $j$ is the specific angular momentum of the displaced body (which is the
conservative quantity). Expansion of the radial acceleration $\ddot{R}$ to first order in 
$R_0$ leads to the expression
\begin{align}\label{epicyclfreq}
\ddot{R}+\left[\frac{\partial^2\Phi}{\partial R^2}\Bigg|_{R_0}+\frac{3j^{\,2}}{R_0^4}\right](R-R_0)=0,
\end{align}
where the term in square bracket represents the square of the epicyclic frequency $\kappa^2$. 

Equation \eqref{mhd19} we can thus consistently rewrite into the simplified form
\begin{align}\label{mhd20}
-\text{i}\omega\delta V_{\phi}+\frac{\kappa^2}{2\Omega}\delta V_R-\frac{\text{i}k_z}{\mu_0\rho}
B_z\delta B_{\phi}=0.
\end{align}
We may write the explicit form of radial component of the Maxwell-Faraday (induction) equation \eqref{mhd8} using the vector identity
$\vec{\nabla}\times(\vec{V}\times\vec{B})=\vec{V}(\vec{\nabla}\cdot\vec{B})+(\vec{B}\cdot\vec{\nabla})\vec{V}-(\vec{V}\cdot\vec{\nabla})\vec{B}-\vec{B}(\vec{\nabla}\cdot\vec{V})$
and involving the Maxwell equation $\vec{\nabla}\cdot\vec{B}=0$, e.g., in the form
\begin{align}\label{mhd21}
\frac{\partial B_R}{\partial t}-\frac{B_{\phi}}{R}\frac{\partial V_R}{\partial {\phi}}-
B_z\frac{\partial V_R}{\partial z}+V_R\frac{\partial B_R}{\partial R}+\frac{V_{\phi}}{R}\frac{\partial B_R}{\partial {\phi}}+
V_z\frac{\partial B_R}{\partial z}+%\nonumber\\
B_R\left(\frac{1}{R}\frac{\partial V_{\phi}}{\partial {\phi}}+
\frac{\partial V_z}{\partial z}+\frac{V_R}{R}\right)=0.
\end{align}
Linearization of perturbations in Eq.~\eqref{mhd21} according to Eq.~\eqref{perturb} and with use of Eq.~\eqref{boussinesq2}, taking into account the above described constraints 
(including the assumption $B_R=0$), leads
to the equation in the form
\begin{align}\label{mhd22}
-\text{i}\omega\delta B_R-\text{i}k_zB_z\delta V_R=0.
\end{align}
Explicit form of vertical component of the Maxwell-Faraday (induction) equation \eqref{mhd8} similarly is
\begin{align}\label{mhd23}
\frac{\partial B_z}{\partial t}-B_R\frac{\partial V_z}{\partial R}-\frac{B_{\phi}}{R}\frac{\partial V_z}{\partial {\phi}}
+V_R\frac{\partial B_z}{\partial R}+\frac{V_{\phi}}{R}\frac{\partial B_z}{\partial {\phi}}+
V_z\frac{\partial B_z}{\partial z}%\nonumber\\
+B_z\left(\frac{\partial V_R}{\partial R}+\frac{1}{R}\frac{\partial V_{\phi}}{\partial {\phi}}
+\frac{V_R}{R}\right)=0.
\end{align}
Linearization of perturbations in Eq.~\eqref{mhd23} according to Eq.~\eqref{perturb} and with use of Eq.~\eqref{boussinesq2}, including the same constraints as in Eq.~\eqref{mhd22}, gives the form
\begin{align}\label{mhd25}
-\text{i}\omega\delta B_z+\text{i}k_RB_z\delta V_R=0,\quad\quad\text{yielding (see Eq.~\eqref{mhd12})}\quad\quad 
-\text{i}\omega\delta B_z-\text{i}k_zB_z\delta V_z=0,
\end{align}
while the explicit form of azimuthal component of the Maxwell-Faraday (induction) equation \eqref{mhd8} consequently is
\begin{align}\label{mhd26}
\frac{\partial B_{\phi}}{\partial t}-B_R\frac{\partial V_{\phi}}{\partial R}-B_z\frac{\partial V_{\phi}}{\partial z}
+V_R\frac{\partial B_{\phi}}{\partial R}+\frac{V_{\phi}}{R}\frac{\partial B_{\phi}}{\partial {\phi}}+
V_z\frac{\partial B_{\phi}}{\partial z}+\frac{B_RV_{\phi}}{R}%\nonumber\\
+B_{\phi}\left(\frac{\partial V_R}{\partial R}
+\frac{\partial V_z}{\partial z}\right)=0.
\end{align}
Linearization of perturbations in Eq.~\eqref{mhd26} according to Eq.~\eqref{perturb} with use of Eq.~\eqref{boussinesq2}
and with use of Eq.~\eqref{mhd12}, including the same constraints as in Eq.~\eqref{mhd22}, gives the expression
\begin{align}\label{mhd27}
-\text{i}\omega\delta B_{\phi}-\frac{\partial V_{\phi}}{\partial R}\delta B_R-
\text{i}k_zB_z\delta V_{\phi}+\frac{V_{\phi}}{R}\delta B_R=0,
\end{align}
which can be further rearranged into the simplified form
\begin{align}\label{mhd28}
-\text{i}\omega\delta B_{\phi}-\frac{d\Omega}{d\,\text{ln}R}\delta B_R-\text{i}k_zB_z\delta V_{\phi}=0.
\end{align}
The described set of Eqs.~\eqref{mhd12}, \eqref{mhd14}, \eqref{mhd16}, \eqref{mhd20}, \eqref{mhd22}, \eqref{mhd25} and \eqref{mhd28} has to be closed
using the condition of entropy of the adiabatic perturbations (where $s$ is the density of the entropy \eqref{int})
\begin{align}\label{mhd29}
\frac{\text{d}s}{\text{d}t}=\frac{\partial s}{\partial t}+\vec{V}\cdot\vec\nabla s=0.
\end{align}
From the first law of thermodynamics ($\text{d}\epsilon=c_V\text{d}T$, $c_V$ is specific heat at constant volume) for
the ideal gas applies
\begin{align}\label{mhd30}
\text{d}s=c_V\frac{\text{d}T}{T}-\mathcal{R}\frac{\text{d}\rho}{\rho}=c_V\left(\frac{\text{d}P}{p}-\gamma\frac{\text{d}\rho}{\rho}\right).
\end{align}
By integrating Eq.~\eqref{mhd30} in case of monoatomic gas ($\gamma=5/3$) between two states ``0'' and ``1'',  we obtain the expression
\begin{align}\label{mhd31}
\Delta s=c_V\,\text{ln}\left[\frac{P_1}{P_0}\left(\frac{\rho_1}{\rho_0}\right)^{-5/3}\right],\quad\text{implying}
\quad s=c_V\,\text{ln}\left(P\rho^{-5/3}\right),
\end{align}
where $s$ is the entropy of a perfect gas with constant specific heats to within an arbitrary constant of integration \citep{zelinar}.

Linearization of isentropic perturbations (i.e.,~of a process that is internally reversible and adiabatic, the entropy of a considered system therefore does not change)
in Eq.~\eqref{mhd31} according to Eq.~\eqref{perturb} with use of Eq.~\eqref{boussinesq2} and 
valid for the Boussinesq approximation gives equation in the form
\begin{align}\label{mhd32}
\text{i}\omega\frac{5}{3}\frac{\delta\rho}{\rho}+\delta V_R\frac{\partial\,\text{ln}\left(P\rho^{-5/3}\right)}{\partial R}
+\delta V_z\frac{\partial\,\text{ln}\left(P\rho^{-5/3}\right)}{\partial z}=0.
\end{align}
Further strategy is to express all perturbations in terms of $\delta V_z$ by eliminating all terms $\delta V_R$.
With use of Eq.~\eqref{mhd12} we can rewrite Eq.~\eqref{mhd32} into the form
\begin{align}\label{mhd33}
\frac{\delta\rho}{\rho}=\frac{3}{5\text{i}\omega}\delta V_z\left(\frac{k_z}{k_R}\frac{\partial\,\text{ln}\left(P\rho^{-5/3}\right)}{\partial R}
-\frac{\partial\,\text{ln}\left(P\rho^{-5/3}\right)}{\partial z}\right).
\end{align}
Combining Eq.~\eqref{mhd16} with Eq.~\eqref{mhd33} one obtains
\begin{align}\label{mhd34}
\frac{\delta P}{\rho}+\frac{B_{\phi}\,\delta B_{\phi}}{\mu_0\rho}=\frac{\delta V_z}{k_z}\left[\omega-\frac{3}{5\omega}\frac{1}{\rho}
\frac{\partial P}{\partial z}\left(\frac{k_z}{k_R}\frac{\partial\,\text{ln}\left(P\rho^{-5/3}\right)}{\partial R}
-\frac{\partial\,\text{ln}\left(P\rho^{-5/3}\right)}{\partial z}\right)\right].
\end{align}
Using the expression for $B_{\phi}$ from Eq.~\eqref{mhd28} we can write
\begin{align}\label{mhd35}
\delta V_{\phi}=\frac{\delta V_z}{\text{i}\omega}\frac{k_z}{k_R}\left(-\frac{\kappa^2}{2\Omega}+\frac{k_z^2V_{Az}^2}{\omega^2}
\frac{d\Omega}{d\,\text{ln}R}\right)\left(1-\frac{k_z^2V_{Az}^2}{\omega^2}\right)^{-1},
\end{align}
where we introduce the definition of Alfv\'en speed \citep{Bittyk} $V_{Az}^2=B_z^2/(\mu_0\rho)$ in SI units.
With use of Eq.~\eqref{mhd12} we can rewrite Eqs.~\eqref{mhd22} and \eqref{mhd25}, respectively, as
\begin{align}\label{mhd36}
\delta B_R=\frac{k_z^2}{k_R}B_z\frac{\delta V_z}{\omega},\quad\quad\delta B_z=-k_zB_z\frac{\delta V_z}{\omega}.
\end{align}
By substituting Eq.~\eqref{mhd35} and Eq.~\eqref{mhd36} into Eq.~\eqref{mhd28} one obtains
\begin{align}\label{mhd38}
\delta B_{\phi}=2\Omega\,\frac{B_z}{\text{i}\omega^2}\frac{k_z^2}{k_R}\left(1-\frac{k_z^2V_{Az}^2}{\omega^2}\right)^{-1}\delta V_z.
\end{align}
Substituting the set of Eqs.~\eqref{mhd33}-\eqref{mhd36} into Eq.~\eqref{mhd13} and making some simplifying rearrangements, we obtain the dispersion relation
\begin{align}\label{mhd39}
\tilde{\omega}^4+\frac{k_z^2}{k^2}\left[\frac{3}{5\rho}\left(\frac{k_R}{k_z}\frac{\partial P}{\partial z}-
\frac{\partial P}{\partial R}\right)\left(\frac{k_R}{k_z}\frac{\partial\,\text{ln}\left(P\rho^{-5/3}\right)}{\partial z}-
\frac{\partial\,\text{ln}\left(P\rho^{-5/3}\right)}{\partial R}\right)-\kappa^2\right]\tilde{\omega}^2%\nonumber\\
-\,4\Omega^2\frac{k_z^4V_{Az}^2}{k^2}=0,
\end{align}
where $\tilde{\omega}^2=\omega^2-k_z^2V_{Az}^2\,\,\,\text{and}\,\,\,k^2=k_R^2+k_z^2$. 
We can further simplify this relation by setting 
\begin{align}\label{mhd40}
\frac{\partial P}{\partial z}\frac{\partial\,\text{ln}\left(P\rho^{-5/3}\right)}{\partial R}=
\frac{\partial P}{\partial R}\frac{\partial\,\text{ln}\left(P\rho^{-5/3}\right)}{\partial z},
\end{align}
which follows from the assumption of rotation on cylinders, or equivalently, that isobaric and
isochoric surfaces coincide \citep{Balbus}. 

The \textit{Brunt-V\"ais\"al\"a} frequency is defined as a frequency of the fluid
parcel with density $\rho_{\text{int}}$ that oscillates due to small displacement $\xi'=\xi-\xi_0$ around equilibrium position $\xi_0$ 
in surrounding medium with density $\rho_{\text{ext}}$, where $\xi$ denotes general
coordinate direction. If the fluid parcel is displaced along the coordinate $\xi$ and the motion is adiabatic without viscous effects, the equation of motion is
\citep{maeder}
\begin{align}\label{brunt1}
\rho_{\text{int}}\,\ddot{\xi}=-g\left[\rho_{\text{int}}-\rho_{\text{ext}}\right].
\end{align}
Expanding the right-hand side of Eq.~\eqref{brunt1} to first order in $\xi$ around the equilibrium position $\xi_0$ and rearranging this, we obtain the equation of harmonic oscillations
\begin{align}\label{brunt2}
\ddot{\xi}'+\left(\frac{g}{\rho_{\text{int}}}\frac{\partial\Delta\rho}{\partial\xi}\Bigg|_{\xi_0}\right)\xi'=0,
\end{align}
where $\Delta\rho=\rho_{\text{int}}-\rho_{\text{ext}}$.
The term in bracket in Eq.~\eqref{brunt2} represents the square of the \textit{Brunt-V\"ais\"al\"a} frequency $N_{\xi}^2$ that corresponds to the oscillations
in the direction of the coordinate $\xi$ (obviously in case of
$\partial(\Delta\rho)/\partial\xi<0$ we obtain unstable solution, diverging to infinity).

% The Brunt-V\"ais\"al\"a frequency we may express in terms of temperature gradient by eliminating $\rho=\rho(P,T)$ by using the equation of state (see Sect.~\ref{statak}). 
We assume the adiabatic behavior of fluid parcel interior \citep{Balbus,stonybal}
% expansion of the frequency gives the expression 
% \begin{align}\label{brunt2a}
% N_{\xi,\,\text{ad}}^2=g\left(\frac{\partial\,\text{ln}\,\rho}{\partial\,\text{ln}\,T}\right)_P\frac{\text{d}\,\text{ln}\,P}{\text{d}\xi}
% \,\,\left[\left(\frac{\text{d}\,\text{ln}\,T_{\text{int}}}{\text{d}\,\text{ln}\,P}\right)_\text{ad}\!\!\!\!-
% \frac{\text{d}\,\text{ln}\,T_{\text{ext}}}{\text{d}\,\text{ln}\,P}\right],
% \end{align}
% where we consider the ratio $\partial\,\text{ln}\,\rho/\partial\,\text{ln}\,T$ being the same for internal and external values, since both the logarithmic functions 
% vary slowly and basically simultaneously , 
as well as we consider that the fluid parcel is in pressure equilibrium with surrounding medium \citep{maeder}, $P_{\text{int}}=P_{\text{ext}}$ 
(this approximation is however valid only when we regard the subsonic motion). Employing the definition of the adiabatic exponent $\gamma=
(\text{d}\,\text{ln}\,P/\text{d}\,\text{ln}\,\rho_{\text{int}})_{\text{ad}}$, given in Sect.~\ref{statak}, we obtain from Eq.~\eqref{brunt2}
the adiabatic expression for the Brunt-V\"ais\"al\"a frequency in the form
\begin{align}\label{brunt2a}
N_{\xi,\,\text{ad}}^2=g\left(\frac{1}{\gamma P}\frac{\partial P}{\partial\xi}-\frac{1}{\rho}\frac{\partial\rho}{\partial\xi}\right),
\end{align}
where $\rho_{\text{ext}}$ we hence denote as $\rho$.
Applying the latter expression for the Brunt-V\"ais\"al\"a frequency for the monoatomic ideal gas in hydrostatic equilibrium, 
we can simplify Eq.~\eqref{mhd40} into the following form (for example for the piece $N_z$),
\begin{align}\label{brunt3}
-\frac{3}{5\rho}\frac{\partial P}{\partial z}\frac{\partial\,\text{ln}\left(P\rho^{-5/3}\right)}{\partial z}=N_z^2.
\end{align}
We thus obtain the expression for the piece (it is not the vector component) of the Brunt-V\"ais\"al\"a frequency that corresponds to vertical
oscillations.
Analogously we obtain the expression for the piece of the Brunt-V\"ais\"al\"a frequency that corresponds to radial oscillations, $N_R$.
The quantities $N_R$ and $N_z$ are the parts (pieces) of overall scalar quantity $N$, one can write
(see Eq.~\eqref{brunt3})
\begin{align}\label{brunt4}
N^2=-\frac{3}{5\rho}\left(\vec\nabla P\right)\cdot\left[\vec\nabla\,\text{ln}\left(P\rho^{-5/3}\right)\right]=N_R^2+N_z^2.
\end{align}
Using the pieces of Brunt-V\"ais\"al\"a frequency, we can rewrite the dispersion relation \eqref{mhd39} into quite simple form
\begin{align}\label{mhd41}
\frac{k^2}{k_z^2}\tilde{\omega}^4-\left[\kappa^2+\left(\frac{k_R}{k_z}N_z-N_R\right)^2\right]\tilde{\omega}^2-
4\Omega^2k_z^2V_{Az}^2=0.
\end{align}

From Eq.~\eqref{mhd39} clearly follows that only $z$-component of magnetic field enters the dispersion relation 
(as a part of term $V_{Az}$ or as a part of term $\tilde{\omega}$), and that it is always multiplied by the wavenumber $k_z$. 
The importance of arbitrarily small magnetic fields can thus be readily understood: strong magnetic tension forces can 
be generated at sufficiently small perturbation wavelengths. We can also see that in case of absence of the magnetic field
the wavenumbers are not scaled: internal waves propagate with a frequency that depends only on the direction of the wavenumber. The presence 
of the magnetic field however enables us to establish the inverse length scale for the wavenumbers, $\Omega/V_{Az}$. 
By normalizing the components of the wavenumber $\vec{k}$ with use of the characteristic value $\Omega/V_{Az}$, 
we can completely scale the magnetic field out of the problem. 
Only the values of the wavenumbers that are relative to the scaled characteristic $\Omega/V_{Az}$ play a role, not directly 
the values of magnetic field intensity (induction).

With reference to \citet{Balbus} we can also briefly analyze the general case with nonzero radial component of magnetic field ($B_R\ne 0$).
Considering the ideal MHD Faraday's law of electromagnetic induction \eqref{mhd8}, assuming axial symmetry ($\partial/\partial\phi=0$) 
with $\Omega=\Omega(R)$, and neglecting $V_R$ and $V_z$
from the component equations \eqref{mhd21}, \eqref{mhd23}, \eqref{mhd26}, we have the only relevant field freezing equation,
\begin{align}\label{brneze1}
\frac{\partial B_{\phi}}{\partial t}=B_R\frac{\partial V_{\phi}}{\partial R}-\frac{B_RV_{\phi}}{R}=
B_R\frac{\text{d}\Omega}{\text{d}\,\text{ln}\,R}.
\end{align}
Since $B_R$ does not change with time, the solution of the equation \eqref{brneze1} is
\begin{align}\label{brneze2}
B_{\phi}(t)=B_{\phi}(0)\left[1+\frac{B_R}{B_{\phi}(0)}\frac{\text{d}\Omega}{\text{d}\,\text{ln}\,R}\,t\right].
\end{align}
We see that the presence of radial field component leads to a linear growth of $B_{\phi}$ with time in the unperturbed disk.
However, since the azimuthal field component is not present in the dispersion relation \eqref{mhd41} and the inclusion of radial field component
does not change that (the $\omega$ frequency is also not explicitly time-dependent), no generality is lost by considering only the special case
$B_R=0$.
\subsection{Analysis of limit of stability of perturbations caused by shears}
\label{limitanalyz}
Since Eq.~\eqref{mhd41} forms a quadratic relation for the squared scaled angular MRI frequency $\tilde{\omega}^2$ (and thus for $\omega^2$), 
it has to be always a real and continuous function of its parameters in that 
dispersion relation. We may further investigate the stability of the weakly magnetized disk by considering conditions in the neighborhood
of the values
$\omega^2=0$ or $\tilde{\omega}^2=-k_z^2V_{Az}^2$. In this limit the equation \eqref{mhd41} may be written as \citep{Balbus}
\begin{align}\label{stabil1}
k_R^2\left(k_z^2V_{Az}^2+N_z^2\right)-2k_Rk_zN_RN_z+k_z^2\left(\frac{\text{d}\Omega^2}{\text{d}\,\text{ln}\,R}+N_R^2+k_z^2V_{Az}^2\right)=0.
\end{align}
We may regard Eq.~\eqref{stabil1} as a quadratic equation for $k_R$, noting that this equation would not allow real solutions for $k_R$ in case
that its discriminant $D$ is negative, 
\begin{align}\label{stabil2}
D\equiv -\left[k_z^4V_{Az}^4+k_z^2V_{Az}^2\left(N^2+\frac{\text{d}\Omega^2}{\text{d}\,\text{ln}\,R}\right)+
N_z^2\frac{\text{d}\Omega^2}{\text{d}\,\text{ln}\,R}\right],
\end{align} and thereby assuring stability, since $\omega^2$ could not then pass through zero. This requirement of stability we may express as
\begin{align}\label{stabil3}
-D>0.
\end{align}
From the assumption $N_z^2>0$ it is obvious that the inequality \eqref{stabil3} applies for all nonvanishing $k_z$ 
by satisfying the inequality  
\begin{align}\label{stabil4}
\frac{\text{d}\Omega^2}{\text{d}R}\geq 0,
\end{align}
which may be regarded as the criterion of stability. In astrophysical disks this criterion is mostly violated, which leads to instability for
values of vertical wavenumber $k_z$ less than the critical value $k_{z,\,\text{crit}}$. 
The value $k_{z,\,\text{crit}}$ we obtain from Eq.~\eqref{stabil2} by setting $D=0$,
\begin{align}\label{stabil5}
\left(k_z\right)_{\text{crit}}^2=\frac{1}{2V_{Az}^2}\left\{\left[\left(N^2+\frac{\text{d}\Omega^2}{\text{d}\,\text{ln}\,R}\right)^2
-4N_z^2\frac{\text{d}\Omega^2}{\text{d}\,\text{ln}\,R}\right]^{1/2}-\left[N^2+\frac{\text{d}\Omega^2}{\text{d}\,\text{ln}\,R}\right]\right\}.
\end{align}
If $N_R^2\ll N_z^2$ (which applies in case of supersonic rotational velocities in Keplerian disks, see Sect.~\ref{isokepldisk}) 
the critical vertical wavenumber becomes
\begin{align}\label{stabil6}
\big|\left(k_z\right)_{\text{crit}}\big|=\frac{1}{V_{Az}}\left|\frac{\text{d}\Omega^2}{\text{d}\,\text{ln}\,R}\right|^{1/2}.
\end{align}
Moreover, if the Brunt-V\"ais\"al\"a frequency $N^2=0$ or if it is quite negligible 
(noting that the square root of the quadratic term $(\text{d}\Omega^2/\text{d}\,\text{ln}\,R)^2$ may become negative), 
the solution of Eq.~\eqref{stabil5}
becomes  \eqref{stabil6}.
If $N_z^2=0$ (which is the important case of the disk midplane), the criterion of stability is
\begin{align}\label{stabil7}
N_R^2+\frac{\text{d}\Omega^2}{\text{d}R}\geq 0.
\end{align}
In case of supersonic Keplerian rotational velocity (where $N_R$ becomes negligible) the relation \eqref{stabil7} practically equals the criterion expressed 
in Eq.~\eqref{stabil4}.
\subsection{Magnetorotational instability in case of thin isothermal Keplerian disk}
\label{isokepldisk}
The full set of equations of disk vertical structure is already introduced in Sect.~\ref{thindisk}. For the brief overview: 
from the hydrostatic equilibrium condition in Keplerian isothermal 
($a=\text{const.}$) case one obtains Eq.~\eqref{thinvertigo}, while
% \begin{align}\label{isoth2}
% \frac{\partial\rho}{\partial z}=-\rho\frac{\Omega^2z}{a^2},\quad\quad\text{yielding}\quad\quad 
% \rho=\rho(R)\text{e}^{-\frac{\Omega^2z^2}{2a^2}}=\rho(R)\text{e}^{-\frac{z^2}{H^2}}.
% \end{align}
Eq.~\eqref{kepscaleheight} defines the disk scale height $H^2=a^2/\Omega^2$. From Eq.~\eqref{brunt3} directly 
follows that at the vertical distance $z=H$ applies
\begin{align}\label{isoth3}
N_z^2=\frac{3}{5\rho}\left(\rho\Omega^2H\right)\left(\frac{\partial\,\text{ln}\,P}{\partial z}\Bigg|_{z=H}-
\frac{5}{3}\frac{\partial\,\text{ln}\,\rho}{\partial z}\Bigg|_{z=H}\right)=\frac{4}{5}\Omega^2.
\end{align}
In Keplerian case applies the equality of epicyclic frequency and angular velocity $\kappa^2=\Omega^2$,
while the expression for the radial piece of the Brunt-V\"ais\"al\"a frequency 
$N_R^2$ can be set from the assumption based on Eq.~\eqref{brunt3},
where we do not apply hydrostatic equilibrium in radial direction, instead we roughly assume the similarity $\rho(R)\sim R^{-2}$ (see Sect.~\ref{radthin}, cf.~also e.g., \citet{okazaki}). We obtain
\begin{align}\label{isoth4}
N_R^2=-\frac{3}{5\rho}a^2\frac{\partial\rho}{\partial R}\left(\frac{1}{\rho}\frac{\partial\rho}{\partial R}-
\frac{5}{3\rho}\frac{\partial\rho}{\partial R}\right)=\frac{2}{5}a^2\left(\frac{1}{\rho}\frac{\partial\rho}{\partial R}\right)^2
=\frac{8}{5}\frac{a^2}{R^2}=2N_z^2\frac{H^2}{R^2}.
\end{align}
Following this equation and rather arbitrarily estimating \citep[cf.][]{Balbus} $H/R=1/10$, we set $N_R^2=0.01N_z^2$.
It is also of interest to relate the ratio of the critical wavelength $\lambda_{\text{crit}}=2\pi/k_{z,\text{crit}}$ over disk thickness to 
the magnetic field strength. From Eq.~\eqref{stabil6} we obtain
\begin{align}\label{isoth5}
\frac{\lambda_{\text{crit}}}{2H}=\frac{\pi}{\sqrt{6}}\frac{V_{Az}}{a}.
\end{align} 
Equation~\eqref{isoth5} can be expressed in terms of the the effective plasma parameter $\beta$, i.e., of the ratio 
$p_{\text{gas}}/p_{\text{mag}}=2\mu_0\rho a^2/B_z^2$. We get
\begin{align}\label{isoth6}
\beta=\frac{\pi^2}{3}\left(\frac{2H}{\lambda_{\text{crit}}}\right)^{2}.
\end{align}

\begin{figure}[t!]
\begin{center}
\includegraphics[width=4.2cm]{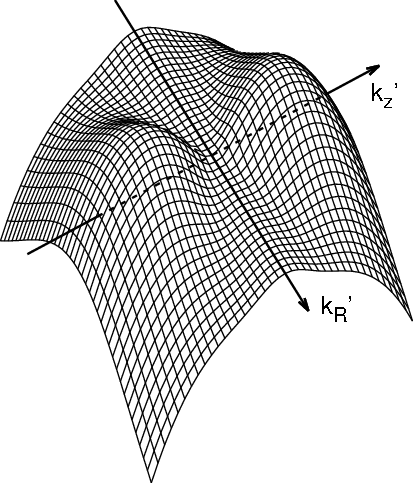}~~~~~~~\includegraphics[width=4.2cm]{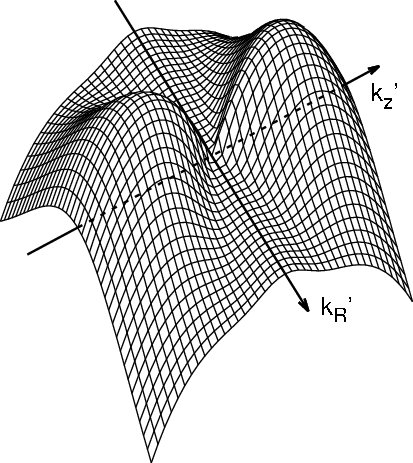}~~~~~~~\includegraphics[width=4.2cm]{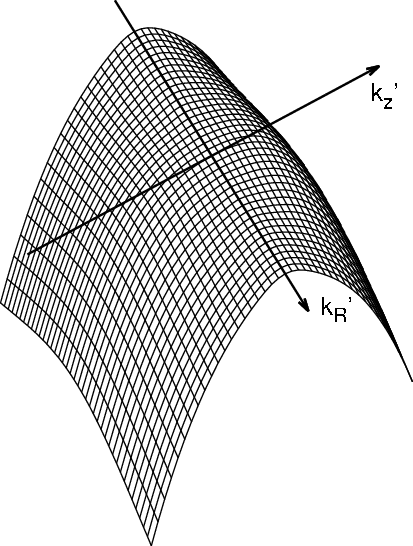}\\
\small{Fig.~\ref{ome1}a:~~$\kappa^2=\Omega^2$}~~~~~~~~~~~~~~~~~~~~~~~~\small{Fig.~\ref{ome1}b:~~$\kappa^2=0$}
~~~~~~~~~~~~~~~~~~~~~~~~\small{Fig.~\ref{ome1}c:~~$\kappa^2=5\Omega^2$}
\caption{\small{(a) 3D-plot of the unstable branch of the dispersion relation \eqref{mhd41}. Relative values of normalized dimensionless magnetorotational angular frequency 
$-\omega'^2$ are plotted in the 
${k_R}',\,{k_z}'$ (the dimensionless scaled wavenumbers) plane for a thin isothermal Keplerian disk with Brunt-V\"ais\"al\"a frequencies $N_z^2=0.8\Omega^2$ and $N_R^2=0.008\Omega^2$. 
Values of the scaled wavenumber ${k_R}',\,{k_z}'$ extend from $-2.0$ to $1.9$ in units of $\Omega/V_{Az}$, grid contours are $0.1$ units apart.
The mounds of positive values (the axes are pictured in zero $xy$ plane) 
represent regions of instability. (b) As in Fig.~\ref{ome1}a but with zero epicyclic frequency, corresponding to 
angular momentum conserving angular velocity curve (where the rotational velocity $V_\phi\sim R^{-1}$). The mounds of positive values, representing regions of instability, are yet higher and
more peaked than in the case of Keplerian rotation. (c) As in Fig.~\ref{ome1}a but with an hypothetic epicyclic frequency equal to $5\Omega^2$, 
corresponding to radially increasing angular velocity. The disk is obviously stable. The idea is adapted from \citet{Balbus}}.}
\label{ome1}
\end{center}
\end{figure}
In the (schematic) plot (Fig.~\ref{ome1}) of the dispersion relation \eqref{mhd41} we use the natural scaling of this problem 
described in Sect.~\ref{sheainst}:
the normalized (dimensionless) components of the wave vector $\vec{k}$ are denoted with prime, i.e.,
$k_z'=k_zV_{Az}/\Omega$, $k_R'=k_RV_{Az}/\Omega$ and the dimensionless magnetorotational angular frequency $\omega'=\omega/\Omega$.
Using this scaling and the values of the pieces of the Brunt-V\"ais\"al\"a frequency $N_R$ and $N_z$
derived in this section, we employ three different values of the epicyclic frequency $\kappa$ in Eq.~\eqref{mhd41}, respectively: the Keplerian value $\kappa^2=\Omega^2$, 
the value $\kappa^2=0$ that corresponds to angular momentum conserving region $\Omega\sim R^{-2}$ 
and the (hypothetic) value $\kappa^2=5\Omega^2$ \citep{Balbus}
that corresponds to outwardly increasing shear (implying the
outwardly increasing angular velocity $\Omega\sim R^{1/2}$).
Fig.~\ref{maxik} plots the section through the region of maximum scaled dimensionless negative magnetorotational instability frequency $-\omega'^2$ 
in dependence on scaled vertical wavenumber $k'_z$, corresponding to $k_R=0$ for the Keplerian case $\kappa^2=\Omega^2$ and the angular momentum 
conserving case $\kappa^2=0$. The plots are calculated using Eq.~\eqref{mhd41} with the corresponding values of $\kappa,\,N_R,\,\text{and}\,N_z$ (cf.
the results demonstrated in Sect.~\ref{magnetousci}).
\begin{figure}[t!]
\begin{center}
\includegraphics[width=4.8cm]{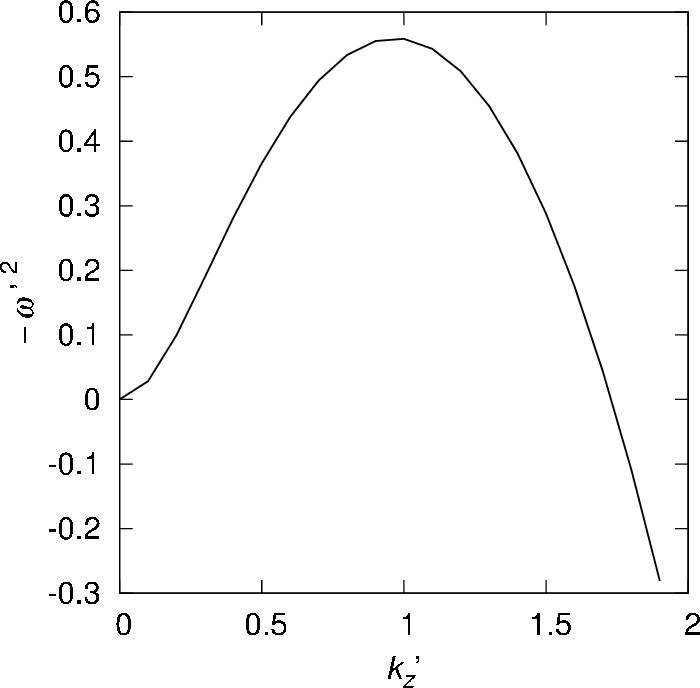}\quad\quad\quad\quad\quad\quad\includegraphics[width=4.8cm]{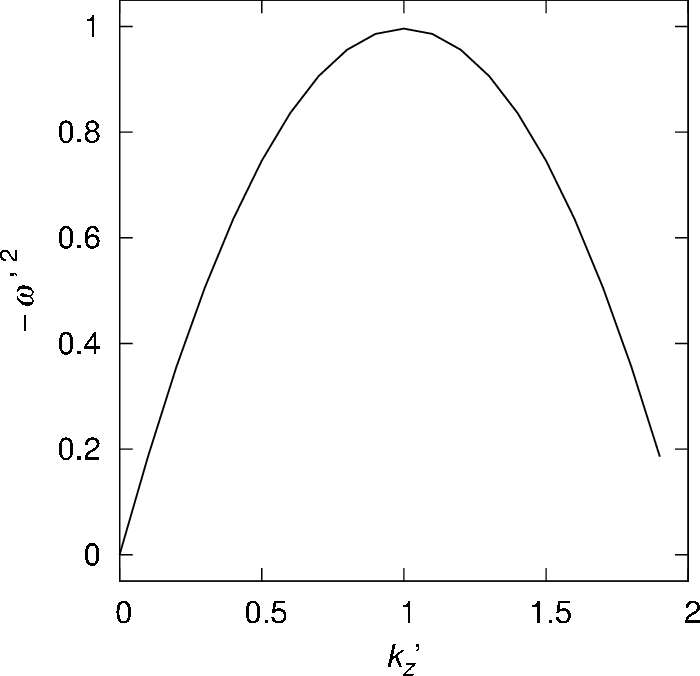}\\
\small{Fig.~\ref{maxik}a:~~$\kappa^2=\Omega^2$}\quad\quad\quad\quad\quad\quad\quad\quad\quad\quad\quad\small{Fig.~\ref{maxik}b:~~$\kappa^2=0$}
\caption{\small{(a) Plot of the section through the region of maximum $-\omega'^2$ in dependence on $k'_z$, 
corresponding to $k_R=0$ in case of Keplerian rotation, calculated using the dispersion relation 
Eq.~\eqref{mhd41}. The largest rate of $\text{i}\omega$, which corresponds to the disk equatorial plane, where $k_z^\prime$,
is approximately $0.75\Omega$. 
(b) Analogous plot of the section through the region of maximum $-\omega'^2$ in dependence on $k'_z$, corresponding to $k_R=0$ 
in case of angular momentum
conserving rotation, calculated using the dispersion relation 
Eq.~\eqref{mhd41} with zero epicyclic frequency (see the description in Sects.~\ref{sheainst} and \ref{isokepldisk}). The largest rate of $\text{i}\omega$ 
in the disk equatorial plane is in this case approximately equal to $\Omega$. The idea is adapted from \citet{Balbus}}.}
\label{maxik}
\end{center}
\end{figure}

\subsection{Local approach to MRI modeling and the rate of the increase of the instability}
\label{nonlinbalbus}
In this section we briefly reproduce the main points of the possible two-dimensional local approach to the modeling of MRI in a selected 
rectangular region (box) with a dimension of a few scale heights. There is a number of papers dealing with the problem, e.g., \citet{balda1,balda2,balda3},
we however demonstrate here the basic principles of a particular simple and instructive simulation provided by \citet{Balbus}, 
who have considered a small box with Keplerian flow in cylindrical coordinates as a computational domain
(the configuration as well as the particular result of the simulation is shown in Fig.~\ref{balbub}).

The grid is radially centered on radius $R=100$ and ranges from $R-1$ to $R+1$ with a global vertical thickness $\Delta z=1$.
They assume only radial gravitational and centrifugal forces with no vertical pressure or density gradients. The assumed Keplerian motion implies 
the initial local balance of gravitational and centrifugal forces, i.e., $\Omega^2=GM_{\!\star}/R^3$. The values of pressure and density are constant through the entire
computational box and the constant value of the specific heats ratio (adiabatic constant) $\gamma=5/3$, assuming the ideal monoatomic gas.
For a computational viability one uses there the periodic boundary conditions 
in the vertical ($z$) direction and reflecting boundary conditions on radial edges (walls) of the box
(see subroutine \textit{boundary} in Appendix~\ref{adjustik}).
\begin{figure}[t!]
\begin{center}
\includegraphics[width=9.6cm]{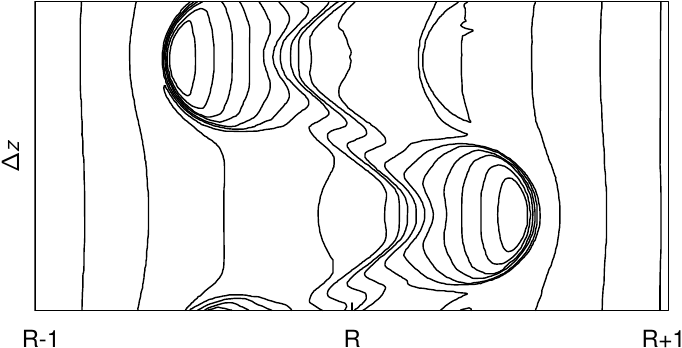}
\caption{\small{Contour plot of the simulation of the angular momentum profile taken after 3.3 orbits with following idealized parameters:
Distance of the center of plot is $R=100$, the radial range of the plot is $R-1,\,R+1$ the vertical range of the plot is $\Delta z=1$
the vertical plasma parameter in the vertical direction $\beta_z=2\mu_0\,P_g/B_z^2=1000$ (see the description in Sect.~\ref{nonlinbalbus}).
The contours denote the angular momentum values, which run from 9.91 on the left side to 10.08 on the right side (the step in one contour therefore is 1.7).
Keplerian value of the angular momentum at the center of the plot is 10. Adapted from \citep{Balbus}}.}
\label{balbub}
\end{center}
\end{figure}

Since the computational domain is a small box centered near the disk's midplane,
the radial and vertical extent of the computational grid is much smaller than its radial distance $R$ and the corresponding disk scale height $H=a/\Omega$
(the scaled sound speed $a=10^{-2}$ and the scaled angular velocity $\Omega=10^{-3}$).
The other essential scaled parameters are $GM_{\!\star}=1$ and $\rho=1$, this implies the gas pressure $P_{\text{gas}}=10^{-4}$, %(there is likely an error in the B\&H paper, 
%setting $p_g=10^{-5}$) and center the grid to $R=100$, this sets the Keplerian velocity at this point equal 
the ratio $a^2/\Omega^2=10^2$, and the local disk scale height $H=10$.
Because there act no forces in the vertical direction, the physically relevant length and time scales are determined by the angular velocity $\Omega$, by the length of the vertical periodicity
$\Delta z$ and by the poloidal component of magnetic field $B_z$ (in this case particularly vertical, since $B_R=0$). This component is 
obtained from Eq.~\eqref{isoth6} by setting the plasma parameter $\beta=1000$. 
We obtain the poloidal component of the Alfv\'en speed $V_{Az}$ consequently from Eq.~\eqref{isoth5}.

The evolution of the magnetic instability is investigated by inserting the weak magnetic field into the hydrodynamically stable initial model.
The simulation starts with the uniform poloidal (vertical) magnetic field. The authors have basically examined 
several initial magnetic field configurations within the simulation.

In Sect.~\ref{sheainst} we have already shown that the instability does not depend directly on the field strength (the field can be scaled out
by normalizing the components of the wavenumber, using the ratio of the angular velocity $\Omega$
over the poloidal Alfv\'en speed $V_{Az}$). It is thus useful to employ
the normalized vector wavenumber parameter $k'_i=k_iV_A/\Omega$ (already introduced in Sect.~\ref{isokepldisk}), where $i$ can be either $R$ or $z$.
It is a priori not known, which component of the wave vector $\vec{k}$ will emerge in the numerical simulations. However, according to the selected 
conditions and parameters and after setting the pieces $N_R$ and $N_z$ of the Brunt-V\"ais\"al\"a 
frequency to zero (which corresponds to the location near the disk equatorial plane and to the region that is relatively close to the star, cf.~the 
constraints used in Sect.~\ref{magnetouscisub1}), 
the dispersion relation \eqref{mhd41} after some rearrangements (after expansion of the resulting fraction by 
the term $2k'^2\left[(\kappa/\Omega)^2+\sqrt{(\kappa/\Omega)^4+16k'^2}\right]$) explicitly is
% \begin{align}\label{nonlinbalb2}
% \omega'^4-2\omega'^2k_z'^2+k_z'^4-\frac{k_z'^2}{k'^2}\left(\frac{\kappa}{\Omega}\right)^2
% \left(\omega'^2-k_z'^2\right)-4\frac{k_z'^4}{k'^2}=0,
% \end{align}
% as a solution of quadratic equation in $\omega'^2$ there is
% \begin{align}\label{nonlinbalb3}
% \omega'^2=k_z'^2\left\{1+\frac{1}{2k'^2}\left[\left(\frac{\kappa}{\Omega}\right)^2-
% \sqrt{\left(\frac{\kappa}{\Omega}\right)^4+16k'^2}\right]\right\},
% \end{align}
% after simple rearrangement (expanding fraction by $2k'^2\left[(\kappa/\Omega)^2+\sqrt{(\kappa/\Omega)^4+16k'^2}\right]$)
% we obtain: 
\begin{align}\label{nonlinbalb4}
\omega'^2=k_z'^2\left\{1-\frac{8}{\left[\kappa^2/\Omega^2+
(\kappa^4/\Omega^4+16k'^2)^{1/2}\right]}\right\}.
\end{align}
Regarding the context of the selected parameters: Eq.~\eqref{nonlinbalb4} shows that in case of a Keplerian disk where $\kappa/\Omega=1$,
the shortest unstable scaled wavelength $k'^2=3$ (the term in brace is equal to zero). By setting this critical wavelength 
equal to the vertical size of the computational domain, $\lambda_{\text{crit}}=\Delta z=1$, we find from Eq.~\eqref{isoth5} 
(where $V_{Az}=k'\Omega/k=k'\Omega\lambda/2\pi$, see the definition of the scaled wavelength $k'$ in Sect.~\ref{isokepldisk}) that this corresponds to the plasma parameter 
$\beta=8\pi^2P_\text{gas}/(\rho k'^2\Omega^2\lambda^2)$. Inserting the values $P_\text{gas}=10^{-4},\,\rho=1,\,k'^2=3,\,\Omega=10^{-3},\,\lambda=1$, the plasma parameter  
$\beta\approx 2630$. 
In case of the selected mode with wavenumbers $k'_z=1,\,k'_R=0$ (therefore $k'=1$) and setting again the critical wavelength $\lambda_{\text{crit}}=\Delta z=1$,
we obtain from Eq.~\eqref{nonlinbalb4} the maximum scaled angular frequency of the magnetorotational instability $\text{i}\omega/\Omega=0.75$ (cf.~Eq.~\eqref{paperref4e}). 
Inserting the values $P_\text{gas}=10^{-4},\,\rho=1,\,k_z'=1,\,\Omega=10^{-3},\,\lambda=1$, one obtains the plasma parameter
$\beta\approx 8000$.

\chapter{Numerical schemes}\label{numerovizci}
\section{Stationary calculations}\label{stationarix}
In the stationary models (see Sect.~\ref{statcalc}) we solve the system of basic hydrodynamic equations that includes the continuity 
equation~\eqref{massconserve} and the radial~\eqref{radmomconserve} and azimuthal~\eqref{phimcon} components of momentum equation in 
their stationary form, using the Newton-Raphson method \citep[see, e.g,][]{tiskarna} supplemented
by the sonic point condition in terms of equation~\eqref{spcon}.
The description of the physical context of the modeling process and its resulting outcomes are given in Sect.~\ref{statcalc}.

The Newton-Raphson method represents very efficient tool for solving the system of (nonlinear) equations in general 
\citep[see, e.g.,][]{krkub,tiskarna} which is the case of our stationary differential equations system.
We write the system $P$ of $n$ equations as
\begin{align}\label{numourix0}
P_i\left(\vec{x}\right)=0,
\end{align}
where $i=1,\,\ldots\,,\,n$ and $\vec{x}$ is the $n$-dimensional vector of variables $x_j$. From Taylor expansion of Eq.~\eqref{numourix0} to first order 
we obtain the general expression for 
the $k$-th iterative step of the system of equations $P_i$ that can be written in the compact form,
\begin{align}\label{numourix1}
J\,^{k}\Delta\vec{x}\,\,^{k}=-\vec{P}\,^{k-1}\big(\vec{x}\,\,^{k-1}\big),
\end{align}
where the vector $\Delta\vec{x}$ is thus the correction of the solution for each variable $x_j$ 
with respect to the previous iterative step.
The vector $\Delta\vec{x}\,\,^{k}$ can be written in the explicit form
\begin{align}\label{numourix2}
\Delta\vec{x}\,\,^{k}=({x}^k_1-{x}^{k-1}_1,\,\ldots\,,\,{x}_n^k-{x}^{k-1}_n)^T.
\end{align}
The expression $\vec{P}\,^{k}$ in Eq.~\eqref{numourix1} is the vector of $k$-th iteration of all the system's equations ${P\,}^k_i$, while
the expression $J\,^{k}$ denotes the corresponding Jacobi matrix whose each term ${J\,}^{k}_{ij}$ can be easily analytically derived from the system of equations ${P\,}^k_i$
by setting
\begin{align}\label{numourix3}
{J\,}^{k}_{ij}=\frac{\partial{P\,}^k_i}{\partial x^k_j}.
\end{align}
To solve the system of linearized equations we use the numerical package LAPACK (Linear Algebra PACKage) \citep{anders}
that provides very efficient and facilitating computational tool for solving most of the problems connected with linear algebra
(i.e., the systems of equations, $n$-diagonal matrices, eigenvalues and eigenvectors of operators, etc.).
For further details and comparisons of numerous applications of this method and its variants in various astrophysical problems see, 
e.g., \citet{henych,auglik,hubis1,hubis2,krkub,krticka1,fialka} and many others.
\section{Time-dependent calculations}\label{timedependix}
\subsection{The equations of (magneto)hydrodynamics}\label{operous}
For the time-dependent calculations we employ the left-hand sides of hydrodynamic equations 
written in conservative form \citep[see, e.g.,][]{Normi, Hirsch, Stony1a, felda, Lev1, kurfek}
\begin{equation}\label{advektik}
\frac{\partial{f}}{\partial t}+\vec{\nabla}\cdot\vec{F}({f})\,, 
\end{equation}
where in the two-dimensional calculations the arbitrary (scalar or vector) conservative quantities
${f}$ $=$ $\rho$, $\rho\vec{V}$, $\vec{R}\times\rho\vec{V}$, $E$ 
and the flux of these quantities $\vec{F}({f})$ $=$ $\rho\vec{V}$, $\rho\vec{V}\otimes{\vec{V}}$, 
$\vec{R}\times\rho\vec{V}\otimes{\vec{V}}$, $E\vec{V}$ 
for the mass, momentum, angular momentum and energy conservation equations, respectively (where the symbol $\times$ denotes the vector product and 
the symbol $\otimes$ denotes the tensor product). The explicit conservative vector form of mass conservation equation is (see Eq.~\eqref{masscylinder001}),
\begin{align}\label{genmasscylinder001}
\frac{\partial\rho}{\partial t}+\vec{\nabla}\cdot\left(\rho\vec{V}\right)=0.
\end{align}
The explicit conservative vector form of momentum conservation equation \eqref{moment} (where the {\textcolor{BlueViolet}{blue terms}} denote the magnetohydrodynamic 
extension of basic hydrodynamic equations according to Eqs.~\eqref{mhd7}, \eqref{mhd8}, \eqref{mhd10} and \eqref{mhd11}) is
\begin{equation}\label{gengenenumermom}
\frac{\partial\big(\rho\vec{V}\big)}{\partial t}+\vec{\nabla}\cdot\big(\rho\vec{V}\otimes\vec{V}\big)
=-\vec{\nabla} P-\rho\vec{\nabla}\Phi{\color{BlueViolet}{\,\,+\frac{1}{\mu}\big(\vec{\nabla}\times\vec{B}\big)\times\vec{B}}}\quad\big(+\vec{f}_{\text{visc}}\,\ldots\,\big),
\end{equation}
where $\Phi$ is the gravitational potential and $\mu$ is the magnetic permeability. The term $\vec{f}_{\text{visc}}$
denotes the general expression for the viscous force, i.e., the divergence of the viscous stress tensor (the nondiagonal terms in tensor $T_{ij}$
from Eq.~\eqref{stresik2}) whose explicit form is given in Appendix~\ref{stressappend}. The left-hand side of Eq.~\eqref{gengenenumermom} represents the 
advection schema (the advection terms are also called the transport terms) and the right-hand side terms are the source terms (see also Eq.~\eqref{oprourix}).
The analogous form of the energy equation (see Eqs.~\eqref{energy} and \eqref{mhd10}) takes the form
\begin{equation}\label{gengenenumerenergy}
\frac{\partial E}{\partial t}+\vec{\nabla}\cdot\left[\left(E+P\right)\vec{V}\right]
=\Psi-\nabla\cdot\vec{q}\,\ldots\,,\quad\quad\text{where}\quad\quad E=\left(\rho\epsilon+\frac{\rho V^2}{2}
{\color{BlueViolet}{\,+\,\frac{B^2}{2\mu}}}\right),
\end{equation}
(cf.~Eqs.~\eqref{mhd10} and \eqref{mhd11}),
where the term $E+P$ is the enthalpy $h$ (cf.~Eq.~\eqref{statix1a}).
The quantity $\epsilon$ in Eq.~\eqref{gengenenumerenergy} is the specific internal energy, 
$\Psi$ is the dissipation function (see Eq.~\eqref{Phi} and the remarks in Sect.~\ref{energyconserve}, cf.~also \citep{Mihalas2}, 
$\vec{q}$ is the vector of the heat flux. Equation \eqref{gengenenumerenergy} can be yet supplemented 
by additional terms, for example by the heat sources other than conduction \citep[see, e.g.,][]{Hirsch}, 
i.e.,~the radiation, chemical reactions, etc. 
The magnetohydrodynamic calculations involve the induction equation, whose explicit conservative form is written in Eq.~\eqref{mhd8}. 
The system of the equations is closed by the equation of state \eqref{statix4} whose explicit form, employed in the numerical codes, is
\begin{equation}\label{gengenenustateeq}
P=\big(\gamma-1\big)\left(E-\frac{\rho V^2}{2}{\color{BlueViolet}{\,-\,\frac{B^2}{2\mu}}}\right).
\end{equation}
Some authors \citep[e.g.,][]{Normi} prefer to use merely the internal energy equation (see Eq.~\eqref{energos5}) in 
order to avoid numerical perturbations (vibrations) that may be caused by the difference of many orders of magnitude between the values of thermal and mechanical energy in some cases. 
Since the internal energy $\rho\epsilon$ is \textit{not} the conservative quantity, we prefer to consistently involve the total energy $E$ into calculations in our codes. 

In order to numerically mimic the conditions in zones with steep negative gradients of density and velocity (e.g., in compressive zones of shock waves) 
we adopt the so-called artificial viscosity (denoted here as $Q$) acting within the precisely limited regions as a bulk viscosity \citep[see also, e.g., \citealt{Caramana}]{Normi}.
Its general one-dimensional form is
\begin{equation}\label{artivis}
Q_{k\,(i)}=\rho_i(V_{k\,(i+1)}-V_{k\,(i)})[-\text{C}_1a_i+\text{C}_2\,\text{min}(V_{k\,(i+1)}-V_{k\,(i)},0)],
\end{equation}
where $V_k$ is the corresponding velocity component (in the one-dimensional case it is obviously radial component, $k\equiv R$), $a$ is the sound speed and the
lower index $i$ denotes the $i$-th spatial grid step (see Sect.~\ref{vanleerix}). The index $k$ in the artificial viscosity term $Q_k$ (since it is the scalar quantity 
with units of pressure) refers to the ``piece'', which is inserted into the corresponding momentum component equation 
(cf.~the nondimensional pieces or parts of the the Brunt-V\"ais\"al\"a frequency in Sect.~\ref{sheainst}).
The second term scaled by a nondimensional constant $\text{C}_2$ is the quadratic artificial viscosity \citep[see][]{Caramana} used in the compressive zones. 
The linear viscosity term (scaled by a constant $\text{C}_1$) is sparingly used for damping 
oscillations in stagnant regions of the flow \citep{Normi}. The artificial viscosity term $Q$ (with the physical dimension of pressure) is included in the momentum equation 
(see Eqs.~\eqref{gennumermom} and \eqref{gennumerang}) and may be also used in
the energy equation \eqref{gengenenumerenergy} (\citet{Normi}, see also subroutine~\textit{artvis} in Appendix~\ref{adjustik} for other details).

In case of one-dimensional calculations in the thin disk approximation (cf.~Sect.~\ref{thindisk}), the analogous forms of the arbitrary (scalar or vector) conservative quantities are 
${f}$ $=$ $\Sigma$, $\Sigma\vec{V}$, $\vec{R}\times\Sigma\vec{V}$.
The flux of these quantities we may express as $\vec{F}({f})$ $=$ $\Sigma\vec{V}$, $\Sigma\vec{V}\otimes{\vec{V}}$, 
$\vec{R}\times\Sigma\vec{V}\otimes{\vec{V}}$
for the mass, momentum, and angular momentum conservation equations, respectively.
Assuming axial symmetry of the thin disk (with $\partial/\partial\phi=0$ and $\partial/\partial z=0$),
all terms in the equations are only time and radially dependent within one-dimensional calculations.
The cylindrical radial part of the mass conservation equation (cf.~Eqs.~\eqref{massfluxtot}, \eqref{massconserve} and \eqref{conticylap}) is 
\begin{equation}\label{genmasik}
\frac{\partial\Sigma}{\partial t}+\frac{1}{R}\frac{\partial}{\partial R}\big(R\Sigma V_R\big)=0.
\end{equation}
The cylindrical radial component of the momentum equation (see Eq.~\eqref{cylmomrad}, cf.~also Eq.~\eqref{radmomconserve}) 
in its conservative form gives 
\begin{equation}\label{gennumermom}
\frac{\partial\big(\Sigma V_R\big)}{\partial t}+\frac{1}{R}\frac{\partial}{\partial R}\big(R\Sigma V_R^2\big)
=\Sigma\frac{V_{\phi}^2}{R}-\frac{\partial P}{\partial R}-\Sigma\frac{GM_{\!\star}}{R^2}+\frac{3}{2}\,\frac{a^2\Sigma}{R}-\frac{\partial Q_R}{\partial R},
\end{equation}
where $P$ is the isothermal gas pressure (Eq.~\eqref{stateq}). The radial piece of the artificial viscosity $Q_R$ is calculated using Eq.~\eqref{artivis},
where the density $\rho$ is replaced by the integrated density $\Sigma$. 
The explicit form of the angular momentum equation (see Eq.~\eqref{cylmomphi}, cf.~also Eqs.~\eqref{angularmomentumflux1} and \eqref{angfl1}) in this case is
\begin{equation}\label{gennumerang}
\frac{\partial}{\partial t}\big(R\Sigma V_{\phi}\big)+\frac{1}{R}\frac{\partial}{\partial R}\big(R^2\Sigma V_RV_{\phi}\big)=
R\Sigma f_{\text{visc}}^{(2)},
\end{equation}
where the term $f_{\text{visc}}^{(2)}$
denotes the density of the viscous force
in the form derived in Eq.~\eqref{angfl}. 
Because we parameterize the disk temperature profile
via Eq.~\eqref{temperature},
we do not employ the energy equation for the calculation of one-dimensional disk models using the thin disk approximation.
The system of the equations is closed by the isothermal form of the equation of state (cf.~Eq.~\eqref{statix6}), 
\begin{equation}\label{stateq}
P=a^2\Sigma.
\end{equation}
For one-dimensional time-dependent calculations we extended the
hydrodynamic code flu.f, developed by \citet{felda}.
Following \citet{Normi}, the angular momentum advection flux acts as the azimuthal component of momentum flow.
We nevertheless do not use the consistent advection 
schema \citep{Norm1}
as it is described in detail in \citet{Normi}, but employ Eq.~\eqref{angularmomentumflux1} in its explicit form (the
key idea of the \citet{Normi} advection schema is to divide 
the conservative quantities $\Sigma$, $\Sigma\vec{V}$ and $\vec{R}\times\Sigma\vec{V}$ by the local density and advect the velocities instead, see \citet{Normi} for 
details, we however use the standard advection of these conservative 
quantities).
Equations~\eqref{massfluxtot},~\eqref{gennumermom}, and~\eqref{angularmomentumflux1} (or equivalently \eqref{genmasik},~\eqref{gennumermom}, and~\eqref{gennumerang})
with use of Eq.~\eqref{angfl} are discretized using
time-explicit operator-splitting (see Eq.~\eqref{oprourix}) and finite volume method on staggered radial grids \citep[see, e.g.,][see also Sect.~\ref{vanleerix}]{Lev1,cune}. 
The advection fluxes are 
calculated on the boundaries of control volumes of these grids \citep[see, e.g.,][]{Roache, Lev2} using van Leer's monotonic interpolation 
\citep{vanleer1, vanleer2}. 

In the source steps, regarding the right-hand sides of Eqs.~\eqref{genmasik}, \eqref{gennumermom} and \eqref{gennumerang}, respectively,
we accelerate the fluid 
by the action of external forces (gravity) and internal pressure forces on the gas momenta \citep[see][]{Normi}. 
Involving the radial artificial viscosity term $Q$ (see Eq.~\eqref{artivis})
we solve the finite-difference approximations 
of the following differential equations \citep[see][]{Normi}, 
\begin{eqnarray}  \label{sourcicek1}
\left.\frac{\text{d}\Sigma}{\text{d}t}\right|_{\text{source}}&=&0,\\
\left.\frac{\text{d}\Pi}{\text{d}t}\right|_{\text{source}}&=&\Sigma\frac{V_{\phi}^2}{R}
-\frac{\partial(a^2\Sigma)}{\partial R}-\Sigma\frac{GM_{\!\star}}{R^2}+\frac{3}{2}\,\frac{a^2\Sigma}{R}-\frac{\partial Q_R}{\partial R},\label{sourcicek2}\\ 
\left.\frac{\text{d}J}{\text{d}t}\right|_{\text{source}}&=&R\Sigma f_{\text{visc}}^{(2)},\label{sourcicek3}%\\
\end{eqnarray}
where $\Pi=\Sigma V_R$ denotes the radial momentum density, $J=\Sigma RV_\phi$ is the angular momentum density,
and $f_{\text{visc}}^{(2)}$ is the second order viscosity term derived in Eq.~\eqref{angfl}.
The radial piece of the artificial viscosity $Q_R$ corresponds to Eq.~\eqref{artivis} 
where the arbitrary velocity component $V_k$ becomes the radial velocity component $V_R$.
The value of the constant we set $\text{C}_2=1.0$. 
We employ the linear viscosity term evaluated as $\text{C}_1=0.5$ only rarely. We use it only if some oscillations occur
near the inner boundary region (near stellar equatorial surface), in disks this usually happens either in case of low $\alpha(R_\text{eq})$ viscosity parameter, 
$\alpha(R_\text{eq})<0.02$, or in case of steeper temperature decrease ($p>0.2$) (see Eqs.~\eqref{temperature} and \eqref{alpvis}).

The schema of the operator splitting (multistep) method is based on the successive addition of different forces, i.e., of the single terms in
the right-hand sides of Eqs.~\eqref{gengenenumermom} and \eqref{gengenenumerenergy} in the vectorial two-dimensional case or
the right-hand sides of Eqs.~\eqref{sourcicek1}-\eqref{sourcicek3} in case of one-dimensional thin disk calculations. 
Each force is successively evaluated using the results of the preceding update \citep{Stony1a}. 
The compact vectorial representation given in Eqs.~\eqref{genmasik}, \eqref{gennumermom} and \eqref{gennumerang} can be schematically written as 
\begin{align}\label{mgui}
\frac{\text{d}f}{\text{d}t}=\mathcal{L}(f), 
\end{align}
where the operator $\mathcal{L}$ represents in general the (negative) mathematical functions of the conservative quantities
$f=\Sigma,\,\Sigma\vec{V}\,\text{and}\,\vec{R}\times\Sigma\vec{V}$, respectively.
The operator $\mathcal{L}(f)$ can be thus split into parts, $\mathcal{L}(f)=\mathcal{L}_1(f)+\mathcal{L}_2(f)+\,\ldots\,$, according to the number of terms
(forces) in right-hand sides of Eqs.~\eqref{sourcicek1}-\eqref{sourcicek3} in one-dimensional thin disk case.
The numerical form of one timestep of the operator split procedure is subdivided into $m$ time substeps, corresponding to $m$
right-hand side terms in each equation ($m$ source terms):
\begin{align}
\begin{array}{llll}\label{oprourix}
(f^1-f^0)/\Delta t & = & L_1(f^0),\\[4pt]
(f^2-f^1)/\Delta t & = & L_2(f^1),\\[4pt]
\,\,\,\,\,\,\vdots &   & \,\,\,\,\,\,\vdots\\[4pt]
(f^m-f^{m-1})/\Delta t & = & L_m(f^{m-1}), 
\end{array}
\end{align}
where $L_i$ represent finite-difference approximations of the operators $\mathcal{L}_i$, the 
superscripts refer to the sequence number of the time substep and $\Delta t$ is the time interval of the corresponding timestep (see subroutine \textit{clock} in Appendix~\ref{adjustik}). 
This method is of course merely a simplification of the exact solution 
of the nonlinear multidimensional operator $\mathcal{L}$. However, numerical experiments and tests have shown that such a multistep procedure is 
much more accurate than a single step integration. Furthermore, there is a little difference between solutions generated using different orders
of finite-difference schemes
\citep[cf.~Sect.~\ref{vanleerix}, see also][]{Hawsma,Normi}.
\subsection{Advection schema}\label{vanleerix}
We use the \textit{staggered} spatial computational grid (see Fig.~\ref{stagis}), which enables us to 
separate numerically the various ranks of physical quantities
onto two meshes which are mutually shifted by one half of a space cell width, we denote them A-mesh and B-mesh. 
For example, the vectors (and all components of tensors of odd rank) are positioned on the A-mesh nodes while scalars and the diagonal components of tensors of even rank are positioned on the 
B-mesh nodes. Different components of these quantities may be calculated on different meshes as well,
for example the radial component of an arbitrary quantity $f$ is located on the A-mesh
while the transverse (azimuthal or vertical) component of the same quantity is located on the B-mesh.
\begin{figure}[t]
\begin{center}
\includegraphics[width=8.25cm]{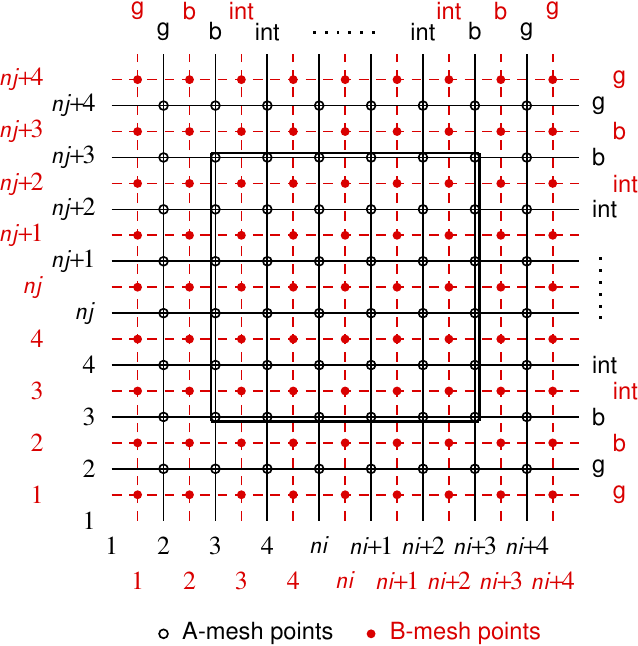}
\caption{Schema of the ~\textit{staggered} computational mesh \citep{felda}. Physical vectors like 
$\rho\vec{V}$ (all \textit{odd} tensors in general) are positioned on the A-mesh points (black colored), 
scalars like $\rho$ (generally all \textit{even} tensors) are positioned on the B-mesh points (red colored). The same quantity, if
advected in different directions, may also be positioned on different meshes (directional splitting). 
The computational domain (interior mesh points) is bound by double line, in the '{b}' zones are applied the physical 
boundary conditions and the '{g}' zones are the so-called~\textit{ghost zones} which are used to apply the boundary conditions in case of
second (or higher) order schemes (i.e.,~the schemes which are of second order accuracy in space and time), or to apply symmetry relations along, e.g.,~the midplane 
(periodic boundary conditions), rotational axis, etc. 
The adjustment of the boundary and ghost zones may in detail differ in accordance to various types of boundary conditions.}
\label{stagis}
\end{center}
\end{figure}

The principle of the calculation of the spatial derivative of the quantity $f$ in advection equation \eqref{advektik}  
(we use a scalar quantity located on B-mesh in Eq.~\eqref{vanlicek}) we demonstrate on the one-dimensional example:
we introduce two difference quotients \citep{felda},
\begin{align}\label{vanlicek}
 \Delta_-^{\text{A}}=\frac{f_{i}^{\text{B}}-f_{i-1}^{\text{B}}}{x^{\text{B}}_{i}-x^{\text{B}}_{i-1}},\quad
 \Delta_+^{\text{A}}=\frac{f_{i+1}^{\text{B}}-f_{i}^{\text{B}}}{x^{\text{B}}_{i+1}-x^{\text{B}} _{i}},
\end{align}
where the symbols $\Delta_-$, $\Delta_+$ are the left-sided and right-sided difference quotients at the location $i$ (or $j$, $k$ in case of another 
coordinate directions) of the grid.
For the calculation of the complete spatial step we employ the \textit{van Leer} derivative \citep{vanleer1,vanleer2,felda}, defined as
\begin{align}\label{valda1}
 \text{d}_{vL}^{\text{B}}=\left\{
\begin{array}{cl}
  \displaystyle\langle\Delta_-\Delta_+\rangle=\frac{2\Delta_-\Delta_+}{\Delta_-+\Delta_+},&\displaystyle\quad\text{if}\quad\quad\Delta_-\Delta_+>0\\[12pt]
  \displaystyle\,\,\,\,\,\,\,\,\,\,\,\,0,&\displaystyle\quad\text{if}\quad\quad\Delta_-\Delta_+<0\,.
\end{array}
  \right.\,
\end{align}
According to the upper row in Eq.~\eqref{valda1}, the van Leer derivative is nonzero if the function $f$ is monotonic while it is
zero in those spatial grid cells where the function $f$ is extremal (lower row in Eq.~\eqref{valda1}). The important property of the van Leer derivative is that it maintains 
the monotonicity of the linearly interpolated function and prevents the creation of the new local extrema before reaching the corresponding 
interface of the neighboring cell: from Eq.~\eqref{valda1} follows that if the left-sided and the right-sided difference quotients are approximately equal,
$\displaystyle\Delta_-\approx\Delta_+\approx\Delta$, then their average product $\displaystyle\langle\Delta_-\Delta_+\rangle\approx\Delta$. 
On the other hand, if there is a sharp inequality
$\displaystyle\Delta_-\ll\Delta_+$ or $\displaystyle\Delta_-\gg\Delta_+$, then the average product of the differences
$\displaystyle\langle\Delta_-\Delta_+\rangle\approx 2\,\text{min}\left(\Delta_-,\Delta_+\right)$.
The zero values of van Leer derivative in case of different signs of the two $\Delta$ 
difference quotients thus prevent the overshooting of the modeled function at local extrema (see Fig.~\ref{vanda}, see also \citet{vanleer1,vanleer2,felda} for details).

For the calculation of time derivatives we use the two-step algorithm, traditionally called in general the 
\textit{predictor-corrector} method \citep[see, e.g.,][among others]{richter,Hirsch}, where the calculation of the timestep is divided
into two substeps: 
the explicitly calculated \textit{predictor} step, followed by the implicit \textit{corrector} step.
The partial result of the first (predictor) step is represented by the quantity $I$ that we call the \textit{interpolant} 
(or interface interpolant), which is thus the product of the advection of the original quantity $f$
on an interface of the other mesh, i.e., from the B-mesh on the A-mesh or vice versa. The explicit one-dimensional form of the predictor
step in case of the scalar quantity $f$ is 
 \begin{align}\label{predator}
   I_{i}^{\text{A},\,n+a}=f_{i-1}^{\text{B},\,n}+\text{d}_{vL}^{\text{B}}\left(x^{\text{A}}_{i}-x^{\text{B}}_{i-1}-\frac{V_{i}^{\text{A},\,n+a}\Delta t}{2}\right),
 \end{align}
where the superscript $n+a$ represents the partially updated timestep $n$ (the quantities are evaluated within the source step, 
see also Eq.~\eqref{multikorektorix} and Sect.~\ref{koropticek} for details).
In case of a vector quantity shifts the point indexed as $\text{A}(i)$ to $\text{B}(i)$ while the point $\text{B}(i-1)$ shifts to $\text{A}(i)$.
The quantity $I^\text{A}$ in Eq.~\eqref{predator} is the interface interpolant of the scalar quantity $f$ advected in partial timestep, 
$x^\text{A}$ and $x^\text{B}$ are the coordinates of the $\text{A}$-mesh and $\text{B}$-mesh points  
and $\text{d}_{vL}$ is the van Leer derivative from Eq.~\eqref{valda1}. The local velocity of advection $V^\text{A}$ is partially updated within the timestep 
via the velocity update which is inserted 
between source and advection steps blocks (see Fig.~\ref{hydroloop}).
The role of particular terms in Eq.~\eqref{predator} is instructively shown in Fig.~\ref{vanda}.
The following corrector step we can describe as
\begin{align}\label{korektorix}
  f_i^{\text{B},\,n+1}=f_i^{\text{B},\,n}-\frac{\Delta t}{x^{\text{A}}_{i+1}-x^{\text{A}}_{i}}
  \left(I_{i+1}^{\,\text{A},\,n+a}V^{\text{A},\,n+a}_{i+1}-I_{i}^{\,\text{A},\,n+a}V^{\text{A},\,n+a}_{i}\right).
\end{align}
After executing the corrector step, the scalar quantity $f$ is advected back on the B-mesh, i.e.,~it is centered back
between the points $\text{A}(i+1)$ and $\text{A}(i)$. We may thus regard Eq.~\eqref{korektorix}
as the numerical form of one-dimensional divergence.
\begin{figure}[t]
\begin{center}
\includegraphics[width=15cm]{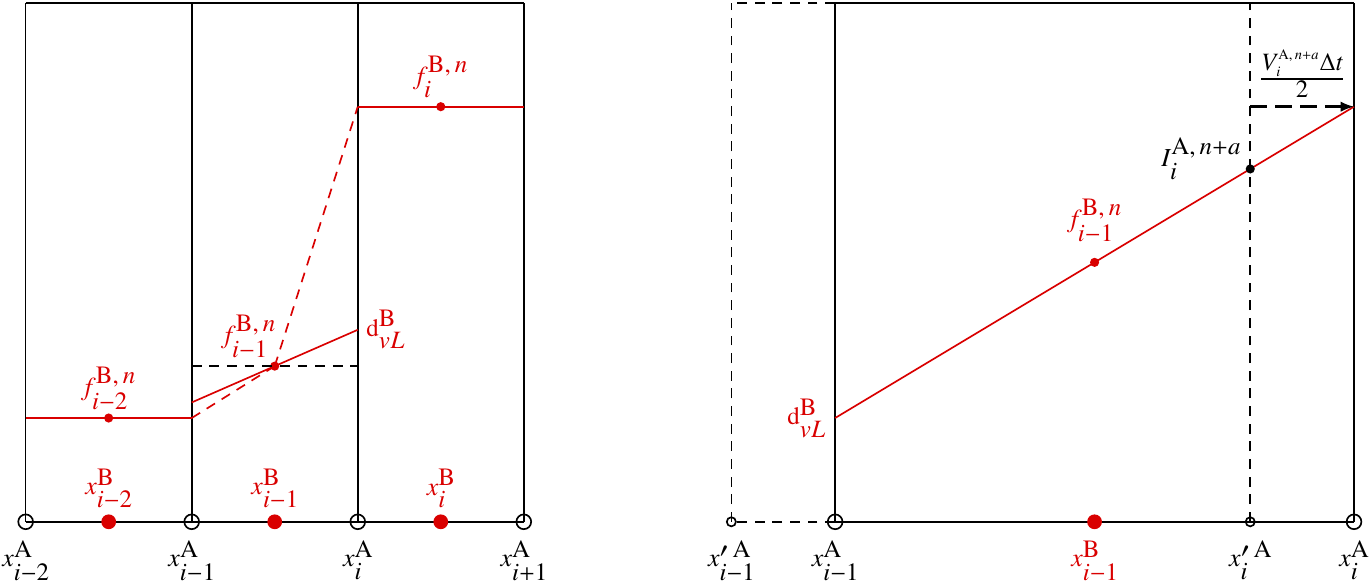}
\caption{Left panel: the schematic picture of the monotonicity condition of the van Leer second order accurate scheme given in Eq.~\eqref{valda1}. 
The slope of the linear distribution in the middle computational cell (dashed lines) is due to van Leer derivative (solid line denoted as $\text{d}_{vL}^{\text{B}}$) reduced so that the values
of a linearly interpolated advected scalar quantity $f$ on a cell interface must fall, over a full cell width, between the volume \textit{averages} of the quantity $f$ at 
the two neighboring zones. Adapted from \citet{vanleer1}. Right panel: the predictor step of 
the advection of the arbitrary scalar quantity $f$, given in Eq.~\eqref{predator}. A quantity $f$ is linearly interpolated (the solid red line, denoted as 
$\text{d}_{vL}^{\text{B}}$, depicts the slope of the 
van Leer derivative) and it is advected on a cell interface in a half-time step $t+\Delta t/2$. The solid cell boundary depicts the fluid parcel advected within the time $\Delta t/2$
while the dashed cell (with primed coordinates) is fixed in space. The position of the linear interpolant $I$ is denoted $I_i^{\text{A},\,n+a}$ (where $n+a$ 
means the partially updated time within the source step 
calculation, see the comments in Sect.~\ref{operous} for details).
The following corrector step, see Eq.~\eqref{korektorix}, advects the quantity on the center of the B-mesh in time $t+\Delta t$. Adapted from \citet{vanleer1}.}
\label{vanda}
\end{center}
\end{figure}

The corresponding two-dimensional (or three-dimensional) numerical schema is called the computational 
\textit{method of finite volumes} or shortly the finite volume method \citep[see, e.g.,][]{Lev1,Lev2,cune}. 
The multidimensional advection is usually solved using the directional splitting (see Fig.~\ref{stagis}) which may employ a series of two one-dimensional partial updates
where each of them is performed in different direction \citep[cf.][]{Stony1a}.
The second (corrector) step is in this case calculated using the partially updated multidimensional quantities from the first (predictor) step 
(e.g., from the interface interpolant $I^{\text{A},\,n+a}_{i,j}$, cf.~Fig.~\ref{hydroloop}).
The multidimensional form of Eq.~\eqref{korektorix} 
(where the advection is calculated in the coordinate direction $i$ while the index $j$ symbolizes here all other coordinate directions that are centered on B-mesh, 
in dependence on dimension of computational grid) thus takes the form
\begin{align}\label{multikorektorix}
f_{i,j}^{\text{B},\,n+1}=f_{i,j}^{\text{B},\,n}-\frac{\Delta t}{\Omega^{\text{B}}_{i,j}}
\left(I_{i+1,j}^{\,\text{A},\,n+a}V^{\text{A},\,n+a}_{i+1,j}S^{\text{A}}_{i+1,j}-
I_{i,j}^{\,\text{A},\,n+a}V^{\text{A},\,n+a}_{i,j}S^{\text{A}}_{i,j}\right),
\end{align}
where the quantity $\Omega^{\text{B}}_{i,j}$ (control volume) denotes the volume of one cell of the numerical computational grid, 
which is centered on the B-mesh. The quantity $S^{\text{A}}_{i,j}$ (control surface) denotes the corresponding area 
of this grid cell surface whose normal vector is parallel with the coordinate direction $i$. 
The surface $S^{\text{A}}_{i,j}$ is thus located on the A-mesh in the $i$-th direction while it is centered on the B-mesh in all other 
directions.
That is, the $i$-th component of the flux of the quantity $f$ 
is perpendicular to the grid cell surface $S$ (see Sects. 
\ref{advecticek} and \ref{magorisadvectis}  as well as Sect.~\ref{flarevolsurfs}, where we describe 
the geometric relations in various coordinate systems;
see also Fig.~\ref{vanda}). In case of modeling the advection equation \eqref{advektik} for a vector quantity $\vec{F}(f)$ (see Eq.~\eqref{advektik})
is the described procedure quite analogous. We start from the A-mesh (instead of B-mesh), the predictor step transports the 
advected quantity on the B-mesh, and the consecutive corrector step returns it back to A-mesh.

The described numerical schema represents the \textit{piecewise linear method}, where the quantity derivatives are linearly
interpolated. Higher order schemes (which may be however more computationally expensive) are developed: e.g.,~the 
piecewise parabolic (third order) algorithm - PPA \citep{kowrda} or the non-grid methods based on different principle - the smooth particle hydrodynamics (SPH) method 
\citep[see, e.g.,][]{monca,cenovka}, etc.
We however developed and tested the time-dependent hydrodynamic and magnetohydrodynamic codes based on the described proven (second order accurate) method, since
it provides quite efficient and accurate computational tool that fits our purpose. 

We demonstrate in Appendix \ref{dvojdimhydrous} the schematic flow charts (including the
ordering of the hydrodynamic and magnetohydrodynamic algorithms) and give the detailed description of 
the computational principles of the numerical schemes of particular routines 
of our codes.
\section{Examples of test problems}\label{testingleerix}
Before we employed the two-dimensional hydrodynamic and magnetohydrodynamic grid time-dependent
code described in Sects.~\ref{operous} and \ref{vanleerix} in modeling the real astrophysical problems leading to serious
scientific results, we tested it properly on the typical test examples such as the Riemann-Sod shock tube
\citep[][among others]{zelinar,Stony1a}, the Rayleigh-Taylor instability \citep[see, e.g.,][]{atas,taknalej}, etc.
Within the testing process of each test example 
we always started with physically as well as geometrically simple arrangements (i.e.,~in Cartesian geometry and with the least necessary number 
of computational subroutines, usually only with
advection terms without 
external or internal forces, etc.) and gradually increased the complexity of the problem until all parts of the code are in use.
We generally use the Cartesian grid with 300-500 grid points in both directions (see subroutine \textit{mesh} in Appendix~\ref{setous}), the computational time does not exceed several minutes.

Within the one-dimensional problems we also checked the behavior of the codes via modeling the essential
spherically symmetric astrophysical problems, 
e.g.,~the Parker solar wind \citep{parkan}, the line-driven (CAK) wind \citep{CAK1}, etc. 
In these models we used 300 grid points, in case of the line-driven (CAK) wind we use the logarithmic grid (see subroutine \textit{mesh} in Appendix~\ref{setous}). 
The computational time in case of these one-dimensional models does not exceed a few tens of seconds.
We have precisely and carefully compared the models obtained from these
tests with previously reported and proven results.
\begin{figure}[t]
\begin{center}
\includegraphics[width=11cm]{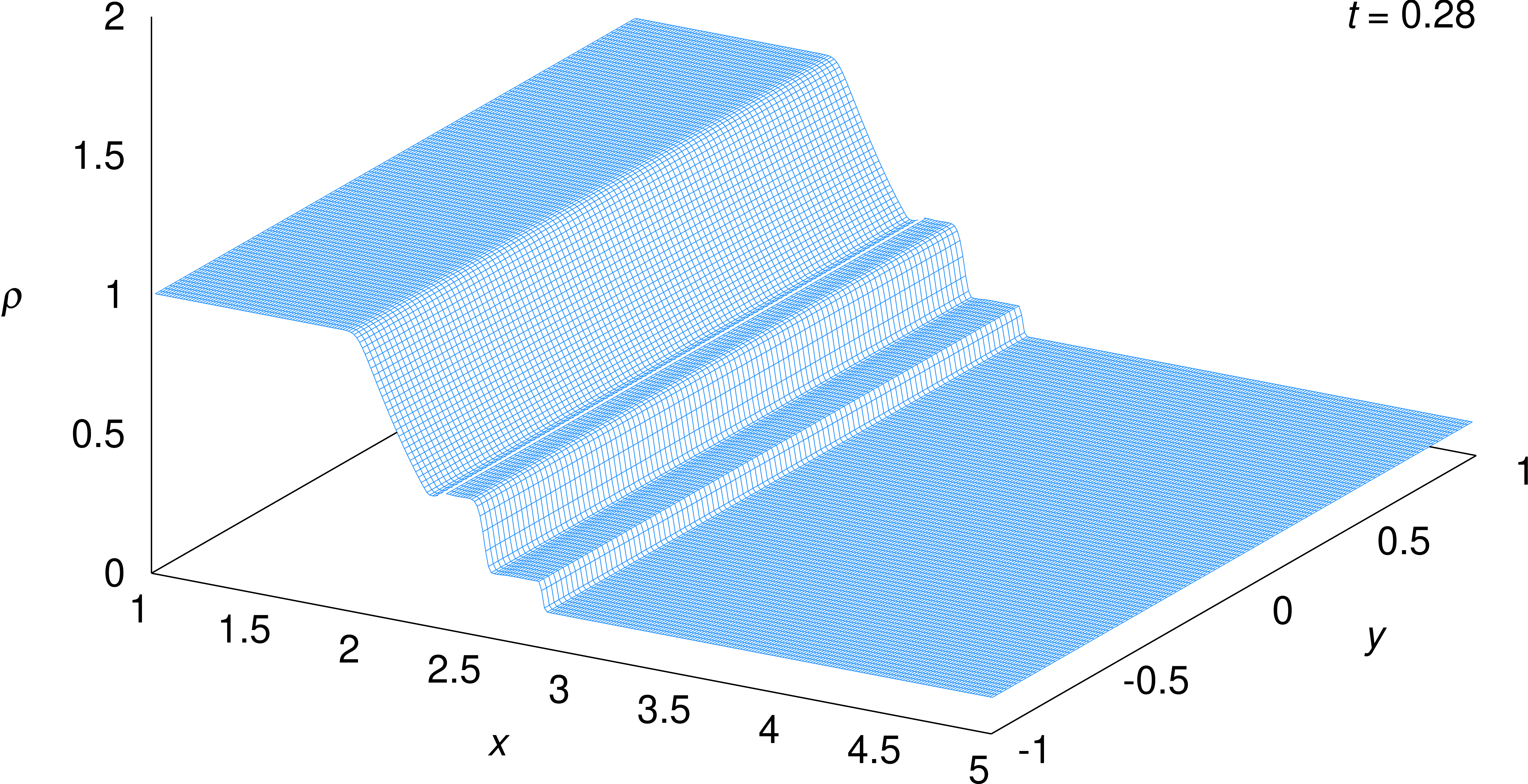}
\caption{The computational result for the density $\rho$ for the Riemann-Sod shock tube inviscid  
problem (with zero Navier-Stokes viscosity) calculated using our 
2D hydrodynamic code in time $t=0.28$ (in arbitrary units).
The static initial state of the gas is fixed by solid 
partition (diaphragm) located in 1/3 of the tube length. The initial values of the density $\rho$ and pressure $P$ on the left side of the diaphragm are 
$\rho_L=1.0$, $P_L=1.0$ while the initial values on its right side are $\rho_R=0.125$, $P_R=0.1$.
The total length $\times$ width of the tube (box) is $4.0\times 2.0$ in arbitrary units and the grid of $300\times 100$ cells is used. 
The boundary values are reflecting, that is, the ``solid walls''.
The three characteristic ``jumps'' in density are, from right to left, the shock wave (which is the fastest and its velocity exceeds the 
gas velocity, see, e.g.,~\citet{zelinar} for details), the contact discontinuity that propagates with the gas velocity and 
the rarefaction wave that propagates in the opposite direction.
The brief description of the process is given in Sect.~\ref{rymashock} \citep[cf.~the graphs of the same test problem, e.g.,~in][]{Stony1a}.}
\label{rimca2}
\end{center}
\end{figure}
\begin{figure}[t]
\begin{center}
\includegraphics[width=11cm]{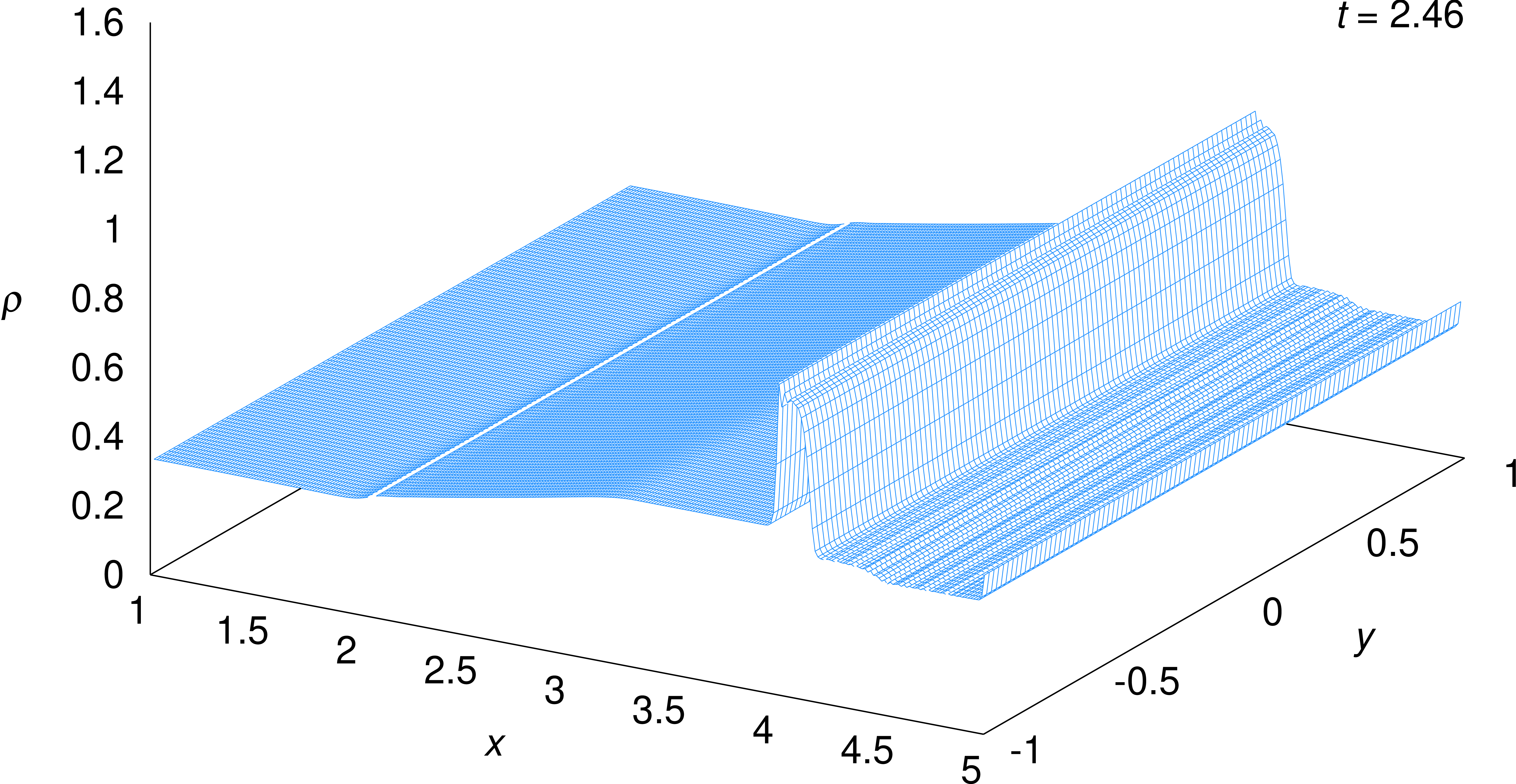}
\caption{As in Fig.~\ref{rimca2}, however in time $t=2.46$. The originally rightward propagating density shock (see Fig.~\ref{rimca2})
has been reflected by the solid wall boundary and currently 
propagates backwards while colliding with the delayed contact discontinuity wave. The originally leftward propagating rarefaction
wave causes the temporary decrease of the density in the left part of the tube.}
\label{rimca3}
\end{center}
\end{figure}
\subsection{Riemann-Sod shock tube}\label{rymashock}
The Riemann-Sod shock tube usually represents the fundamental test problem to prove and benchmark all hydrodynamic computational algorithms, first used by \citet{soda}, which 
is studied in detail with widely published and easily verifiable and comparable results. 
It is a closed tube or box divided into two compartments by a solid wall or partition called \textit{diaphragm} \citep{zelinar}, where
both compartments are filled with gas with significantly different pressure and density. Suddenly the partition disappears,
which produces the shock wave propagating in the gas perpendicularly to the original diaphragm in the direction towards the lower gas density
region accompanied by a rarefaction wave that propagates in the opposite direction.
We perform the shock tube hydrodynamic tests in two (Cartesian) modes
- with the flow characteristics (i.e.,~the density, velocity, etc.) varying only in one coordinate direction 
while in the second direction they are constant (it is in
fact only formally two-dimensional modeling), and with flow characteristics varying in both coordinate directions. The solution involves the full set of hydrodynamic equations,
i.e., the continuity equation \eqref{masscylinder001}, momentum equation \eqref{moment}, energy equation \eqref{energy} (where the thermal energy flux $\vec{q}$ 
and the influence of external forces are omitted) and equation of state \eqref{statix4}, in Cartesian coordinates.

\begin{figure}[t]
\begin{center}
\includegraphics[width=11cm]{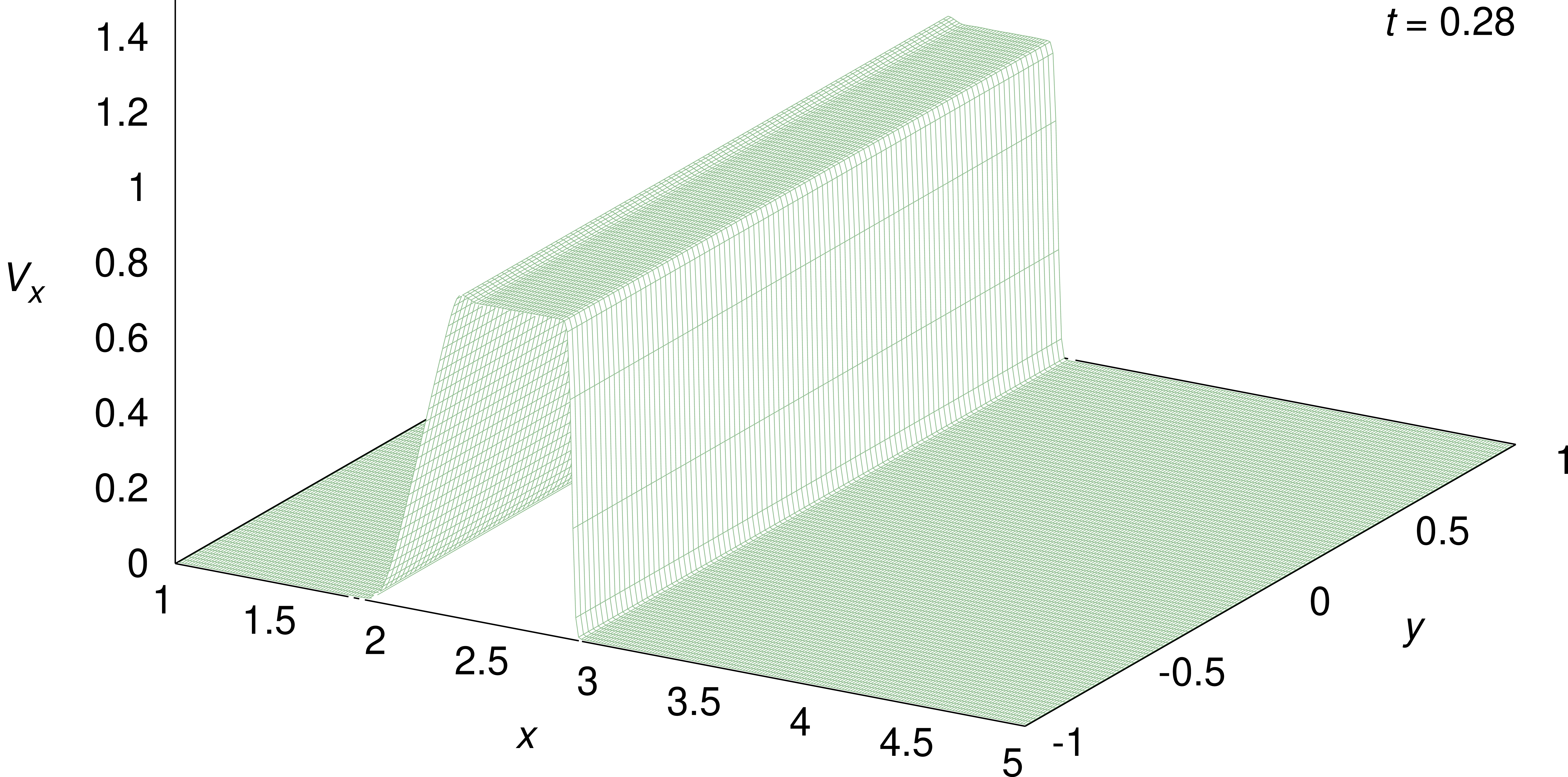}
\caption{The graph of the $x$-velocity component corresponding to the density map in Fig.~\ref{rimca2}, in time $t=0.28$ (in arbitrary units).}
\label{rimca3a}
\end{center}
\end{figure}
\begin{figure}[h!]
\begin{center}
\includegraphics[width=10.5cm]{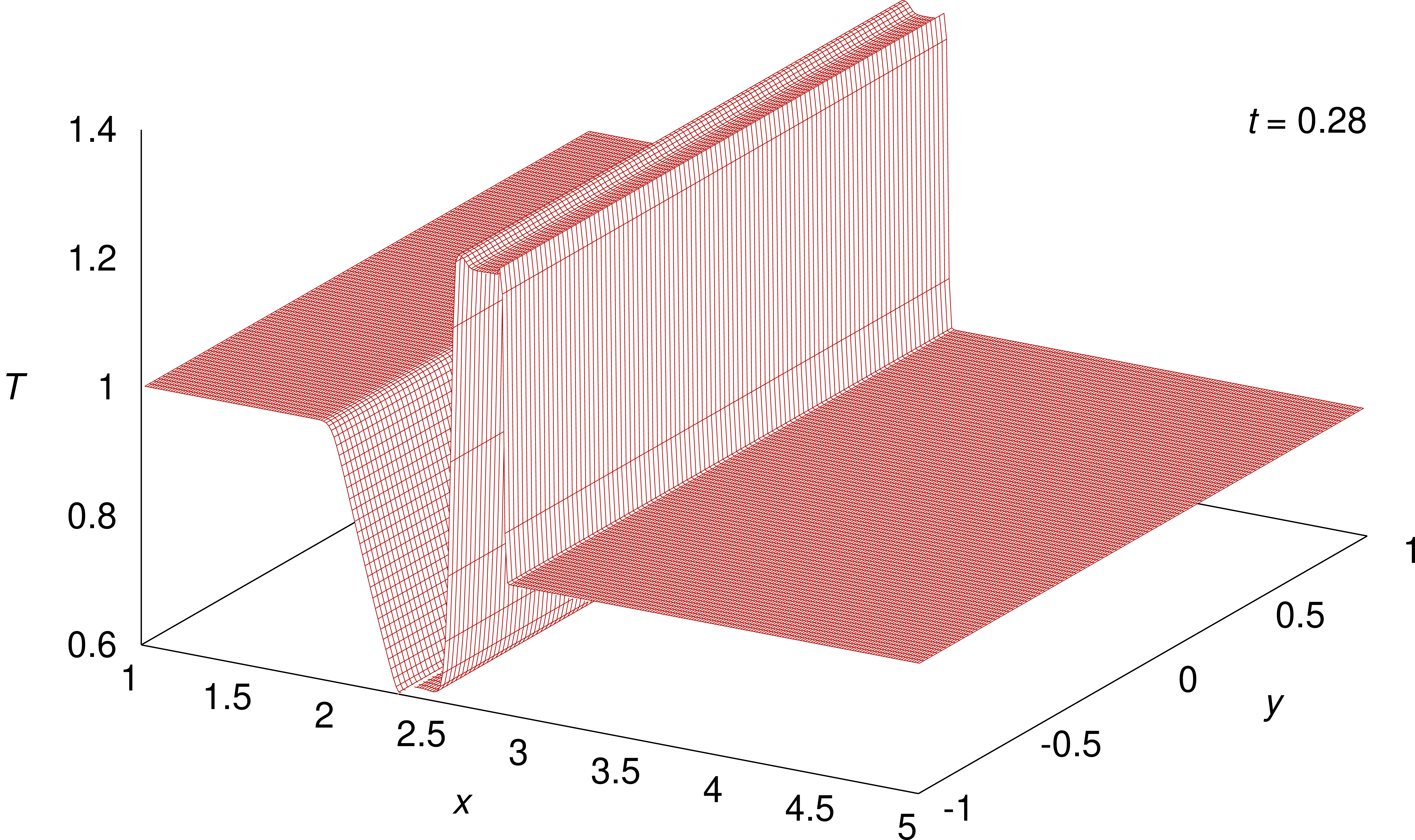}
\caption{The graph of the temperature profile of the same problem as in Figs.~\ref{rimca2} and \ref{rimca3} in the same time as in Fig.~\ref{rimca2}.
The temperature of the gas is calculated using
Eq.~\eqref{statix5} with unit values of the constants $\mu,\,m_u$ and $k$. The characteristic significant increase of the temperature
in the zone of the shock compression \citep[see, e.g.,][]{zelinar} is followed by region of the temperature drop behind this zone
\citep[cf.~the graph of the evolving temperature profile in Riemann-Sod shock tube, e.g.,~in][]{Stony1a}.}
\label{rimtempca1}
\end{center}
\end{figure}
\begin{figure}[h!]
\begin{center}
\includegraphics[width=10.5cm]{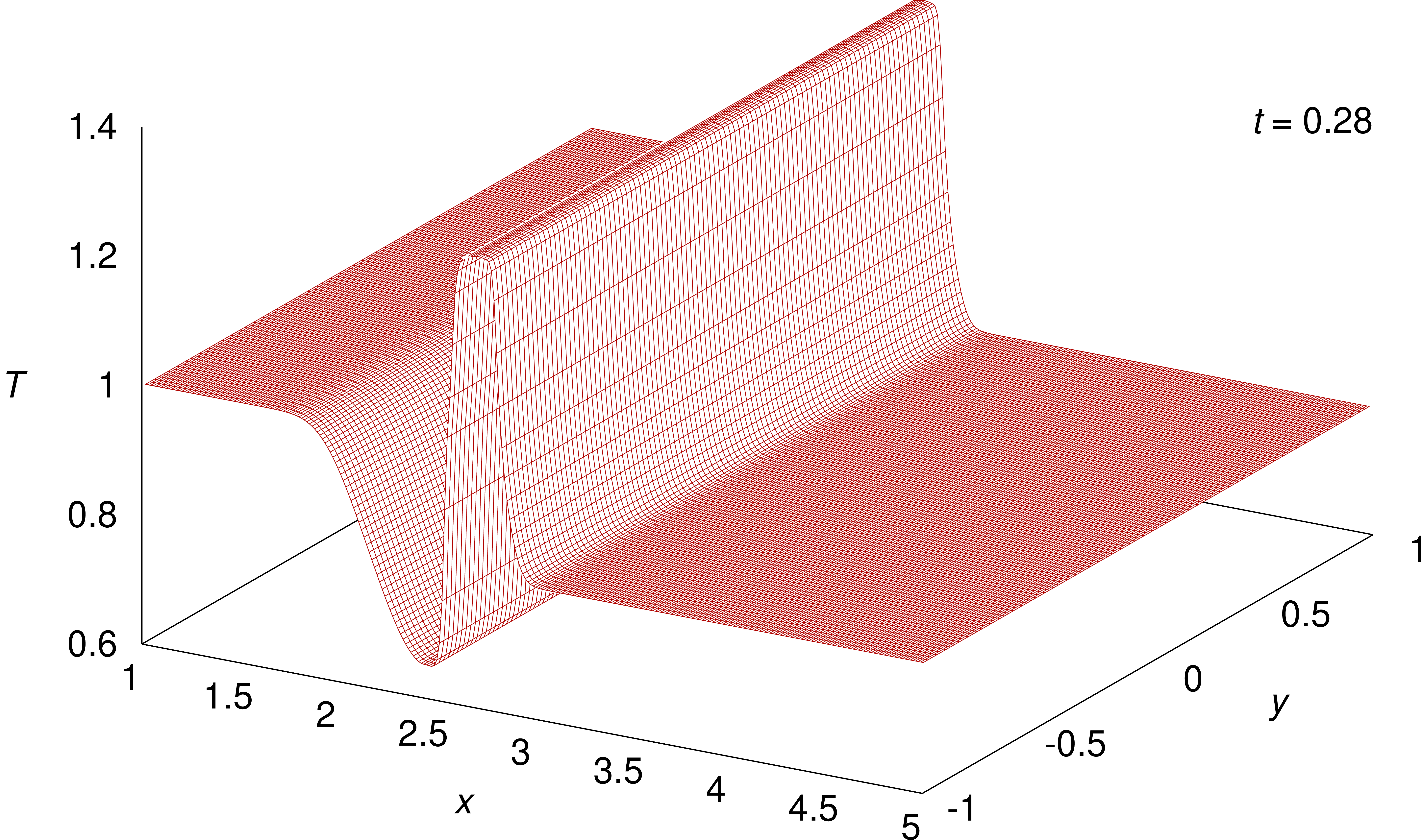}
\caption{As in Fig.~\ref{rimtempca1}, with the same values of parameters and in the same time, however 
with included Navier-Stokes viscosity and energy dissipation and with higher values of the artificial viscosity constants.
The region of increased temperature just behind the compressive zone of the shock is slightly lower and the 
temperature drop zone is shallower than in the inviscid case. The profiles of the discontinuities are 
less sharp than those in Fig.~\ref{rimtempca1}.}
\label{rimtempcavis}
\end{center}
\end{figure}

In Figs.~\eqref{rimca2}-\eqref{rimtempca1} we demonstrate two selected time snapshots of the propagating shock wave in the closed Riemann-Sod shock tube filled
with gas regarded as inviscid. 
The flow characteristics are in this case variable along the coordinate axis $x$ and constant along the coordinate axis $y$. 
The subscript $L$ denotes the left side of the tube with higher initial gas pressure and density
while the subscript $R$ denotes the right side of the tube with lower initial gas pressure and density, we keep this notation in 
all the following performed shock tube tests.
The initial state of the gas is selected (in arbitrary units) as follows:
$\rho_L=1.0$, $\rho_R=0.125$, $P_L=1.0$, $P_R=0.1$, $\gamma=5/3$, where $\rho$ is the density, $P$ is the pressure and $\gamma$ is the 
adiabatic constant. This implies the initial (squared) sound speed $a^2_L=5/3$ in the left compartment and $a^2_R=4/3$ in the right 
compartment. The total longitudinal dimension of the tube along the $x$ axis (the tube's length) is $4.0$, 
its latitudinal dimension along the $y$ axis (the tube's width) is $2.0$ in the same arbitrary units and the position 
of the diaphragm is located in $1/3$ of the tube's length. 
Since the initial gas is static in both compartments, the initial gas momenta $\rho V_x=\rho V_y=0$. The total energy $E$
of the gas is calculated using Eq.~\eqref{gengenenustateeq} omitting the magnetic pressure term (the initial values are $E_L=1.5$, $E_R=0.15$). The gas temperature is calculated from 
Eq.~\eqref{statix5}, where we set the values of constants $\mu,\,m_u$ and $k$ equal to $1$ (which implies the initial $T_L=1.0$, while the initial $T_R=0.8$ in 
corresponding arbitrary units). 
The Navier-Stokes viscosity is suppressed in this example and 
the coefficients of the artificial viscosity $Q$ (see Eq.~\eqref{artivis}) are $\text{C}_1=0.1,\,\text{C}_2=0.5$.
The boundary values are reflecting (i.e.,~the ``solid walls'', see subroutine \textit{boundary} in Appendix~\ref{adjustik}). In time $t=0.28$ (see Fig.~\ref{rimca2}), 
which is however scaled in arbitrary
units (see the description below within this section) evolves the density wave with the three 
characteristic ``jumps''. The fastest of them is the rightmost shock wave, followed by a contact discontinuity that 
propagates with the velocity of the gas, and by the leftmost rarefaction wave that propagates in the opposite direction 
\citep[see, e.g.,][]{zelinar,Stony1a} towards the solid wall that closes the shock tube (box). 
\begin{figure}[t]
\begin{center}
\includegraphics[width=11cm]{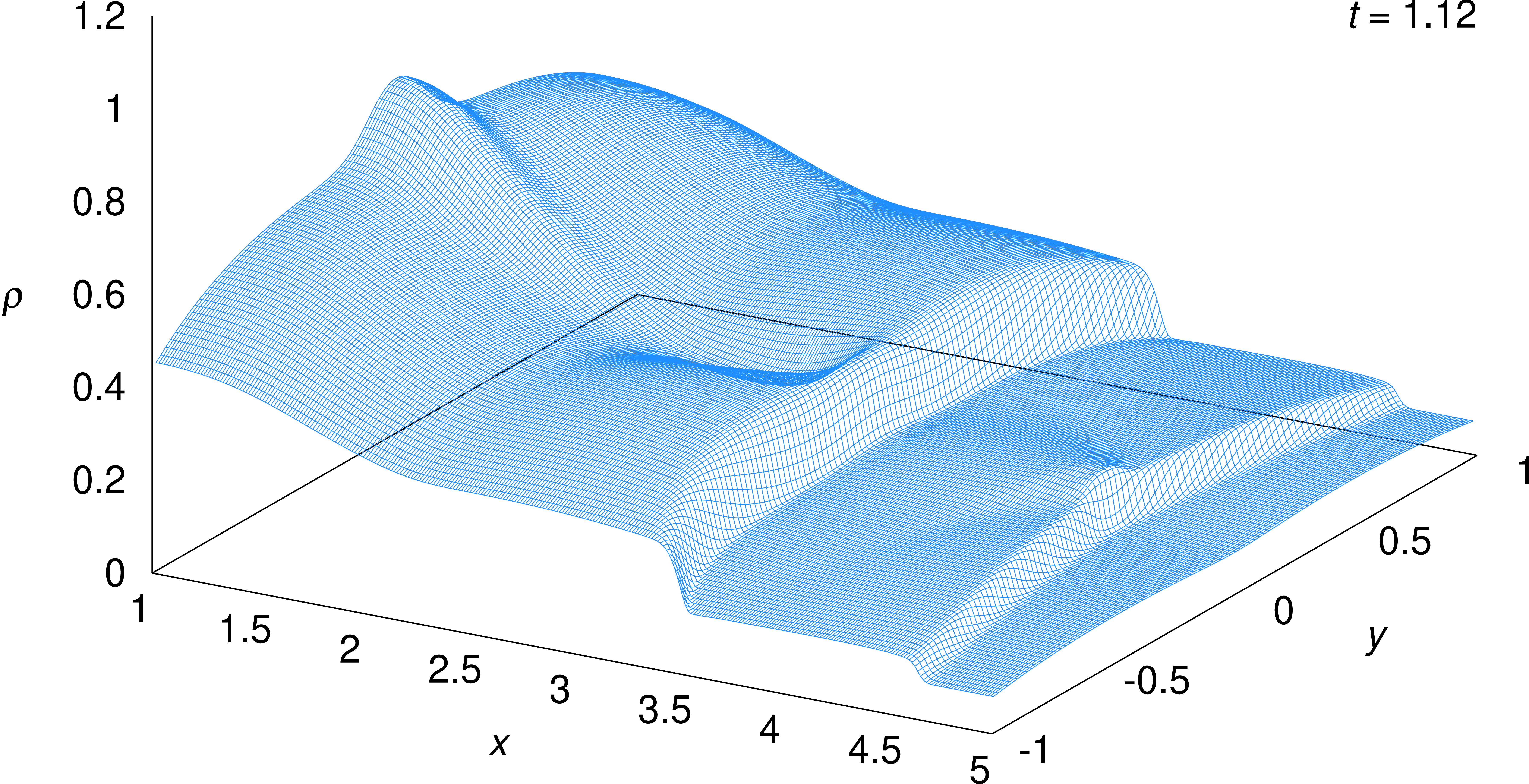}
\caption{The two-dimensional graph of the density evolution in Riemann-Sod shock tube (cf.~Fig.~\ref{rimca2}) 
where the initial density (together with other quantities) is set as varying in both coordinate directions.
The snapshot shows the flow in the moment just before the shock wave hits the ``front'' solid wall and is reflected backwards.
We use in this case the Cartesian grid with $300\times 200$ points.}
\label{rimdenwrap60}
\end{center}
\end{figure}
\begin{figure}[h!]
\begin{center}
\includegraphics[width=11cm]{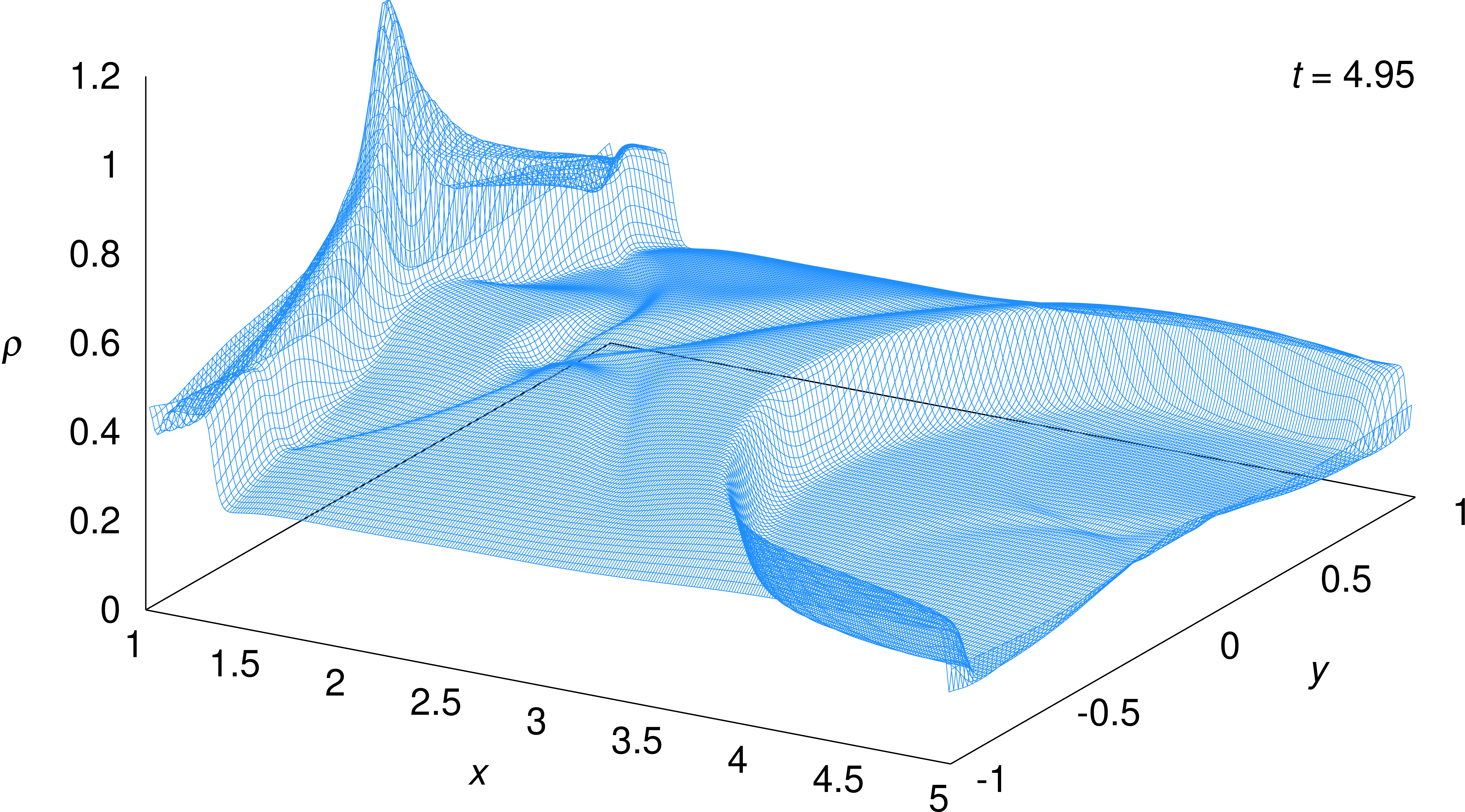}
\caption{As in Fig.~\ref{rimdenwrap60}, now in more advanced time when the density shock wave bounces off the ``back'' solid wall
and is starting for the second time to propagate rightwards.}
\label{rimdenwrap265}
\end{center}
\end{figure}

We can however briefly (since we do not introduce here the complete derivations) 
compare the numeric solution with the analytic Rankine-Hugoniot relations given by the gas dynamics of shock waves. Following \citet{zelinar} we denote $D$ the 
propagation speed of the shock wave. We use the subscripts 0,1,2,3, respectively (from right to left in the graph \ref{rimca2}), 
for the regions of the unperturbed gas in front of the shock wave ($0\equiv R$), 
between the shock wave and the contact discontinuity (1), between the contact discontinuity and the rarefaction wave (2), and the left initial state ($3\equiv L$). The analytic relations 
for the gas density and for the gas velocity in the region (1) (i.e., behind the shock front) and for the shock front propagation velocity are
\begin{align}\label{rahug1}
\frac{\rho_1}{\rho_0}=\frac{P_1+[(\gamma-1)/(\gamma+1)]P_0}{P_0+[(\gamma-1)/(\gamma+1)]P_1},\quad 
\rho_0V_1^2=\frac{2\left(P_1-P_0\right)^2\!\!/(\gamma+1)}{P_1+[(\gamma-1)/(\gamma+1)]P_0},\quad D^2=\frac{\rho_1\left(P_1-P_0\right)}{\rho_0\left(\rho_1-\rho_0\right)}.
\end{align}
We also note that the contact discontinuity separates two fluids with different density and temperature but with the same pressure
and the same velocity. The direct analytic solution is due to the nonlinearity of the problem quite intricate, if we however adopt, 
e.g., the calculated pressure $P_1\approx 0.295=P_2$, we derive from Eq.~\eqref{rahug1} the following values: $\rho_1\approx 0.23$, $V_1\approx 0.84$, $D\approx 1.85$.
Employing the equation (cf.~Sect.~\ref{statak}) for adiabatic ideal gas $P=K\rho^\gamma$ (and assuming that $K$ is the same
at least everywhere left of the contact discontinuity), we can write $\rho_2=\rho_3(P_2/P_3)^{1/\gamma}$, obtaining the value $\rho_2\approx 0.48$. 
Using the equation of state \eqref{statix5} for calculation of the temperature, we get $T_1\approx 1.23$ and $T_2\approx 0.62$. We may thus confirm that all the analytical results are 
in good agreement with the numerical solution (cf.~Figs.~\ref{rimca2} and \ref{rimtempca1}).

In Fig.~\ref{rimca3} (time $t=2.46$) the density shock wave has been reflected and is propagating leftwards. The test within the described conditions 
continues until the time $t=12$ during which the shock wave is 5 times reflected 
from the opposite boundary walls. Since the gas is inviscid (with merely the artificial viscosity included) and the energy equation
\eqref{gengenenumerenergy}
does not include the dissipation term $\Psi$, the total energy of the shock wave remains constant during the whole test.
Fig.~\ref{rimca3a} shows the $x$-velocity profile of the gas in the time $t=0.28$ while
Fig.~\ref{rimtempca1} shows the analogous snapshot of the temperature 
profile of the propagating gas in the time $t=0.28$, demonstrating the characteristic zone of
significant temperature increase in the compressive zone of the shock wave \citep[see, e.g.,][]{zelinar}
followed by region of the temperature drop \citep[cf.][]{Stony1a} in the rarefied gas behind the 
shock front, which is simultaneously the region of increasing gas velocity.
\begin{figure}[t]
\begin{center}
\includegraphics[width=11cm]{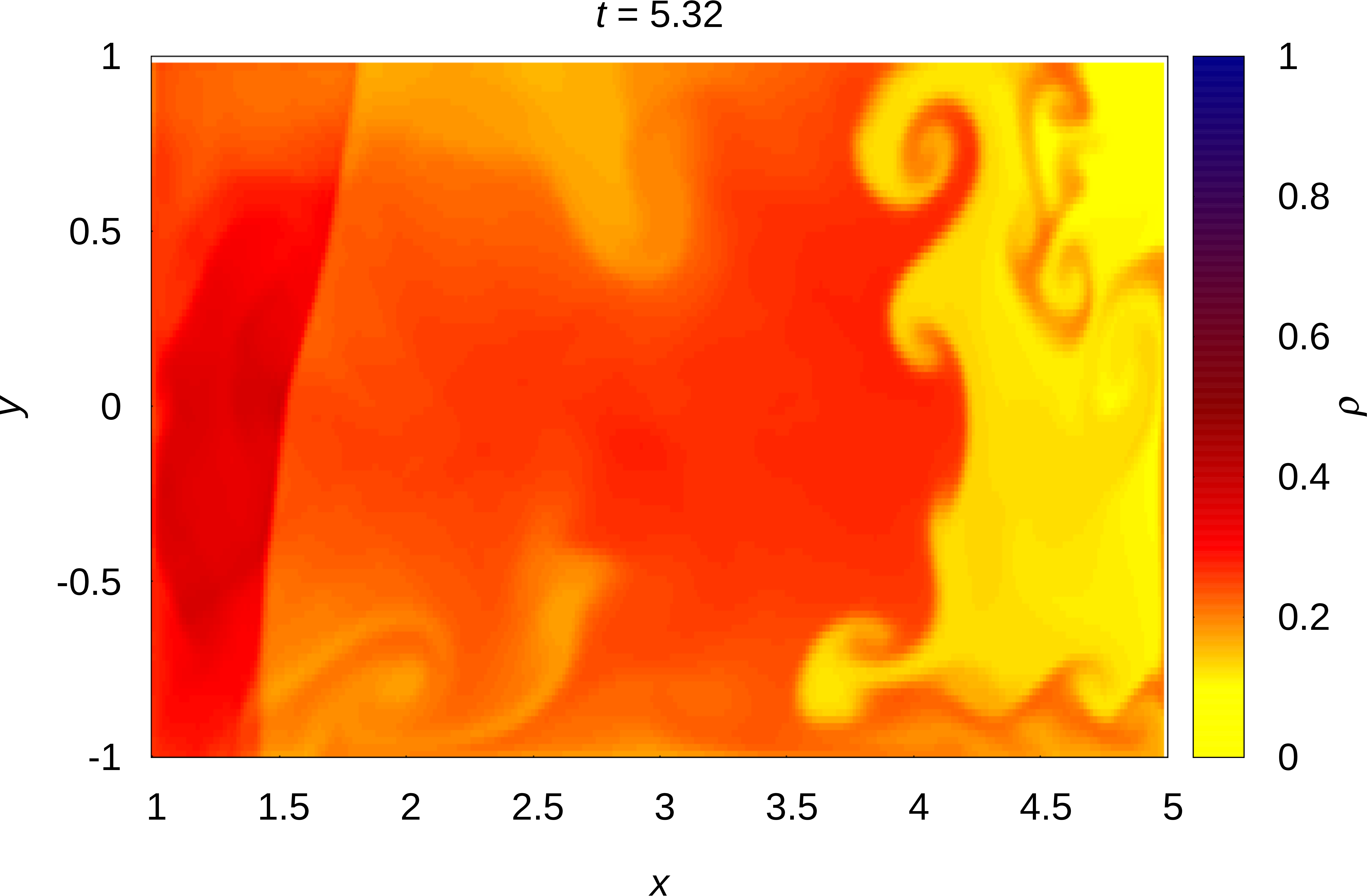}
\caption{The color map of the density profile in the Riemann-Sod shock tube in slightly more advanced time than in Fig.~\ref{rimdenwrap265}
where the small initial $y$-velocity component $V_y=0.05$ is added. This ``perturbation'' produces the latitudinal flow distortion with 
Kelvin-Helmholtz and Rayleigh-Taylor instabilities.}
\label{rimcolor285}
\end{center}
\end{figure}
\begin{figure}[t]
\begin{center}
\includegraphics[width=11cm]{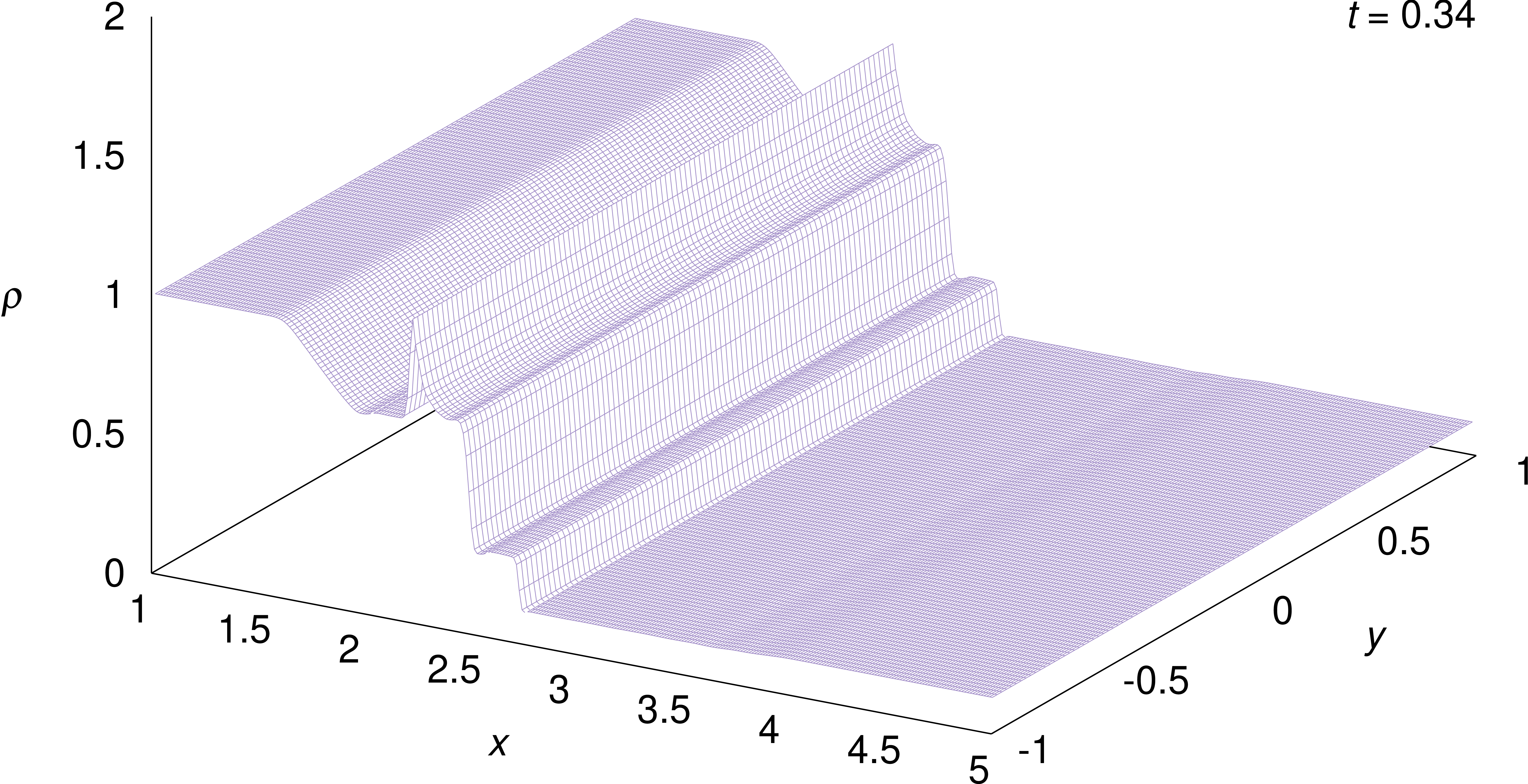}
\caption{The graph of the density profile for the same problem and approximately in the same time as in Fig.~\ref{rimca2}
with included initial internal magnetic field with following parameters (see also the more detailed description in Sect.~\ref{rymashock}):
$B_x=0.75\sqrt{\mu}$, $B_{y,L}=1.5\sqrt{\mu}$, $B_{y,R}=-1.5\sqrt{\mu}$ and $B_z=0$, producing the characteristic
density peak which moves along with the contact discontinuity zone \citep[cf.][]{Stony1b}.}
\label{magnetodensity}
\end{center}
\end{figure}
\begin{figure}[h!]
\begin{center}
\includegraphics[width=11cm]{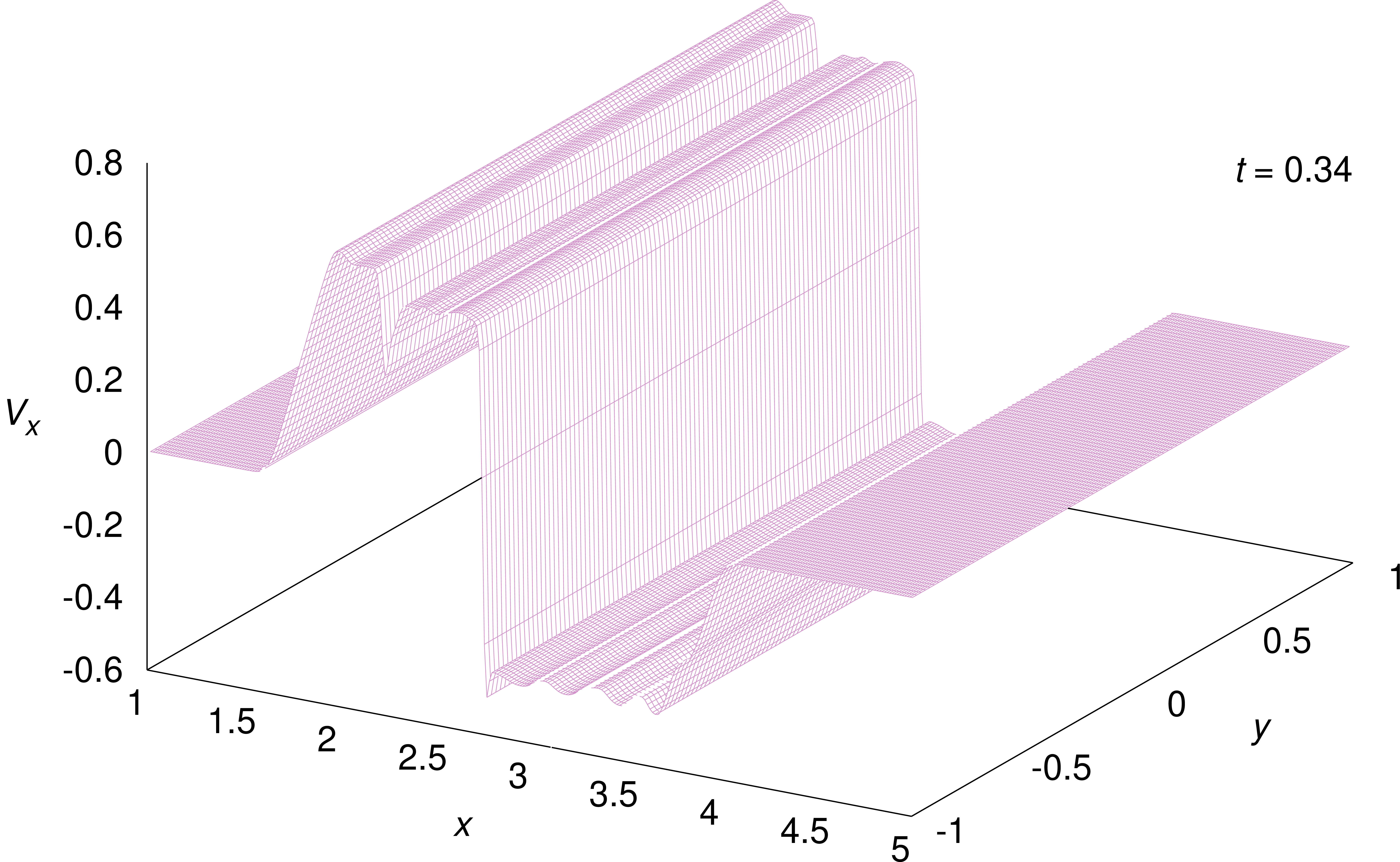}
\caption{The graph of the longitudinal flow velocity component $V_x$ for the same problem and with the same parameters as in Fig.~\ref{magnetodensity}.
The magnetic field produces the significant peak of the $x$-velocity drop, moving along with the contact discontinuity zone \citep[cf.][]{Stony1b}.}
\label{magnetoradvelo}
\end{center}
\end{figure}
\!\!\!\,Fig.~\ref{rimtempcavis} illustrates the 
temperature distribution within the same process in case of the viscous gas. 
The coefficient of the quadratic (compressive) artificial viscosity in Eq.~\eqref{artivis} 
was increased to the value $\text{C}_2=3$ and the Navier-Stokes viscosity according to Eq.~\eqref{stresik2} is involved. The value of
the coefficient of the dynamic viscosity $\eta=1$ and the coefficient 
of the bulk viscosity $\zeta=1$. The dissipation function $\Psi$ in energy equation \eqref{gengenenumerenergy} is included as well.
The sharp edges of the density and temperature discontinuities in the evolving shock front profile are clearly smoothed in this case and
the height of the temperature increase within the compressive zone is lower than in the inviscid~case.

To obtain the picture of the behavior and efficiency of the code in more complex situation, we arranged the Riemann-Sod shock tube
test example with variable characteristics in both directions $x$ and $y$, involving the following initial parameters (where we keep
the already introduced notation of the selected quantities):
$\rho_L=\text{e}^{-y^2}$, $\rho_R=0.125\,\text{e}^{-y^2}$, $P_L=\text{e}^{-y^2}$, $P_R=0.1\,\text{e}^{-y^2}$,
$\gamma=5/3$, the tube length is $4.0$, the tube width is $2.0$, the position 
of the diaphragm is in $1/3$ of the tube's length. The profiles of the density and pressure
in the lateral $y$ direction is thus ``Gaussian''; from the physical point of view this may be regarded 
as an artificial initial configuration but for the purpose of the numerical code testing 
it is however an effective arrangement that consequently leads to very complex two-dimensional morphology 
of the evolving density (together with other characteristics) shock. The solid wall boundary conditions (see Appendix~\ref{adjustik}) are 
in this case selected as well. Since the image of the test, produced in the inviscid case with 
more contrast amplitudes and discontinuities of the density perturbations, is visually more impressive, we show here exclusively the test example 
with the suppressed viscosity (except the artificial viscosity, necessary for the stability of the numerical solution). However, we have properly 
tested also cases within the range of viscosity parameters up to the values of the dynamic viscosity $\eta=2$ and 
the bulk viscosity $\zeta=1$. 

In Figs.~\ref{rimdenwrap60} and \ref{rimdenwrap265} we demonstrate the model of the evolving density
in two different times in the line contours graphs. Fig.~\ref{rimdenwrap60} shows the rightwards evolving density 
shock wave just before it hits the opposite wall, while the reflection of the ``density slope'' from side walls (which are much nearer) forms the
significant density enhancement close to the longitudinal axis. Fig.~\ref{rimdenwrap265} gives the snapshot of the evolving density wave after
the second reflection from the left closing wall.
In Fig.~\ref{rimcolor285} we show the same Riemann-Sod shock tube model in the slightly more advanced time, plotted as the density color map. 
In this model we however add another ``perturbation'' caused by small initial
latitudinal velocity component, $V_y=0.05$, in the whole computational domain. The latitudinal density profile is slightly distorted and there are obvious Kelvin-Helmholtz
(cf.~Fig.~\ref{khcolor408}) as well as Rayleigh-Taylor instabilities appearing in the flow. The numerical stability of the code is however excellent in this 
case during the whole time of computation.

As a successful test of the magnetohydrodynamic extension of our code (see Appendix~\ref{dvojdimmagnetohydrous})
we demonstrate once again the Riemann-Sod shock tube 
with the following initial internal magnetic field parameters adapted from \citet{Stony1b}: 
$B_x=0.75\sqrt{\mu}$ in the whole tube (where the numerical value of $\mu=4\pi\cdot 10^{-7}$, however, within this section we scale it in arbitrary units 
while in all other sections we scale it in SI units, $\text{Hm}^{-1}$),
$B_y=1.5\sqrt{\mu}$ in the left tube compartment while $B_y=-1.5\sqrt{\mu}$ in the right tube compartment (and $B_z=0$ everywhere).
The solution that includes the magnetic field involves the full set of magnetohydrodynamic equations,
i.e., the continuity equation \eqref{genmasscylinder001}, momentum equation \eqref{gengenenumermom}, energy equation \eqref{gengenenumerenergy} (where the thermal energy flux $\vec{q}$ 
is omitted) and equation of state \eqref{gengenenustateeq} in Cartesian coordinates.
Within the test we modeled the profiles of the gas density, pressure, temperature and velocity components with varying viscosity and 
with initial density conditions that correspond to the hydrodynamic test demonstrated in Fig.~\ref{rimca2}. We introduce here however only the density and longitudinal velocity graphs with zero 
(Navier-Stokes) viscosity with the significant density peak and with the significant peak of the $x$-velocity drop behind the contact discontinuity area  
\citep[cf.~][]{Stony1b}.
\subsection{Kelvin-Helmholtz instability}\label{Keshock}
Another frequently used test problem that takes account of the turbulent motion of the gas is the 
Kelvin-Helmholtz instability \citep[see, e.g.,][see also Fig.~\ref{khcolor408}]{chandrous,frankiea}. We modeled this instability using our time-dependent hydrodynamic code 
with the following setup.
The square domain (box) with the dimension $L\times L=1.0\times 1.0$ in arbitrary units, $0\leq x\leq 1.0$, $0\leq y\leq 1.0$, 
is filled with gas where the two oppositely directed streams are divided by a discontinuity. 
We consider the interface between the streams as a "slip surface" \citep[e.g.,][]{atas,skokis}. 
The boundary conditions are periodic on the ``upstream'' and ``downstream'' edges with $x=0$ and $x=1.0$ 
(left and right side of the depicted domain in Fig.~\ref{khcolor408}), while at the two remaining edges they are reflecting
(see Appendix~\ref{adjustik} for the detailed description of the boundary conditions).
From the problem parameters used by \citet{skokis} in instructions to the ATHENA hydrodynamic code \citep{atas} we adapted the following initial conditions:
for $y>0.5$ we set the longitudinal flow velocity $V_{x,1}=0.3$ and the gas density $\rho_1=1$, while for $y\leq 0.5$ we set $V_{x,2}=-0.3$ and $\rho_2=2$. 
The pressure $P_1=P_2=1.0$ within the whole computational domain and the adiabatic constant $\gamma=5/3$. The gas is therefore clearly subsonic, 
giving a Mach number approximately $0.232$ in the $\rho_1=1$ gas density region and approximately $0.329$ in the $\rho_2=2$ gas density region.
In order to avoid a perfectly sharp boundary in the
initial state between these two phases of the flow, we introduce a small transition region that smoothly connects them. 
It is described by the equations \citep{skokis}:
\begin{align}\label{inidenperty}
\rho(x,y)=\rho_1+(\rho_2-\rho_1)\left(1+\text{e}^{\frac{y-0.5}{\sigma}}\right)^{-1}\!\!\!\!,\quad\quad V_x(x,y)=V_{x,1}+(V_{x,2}-V_{x,1})\left(1+\text{e}^{\frac{y-0.5}{\sigma}}\right)^{-1},
\end{align}
that characterize the initial perturbation of the density field and 
% similarly
% \begin{align}\label{iniveloxperty}
% V_x(x,y)=V_{x,1}+(V_{x,2}-V_{x,1})\left(1+\text{e}^{\frac{y-0.5}{\sigma}}\right)^{-1}
% \end{align}
% that characterizes 
the initial perturbation of the $x$-velocity field in the $y$ direction,
where the dispersion of the perturbation $\sigma=0.01$. In these unperturbed initial state we now impose a so-called ``seed perturbation'' of
the velocity in the $y$-direction of the form
\begin{align}\label{iniperturbix}
V_y(x,y)=A\,\text{cos}\,(kx)\,\text{e}^{-k\,|y-0.5|}, 
\end{align}
with the wavenumber $k=2\times(2\pi/L)$ and the amplitude of perturbation $A=0.05$. 
\begin{figure}[t]
\begin{center}
\includegraphics[width=11.5cm]{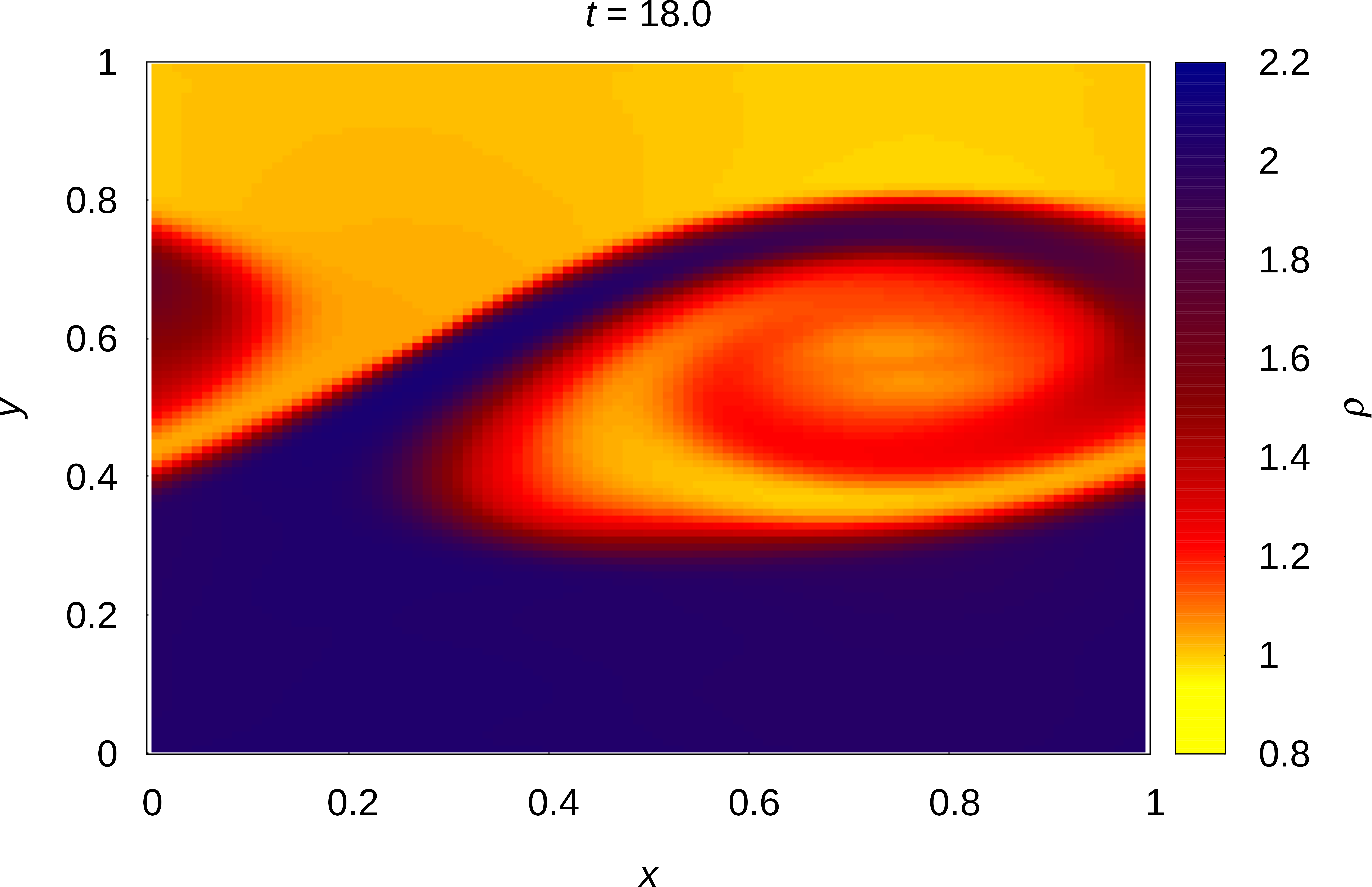}
\caption{The color map of the density profile for the Kelvin-Helmholtz instability problem (see Sect.~\ref{Keshock}). The snapshot 
shows the flow at late time, once the instability has gone fully nonlinear (the turbulences are fully developed).}
\label{khcolor408}
\end{center}
\end{figure}

The significance of the test consists mainly in the fact that it shows that 
at early time of the evolution of the perturbation one can visually check that the growth rate of the transverse component of the velocity is linear, while 
at late times, once the instability has gone fully nonlinear, 
it is then difficult to make any quantitative comparisons. 
The sharpness of the boundary between the two streams may become an indication of the numerical diffusion of the computational scheme.
The solution (see Fig.~\ref{khcolor408}) shows a fine-scale structure, comparable with other
methods \citep[e.g.,][]{atas}, which tends not to be disintegrated in much more advanced time of simulation (the maximum time of this test under the same conditions was 
$t=100.0$). %Future experiments including e.g.,~\textit{Rayleigh-Taylor} instability test or different arrangements of shock tube, are urgently planned.
\subsection{Simple one-dimensional astrophysical simulations}\label{astrochecks}
We briefly mention also the tests of the proper convergence of the code 
by performing a few simple one-dimensional (spherically symmetric) 
astrophysical simulations, where the final stationary state is known or can be relatively easily analytically verified.
In this point we modeled e.g., the simplified Parker solar wind velocity distribution \citep{parkan}
described in Eq.~\eqref{parkousek} in the spherical form. 
We employed the two different initial density functions, respectively, in the following way: first we involved the inner boundary density at the schematic solar radius $R_{\odot}$, 
corresponding to the solar photosphere density $\rho_0$ of the order of 
$10^{-4}\,\text{kg}\,\text{m}^{-3}$ \citep{alalik}. The density distribution in the whole computational domain obeys the power law $\rho=\rho_0\,(R_{\odot}/r)^2$,
where $r$ is the radial distance ranging from $R_{\odot}$ to the outer boundary of the problem, $R_{\text{max}}=10\,R_{\odot}$. In the second variant the 
initial density up to the sonic point $r_\text{s}$ (i.e., approximately up to 2 solar diameters, cf.~Eq.~\eqref{parkousek2}) corresponds to the hydrostatic equilibrium, giving 
the relation 
\begin{align}\label{solaratratous}
\rho=\rho_0\,\text{exp}\left[\frac{GM_{\odot}}{a^2}\left(\frac{1}{r}-\frac{1}{R_{\odot}}\right)\right],
\end{align}
while it obeys the power law $\rho=\rho_\text{s}\,(r_{\text{s}}/r)^2$ outside the sonic point (where $\rho_\text{s}$ is the density at the sonic point calculated via Eq.~\eqref{solaratratous}).
The initial (radial) velocity is set either as a constant, $\varv_r=\dot{M}_{\odot}/(4\pi\,R_{\odot}^2\,\rho_0)$ (where $\dot{M}_{\odot}=2\times 10^{-14}\,M_{\odot}\,\text{yr}^{-1}$
is the solar wind mass loss rate, 
see~\citet{polnice,vudy}), or it obeys the linear law in the form
\begin{align}\label{solarvelostratous}
\varv_r=\frac{\dot{M}_{\odot}}{4\pi\,R_{\odot}^2\,\rho_0}\frac{r-R_{\odot}}{R_\text{max}-R_{\odot}}.
\end{align}
\begin{figure}[t]
\begin{center}
\includegraphics[width=11cm]{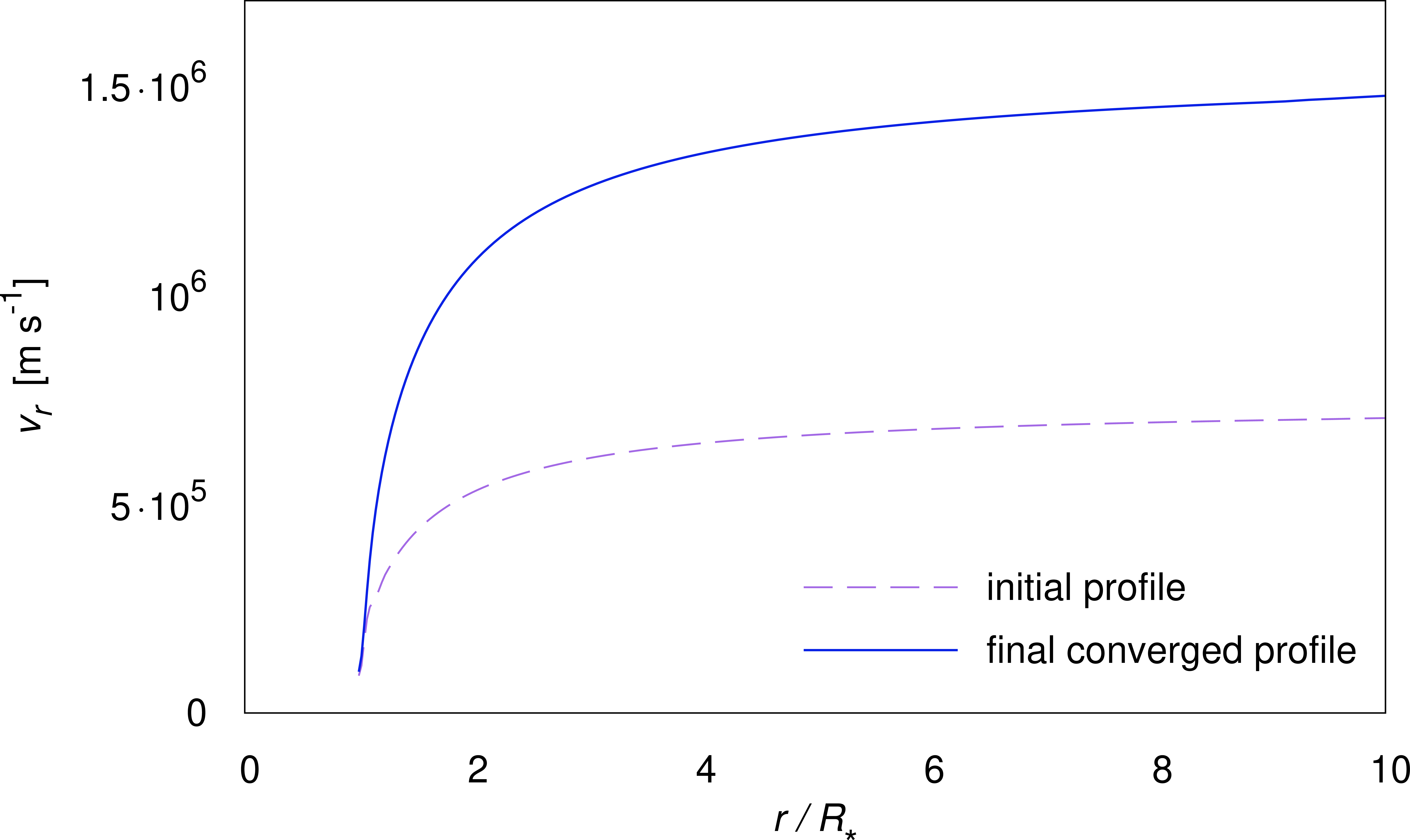}
\caption{Demonstration of the 
proper convergence of the code in case of the spherically symmetric line driven (CAK) wind velocity simulation with the 
significantly underestimated terminal wind velocity 
in the initial function (see Sect.~\ref{astrochecks}).}
\label{cakisevolve}
\end{center}
\end{figure}
The temperature is schematically set as constant, $T=3\times 10^6\,\text{K}$ (solar corona temperature), in the whole computational domain. 
The inner boundary conditions are in this case fixed for the density and free for the radial momentum $\rho\varv_r$ while the outer boundary conditions are 
free (outflow) for the both quantities (see Appendix~\ref{adjustik} for the detailed description of the boundary conditions).

We also modeled the distribution of physical quantities (density, velocity) in line driven wind of hot stars 
\citep[][see also Appendix~\ref{lajncakis}]{CAK1,Lamers1} with various initial and boundary 
conditions. We examined namely the limits of convergence of the code 
in case that the selected initial conditions significantly differ from the assumed final state. Fig.~\ref{cakisevolve} illustrates the 
excellent convergence of the code in case of the spherically symmetric line driven (CAK) wind simulation for averaged 
O-type star with the idealized parameters $M_{\star}=30\,M_{\odot}$, $R_{\star}=9\,R_{\odot}$
and $T_{\text{eff}}=30\,000\,\text{K}$. We use here simplified CAK equation of the (time-dependent) form of Eq.~\eqref{CAK10}
including also the correction for the finite disk of central star (“finite disk correction factor” -
FDCF) described by Eq.~\eqref{CAK39} with the CAK parameter $\alpha=0.62$ \citep{CAK1}.
We employed two variants of the initial velocity functions. In the first variant we 
use a $\beta$-law according to Eq.~\eqref{CAK41} for the analytical approximation of the initial velocity profile with the parameter $\beta=0.8$. 
In the second variant the initial wind velocity up to the sonic point $r_\text{s}$ (where, assuming the simplified  
isothermal solution, $r_\text{s}$ is approximately $1.005 R_{\star}$) obeys the exponential relation 
\begin{align}\label{starvelotratous}
\varv_r=\varv_0\,\text{exp}\left[\frac{GM_\star}{a^2}\left(\frac{1}{r}-\frac{R_\star}{r}\right)\right],
\end{align}
while outside the sonic point the wind velocity obeys the $\beta$-law \eqref{CAK41}. We set the inner boundary velocity $\varv_0$ in Eq.~\eqref{starvelotratous} as an arbitrary small value 
(e.g., $\varv_0=10^{-6}\,\text{m}\,\text{s}^{-1}$).
However, we tested also the models where the terminal wind velocity $\varv_\infty$ (see Eq.~\eqref{CAK41})  
is in the initial function intentionally significantly underestimated, the case is demonstrated in Fig.~\ref{cakisevolve}. 
We set the initial density distribution in the whole computational domain by using the stationary spherical mass conservation equation \eqref{masscylinder002} as $\rho=\dot{M}/(4\pi\,r^2\,\varv_r)$,
where $\dot{M}=0.2\times 10^{-6}\,M_{\odot}\,\text{yr}^{-1}$ \citep{pulspauld} is the O-star's mass loss rate.
Similarly to the solar wind models, the inner boundary conditions are in this case fixed for the density and free for the radial momentum $\rho\varv_r$ while the outer boundary conditions are 
free (outflow) for the both quantities (see Appendix~\ref{adjustik} for the detailed description of the boundary conditions).

In the very inner regions of the computational domain (close to the sun or the star) are the detailed solutions of the referred models very sensitive 
to selected initial and boundary conditions,
while the global solution does not depend on them. The computations exhibit a fast convergence and the final convergent states
show a good agreement with analytical predictions.
(e.g.,~the terminal velocity given by Eq.~\eqref{CAK41} is in this case $\varv_\infty\approx 1.44\cdot 10^6\,\text{m}\,\text{s}^{-1}$, cf.~the profile in Fig.~\ref{cakisevolve}).
However, the overall picture may be incomplete due to considerably simple nature of the performed one-dimensional models (smooth radial winds). We currently plan to use our code also for
hydrodynamic modeling of much more complex tasks, e.g.,~of the clumped stellar wind \citep{branka12,branka13}. 

\chapter[One-dimensional modeling of extended disk radial structure]
{One-dimensional modeling of extended disk radial structure}\label{largemodel}
\section{The parameterization of hydrodynamic equations used in one-dimensional models}\label{largemodelix}
Evolutionary models of rotating massive stars show that the stellar equatorial surface velocity
may reach the critical velocity
during the main-sequence evolution \citep{Chiappi,Meynet2,Meynet3,granada} as a consequence 
of the outward vertical transport of angular momentum 
in the stellar interior, primarily driven 
by the convection and meridional circulation (see Sect.~\ref{diskfome}).
Following the viscous disk scenario (see Sect.~\ref{viskodisks}) we assume the time-independent solution of the disk being
isothermal and in vertical hydrostatic equilibrium \citep[e.g.,][]{okazaki,Karfiol08}. Due to highly
supersonic disk rotational velocity in the inner radial regions (approximately up to the sonic point distance) the disks are geometrically
very thin while the disk opening angle is there only a few degrees.
The scenario of the viscous decretion disk model naturally leads to the idea of the formation of 
near-Keplerian disks around critically rotating stars.
Since the specific angular momentum in the 
Keplerian (or near-Keplerian) disk increases with radius, we expect that far from the star a disk
becomes angular-momentum conserving. 
This transitional feature of stellar decretion disks
is however not very well understood theoretically nor has it still been observationally confirmed.
Most of the current models analyze 
the inner parts of the disk, whereas the evolution close to the sonic point or even in 
the supersonic regions up to the possible outer disk edge has not been, to our knowledge,
very well studied \citep{kurfek}.

We study the characteristics of the outflowing disks of
critically (or near-critically) rotating stars.
The mass loss rate is determined by the angular momentum loss rate needed to keep the star at critical rotation 
\citep[see also \citealt{Kurf2}, \citealt{kurfek}]{Krticka}.
The model of the viscous decretion disk proposed by 
\citet[see also \citealt{okazaki}]{Lee} outlined the basic scenario where one obtains 
a steady structure of viscous decretion disks around Be stars in thermal and radiative equilibrium. 
Since the physics of accretion disks is quite similar, we follow the main principles \citep{Pringle,Frank} 
of accretion disk theory for the description of decretion disks.
As the major uncertainties we consider the viscous coupling and the temperature distribution. Most of recent
models indicate a constant value of the viscosity throughout the inner region of 
such disks \citep{Penna}, 
we investigate however also the cases when the viscous coupling varies outward as a certain power law. The disk temperature 
is mainly affected
by the irradiation from the central star \citep{Lee}. 
Motivated by the non-local thermodynamic equilibrium (NLTE) simulations \citep{Karfiol08} 
we extrapolate the temperature distribution up to quite distant regions as 
a power law \citep{kurfek}.

To summarize the key cylindrical axisymmetric hydrodynamic equations of the disk structure 
introduced in Sects.~\eqref{eqraddisk}, \eqref{vertikalekdisk} and \eqref{thindisk} (using the notation introduced therein) in the 
one-dimensional form: the mass conservation (continuity) equation 
(cf.~Eqs.~\eqref{masscylinder1}, \eqref{angularmomentumfluxix0} and \eqref{conticylap}) is
\begin{align}
\label{massconserve}
R\frac{\partial\Sigma}{\partial t}+\frac{\partial}{\partial R}\left(R\Sigma V_{\scriptscriptstyle{R}}\right)=0.
\end{align}
The radial momentum conservation equation in the thin disk approximation 
(cf.~Eq.~\eqref{cylmomrad} supplemented by Eq.~\eqref{gravvertixosix}) is
\begin{align}
\label{radmomconserve}
\frac{\partial V_R}{\partial t}+V_R\frac{\partial V_R}{\partial R}=
\frac{V_{\phi}^2}{R}-\frac{GM_{\!\star}}{R^2}-\frac{1}{\Sigma}\,\frac{\partial (a^2\Sigma)}{\partial R}+\frac{3}{2}\,\frac{a^2}{R},
\end{align}
while the conservation of azimuthal component of momentum in the compact form 
(cf.~Eqs.~\eqref{anabjk1} and \eqref{cylmomphi}) is
\begin{align}
\label{phimcon}
\frac{\partial V_\phi}{\partial t}+V_R\frac{\partial V_\phi}{\partial R}+
\frac{V_{R}V_{\phi}}{R}=f_{\text{visc}},
\end{align}
where $f_{\text{visc}}$ shortly denotes the viscous force per unit volume,
exerted by the outer disk segment on the inner disk segment \citep[cf.]{kurfek}. In the axisymmetric
($\partial/\partial\phi=0$)
and geometrically thin case
the viscous shear force density is represented by differentiation of the viscous tensor \eqref{stresikvalecexplirphi}
$f_{\text{visc}}=\partial T_{R\phi}/\partial R+2T_{R\phi}/R$.
The viscous force density $f_{\text{visc}}$ is
usually represented by using merely the first order linear viscosity term (see Eq.~\eqref{viscon}),
we examine the viscous force density including the full second order viscosity
(second radial derivatives and the products of first radial derivatives in Eq.~\eqref{appendphimomcylinder1}) 
with the adopted Shakura-Sunyaev $\alpha$~parameter \citep{Shakura}. 
We also examine the cases with non-constant $\alpha$ parameters \citep{kurfek}.
Taking into account the turbulent motion of the gas, 
we can relate the kinematic viscosity $\nu$ and the Shakura-Sunyaev $\alpha$~parameter \citep[e.g.,][cf.~also Eq.~\eqref{Shakurakos}]{Frank} via
\begin{align}
\label{shak}
\nu=\alpha\frac{a^2R}{V_{\phi}}\approx\alpha aH,
\end{align}
where $H$ denotes the typical vertical scale height of the disk, $H^2=a^2R^3/(GM_{\!\star})$ (see Eq.~\eqref{kepscaleheight})
in Keplerian case.

NLTE simulations \citep[e.g.,][]{Karfiol08} basically conclude that the radial temperature distribution 
in the very inner regions (up to few stellar radii)
corresponds to a flat blackbody reprocessing disk due to the optically thick nature of this inner part,
$T(R_\text{eq})\approx\frac{1}{2}T_{\mathrm{eff}},\,T(R)\sim R^{-0.75}$,
where $T(R_\text{eq})$ is the inner boundary disk temperature at $R=R_{\mathrm{eq}}$. 
As the disk in larger distance becomes vertically optically thin, the disk temperature rises to the
optically thin radiative equilibrium temperature with the average of about $60\%$ of $T_{\mathrm{eff}}$. 
The temperature radial profile at larger radii is nearly isothermal with
very gradual temperature decrease (i.e., about $1000~\text{K}$ decrease within the distance from $10$ to $50$ stellar radii).
\citet{Millar1, Millar2} consistently found that the radial temperature distribution up to approximately $100$ stellar radii is nearly isothermal.
We approximate these dependencies
by a radial power law \citep{Krticka,Kurf1,Kurf2,kurfek}
\begin{align}\label{temperature}
T=T(R_\text{eq})\left(\frac{R_{\text{eq}}}{R}\right)^p,
\end{align}
where $p$ is a free parameter that ranges within the interval $0\leq p<0.5$. 
We use the same power law temperature decline for the extrapolation of the radial temperature structure of the 
outer distant part of the disk.

The disk temperature is also influenced by the viscosity, however, the
contribution of the viscous heating production in the disk is very
small (practically negligible) compared
with the heating that comes from
radiative flux from the central star \citep[e.g.,][]{smak,Karfiol08}. 
The viscous heating may dominate
in the inner disk regions \citep{Lee} only in
case of enormous value of mass loss rate ($\dot{M}\geq 10^{-5}\,M_{\odot}\text{yr}^{-1}$, cf.~also Eq.~\eqref{smakratio}).
This is the reason 
why we parameterize the viscosity via the temperature independent $\alpha$ parameter \citep{kurfek}.

The radial profile of $\alpha$ parameter is not quite certainly known. There is however a common agreement that the value of
the dimensionless $\alpha$ parameter of turbulent viscosity is less than $1$, 
since in case that $\alpha>1$, the rapid thermalization caused by shocks
would lead again to $\alpha\le 1$ \citep{Shakura}. Most authors use the value about $0.1$; some of the most recent
works focused on the disk viscosity problem \citep[e.g.,][]{Penna}
find the $\alpha$ viscosity coefficient as constant, $\alpha\approx 0.025$. 
However, we regard as an
open question if $\alpha$
should be considered as constant throughout the whole disk or not \citep{kurfek}.
We therefore introduce the power law
\begin{align}
\label{alpvis}
\alpha=\alpha(R_\text{eq})\left(\frac{R_{\text{eq}}}{R}\right)^n,
\end{align}
where $\alpha(R_\text{eq})$ is the viscosity in the inner region of the disk near the stellar surface, 
$n$ is a free parameter (power) of the radial viscosity dependence, $n>0$.
The kinematic viscosity $\nu(R)$ is related to 
temperature via Eq.~\eqref{shak}.

Following the radial derivative of Eq.~\eqref{appendphimomcylinder1}, the first order linear viscosity term in Eq.~\eqref{phimcon} takes the form \citep[see, e.g.,][]{Krticka}
\begin{align}
\label{viscon}
f_{\text{visc}}^{(1)}=-\frac{1}{R^2\Sigma}\,\frac{\partial}{\partial R}(\alpha a^2R^2\Sigma),
\end{align}
while including the full second order viscosity (employing the second radial derivatives and 
corresponding nonlinear terms from Eq.~\eqref{appendphimomcylinder1}) 
gives the right-hand side of Eq.~\eqref{phimcon} in the form \citep[][see subroutine \textit{nav} in Appendix~\ref{koropticek} for its explicit numerical 
expression]{kurfek}
\begin{align}\label{angfl} 
f_{\text{visc}}^{(2)}=\frac{1}{R^2\Sigma}\,\frac{\partial}{\partial R}
\left(\alpha a^2R^3\Sigma\,
\frac{\partial\,\text{ln}\,V_{\phi}}{\partial R}-
\alpha a^2R^2\Sigma\right).
\end{align}
The identical relation is obtained by differentiation of the stationary part of angular momentum equation \eqref{angularmomentumflux1} \citep{Pringle,Frank},
whose form there is
\begin{align}\label{angfl1} 
\frac{\partial}{\partial t}\left(\Sigma R^2\Omega\right)+\frac{1}{R}\frac{\partial}{\partial R}\left(\Sigma V_RR^3\Omega\right)
=\frac{1}{2\pi R}\frac{\partial\mathcal{G}}{\partial R}, 
\end{align}
(we note that we use angular momentum in the numerical scheme instead of $\phi$-momentum component; see Sect.~\ref{timedependix})
where $\Omega$ is the angular velocity ($\Omega=V_{\phi}/R$) and $\mathcal{G}$ is the viscous torque acting between two neighboring disk segments
(cf.~Eqs.~\eqref{viscoustorque} and \eqref{shak}),
\begin{align}\label{vist}
\mathcal{G}=2\pi\alpha a^2\Sigma R^3\frac{\partial\,\text{ln}\,\Omega}{\partial R}.
\end{align}
We can thus relate $\mathcal{G}$ and $f_{\text{visc}}^{(2)}$ from Eq.~\eqref{angfl}, $\partial\mathcal{G}/\partial R=2\pi R^2\Sigma\,f_{\text{visc}}^{(2)}$.

Most authors use the concept of some outer disk radius 
$R_\text{out}$ regarded as free parameter (see Sect.~\ref{thindisk}) where, due to e.g., the 
radiation pressure, the disk matter may be completely driven outward \citep[e.g.,][among many others]{Lee,Houba12}.
Using the reasonable parameterization of $T$ and $\alpha$, we are not 
dependent on rather arbitrary choice of $R_{\text{out}}$ and we can therefore calculate the radial profiles of the characteristics as a direct 
solution of the hydrodynamic equations \citep{kurfek}.
\section{Analytical approximations of the radial thin disk structure}
\label{radthin}
To obtain an instructive overall picture of the behavior of the main characteristics of the disk system,
we analytically expand the stationary form of the hydrodynamic equations in a way that is basically described, e.g.,~in \citet{okazaki,Krticka}, involving
however the term $f_{\text{visc}}^{(2)}$ from Eq.~\eqref{angfl}).
Integrating the stationary form of Eq.~\eqref{phimcon}, multiplying it by $R^2\Sigma$ and dividing it by the integrated stationary mass conservation equation 
\eqref{massconserve} $R\Sigma V_R=\text{const.}$, in case of $\alpha=\text{const.}$ we obtain \citep[see][]{kurfek}
\begin{equation}\label{angestim2} 
RV_{\phi}+\frac{\alpha a^2R}{V_R}\left(1-R\frac{\partial\,\text{ln}\,V_{\phi}}{\partial R}\right)=\text{const.}
\end{equation}
If we include only the first order linear viscosity $f_{\text{visc}}^{(1)}$ (Eq. \eqref{viscon}) 
we get a relation similar to Eq.~\eqref{angestim2} where however the second term in the bracket drops.
Assuming $V_R\ll a$ in the innermost part of the isothermal disk (see also the considerations referring to the radial velocity in Sect.~\ref{thindisk}), the second term on
the left-hand side of Eq.~\eqref{angestim2} dominates, therefore $V_R\sim R$ and
$\Sigma\sim R^{-2}$ from the continuity equation \citep[see, e.g,][]{okazaki}.
Since in the inner region of the disk the radial pressure gradient is negligible
compared with the gravitational force,
from the momentum equation \eqref{radmomconserve} follows the the Keplerian rotation
$V_\phi\sim R^{-1/2}$. 
The constant mass loss rate $\dot{M}=\text{const.}$ (see Eq.~\eqref{angularmomentumfluxix0}) implies in this region the relation for the angular momentum loss rate
$\dot{J}(\dot{M})\sim RV_{\phi}\sim R^{1/2}$.
Eq.~\eqref{angestim2} may be simplified into the form \citep{kurfek}
\begin{equation}\label{angestim3} 
RV_{\phi}+\gamma\frac{\alpha a^2R}{V_R}=\text{const.},
\end{equation}
where the numerical factor $\gamma=3/2$ for the Keplerian rotational velocity 
(in this point we note that the second order viscosity term in the Keplerian case represents 
one half of the corresponding first order viscosity term, 
assuming the same $\alpha$ parameter, see~\citet{kurfek}).
In the distant region near the sonic point (where $V_R\approx a$) the radial velocity linearly increases;
the first term on the left-hand side of Eqs.~\eqref{angestim2} and \eqref{angestim3} then fully
dominates, hence $RV_{\phi}=\text{const.}$,
and the disk becomes above this region angular momentum conserving.

If we go further outward to very distant supersonic regions with nearly flat disk temperature distribution 
($V_R\gg a$, $V_R\gg V_{\phi}$, $\partial a^2/\partial R\approx 0$)
and negligible gravity and density, we may take into account only the last terms on both sides of the radial momentum equation \eqref{radmomconserve}.
Integration of these terms implies logarithmic
radial dependence, $V^2_R\sim \text{ln}\,R$ \citep{okazaki,Krticka,kurfek}. 
Consequently the second term on the left-hand side of Eq.~\eqref{angestim3} grows and
simultaneously the azimuthal velocity substantially decreases while it may in case of including merely the viscosity term $f_{\text{visc}}^{(1)}$
become even negative. 
However, its fall to negative values is impossible due to the logarithmic term in Eq.~\eqref{angestim2}.
This is also actually proved by the numerical simulations (see Figs.~\ref{B0pt0}-\ref{ptI}), where the  
azimuthal velocity even in extremely distant regions does not change its sign (i.e.,~its direction).
The numerical models (Sect.~\ref{onedimtimemod}) exhibit the rapid rotational velocity drop in very outer disk region whose location is radially dependent of 
the viscosity and temperature profiles.
Equation \eqref{angestim3} consistently indicates that for steeper viscosity and temperature decrease (lower $\alpha$ and $a^2$ in
distant regions) the region of that azimuthal velocity drop moves outwards.

From the stationary form of the radial momentum equation \eqref{radmomconserve} and the mass conservation equation
\eqref{massconserve}, using the equation of the radial temperature parameterization \eqref{temperature},
we obtain the expression of the sonic point condition $V_R=a$ that applies at the sonic point radius $R_{\text{s}}$ \citep[cf.][]{okazaki,Krticka},
\begin{equation}\label{spcon}
R_{\text{s}}=GM_{\!\star}\left[\left(\frac{5}{2}+p\right)a^2+V_{\phi}^2\right]^{-1}.
\end{equation}
From previous stationary models calculated by \citet{Krticka} we can estimate that the azimuthal velocity at the sonic point roughly equals one half
of the corresponding Keplerian velocity at the same point.
Substituting $V_{\phi}(R_\text{s})\approx\frac{1}{2}V_K(R_\text{s})$, where
$V_{K}(R)=\sqrt{GM_{\!\star}/R}$ is the Keplerian velocity at the radius $R$ \citep[see also
Figs.~\ref{B0pt0}-\ref{ptI}]{Krticka,kurfek}, we obtain an approximate sonic point
radius
\begin{equation}\label{sonicestR}
\frac{R_\text{s}}{R_{\text{eq}}}\approx
\left[\frac{3}{10+4p}\left(\frac{V_{K}(R_{\text{eq}})}{a(R_{\text{eq}})}\right)^2\right]^{\frac{1}{1-p}}.       
\end{equation}
In a similar way we can approximate the radius of the disk angular momentum loss rate maximum $\dot{J}_{\text{max}}$, 
it however roughly corresponds to the sonic point radius 
(see Figs.~\ref{B0pt0}-\ref{ptI}).
Since the maximum $\dot{J}_{\text{max}}\approx\dot{M}R_\text{s}V_{\phi}(R_\text{s})$, we write \citep[cf.][]{Krticka,kurfek}
\begin{equation}\label{sonicestL}
\dot{J}_{\text{max}}(\dot{M})\approx
\frac{1}{2}\left[\frac{3}{10+4p}\left(\frac{V_{K}(R_{\text{eq}})}{a(R_{\text{eq}})}\right)^2\right]^{\frac{1}{2-2p}}
\dot{M}R_{\text{eq}}V_{K}(R_{\text{eq}}).       
\end{equation}
From Eqs.~\eqref{sonicestR} and \eqref{sonicestL} follows that adding cooling (higher parameter $p$) implies substantial increase of the sonic point distance and
consequently also the increase of the angular momentum loss of the disk for a fixed $\dot M$.
For example, for $p=0.2$ the maximum loss rate of angular momentum
increases roughly 2 times, while for $p=0.4$ it increases roughly 5-6 times in comparison with the isothermal case, 
according to type of star \citep{kurfek}.
\section[One-dimensional hydrodynamic models]{Stationary models of one-dimensional disk hydrodynamic structure}\label{statcalc}
\begin{figure} [t]
\centering\resizebox{0.6\hsize}{!}{\includegraphics{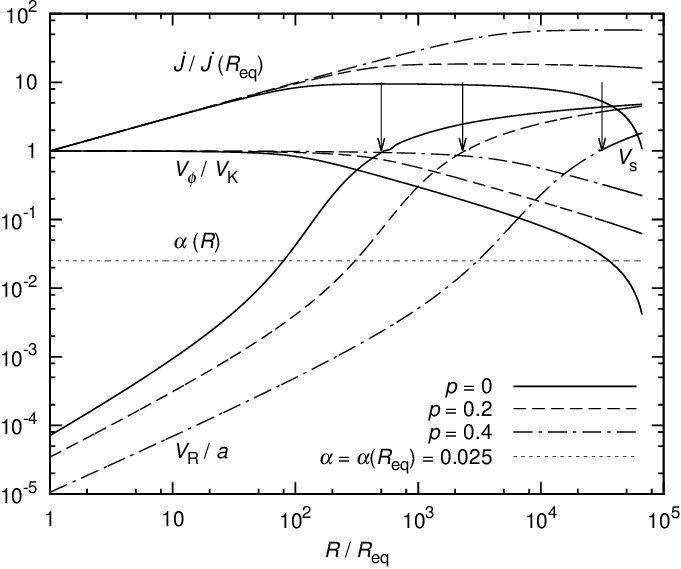}}
\caption{Dependence of the scaled radial and azimuthal velocities and the angular momentum loss rate 
${\dot{J}/\dot{J}(R_{\text{eq}})}$ on radius
for various temperature profiles calculated by the method described in Sect.~\ref{statcalc}. 
Constant viscosity ${\alpha=0.025}$ is assumed. 
Arrows denote the sonic point. Adapted from \citet{kurfek}.}
\label{fig1}
\end{figure}
\begin{figure} [h!]
\centering\resizebox{0.6\hsize}{!}{\includegraphics{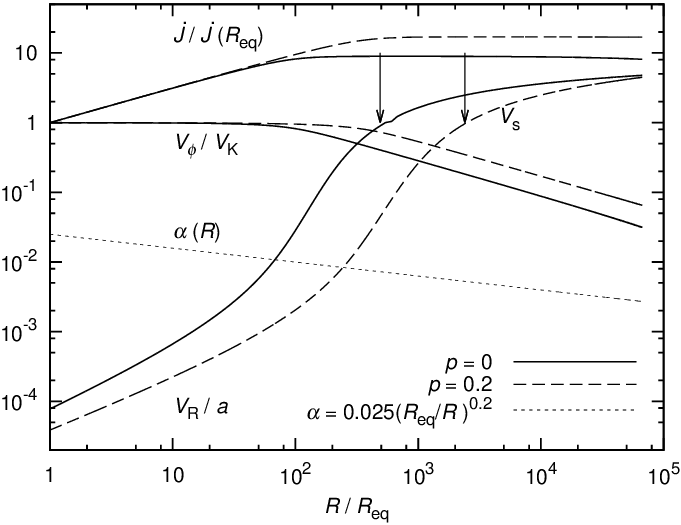}}
\caption{As in Fig.~\ref{fig1}, but
with variable ${\alpha}$ parameter, ${\alpha\sim R^{-0.2}}$. Adapted from \citet{kurfek}.}
\label{fig2}
\end{figure}
Before we developed the time-dependent code (Sect.~\ref{timedependix}) we calculated the stationary models \citep{Kurf2,kurfek} solving the (stationary)
equations~\eqref{massconserve}, \eqref{radmomconserve}, and \eqref{phimcon}, using the Newton-Raphson method described in Sect.~\ref{stationarix}.
As a central object we selected the main-sequence star of spectral type B0 with the following parameters \citep{Harmanec}:
$T_{\text{eff}}=30\,000\,\text{K},\,M_{\!\star}=14.5\,M_{\odot},\,R_{\star}=5.8\,R_{\odot}.$
We solved the system of hydrodynamic equations \eqref{massconserve}-\eqref{phimcon} supplemented by the sonic point condition \eqref{spcon} 
using a kind of shooting method (i.e., the radial velocity curve has to target
and smoothly pass through the sonic point position $R_\text{s}$), based on changing the inner boundary (photospheric) 
radial velocity in order to find a properly converging branch of the solution.
The azimuthal velocity at the inner disk boundary 
(stellar equatorial surface) is fixed as the stellar equatorial critical rotation velocity. 
The mass loss rate $\dot{M}$ is regarded in the calculations as a free parameter.
The solution is completely independent of the value of
the surface density $\Sigma$ \citep[cf.][]{Krticka}. For the numerical calculation we selected a radial grid consisting of 300-1000 grid points according to various initial conditions.

Despite the efficiency of this method, some 
numerical problems occurred in our stationary thin disk calculations, above all in case of very radially extended models. The models are therefore cut off at approximately $10^5$ 
stellar equatorial radii (see Figs.~\ref{fig1}-\ref{fig2}). 
In the stationary calculations we used therefore merely the first order linear viscosity (Eq.~\eqref{viscon} inserted into Eq.~\eqref{phimcon}). 
The reason is that despite its complete analytic linearization in the Jacobi matrix (Eq.~\eqref{numourix3}), 
the inclusion of the second order viscosity term (see Eq.~\eqref{angfl}) have caused severe numerical vibrations 
near the sonic point and further out, which we did not succeed to eliminate.

The model with constant viscosity profile and with $p=0$
(see Fig.~\ref{fig1})
obviously shows a rapid decrease in the rotational velocity and the angular momentum loss at large radii. These quantities 
may even drop to negative values in case of the adopted 
first order linear viscosity prescription (see Sect.~\ref{radthin}). 
From the analytical point of view is this drop caused by the increase of the second term in
Eq.~\eqref{angestim3} at large radii. We however consider this drop as clearly unphysical.
As a solution to this problem we introduce the models with
higher $p$ and $n$ parameters (power law temperature and viscosity decrease, Eqs.~\eqref{temperature} and \eqref{alpvis}) that avoid the rapid 
rotational velocity drop (providing a
constant angular momentum loss rate) in supersonic region (see Fig.~\ref{fig2}) \citep{kurfek}. 
\begin{figure} [t]
\centering\resizebox{0.5\hsize}{!}{\includegraphics{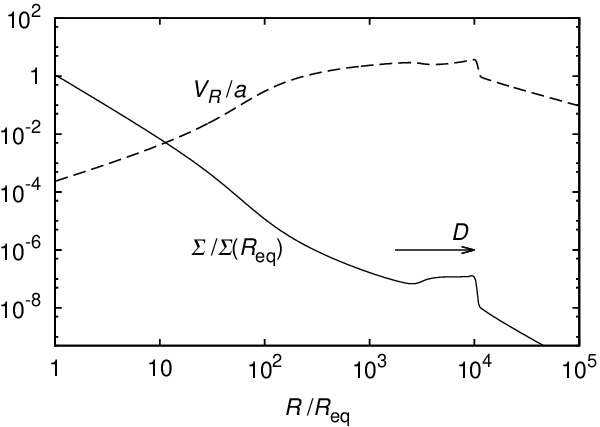}}
\caption{The snapshot of the radial velocity and the surface density transforming wave in the case of 
a B0-type star isothermal constant viscosity model (see Fig.~\ref{B0pt0}),
the time $t_{\text{dyn}}\approx 40\,\text{yr}$.
The wave propagation velocity is denoted as $D$.
In the model, the ratio $\Sigma_1/\Sigma_0$ (the surface densities behind and in front of the wavefront) 
is about one order of magnitude and slightly increases with the distance.
Adapted from \citet{kurfek}.}
\label{shock}
\end{figure}
\begin{figure} [t]
\centering\resizebox{0.5\hsize}{!}{\includegraphics{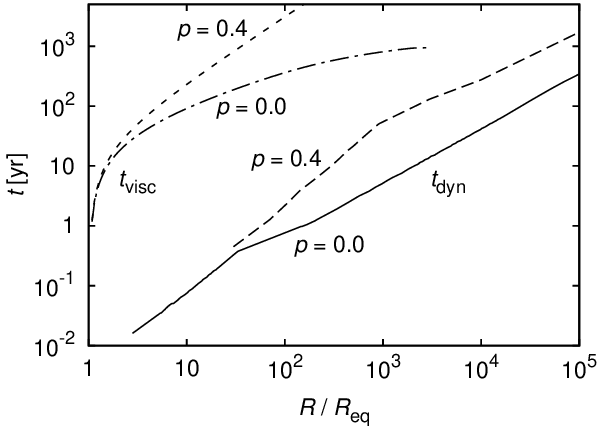}}
\caption{Comparison of the density wave propagation time (lower two branches denoted as $t_{\text{dyn}}$)
with the disk viscous time (upper two branches denoted as $t_{\text{visc}}$),
for B0-type star isothermal ($p=0$) constant viscosity model and the model with decreasing ($p=0.4$) temperature 
profile (see Fig.~\ref{B0pt0n04}), in dependence on radius.
The disk viscous time is calculated from Eq.~\eqref{visctime}. Since the rotational 
velocity $V_{\phi}$ is adopted from the models, the graph of $t_{\text{visc}}$ is cut off
in the region of the rapid rotational velocity drop. The plotted values of $t_{\text{dyn}}$ are adopted from the models.
Adapted from \citet{kurfek}.}
\label{shortlong}
\end{figure} 
\begin{figure} [t]
\centering\resizebox{0.56\hsize}{!}{\includegraphics{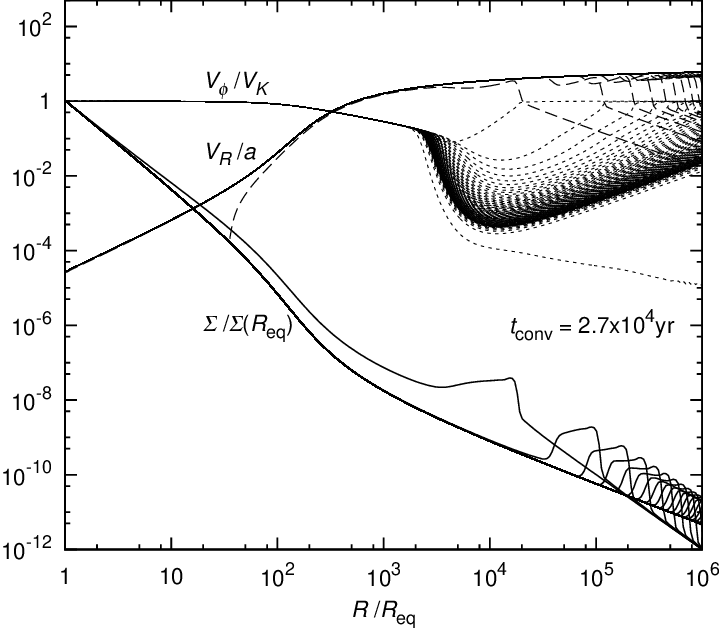}}
\caption{The composite graph of time snapshots of surface density, radial and azimuthal velocity. It demonstrates the propagation of the transforming wave described in Sect.~\ref{chartime} 
in the dynamical timescale
$t_\text{dyn}$ of the isothermal constant viscosity ($\alpha=0.025$) model of B0-type star. The location of the first wave contour (at the distance of approximately $10^4 R/R_{\text{eq}}$) 
represents the dynamical time of approximately $65$ years. The corresponding contour of scaled radial velocity (dashed line) 
is truncated inwards since it drops in the inner region to negative values.
The next shifts of the transforming wave represent the time interval of approximately $250$ years. 
The dynamical time is calculated directly in the model. The time $t_{\text{conv}}$, denoted in the graph, is the minimum time needed for the model convergence into the final stationary state.}
\label{slozenka1}
\end{figure} 
\begin{figure} [h]
\centering\resizebox{0.56\hsize}{!}{\includegraphics{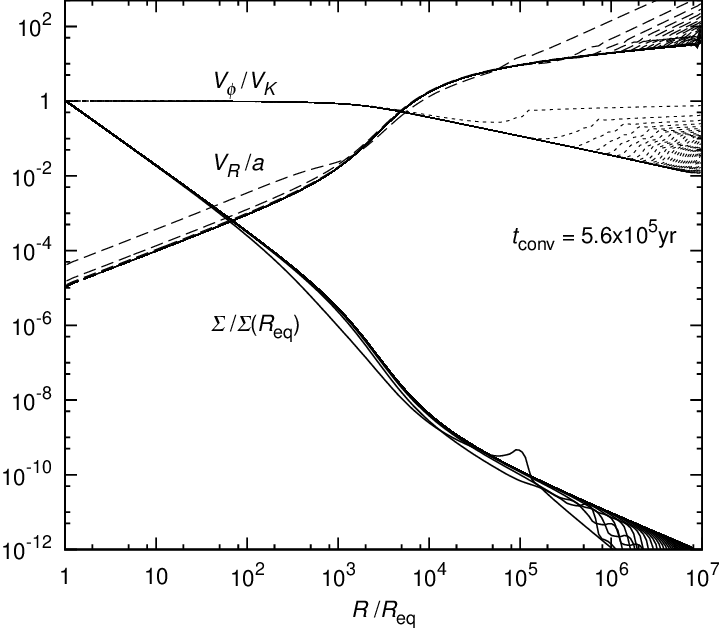}}
\caption{As in Fig.~\ref{slozenka1}, however for the model with decreasing viscosity profile ($\alpha(R_\text{eq})=0.025,\,n=0.2$) and with decreasing temperature profile ($p=0.4$). 
The location of the first wave contour (at the distance of approximately $10^5 R/R_{\text{eq}}$) 
in this case represents the dynamical time of approximately $3\,800$ years, the next shifts of the transforming wave (see the description in Sect.~\ref{chartime}) 
represent the time interval of approximately 
$15\,000$ years. The time $t_{\text{conv}}$ is the time needed for the model convergence into the final stationary state within
the range of the computational domain (up to $R=10^7\,R_{\text{eq}}$).}
\label{slozenka2}
\end{figure} 
\section[One-dimensional hydrodynamic models]{Time-dependent models of one-dimensional disk hydrodynamic structure}\label{onedimtimemod}
We calculate the time-dependent disk models using our time-dependent hydrodynamic code (see Sect.~\ref{timedependix}) involving 
Eqs.~\eqref{genmasik}, \eqref{gennumermom}, \eqref{gennumerang}, \eqref{stateq} and Eq.~\eqref{angfl}. The temperature and viscosity profiles are parameterized 
according to Eqs.~\eqref{temperature} and \eqref{alpvis}.
\subsection{Disk evolution time}
\label{chartime}
Within the process of time-dependent 
modeling we recognize the wave that converges the initial state of the calculated quantities to their final stationary state.
Since the wave is connected above all with the transformation of the initial density distribution,
we may regard the wave as physical \citep[not only numerical artifact, see,][]{kurfek}.
We assume that during the disk developing phase a similar transforming wave occurs and
its amplitude and velocity depends on physical conditions (namely on the distribution of the density) 
in the stellar neighborhood. We regard the velocity of the wave as the velocity of the spreading of the disk. There might be a possibility to observe some 
bow shocks at the boundary between the developing disk and the interstellar medium.  
Similar bow shocks are created as a result of the interaction of the stellar wind and interstellar medium, however, 
the disk radial velocity in the distant regions is about one order of magnitude lower 
than in the case of line-driven stellar winds 
\citep[cf.~Appendix~\ref{lajncakis},][]{kurfek}.
The wave may also basically determine the timescale of the Be star disk growth and
dissipation phases \citep[e.g.,][]{guhaj,stebar}.
Within the subsonic region the wave establishes nearly hydrostatic equilibrium in
the radial direction (Eq.~\eqref{radmomconserve}) where the wave speed approximately equals the sound speed.
In the distant (supersonic) region this wave propagates as a shock wave. 

The Rankine-Hugoniot relations (cf.~Eqs.~\eqref{rahug1}, see also \citet{zelinar}) establish the (square of) the shock wave propagation velocity 
$D^2=[\rho_1/(\Gamma_1\,\rho_0)](a_1^2\rho_1-a_0^2\rho_0)/(\rho_1-\rho_0)$ where $\Gamma_1$ is the general adiabatic exponent (see Sect.~\ref{statak}),
and the subscripts
$0,1$ denote the quantities before and behind the shock front, respectively. In the (nearly) isothermal case we assume $\Gamma_1\approx 1$ and 
$a_0\approx a_1=a$. We thus approximate
$D=a\sqrt{\Sigma_1/\Sigma_0}$ while from the models the ratio $\Sigma_1/\Sigma_0$ is one order of magnitude.
We regard 
the shock propagation time as the dynamical time
\begin{equation}\label{tshock}  
t_{\text{dyn}}\approx R/D\approx 0.3R/a. 
\end{equation}
For example, for the distance 
$10^4\,R_{\text{eq}}$ the isothermal constant viscosity B0-type star disk model gives $t_{\text{dyn}}\approx 40\,\text{yr}$ 
\citep[][cf.~also Figs.~\ref{shock} and \ref{slozenka1}]{kurfek}.
Analogous model with decreasing viscosity ($n=0.2$) gives for the same distance 
approximately the same value of $t_{\text{dyn}}$ and the model with decreasing temperature
profile ($p=0.4$) and decreasing viscosity ($n=0.2$) gives for the same distance $t_{\text{dyn}}\approx 275\,\text{yr}$
(while for the distance of approximately $10^5\,R_{\text{eq}}$ it gives $t_{\text{dyn}}\approx 3\,000\,\text{yr}$, cf.~Fig.~\ref{slozenka2}).
The dynamical time is thus almost independent of the viscosity while it significantly increases with decreasing temperature.

We associate the disk evolution time with disk viscous time $t_{\text{visc}}$ \citep[see Eq.~\eqref{visctime}, see also][]{okazaki,maeder,kurfek}, 
i.e., with the timescale on which matter diffuses through the disk under the effect of the viscous torques \citep{Frank}.
We can analytically estimate $t_{\text{visc}}$ from the similarity \citep{maeder, granada} $t_{\text{visc}}\sim R^2/\nu$, 
where $\nu$ is the kinematic viscosity. According to Eq.~\eqref{shak} we write,
\begin{equation}\label{visctime}
t_{\text{visc}}=\int_{R_{\text{eq}}}^RV_{\phi}\,\text{d}R/(\alpha a^2).
\end{equation}
In the isothermal constant viscosity case the same model gives 
$t_{\text{visc}}\sim 10^2\,\text{yr}$ for the sonic point radius \citep{kurfek}.
The viscous time significantly grows with temperature and viscosity.
The comparison of the two times $t_{\text{dyn}}$ and $t_{\text{visc}}$ is shown in Fig.~\ref{shortlong}.

\subsection{Stationary state reached by time-dependent models}  
\label{timemod}
The properties and the technical details of the time-dependent hydrodynamic code that we have developed for this purpose over nearly past four years
are fully described in Sect.~\ref{timedependix} and in Appendix~\ref{dvojdimhydrous}.
For the time-dependent modeling we selected following two types of stars: main sequence star of spectral Type B0 \citep{Harmanec} 
with the parameters already introduced in Sect.~\ref{statcalc}, used in stationary calculations,  
and a Pop III star (see Sect.~\ref{poptri} for general description of these stars' type link to the disk phenomenon) with the following parameters
\citep{marigo, ekstrom}: ${T_{\mathrm{eff}}=30\,000\,\mathrm{K},M_{\!\star}=50M_{\odot},R_{\star}=30R_{\odot}}$.
The calculations were extended up to a considerable distance from the parent star \citep{kurfek}, although this may in many cases be regarded as
a purely hypothetical issue due to the extremely low density of the disk material in such distance \citep[cf.~e.g.,][]{misirio}
as well as due to the nearby objects or fields (various types of stellar companions,~etc.) whose existence prevents development of such large disks
due to the mechanical or radiative ablation, tidal truncation, etc., of their outer layers \citep[][etc.]{milyartycok,negrokac,okacnegr,Krticka}.

We solve the system of fundamental hydrodynamic equations
\eqref{massconserve}, \eqref{radmomconserve} and \eqref{angfl1}, supplemented by Eq.~\eqref{angfl} or Eq.~\eqref{vist}.
\begin{figure} [t]
\centering\resizebox{0.6\hsize}{!}{\includegraphics{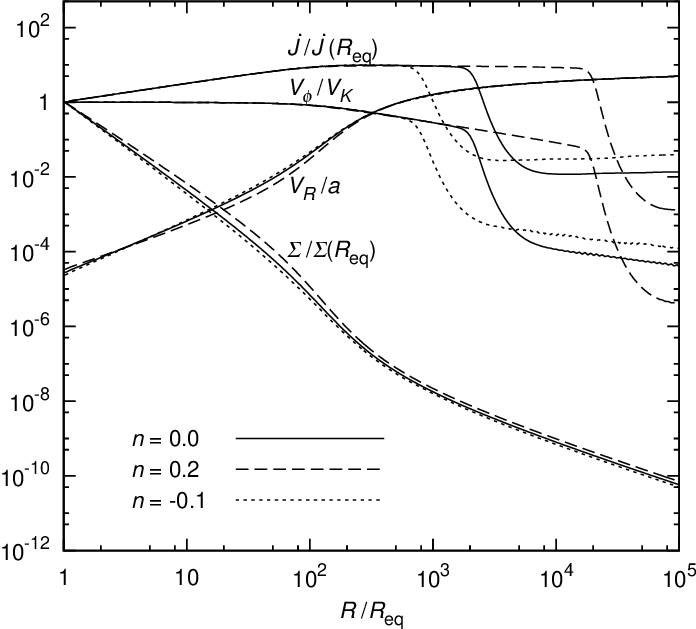}}
\caption{The dependence of scaled surface density and radial and azimuthal disk velocities and the scaled 
angular momentum loss rate  
$\dot{J}/\dot{J} (R_{\mathrm{eq}})$ on radius in case of the isothermal disk ($p=0$) of selected B0-type star
for various radial viscosity profiles (various $n$) in a final stationary state of
time-dependent models (the rapid drop of rotational velocity and the angular momentum loss rate in the outer disk region is 
a finally converged stationary jump, not a propagating shock wave). Adapted from \citet{kurfek}.}
\label{B0pt0}
\end{figure}
We have adopted the inner boundary (stellar equatorial) values  
on the equatorial radius of the critically rotating star ($R_{\text{eq}}=3/2\,R_{\star}$)
in the following way \citep{kurfek}: the estimation of inner boundary surface density $\Sigma(R_{\text{eq}})$ is 
implemented as a fixed boundary value (the technical description of various types of boundary conditions is given in subroutine \textit{boundary} in Appendix~\ref{dvojdimhydrous});
in the case of a critically rotating B0-type 
star the isothermal disk surface density that roughly corresponds to
$\dot{M}\approx 10^{-9} M_{\odot}$ yr$^{-1}$ (the mass loss rate of this type of star proposed by
\citet{granada}) is $\Sigma(R_{\text{eq}})=1.6\times 10^2$ g cm$^{-2}$. In the case of a critically rotating Pop III type star the isothermal disk surface density is
$\Sigma(R_{\text{eq}})=1.6\times 10^5$ g cm$^{-2}$, which roughly corresponds to
$\dot{M}\approx 10^{-6} M_{\odot}$ yr$^{-1}$ \citep{ekstrom}.
Similarly to the stationary calculations described in Sect.~\ref{statcalc}, the
time-dependent models are totally independent of the scaling of the surface density $\Sigma$.
The inner boundary condition for $V_R$ is set as outflowing, i.e.,~the radial velocity is extrapolated 
from mesh interior values as a 0-th order extrapolation (the boundary value is simply 
identical with the value in neighboring computational cell in the interior mesh, cf.~Fig.~\ref{stagis} and see subroutine \textit{boundary} in Appendix~\ref{dvojdimhydrous}).
We assume the inner boundary condition for $V_{\phi}$ as the fixed rotational (Keplerian) velocity of the critically rotating 
star.
The outer boundary conditions are considered as outflowing for all these quantities \citep{kurfek}.

The initial conditions (initial state) we set in the following way: the initial scaled surface density profile $\Sigma/\Sigma(R_\text{eq})=R^{-2}$ 
\citep[see, e.g.,][see also Sect.~\ref{radthin}]{okazaki} while the initial azimuthal velocity is exactly Keplerian ($V_\phi=V_K$) everywhere.
We start the numerical calculations with zero initial gas radial velocity 
throughout the whole disk. 
Our numerical tests have confirmed that the final solution is fully independent of any
selected initial radial velocity profile, we might as well set a zero initial radial velocity $V_R$ within the entire disk \citep{kurfek}.

The calculations clearly prove
that the profiles of the surface density $\Sigma$ and the radial velocity $V_R$ as well as the sonic point distance
(where $V_{R}/a=1$) very weakly depend on the viscosity slope parameter $n$ \citep[Eq.~\eqref{alpvis}, see also][]{Kurf2,Kurf4,kurfek}.
The outer limit of Keplerian rotation velocity region ($V_{\phi}\sim R^{-0.5}$) is almost independent of the viscosity parameter. 
The calculations nevertheless show strong dependence of the outer edge of the region where the rotational velocity behaves as
angular momentum conserving ($V_{\phi}\sim R^{-1}$) (i.e., of the distance where the rotational velocity begins to 
rapidly drop) on the viscosity parameter (for a given temperature profile).  
For a selected range of the viscosity parameter $n$ the distance of this region differs approximately by
one order of magnitude (see, e.g., Fig.~\ref{B0pt0}, see also Table~\ref{tablicek}) \citep{kurfek}. 
However, in the models the location of this rapid rotational velocity drop
does not anywhere exceed the radius where the disk equatorial density falls to 
averaged interstellar medium density (where we assume its average density as $10^{-23}\,\text{g}\,\text{cm}^{-3}$, see, e.g.,~\citet{misirio, maeder}).
At this distance a kinetic plasma modeling
would likely be required; moreover, the interaction of the disk with interstellar medium 
has to be taken into account (cf.~Sect.~\ref{zerodens}).

Within the time-dependent calculations we have examined the differences in the
numerical results between the two different prescriptions for viscous torque (Eqs.~\eqref{viscon} and \eqref{angfl}) \citep{kurfek}.
Similarly to the stationary disk calculations (see Sect.~\ref{statcalc}), the
first order linear viscosity calculations (Eq.~\eqref{viscon}) show in distant regions the rapid rotational velocity 
drop to negative values and consequently indicate very slow convergence back to zero. 
The rotational velocity profiles calculated involving
the second order linear viscosity term (Eq.~\eqref{angfl}) however consistently confirm the analytical 
implication $V_{\phi}>0$, resulting from Eq.~\eqref{angestim2}, through the entire disk range (see Sect.~\ref{radthin}) \citep{kurfek}.

Figure \ref{B0pt0} illustrates the isothermal case ($p=0$) of a B0-type star 
where we assume $\dot{M}\approx 10^{-9}\,M_{\odot}$ yr$^{-1}$ \citep{granada}
and where the inner boundary value of the viscosity parameter $\alpha(R_{\text{eq}})=0.025$ \citep{Penna}.
The calculated distance of the sonic point, $R_\text{s}\approx 550\,R_{\mathrm{eq}}$, roughly corresponds to
the analytical prediction from Eq.~\eqref{sonicestR} with
$R_\text{s}$ being approximately $480\,R_{\mathrm{eq}}$.
The value of maximum angular momentum loss rate $\dot{J}_{\text{max}}$ (see Eq.~\eqref{sonicestL}) is independent of viscosity 
while it strongly depends on the profile of temperature.
Since $\dot{J}_{\text{max}}$ roughly equals the angular momentum loss rate at the sonic point 
radius, we assume that the total angular momentum contained in the disk is \citep{kurfek}
\begin{equation}\label{angtot}
J_{\text{disk}}=\int_{R_{\text{eq}}}^{R_{\text{s}}}2\pi R^2\Sigma V_{\phi}\,\text{d}R.
\end{equation} 
We therefore regard the sonic point distance $R_\text{s}$ as the disk outer edge (cf.~Eq.~\eqref{rsrout}). 
Analogously this also determines the mass of the disk
\begin{equation}\label{masstot}
M_{\text{disk}}=\int_{R_{\text{eq}}}^{R_{\text{s}}}2\pi R\Sigma\,\text{d}R.
\end{equation}
If we compare for example
the total disk angular momentum $J_{\text{disk}}$ with the total stellar angular momentum $J_{\star}=\eta 
M_{\!\star}R_{\star}^2\Omega_\text{crit}$ where 
a dimensionless parameter $\eta\approx 0.05$ (for a main sequence star with $M_{\!\star}=9M_\odot$, \citet{Meynet1}), 
the ratio $J_{\text{disk}}/J_{\star}$ in this case takes the value $1.2\times 10^{-6}$ \citep{kurfek}.
\begin{figure} [t]
\centering\resizebox{0.6\hsize}{!}{\includegraphics{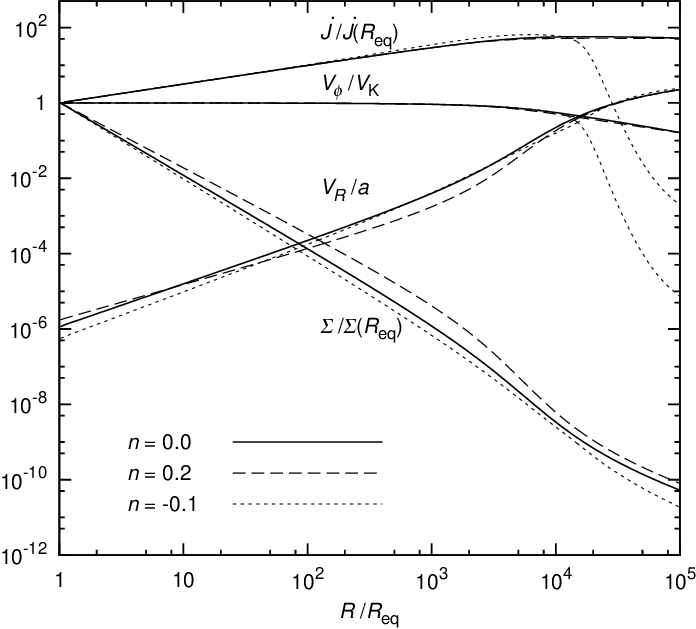}}
\caption{As in Fig.~\ref{B0pt0}, however for temperature decreasing with a power
law with $p=0.4$.
Inner boundary viscosity $\alpha(R_{\text{eq}})=0.025$ is considered. 
The characteristic radii (sonic point distance, 
outer disk radius) are in this case significantly larger. Adapted from \citet{kurfek}.} 
\label{B0pt0n04}
\end{figure}
\begin{figure} [t]
\centering\resizebox{0.6\hsize}{!}{\includegraphics{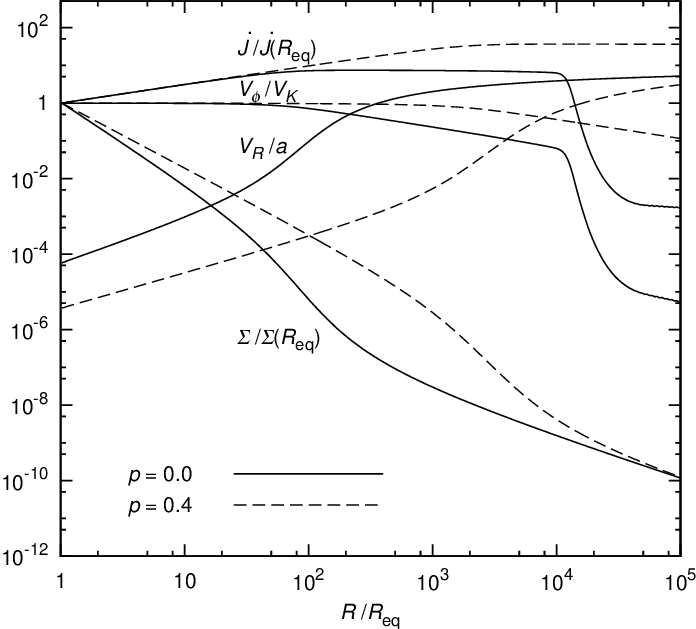}}
\caption{Comparison of the radial profiles of relative surface density and relative velocities and the radial
profiles of angular momentum loss rate in case of decreasing viscosity ($n=0.2$) for isothermal disk ($p=0$)
and for outward decreasing temperature profile ($p=0.4$) in a final stationary state of time-dependent disk models
for Pop III star. Adapted from \citet{kurfek}.}
\label{ptI}
\end{figure}
\begin{table}[h!]
\caption{Comparison of the distances of sonic point ($R_{\text{s}}$), maximum angular momentum loss rate 
($\dot{J}_{\text{max}}$) and the beginning of the rotation velocity drop
for B0-type star and Pop III star for selected temperature parameters $p$ and viscosity parameters $n$.}
\label{tablicek}
\begin{center}
\begin{tabular}{lrlccc}
\multicolumn{5}{l}{B0-type star:}\\
\hline
$p$  &\!\! & \!\!\!\!\!\!$n$   &$R_{\text{s}}$ $(R/R_{\text{eq}})$&
$\dot{J}_{\text{max}}$ $(R/R_{\text{eq}})$&\!\!\!\!\!$V_\phi$ drop $(R/R_{\text{eq}})$\\
\hline
   &\!\! & \!\!\!\!\!\!0   &$5.5\cdot 10^2$&$3.1\cdot 10^2$&\!\!\!\!\!$3.3\cdot 10^3$\\
0  &\!\! & \!\!\!\!\!\!0.2 &$5.6\cdot 10^2$&$3.6\cdot 10^2$&\!\!\!\!\!$2.0\cdot 10^4$\\
   &\!\!-& \!\!\!\!\!\!0.1 &$5.6\cdot 10^2$&$2.7\cdot 10^2$&\!\!\!\!\!$3.0\cdot 10^3$\\
\hline
   &\!\! & \!\!\!\!\!\!0   &$3.1\cdot 10^4$&$1.7\cdot 10^4$&\!\!\!\!\!$5.0\cdot 10^5$\\
0.4&\!\! & \!\!\!\!\!\!0.2 &$3.2\cdot 10^4$&$2.1\cdot 10^4$&\!\!\!\!\!$-$\\
   &\!\!-& \!\!\!\!\!\!0.1 &$3.2\cdot 10^4$&$1.3\cdot 10^4$&\!\!\!\!\!$1.5\cdot 10^4$\\
\hline\\
\multicolumn{5}{l}{Pop III star:}\\
\hline
$p$  &\!\! & \!\!\!\!\!\!$n$   &$R_{\text{s}}$ $(R/R_{\text{eq}})$&
$\dot{J}_{\text{max}}$ $(R/R_{\text{eq}})$&\!\!\!\!\!$V_\phi$ drop $(R/R_{\text{eq}})$\\
\hline
   &\!\! & \!\!\!\!\!\!0   &$3.6\cdot 10^2$&$2.1\cdot 10^2$&\!\!\!\!\!$3.0\cdot 10^3$\\
0  &\!\! & \!\!\!\!\!\!0.2 &$3.7\cdot 10^2$&$2.3\cdot 10^2$&\!\!\!\!\!$1.5\cdot 10^4$\\
   &\!\!-& \!\!\!\!\!\!0.1 &$3.7\cdot 10^2$&$1.9\cdot 10^2$&\!\!\!\!\!$2.0\cdot 10^3$\\
\hline
   &\!\! & \!\!\!\!\!\!0   &$1.9\cdot 10^4$&$1.0\cdot 10^4$&\!\!\!\!\!$3.0\cdot 10^5$\\
0.4&\!\! & \!\!\!\!\!\!0.2 &$2.0\cdot 10^4$&$1.2\cdot 10^4$&\!\!\!\!\!$-$\\
   &\!\!-& \!\!\!\!\!\!0.1 &$2.0\cdot 10^4$&$8.3\cdot 10^3$&\!\!\!\!\!$1.2\cdot 10^4$\\
\hline
\end{tabular}
\end{center}
\end{table} 

Figure \ref{B0pt0n04} shows the case of decreasing temperature profile ($p=0.4$) of the 
B0-type star
with the same viscosity profiles as in Fig.~\ref{B0pt0}.
The sonic point distance is in this case roughly
$R_{\text{s}}\approx 31\,500\,R_{\mathrm{eq}}$ for all the viscosity profiles, which is about two orders of magnitude larger
than in the isothermal case. 
In this model the ratio $J_{\text{disk}}/J_{\star}$ becomes $7.9\times 10^{-5}$,
which is similarly 
about two orders of magnitude larger than in the isothermal case.

Figure \ref{ptI} shows the calculated profiles of the Pop III star's disk with assumed
$\dot{M}\approx 10^{-6}\,M_{\odot}$ yr$^{-1}$ \citep{ekstrom} and with various temperature parameters 
for fixed decreasing viscosity ($n=0.2$).
The graph clearly shows a strong dependence of radii of the sonic point and of 
the rapid rotational velocity drop, as well as of the slopes 
of surface density and radial velocity on temperature profile.
The sonic point radius is now located at the distance
$R_{\text{s}}\approx 360\,R_{\mathrm{eq}}$ in the isothermal case while its position 
$R_{\text{s}}$ is approximately $16\,500\,R_{\mathrm{eq}}$ in case of radially decreasing temperature with the selected parameter $p=0.4$. 
The same dimensionless stellar angular momentum parameter $\eta\approx 0.05$ 
gives the ratio $J_{\text{disk}}/J_{\star}=1.2\times 10^{-3}$ for the isothermal model and
$J_{\text{disk}}/J_{\star}=0.2$ for the model with decreasing temperature \citep{kurfek}.
In the latter case the disk carries away a significant fraction of stellar angular momentum
and the star may not have enough angular momentum to develop the disk fully.
In this case the stellar evolution has to be calculated together with the disk evolution.
All the performed calculations however support the idea that the unphysical drop of the
rotational velocity at large radii can be avoided in the models with radially decreasing
viscosity parameter and temperature \citep{kurfek}.
\begin{figure} [t]
\centering\resizebox{0.485\hsize}{!}{\includegraphics{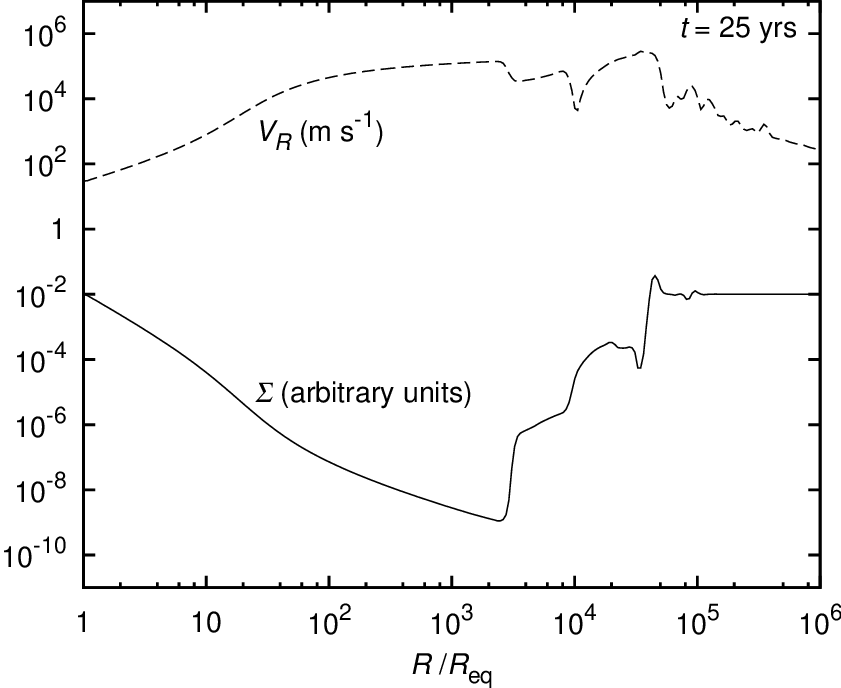}}\hfill\resizebox{0.485\hsize}{!}{\includegraphics{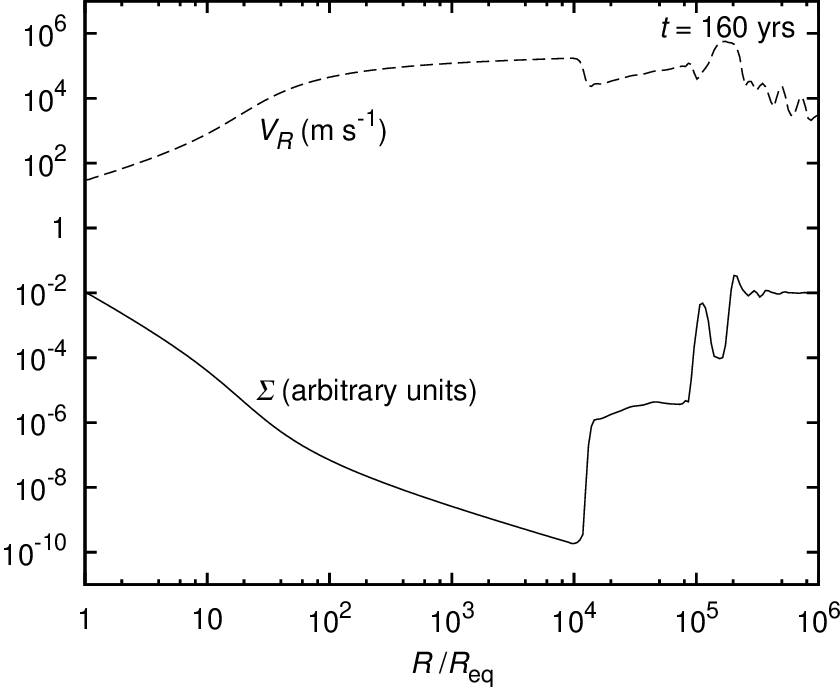}}
\caption{Snapshots of evolved radial profiles of the disk integrated density $\Sigma$ and the disk radial velocity $V_R$. 
The initial integrated density profile is set as constant value $\Sigma_{\text{ini}}=10^{-2}\,\text{kg}\,\text{m}^{-2}$ 
throughout the entire isothermal disk range, whose temperature corresponds to B0-type star's effective 
temperature $T_{\text{eff}}=30\,000\,\text{K}$. Since no constraint applies to the inner boundary value of the surface density,
the values of $\Sigma$ correspond to arbitrary units (i.e., to kg m$^{-2}$ multiplied by an arbitrary constant). The left panel shows these profiles in time $t=25\,\text{yrs}$ 
while the right panel shows the evolution of the same disk
in more evolved time $t=160\,\text{yrs}$ (see Sect.~\ref{zerodens} for other details).}
\label{zerovcak}
\end{figure}
\subsection[Models with very low constant initial density]{Models with very low constant initial density}\label{zerodens}
We have also investigated the case with arbitrarily low nonzero initial surface density values $\Sigma_{\text{ini}}$ that are constant within  
the whole computational domain (we do not presume in the initial state any similarity with the supposed final surface density profile).
We define the initial (Keplerian) value of $V_{\phi}$ of such very rare circumstellar matter only up to a few tens of stellar radii while it is 
followed by a discontinuous jump of the rotational velocity down to zero.
The initial profile of the radial velocity is zero everywhere.
Even using these initial conditions the disk evolves to a large distance and tends to converge
to the proper final state \citep{kurfek}. 

Fig.~\ref{zerovcak} demonstrates two snapshots of the propagating converged state of the vertically integrated density $\Sigma$
and the radial velocity $V_R$. The initial state of the surface density was in this case set to a value $\Sigma_{\text{ini}}=10^{-2}\,\text{kg}\,\text{m}^{-2}$ 
(this was however selected rather arbitrarily as a very low constant number)
within the whole computational domain. No constraint applies to the inner boundary value of the surface density, 
it may freely evolve (outflowing boundary condition). The outer boundary condition for the surface density is outflowing in all the presented models. 
The boundary conditions for the radial velocity are set as outflowing as well (cf.~the boundary conditions 
described in Sect.~\ref{timemod} and in the subroutine \textit{boundary} in Appendix \ref{adjustik}). The profiles of the calculated characteristics establish in a relatively short time 
the expected shape in the inner disk region, corresponding to the converged shape in Sect.~\ref{timemod}. The only yet unresolved problem is the proper scaling of the surface density during the calculation. 
The scaling of $\Sigma$ however does not affect the calculations as well as the values of 
other quantities at all, see the notes in Sects.~\ref{statcalc} and \ref{timemod}
referring to this point. 

Left panel of Fig.~\ref{zerovcak} shows the evolution of $\Sigma$ and $V_R$ in the isothermal disk with the stellar effective 
temperature $T_{\text{eff}}=30\,000\,\text{K}$ and with constant viscosity ($\alpha=0.025$) in time $t=25\,\text{yrs}$.  
Right panel of Fig.~\ref{zerovcak} shows the evolution of the same disk
in more evolved time $t=160\,\text{yrs}$. 
The density in the region of an outer disk radius drops below the initial state density (which we may regard as an interstellar medium density, however,
the integrated density $\Sigma$ of interstellar medium is questionable), forming a rarefaction wave that extends
radially with time \citep{kurfek}. The matter from this rarefaction wave region is moved to the area of an excess of matter on the edge of the unperturbed interstellar medium.

When examining higher disk temperatures, corresponding to $T_{\text{eff}}=50\,000\,\text{K}$ or more (which implies higher gas pressure),
the expanding wave may at some time begin to return back, which may again change to expansion, etc. We have also examined the 
models with decreasing temperature and viscosity. The propagating speed is for the decreasing temperature profile lower, which corresponds to conclusions made in Sect.~\ref{chartime}. 
The wavefront of the propagating converging wave is however not smooth in this case 
like it is in Sect.~\ref{chartime} (see Fig.~\ref{shock}), instead it exhibits several density as well as radial velocity bumps whose evolution 
is relatively stable in time. Further investigation is needed for the explanation of their nature (random or numerical artifacts vs. 
possible link to bow shocks in the contact zone
between the expanding disk material and interstellar medium, see also notes in Sect.~\ref{chartime}). 

\subsection[Models with subcritically rotating central star]{Models with subcritically rotating central star}\label{subcritic}
Within the time-dependent calculations we have also examined 
subcritically rotating stars modifying the inner boundary condition for disk rotation velocity $V_{\phi}$.
Following the prescription for angular momentum loss at radius $R$ in the steady Keplerian circumstellar disk, 
\begin{align}\label{slowdown1}
\dot{J}=\dot{M}RV_\phi,                          
\end{align}
we regard the quantity $\dot{M}$ as the radial mass flux through the arbitrary circular segment of the disk 
(cf.~Eq.~\eqref{angularmomentumfluxix0}). We however consider $\dot{M}=\dot{M}_{\!\star}=\text{const.}$ (at least within sufficiently 
long period of stellar as well as of the disk evolution time). Approximating the loss of the stellar mass ${M}_{\!\star}$ (of the B0-type star described in Sect.~\ref{statcalc}) due to 
$\dot{M}$ as negligible, we consider also ${M}_{\!\star}=\text{const.}$ We also neglect in this simple model
the stellar oblateness, considering merely the radius $R_\star$ of the spherical stellar body. Following these constraints, 
the effect of $\dot{J}$ on the left-hand side of Eq.~\eqref{slowdown1} may be regarded as slowing down of the stellar rotation,
$\dot{J}=-{M}_{\!\star}R_\star\dot{V}_{\phi}(R_\star)$.
Combining this with Eq.~\eqref{slowdown1}, we obtain the rate of braking of the stellar rotation in the quite simple form,
\begin{align}\label{slowdown2}
V_{\phi}(R_\star)=V_K(R_\star)\,\text{e}^{-\frac{\dot{M}}{{M}_{\!\star}}t},                     
\end{align}
where $V_K(R_\star)$ denotes the Keplerian (critical) rotational velocity at stellar equatorial surface which enters the calculation as the initial state.
Equation \eqref{slowdown2} was incorporated in the hydrodynamic code as a time-dependent inner boundary condition for the disk azimuthal velocity. We
successively examined the calculation for various values of $\dot{M}$ ranging from $10^{-9}\,M_\odot\,\text{yr}^{-1}$ up to hypothetical $10^{-4}\,M_\odot\,\text{yr}^{-1}$.

In the initial phase of the calculation the central star rotates critically (does not slow down) until the radial velocity of the disk converges in the very inner disk region
(cf.~the radial velocity contours in Fig.~\ref{slozenka1}). Consequently is the stellar rotational slowdown switched on.
For the inner boundary value of the azimuthal velocity
$V_{\phi}(R_\star)\gtrapprox 0.97 V_K(R_\star)$
the models (regardless of the selected $\dot{M}$) precisely converge in the supersonic region. However, in the case
when the boundary rotational velocity is only slightly higher than the above limit 
($0.97 V_K(R_\star)\lessapprox V_{\phi}(R_\star)\lessapprox 0.98 V_K(R_\star)$), 
there occur (more or less regular) pulsations in density and radial velocity profiles in the region close to the star, i.e., up to approximately $2$-$3$ stellar radii.
\begin{figure} [t]
\centering\resizebox{0.525\hsize}{!}{\includegraphics{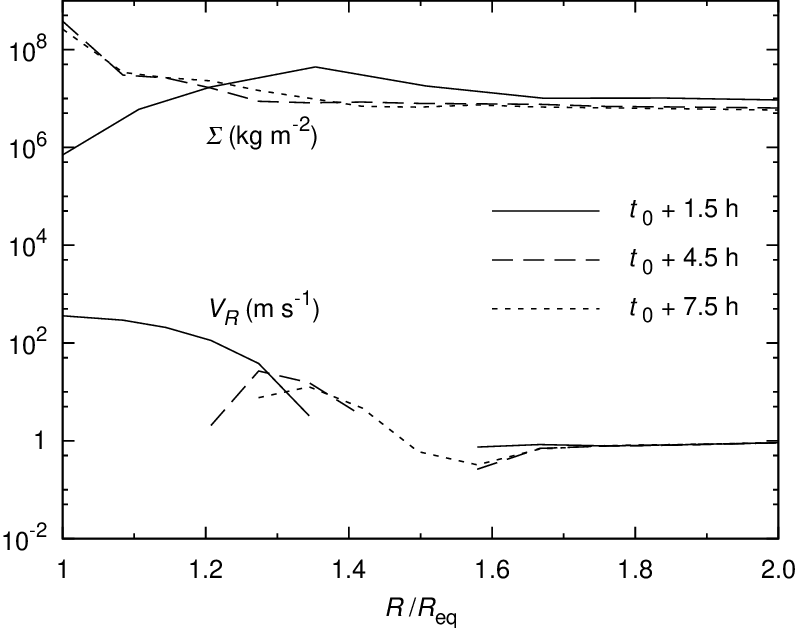}}
\caption{Example of three snapshots of the propagation of the (associated) waves of integrated density $\Sigma$ and the radial velocity $V_R$, 
generated by the pulsations of central star (see Sect.~\ref{subcritic} for details of the parameters of pulsation). 
The wave is showed in three successive times, $t_0+1.5\,\text{hour}$ (corresponding to fastest rotation and minimum radius), 
$t_0+4.5\,\text{hour}$ (slowest rotation and maximum radius), $t_0+7.5\,\text{hour}$, etc., after the time $t_0$ of the equilibrium position of the stellar pulsation. 
The missing parts of the radial velocity curves correspond to negative values of $V_R$ in this regions (the matter infall).
Similar wave is produced by each successive stellar pulse. The stationary (not affected by the stellar pulsations) 
surface density at the stellar surface, which corresponds to selected hypothetical stellar $\dot{M}=10^{-5}\,M_\odot\,\text{yr}^{-1}$, 
is in this case approximately $4\cdot 10^7\,\text{kg}\,\text{m}^{-2}$, the stationary radial velocity is approximately $0.4\,\text{m}\,\text{s}^{-1}$.}
\label{ubwave}
\end{figure}
For $V_{\phi}(R_\star)\lessapprox 0.97 V_K(R_\star)$ the density (and consequently the radial velocity) 
profile is unstable and gradually declines; for lower $V_{\phi}(R_\star)$ the decrease in density is faster \citep{kurfek}. In all these cases there forms a significant gap in 
the density and radial velocity in the proximity of the star (which fall in this region into negative values) that may indicate the infall of the material that is not supported by fast rotation.

We have calculated also the basic introductory model with the inner disk characteristics affected by the stellar pulsations. We involved merely 
radial pulsations in order to yield a simple kinematic relations. We have considered small radial pulsations
of critically rotating star,
where the amplitude of the spatial variations of the stellar equatorial radius $\delta R_{\text{eq}}\ll R_{\text{eq}}$. 
We do not model the spatial variations of the stellar equatorial radius $R_{\text{eq}}$ and involve only the variations of the stellar rotational velocity.
The pulsations are linearized \citep{maeder}, i.e., we consider
the time dependence of the pulsations as a harmonic motion with angular frequency $\omega=2\pi/P$ (where $P$ is the period). The time dependent 
stellar equatorial surface rotation velocity therefore is
\begin{align}\label{pulsacek}
V_\phi(R_{\text{eq}})(t)=\frac{\sqrt{GM_{\!\star} R_{\text{eq}}}}{R_{\text{eq}}+(\delta R_{\text{eq}})\,\text{sin}\,\omega t}. 
\end{align}
We employed the period of the pulsations $P=0.25\,\text{d}$ (6 hours), 
which roughly corresponds to typical period of Be stars' non-radial $g$-modes \citep[][cf.~also Sect.~\ref{Bephen}]{sako}. We have assumed the amplitude
of the radial oscillations $\delta R_{\text{eq}}=0.05 R_{\text{eq}}$. Figure~\ref{ubwave} shows a few snapshots of the (associated) density and radial velocity waves propagation,
produced by each stellar pulse. The waves propagate in the model
outwards up to the distance of about 2 stellar equatorial radii where they are evidently damped. 

The inner boundary disk surface density ranges 
from approximately $\Sigma(R_{\text{eq}})_{\text{min}}\approx 8\cdot 10^{5}\,\text{kg}\,\text{m}^{-2}$ (fastest rotation)
to $\Sigma(R_{\text{eq}})_{\text{max}}\approx 3.8\cdot 10^{8}\,\text{kg}\,\text{m}^{-2}$ (slowest rotation), while the stationary (not affected by pulsations) value,
corresponding via the mass conservation equation to the selected hypothetical stellar $\dot{M}=10^{-5}\,M_\odot\,\text{yr}^{-1}$, is 
$\Sigma(R_{\text{eq}})\approx 4\cdot 10^{7}\,\text{kg}\,\text{m}^{-2}$. The corresponding inner boundary disk radial velocity ranges within the process
from approximately $V_R(R_{\text{eq}})_{\text{min}}\approx -400\,\text{m}\,\text{s}^{-1}$ (infall) to $V_R(R_{\text{eq}})_{\text{max}}\approx 350\,\text{m}\,\text{s}^{-1}$, 
while the stationary value is $V_R(R_{\text{eq}})\approx 0.4\,\text{m}\,\text{s}^{-1}$ (see Fig.~\ref{ubwave}). However, the study has to be yet properly verified by 
a large number of similar simulations.

Much more work on the models with variations of the stellar rotation 
is however needed. We have not yet performed for example the proper investigation of the (possible) influence of the temperature and viscosity
as well as of the absolute values of various physical quantities and parameters, i.e., of $T_\text{eff}$, $\alpha(R_\text{eq})$, mass and radius of the central star, on the described process. 

Another similar idea is to examine the modeling 
of the disk profiles evolution if the subcritical rotation is already entered as the initial state, such attempts have 
been however not successful so far.
It is also necessary to improve the calculations of the models with pulsations, e.g.,~by
using a spatial grid with much more steeper concentration of grid points towards the inner boundary,
achieving an optimal time coordination between pulsation period and the time step of the model (while not exceeding a reasonable limit of the computational cost)
as well as by including various parameters of pulsations, etc.

\section[One-dimensional modeling of disk magnetorotational instabilities]{One-dimensional calculations of disk magnetorotational instabilities}\label{magnetousci}
Magnetorotational instabilities (hereafter MRI) generated in accretion disks threaded by sufficiently weak magnetic fields 
are considered to be the main source 
of anomalous viscosity that plays a key role in the angular momentum transport in such disks \citep[][see also Sect.~\ref{sivikis}]{Balbus}. 
The driving mechanism that is responsible for the material transport from the central star into the decretion (outflowing) disk 
is still not fully resolved (see Sect.~\ref{diskfome}), it is however believed that the outward transporting of the material and angular momentum in the 
disk is controlled by the same anomalous viscosity. Although the theoretical as well as observational studies indicate
that the stellar decretion disks may extend up to quite large distance of several hundreds of stellar radii 
\citep[e.g.,][see also Sect.~\ref{onedimtimemod}]{Krticka,kurfek}, we do not still know exactly the quantitative behavior 
of magnetorotational instabilities in various disk regions, namely whether their importance in the disk outer regions is the 
same or comparable to that in the inner disk regions, etc. We provide the analytical study \citep{Krticka_2014} that is accompanied by one-dimensional 
hydrodynamic modeling of the disk MRI radial profiles and their dependence on various temperature and viscosity parameters
(which is however expected to be followed by the two-dimensional modeling of this 
problem in an oncoming study).

\subsection[Radial structure of MRI]{Radial structure of MRI}\label{magnetouscisub1}
We describe the basic magnetohydrodynamics as well as the basic equations that rule the magnetorotational instabilities in Sect.~\ref{sivikis}.
The geometrical condition for the vertical component of magnetic field that may develop an instability, 
\begin{align}\label{paperref1}
\lambda_{z,\text{max}}<2H,\quad\text{i.e.,}\quad\frac{2\pi}{k_{z,\text{max}}}<2H,
\end{align}
indicates that the maximum wavelength of any magnetically driven perturbation (see Eq.~\eqref{perturb})
should be shorter than the disk (scaled) thickness at arbitrary radius of the disk. Inserting the dispersion relation $a/\Omega=H$  
(Eq.~\eqref{slightvertigo}) into equation of the disk equatorial plane density $\rho_0=\Sigma/(\!\!\sqrt{2\pi}H)$ \eqref{Sigmasegva} where we expand 
the fraction using the mass loss rate $\dot{M}=2\pi R\Sigma V_R$ (Eq.~\eqref{angularmomentumfluxix0}), and following the criterion of stability given in
Eq.~\eqref{stabil6} (where we note that for 
values of vertical wavenumber $k_z$ less than the critical value $k_{z,\,\text{crit}}=k_{z,\,\text{max}}$ this criterion leads to instability),
we obtain the condition for the development of this instability rewritten as the condition for the vertical component of magnetic induction $B_z$ for general rotational velocity and $B_{z,K}$ for 
Keplerian regions (where $V_{\phi}=V_K=\sqrt{GM_\star/R}$), respectively,
\begin{align}\label{paperref3}
B_z<\frac{\sqrt{\mu_0}}{\pi}\left(\frac{2\dot{M}a}{(2\pi)^{3/2}V_RV_{\phi}}
\left|\frac{V_{\phi}}{R}\frac{\text{d}V_{\phi}}{\text{d}R}-\frac{V_{\phi}^2}{R^2}\right|\right)^{1/2}\!\!\!,\quad\quad
B_{z,K}<\frac{\sqrt{3\mu_0}}{\pi}\left(\frac{\dot{M}a V_K}{(2\pi)^{3/2}V_R R^2}\right)^{1/2}\!\!\!.
\end{align}
Physically this relation represents
the gas pressure dominance over the magnetic one, i.e.,~the plasma
parameter $\beta>\pi^2/3\approx3$ \citep[cf.~Eq.~\eqref{isoth6}, see also,][]{Balbus,horal}.
We note that for lower values of $\beta$ parameter only stable modes ($k_z>k_{z,\text{max}}$) can exist
in the disk.
The Keplerian relation in Eq.~\eqref{paperref3} can be rewritten, using the scaled quantities, as 
\begin{equation}
\label{paperref3a}
B_z<0.025\,\text{T}\left[\left(\frac{a/V_R}{10^{3}}\right)
\left(\frac{V_\text{K}}{100\,\text{km}\,\text{s}^{-1}}\right)
\left(\frac{\dot M}{10^{-9}\,M_\odot\,\text{yr}^{-1}}\right)\right]^{1/2}
\left(\frac{R}{1\,R_\odot}\right)^{-1}
\end{equation}
(where we express the magnetic field induction in tesla (T) units in the system SI).
In stars with magnetic field stronger than that given by Eq.~\eqref{paperref3a}
the MRI does not operate in stationary conditions. Such stars may possibly
accumulate the material close to their equator until the condition for the MRI
development $k_z<k_{z,\text{max}}$ is fulfilled. The stars may show episodic
ejections of matter.

Considering only the disk equatorial plane (or the region that is vertically very close), we neglect the vertical piece $N_z\approx 0$ 
of the Brunt-V\"ais\"al\"a frequency (see Sect.~\ref{sheainst}). The MRI frequency dispersion relation \eqref{mhd41}, where $\kappa$ is 
the epicyclic frequency from Eq.~\eqref{epicyclfreq}, thus modifies to
\begin{align}\label{paperref4}
\frac{k_R^2+k_z^2}{k_z^2}\left(\omega^2-k_z^2V_{Az}^2\right)^2-\left(\kappa^2+N_R^2\right)\left(\omega^2-k_z^2V_{Az}^2\right)-4\Omega^2k_z^2V_{Az}^2=0,
\end{align}
giving the quadratic relation for squared MRI frequency $\omega^2$ whose solution we write as
\begin{align}\label{paperref4a}
\omega^2_{1,2}=\frac{k_z^2}{2\left(k_R^2+k_z^2\right)}\left[2\left(k_R^2+k_z^2\right)V_{Az}^2+\kappa^2+N_R^2
\pm\sqrt{\left(\kappa^2+N_R^2\right)^2+16\left(k_R^2+k_z^2\right)V_{Az}^2\Omega^2}\right].
\end{align}
This however leads to MRI development in the case that for some wavenumbers the root of Eq.~\eqref{paperref4a} becomes negative, $-\omega^2>0$, 
which implies that the negative branch of Eq.~\eqref{paperref4a} is the desired solution.
By setting $\partial\omega^2/\partial k_R=0$ ($k_R$ and $k_z$ are positive numbers) we find the extremum of $\omega^2$ in dependence on the 
radial wavenumber for $k_R=0$.
The extremum of MRI frequency $\omega^2$ 
in dependence on vertical wavenumber $k_z$ we find analogously by setting $\partial\omega^2/\partial k_z=0$ in Eq.~\eqref{paperref4a}.
Inserting the solution $k_R=0$ we obtain the relation
\begin{align}\label{paperref4c}
\sqrt{\left(\kappa^2+N_R^2\right)^2+16\,k_z^2V_{Az}^2\Omega^2}=4\Omega^2,\quad\quad\text{yielding}\quad\quad 
k_z=\frac{1}{4V_{Az}\Omega}\sqrt{16\,\Omega^4-\left(\kappa^2+N_R^2\right)^2}.
\end{align}
Inserting $k_z$ from Eq.~\eqref{paperref4c} together with the constraint $k_R=0$ into Eq.~\eqref{paperref4a}, 
we get the disk equatorial plane solution of the MRI frequency,
\begin{align}\label{paperref4d}
\omega=\left[\frac{\kappa^2+N_R^2}{2}\left(1-\frac{\kappa^2+N_R^2}{8\,\Omega^2}\right)-\Omega^2\right]^{1/2},\quad\quad\text{where}\quad\quad
N_R^2=0.4\left(\frac{a}{\rho}\frac{\text{d}\rho}{\text{d}R}\right)^2
\end{align}
\begin{figure}[t]
\centering
\includegraphics[width=6.8cm]{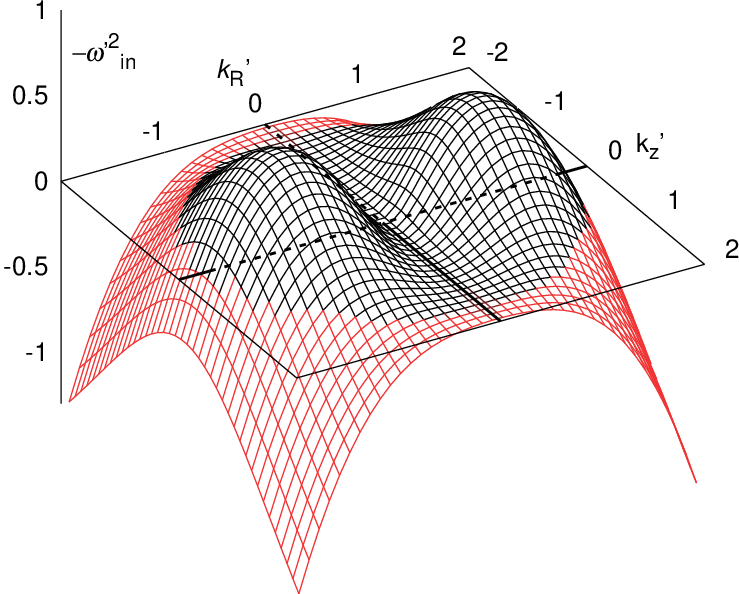}\quad\quad\quad\includegraphics[width=6.8cm]{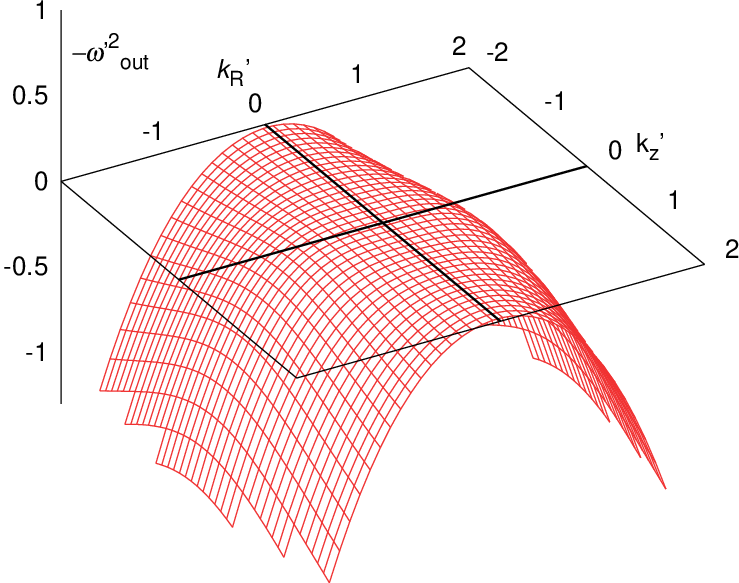}
\small{Fig.~\ref{inthrbol}a}\quad\quad\quad\quad\quad\quad\quad\quad\quad\quad\quad\quad\quad\quad\quad\quad\quad\quad\small{Fig.~\ref{inthrbol}b}
\caption{Graph of the root $-{\omega^\prime}^2=-\omega^2/\Omega^2$ of the negative branch of the dispersion relation Eq.~\eqref{paperref4a}
in the equatorial plane of the disk as an
analytical function of scaled radial and vertical wavenumbers $k_R'=k_RV_{Az}/\Omega$ and $k_z'=k_zV_{Az}/\Omega$ 
in the inner Keplerian disk region (Fig.~\ref{inthrbol}a) described in Eq.~\eqref{paperref4g}
and in the outer disk region (Fig.~\ref{inthrbol}b). The plane $\omega^\prime\!\!=\omega=0$ divides the stable perturbations
$-{\omega}^2<0$ (the red domain) from the unstable ones that may lead to an instability $-{\omega}^2>0$ (the black domain).}
\label{inthrbol}
\end{figure}
in the isothermal case (cf.~Eq.~\eqref{mhd40} and the radial analog of Eq.~\eqref{brunt3}). 
In the region close to the star where the disk is vertically very thin (see Sect.~\ref{thindisk}) 
we set the radial piece (see Eqs.~\eqref{isoth3} and \eqref{isoth4}) 
of the Brunt-V\"ais\"al\"a frequency $N_R\approx 0$.
Assuming also the Keplerian rotation where $\kappa=\Omega$ (cf.~Eq.~\eqref{epicyclfreq} and accompanying explanations), 
we obtain from Eq.~\eqref{paperref4c} the vertical wavenumber $k_z=\sqrt{15}\Omega/(4V_{Az})$ and 
the MRI frequency in the equatorial plane of the Keplerian disk region is
\begin{align}\label{paperref4e}
\omega=\frac{3}{4}i\Omega.
\end{align}

Within the Keplerian isothermal conditions where the sound speed $a=\text{const.}$, $V_R\sim R$ and $V_\phi\sim R^{-1/2}$ (see Sect.~\ref{radthin}), the radial dependence 
of the limiting vertical magnetic field component for the MRI development (which is given in general by Eq.~\eqref{paperref3a}) is $B_{z,\text{max}}\sim R^{-7/4}$.
Comparing this with the dependence of the decrease of the stellar dipolar magnetic field intensity $B_{z,\text{dip}}\sim R^{-3}$ (cf.~Eq.~\eqref{rigid1corot}),
it indicates that the conditions that may enable to develop the MRI close to the star yet improve with increasing distance.
This conclusion is however violated in large distance from the central star due to different physical conditions that govern there the disk behavior 
(see also Sect.~\ref{timemod}). Within the transitional region from the Keplerian to angular momentum conserving rotational velocity
($V_\phi\sim R^{-1}$) the epicyclic frequency (Eq.~\eqref{epicyclfreq}) diminishes while in the latter region it becomes completely zero
(unlike the Keplerian region where $\kappa^2=\Omega^2$). 
We find a certain radius $R_{\omega,0}$ where the squared MRI frequency $\omega^2$ transits from negative to positive sign and the MRI may vanish at larger radii,
i.e., where $R>R_{\omega,0}$. Setting $\omega^2=0$, we find from Eq.~\eqref{paperref4d} the zero MRI frequency radius $R_{\omega,0}$ located at the point where
(cf.~also Eq.~\eqref{mhd40})
\begin{align}\label{paperref4f}
\Omega=\frac{1}{2}N_R=\frac{1}{\sqrt{10}}\frac{a}{\rho}\left|\frac{\text{d}\rho}{\text{d}R}\right|.
\end{align}
In Fig.~\ref{inthrbol}a we plot the dispersion relation for the disk midplane MRI in the inner Keplerian disk region (Eq.~\eqref{paperref4a}), setting $\kappa^2=\Omega^2$
and $N_R=0$ (see the description above in this section),
while in Fig.~\ref{inthrbol}b we demonstrate the same prescription for the outer disk region,
regarding the conditions that correspond to the radius $R_{\omega,0}$, i.e.,~$\kappa=0$ (angular momentum conserving region, see Fig.~\ref{ome1} in Sect.~\ref{sheainst}) 
and $N_R=2\Omega$ (see Eq.~\eqref{paperref4f}). 
Using the same natural scaling formalism introduced in Sect.~\ref{isokepldisk},
where $k_R'=k_RV_{Az}/\Omega$, $k_z'=k_zV_{Az}/\Omega$, ${k^\prime}^2={k_R^\prime}^2+{k_z^\prime}^2$ and $\omega'=\omega/\Omega$, we obtain the inner ($R<R_{\omega,0}$) disk scaled frequency
$\omega^\prime_{\text{in}}$ and the outer ($R=R_{\omega,0}$) disk scaled frequency $\omega^\prime_{\text{out}}$, respectively,
\begin{align}\label{paperref4g}
{\omega^\prime}^2_{\text{in}}=\frac{{k_z^\prime}^2}{2{k^\prime}^2}\left(2{k^\prime}^2+1-\sqrt{1+16{k^\prime}^2}\right),\quad\quad\quad
{\omega^\prime}^2_{\text{out}}=\frac{{k_z^\prime}^2}{{k^\prime}^2}\left({k^\prime}^2+2-2\sqrt{1+{k^\prime}^2}\right).
\end{align}
These equations show that $-{\omega^\prime}^2_{\text{in}}>0$ for $k^\prime<\sqrt{3}$
while $-{\omega^\prime}^2_{\text{out}}\leq 0$ everywhere.
The red region of the plot of the lower root (that
may cross zero) of the dispersion relation Eq.~\eqref{paperref4a} in the midplane of the disk thus denotes the stable domain of the perturbations 
while the black region denotes the perturbations that are unstable and where the MRI may develop. Since all the roots of the disk midplane dispersion relation in the outer disk region are positive, 
the MRI cannot develop there. We note that this happens
regardless of the strength of the magnetic field. The same effect
appears also above the midplane, but for slightly larger radii.
\subsection{Time-dependent models with variable viscosity profile}\label{lkajsd}
\begin{figure}[t]
\centering
\includegraphics[width=0.66\hsize]{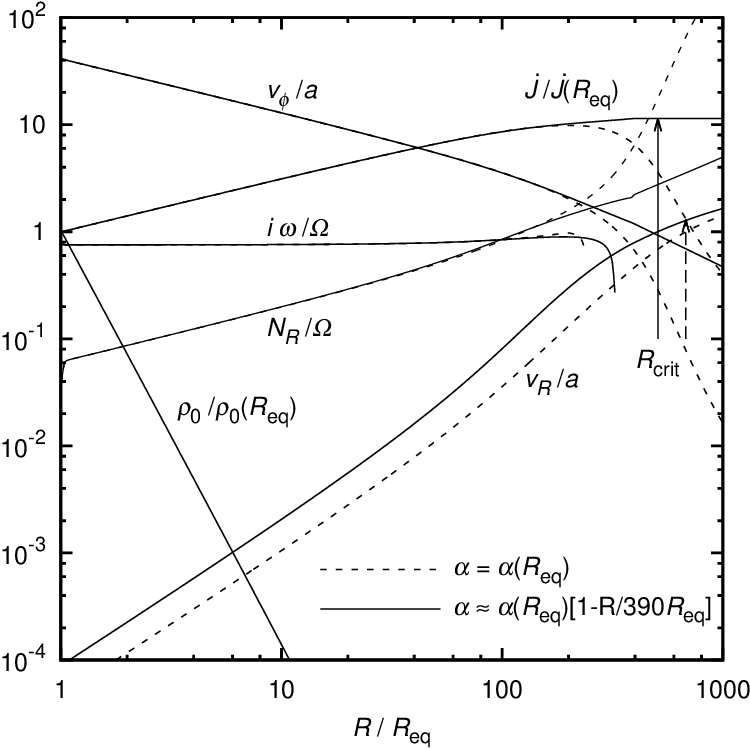}
\caption{Final stationary state of time-dependent disk models
with constant as well as with variable viscosity parameter $\alpha$.
In the constant viscosity model (dashed line) the viscosity parameter 
$\alpha=\alpha(R_{\text{eq}})$, while in the decreasing viscosity model 
(solid line) the parameter $\alpha=\alpha(R_{\text{eq}})[(391\,R_{\text{eq}}-R)/(390\,R_{\text{eq}})]$
below the radius $R=390\,R_{\text{eq}}$ and $\alpha=\alpha(R_{\text{eq}})/390$ elsewhere.
Inner boundary viscosity parameter $\alpha(R_{\text{eq}})= 0.1$.
We consider the isothermal disk of the main-sequence B2-B3 type star with the parameters described in
Sect.~\ref{lkajsd}. Arrows denote the location of critical point. Adapted from \citet{Krticka_2014}.}
\label{alp0.100}
\end{figure}
%In our time-dependent models we study the radial profile of the magnetorotational instability, which is considered to be the origin of anomalous viscosity in outflowing disks.
We calculate the radial variations of the growth rate of the magnetorotational instability using the hydrodynamic structure of our disk models.
We assume an circumstellar outflowing disk that transports an excess of angular momentum from
critically rotating star \citep{Krticka,kurfek,Krticka_2014} whose equatorial radius $R_{\text{eq}}=3/2\,R_\star$ (where $R_\star$ is the stellar polar radius, see, e.g., Sect.~\ref{timemod}). 
Analogously to hydrodynamic studies introduced in Sects.~\ref{statcalc}-\ref{timemod} we calculate the stationary solution based on 
solving the basic hydrodynamic equations ~\eqref{massconserve}, \eqref{radmomconserve}, and \eqref{phimcon}, 
using the Newton-Raphson method described in Sect.~\ref{stationarix} which
provide a radial dependence of the integrated disk
density $\Sigma$ and radial $V_R$ as well as azimuthal $V_\phi$ components of the
velocity. For a simplicity we assumed here an isothermal disk with
$T=T(R_\text{eq})\equiv 1/2\,T_{\text{eff}}$ (cf.~Sects.~\ref{largemodelix} and \ref{statcalc}).
The stellar parameters selected for the modeling correspond
to the main-sequence B2-B3 type star with the effective temperature
$T_{\text{eff}}=20\,000\,\text{K}$, mass $M_{\!\star}=6.60\,M_\odot$, and radius
$R_\star=3.71\,{R}_\odot$ \citep{Harmanec}. We also calculated alternative models for main-sequence stars with effective temperatures
$T_{\text{eff}}=30\,000\,\text{K}$ and $T_{\text{eff}}=14\,000\,\text{K}$ as well as with radially decreasing
disk temperature $T=T(R_\text{eq})(R_{\text{eq}}/R)^p$ with $p=0.1$ and $p=0.2$
(see Eq.~\eqref{temperature}, see Sect.~\ref{largemodelix} for the description). %We do not introduce here the stationary models, since the results do not 
%practically differ from the time-dependent model described below. 
The results of these models however show that their properties in general do not significantly depend on the particular choice of the disk and stellar parameters. 
Consequently, further in this section we concentrate mainly on the examination of a model
with $T_{\text{eff}}=20\,000\,\text{K}$.

Using the final stationary state of our time-dependent hydrodynamic models we study the consequences of 
MRI disappearance at large radii
(see Sect.~\ref{largemodelix} for details). For these
models we used the same set of basic hydrodynamic equations as well as the same
disk initial and boundary conditions described in
Sect.~\ref{timemod} and the same stellar parameters for B2-B3 type star with the effective temperature
$T_{\text{eff}}=20\,000\,\text{K}$, described within the stationary models in this section. We calculated the models of
isothermal disk in case of constant viscosity parameter
$\alpha=\alpha(R_{\text{eq}})$ and with the decreasing viscosity radial dependence
$\alpha=\alpha(R_{\text{eq}})[(391\,R_{\text{eq}}-R)/(390\,R_{\text{eq}})]
\approx\alpha(R_{\text{eq}})[1-R/(390\,R_{\text{eq}})]$, i.e., up to the radius
$390\,R_\text{eq}$ corresponding to the radius $R_{\omega,0}$ from Sect.~\ref{magnetouscisub1}, where the MRI vanishes (i.e., 
where the root of the dispersion relation \eqref{paperref4a} changes
its sign, see also Eqs.,~\eqref{paperref4g}). Below this radius the viscosity $\alpha$ parameter decreases relatively steeply
and beyond this radius it is fixed as the constant value
$\alpha=\alpha(R_{\text{eq}})/390$.
In the time-dependent models we involve
the full second-order Navier-Stokes prescription of the viscous torque (see the description in Sects.~\ref{largemodelix} and \ref{timemod}). This
provides physically more relevant distribution of rotational velocity (and
consequently of the specific angular momentum loss rate) even in very distant
disk regions \citep[see ][for the details]{kurfek}. The results of the models we
obtain as a final stationary state of the converging time-dependent
calculations.

Figure \ref{alp0.100} compares the radial profiles of the relative disk midplane
density $\rho_0/\rho_0(R_{\text{eq}})$ and of the scaled radial and azimuthal
velocities $V_R/a$ and $V_{\phi}/a$ (where $a$ is the speed of sound) as
well as of the scaled specific angular momentum loss rate
$\dot{J}/\dot{J}(R_{\text{eq}})$ (cf. Figs.~\ref{B0pt0}, \ref{B0pt0n04} and \ref{ptI} in
Sect.~\ref{timemod}) for the two cases of radial viscosity distribution.
The depicted profile of the relative midplane maximum MRI growth rate scaled as $i\omega/\Omega$ is
calculated using Eq.~\eqref{paperref4a}, where we involve the full prescription
of the epicyclic frequency $\kappa$ (see Eq.~\eqref{epicyclfreq} and its comments) and the radial
piece of Brunt-V\"ais\"al\"a frequency $N_R$ (Eq.~\eqref{paperref4f}).
The time-dependent calculations show that the decretion disks of critically rotating stars may spread up to very
large radii despite the vanishing MRI instability. From Fig.~\ref{alp0.100} we can see that in the case
of constant viscosity distribution the rotational velocity (and and the angular
momentum loss rate $\dot{J}$) begins to rapidly drop even below the corresponding sonic
(critical) point distance. This can be avoided in the model with decreasing
viscosity coefficient \citep{kurfek}. Moreover, Eq.~\eqref{paperref4e} is also valid
in the cases with decreasing viscosity in Keplerian region, however, the radius
where the MRI instability vanishes increases in case of decreasing viscosity.  
This one-dimensional analytical study of the MRI will be supported
by a future MHD simulations using the MHD code described in Appendix \ref{dvojdimmagnetohydrous}, but this does not seem to alter the
results of this MRI modeling:
\begin{itemize}
\item[-] The MRI does not develop in stars with strong surface magnetic field. In such stars the plasma parameter $\beta$ is too low
and the material is not transported outwards by MRI.
\item[-] The MRI disappears at a certain distance from the star, where the
disk rotational velocity $V_\phi$ and the
disk radial velocity $V_R$ roughly approach the sound speed. In isothermal disks this point is located at the distance of several hundreds of 
stellar radii (see Fig.~\ref{alp0.100}). 
The radial profile of the
angular momentum loss rate flattens there and consequently the disk mass loss rate can be estimated from the maximum angular momentum loss rate at the sonic point radius (see Eq.~\eqref{sonicestL}) 
that can be therefore roughly regarded as an
effective outer disk radius.
The time-dependent simulations show that the
disks may disseminate to the infinity even in the case of vanishing MRI \citep{kurfek}.
We assume that the reason is the very high gas radial velocity that determines the 
disk behavior in the distant supersonic regions \citep{Krticka_2014}.
\end{itemize}

\chapter[Two-dimensional (radial-vertical) time-dependent modeling]{Two-dimensional (radial-vertical) time-dependent modeling}\label{twodajmmodel}
\section{Hydrodynamic models with vertical hydrostatic equilibrium}\label{hydrostatous}
\begin{figure}[t]
\begin{center}
\includegraphics[width=12cm]{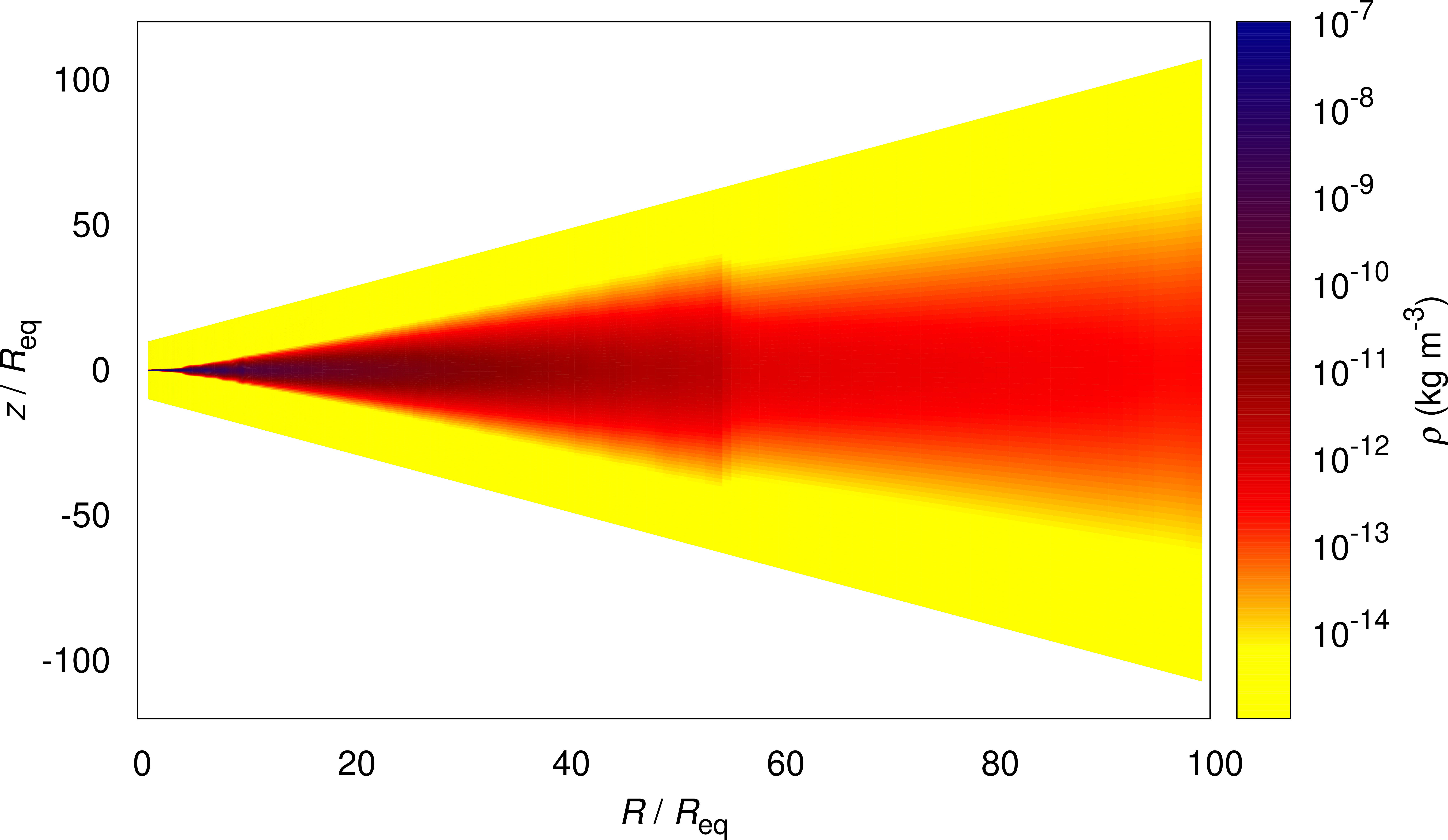}
\caption{The snapshot of the propagating density wave in the distance $R\approx 54 R_{\text{eq}}$ in the converging 
two-dimensional model of the isothermal outflowing disk (cf.~Fig.~\ref{shock} in Sect.~\ref{chartime}), calculated in the $R$-$z$ plane. 
The time of the snapshot is approximately 80 days. The parameters 
of the parent star are the same as those given for B0-type star introduced, e.g., in Sect.~\ref{timemod}. The sonic point is however far beyond the 
computational range ($R_\text{s}\approx 5.5\cdot 10^2\,R_{\text{eq}}$). The model is calculated using the flaring 
grid whose analytical prescription we give in Sect.~\ref{flarecoords}. The initial midplane density is determined by the stellar mass loss rate $\dot{M}=10^{-10}\,M_{\odot}\,\text{yr}^{-1}$.}
\label{flerousekhustotni}
\end{center}
\end{figure}
The models of the disk with the vertically integrated density $\Sigma$ may not be realistic close to the sonic point.
As a first step towards the more realistic two-dimensional models we will study viscous outflowing disk in vertical hydrostatic equilibrium.
The radial parameterization 
of the temperature and viscosity remains the same as in Eqs.~\eqref{temperature} and \eqref{alpvis}.
We assume the vertical profiles of the temperature and viscosity as constant.
The mass conservation (continuity) equation 
(cf.~Eqs.~\eqref{masscylinder1} and \eqref{conticylap}) in the two-dimensional axisymmetric ($\partial/\partial\phi=0$) form is
\begin{align}
\label{massconserve2D}
\frac{\partial\rho}{\partial t}+\frac{1}{R}\frac{\partial}{\partial R}\left(R\rho V_{{R}}\right)=0.
\end{align}
The two-dimensional radial momentum conservation equation (cf.~Eqs.~\eqref{cylmomrad} and \eqref{gravcylexpli}) is
\begin{align}
\label{radmomconserve2D}
\frac{\partial V_R}{\partial t}+V_R\frac{\partial V_R}{\partial R}=
\frac{V_{\phi}^2}{R}-\frac{1}{\rho}\,\frac{\partial(a^2\rho)}{\partial R}-\frac{GM_{\!\star}R}{\left(R^2+z^2\right)^{3/2}},
\end{align}
while the explicit two-dimensional form of the conservation equation of the angular momentum 
(cf.~Eqs.~\eqref{angfl1}, \eqref{vist} and the full second order Navier-Stokes viscosity term in Eq.~\eqref{appendphimomcylinder1}) is
\begin{align}
\label{phimcon2D}
\frac{\partial}{\partial t}\left(R\rho V_{\phi}\right)+\frac{1}{R}\frac{\partial}{\partial R}\left(R^2\rho V_RV_{\phi}\right)
=\frac{1}{R}\frac{\partial}{\partial R}\left(\alpha a^2R^3\rho\,\frac{\partial\,\text{ln}\,V_{\phi}}{\partial R}-\alpha a^2R^2\rho\right).
\end{align}
The explicit expression for the angular momentum (per unit volume) is $J=\rho RV_{\phi}$. We may write the fully explicit form of the equation of the vertical hydrostatic equilibrium 
(using Eqs.~\eqref{hydroequix}, \eqref{fullvertigo} and \eqref{gravcylexpli}) as
\begin{align} \label{fullvertigo2D}
\frac{\text{d}P}{\text{d}z}=-\frac{GM_{\star}z}{\left(R^2+z^2\right)^{3/2}}\,\,\rho(R,0)\,\text{e}^{{\textstyle{-\frac{GM_{\star}}{a^2R}+\frac{GM_{\star}}{a^2\sqrt{R^2+z^2}}}}},
\end{align}
where we obtain the disk midplane density $\rho(R,0)=\rho_0$ from Eq.~\eqref{Sigmasegva}, assuming $H=a/\Omega$ (see Eq.~\eqref{kepscaleheight}). 
We can approximate the radial dependence of the integrated disk density in Eq.~\eqref{Sigmasegva} 
(in the initial state), e.g., as $\Sigma(R)\sim R^{-2}$ (see, e.g., Eq.~\ref{anabjk8}). However, most authors, combining Eqs.~\eqref{kepscaleheight} and \eqref{Sigmasegva}, 
directly assume $\rho(R,0)\sim R^{-3.5}$. 

From the computational point of view, the fundamental problem of the two-dimensional modeling in the $R$-$z$ plane is the unproportional increase 
of the disk vertical scale height $H$ within 
a distance of a few stellar radii (we note that $H\sim R^{3/2}$ in the Keplerian region, while there is 
$H\sim R^2$ in the outer angular momentum conserving region, cf.~Eq.~\eqref{shak}).
To overcome this difficulty, we prepared two variants of the two-dimensional radial-vertical cylindrical computational grid. 
Either we use the standard orthogonal cylindrical grid, which is however logarithmically scaled in both directions of computation
(cf.~Eq.~\eqref{cylmesh}, see also Figs.~\ref{lajzlik}a and \ref{lajzlik}b with schematically shown parts of two-dimensional logarithmic grids), 
or we alternatively proposed a specific type of nonorthogonal (we call it ``flaring'') grid, which is kind of a hybrid of cylindrical and spherical 
spatial mesh. We analyze the grid geometry in detail in Sect.~\ref{flarecoords}. The motivation which has led for such grid development was 
to exclude the vertically eccentric locations in the stellar vicinity (i.e.,~the points with 
high $z$ and low $R$ in Fig.~\ref{conus2}), which are clearly unphysical from the kinematical point of view, while we have to involve
these regions in greater distance from the star. The advantage of the logarithmic orthogonal system is its relative mathematical simplicity while the basic disadvantage 
lies in the fact that it can be used (or at least we have still successfully used it) merely up to approximately 50 stellar radii. Conversely, 
the gas hydrodynamic computations on the flaring grid successfully run up to (at least) 1000 stellar radii, on the other hand, 
the use of the grid for MRI or radiative computations is not yet feasible for its excessive mathematical complexity.
Another possibility might be to use a kind of structured quadrilateral mesh presented, e.g., in \citet{cune} with 
multigrid hierarchy of discretization where the finest grid structure corresponds to dense region of the thin disk near the star while the 
outer regions of the disk are solved on coarser grids.
The viability of these systems for various purposes and conditions will be yet however a matter of further proper testing.

In Fig.~\ref{flerousekhustotni} we demonstrate the snapshot of the converging two-dimensional model of the disk density calculated 
using the flaring grid. The parent star is the B0-type star with the parameters 
given in one-dimensional models introduced, e.g., in Sect.~\ref{timemod}. 
The boundary conditions are identical to those adopted in Sect.~\ref{timemod} for one-dimensional model.
The initial conditions are identical for the corresponding quantities: the 
zero radial velocity ($V_{R,\,\text{ini}}=0$) while the initial (Keplerian) rotation velocity profile is defined as $V_{\phi,\,\text{ini}}=\sqrt{Rg_R}$
where $g_R$ is the radial component of gravitational acceleration (Eq.~\eqref{gravcylexpli}).
The initial density profile $\rho(R,z)_\text{ini}$ is derived from the initial integrated surface density profile $\Sigma(R)\sim R^{-2}$  
using Eqs.~\eqref{kepscaleheight}, \eqref{thinvertigo} and \eqref{Sigmasegva}.
The transforming density wave described in Sect.~\eqref{chartime} propagates near the center of the radial domain.
The initial disk midplane density is determined by the mass loss rate $\dot{M}=10^{-10}\,M_{\odot}\,\text{yr}^{-1}$ (selected as a free parameter).

In this point we note that we have also attempted to calculate the models where the vertical hydrostatic equilibrium is
included merely as an initial state, while in the further evolution the hydrostatic equilibrium is numerically switched off. However, such calculations
have not yet been successful due to a violent computational instability after a few timesteps.
On the other hand, we are not still entirely sure about the effect of the full two-dimensional calculations 
of the vertical disk hydrodynamic (density) structure within the vertical hydrostatic equilibrium. To verify this, we have yet to examine very carefully 
the difference between the numerically consistently calculated vertical profile of the disk density and that which results 
from the analytical relation \eqref{fullvertigo}. Our current experience  
(resulting from a large number of time-dependent hydrodynamic calculations at least up to the radial distance $10^3\,R_{\text{eq}}$ 
which is the maximum radial distance of our consistent two-dimensional models achieved so far) 
is that it differs very little in case of radially parameterized, vertically isothermal temperature structure.

\section{Basic inclusion of the effects of stellar irradiation}\label{teplous}
As a second step to more realistic models we include the radiative heating of the disk by the irradiation from the central star.
The basic theoretical background of the effects of the impinging stellar irradiation is described in Sect.~\ref{kolobrik}.
In the simple (preliminary) model we assume the Roche model of a rigidly rotating gravitationally oblate star 
(where most of the stellar mass is concentrated to the center which is spherically symmetric). The stellar surface thus forms the equipotential according to Eq.~\eqref{fomis1}.
In this point we note that we have properly examined also the gravitational field of the critically rotating star, induced by
realistically modeled stellar internal radial density distribution \citep[e.g.,][]{Meynet1}. 
We mapped the gravitational field of the nonspherically three-dimensionally distributed stellar mass to the two-dimensional disk $R$-$z$ plane. We however found that 
the difference in the gravity distribution obtained by such complex model is very small. Very near to the star the radial component of the gravity is 
slightly lower than in the spherical case, but the difference does not exceed 2\%. The nonradial component of gravity (which in spherical case has to be zero) might play a nonnegligible
(but not very important) role only very close to the star in vertically eccentric (however physically irrelevant) regions (see Sect.~\ref{hydrostatous}).
\subsection{The geometry of the irradiation of the flaring disk by the oblate central star}\label{vikvous}
In order to define the maximally uniform mesh covering the stellar surface (i.e., avoiding the significant concentration of the mesh points 
towards poles or equator, which arises from the cylindrical system), we use the spherical coordinates $\theta$ and $\phi$ (see Appendix \ref{diffsphere})
and divide the stellar surface in the both angular directions almost uniformly (see Appendix \ref{dvojdimradous} for details of the computational schema).
The spherical $\theta$ direction ranges maximally from $0$ to $\pi/2$ (since we assume that the irradiation comes to the disk ``surface'' only from the ``visible'' stellar hemisphere).
The azimuthal $\phi$ direction ranges maximally from $-\pi$ to $\pi$ in order to take into account also the stellar irradiation 
that emerges from behind the $-\pi/2,\,\pi/2$ stellar meridional circle, which impinges the disk regions with high $z$.
\begin{figure}[t]
\begin{center}
\includegraphics[width=6.3cm]{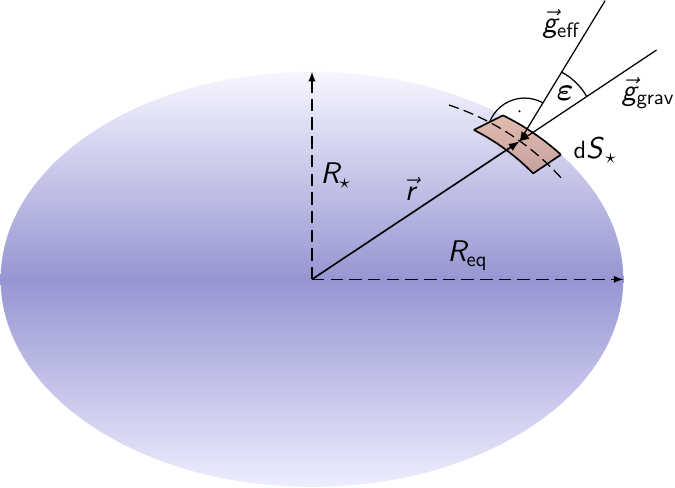}
\caption{Schematic picture of the rotationally oblate star with polar radius $R_{\!\star}$, equatorial radius $R_{\text{eq}}$ and stellar surface element denoted as $\text{d}S_{\!\!\star}$
whose position vector is $\vec{r}$ (see the description in Sect.~\ref{vikvous}). Vector of gravitational force $\vec{g}_{\text{grav}}=-\vec\nabla\Phi$ 
(Eq.~\eqref{fomis1}) is antiparallel to the vector $\vec{r}$ while the vector $\vec{g}_{\text{eff}}=-\vec\nabla\Psi$ is normal to the stellar surface element $\text{d}S_{\!\!\star}$.
The angle $\varepsilon$ denotes the deviation between the vectors $\vec{g}_{\text{grav}}$ and $\vec{g}_{\text{eff}}$ (cf.~Eq.~\eqref{effcoseps}.}
\label{epsideviatik}
\end{center}
\end{figure}

Following Eq.~\eqref{fomis1}, we obtain the spherical radial distance $r$ of each grid point on the stellar surface for each given $\theta$ by finding the root 
of the equation (cf.~\citet{maeder})
\begin{align}\label{effpotroot}
1-\frac{r}{R_\star}+\frac{4}{27}\left(\frac{\Omega}{\Omega_\text{crit}}\right)^2\left(\frac{r}{R_\star}\right)^3\text{sin}^2\theta=0, 
\end{align}
where $R_\star$ is the stellar polar radius and $\Omega_\text{crit}$ is the angular velocity of critically rotating star.
We consequently easily find the cosine of the deviation angle $\varepsilon$ between the radius vector $\vec{r}=\left(r,\,
0,\,0\right)$ and the vector of effective gravity $\vec{g}_\text{eff}=
\left(-GM_\star/r^2+\Omega^2r\,\text{sin}^2\theta,\,\Omega^2r\,\text{sin}\,\theta\,\text{cos}\,\theta,\,0\right)$ in the coordinate directions $r,\,\theta,\,\phi$
(cf.~Sect.~\ref{diffsphere}). Vector $\vec{g}_\text{eff}$ is 
normal to the oblate stellar surface (see Fig.~\ref{irrad}) in each given surface point. From the vector dot product rule we get (cf.~\citet{maeder})
\begin{align}\label{effcoseps}
\text{cos}\,\varepsilon=\frac{\left(\frac{R_\star}{r}\right)^2-\frac{8}{27}\left(\frac{\Omega}{\Omega_\text{crit}}\right)^2
\frac{r}{R_\star}\text{sin}^2\theta}{\left\{\left[-\left(\frac{R_\star}{r}\right)^2+\frac{8}{27}\left(\frac{\Omega}{\Omega_\text{crit}}\right)^2
\frac{r}{R_\star}\text{sin}^2\theta\right]^2+\left[\frac{8}{27}\left(\frac{\Omega}{\Omega_\text{crit}}\right)^2
\frac{r}{R_\star}\text{sin}\,\theta\,\text{cos}\,\theta\right]^2\right\}^{1/2}}. 
\end{align}
We may express the denominator of Eq.~\eqref{effcoseps} as $g_{\text{eff}}\,R_\star/(GM_\star)$ where $g_{\text{eff}}$ is the magnitude of the vector 
of the local effective gravity $\vec{g}_\text{eff}$.
Using Eq.~\eqref{effcoseps}, we may write the stellar surface element $\text{d}S_{\!\!\star}$ and the area of the stellar surface grid cell $\Delta S_{\!\!\star}$, respectively, 
in spherical coordinates as
\begin{align}\label{effsurarea}
\text{d}S_{\!\!\star}=\frac{r^2\text {sin}\,\theta\,\text{d}\theta\,\text{d}\phi}{\text{cos}\,\varepsilon},\quad\quad
\Delta S_{\!\!\star}=\frac{r^2\text {sin}\,\theta\,\Delta\theta\Delta\phi}{\text{cos}\,\varepsilon}.
\end{align}
Following the vector subtraction $\vec{r}_A-\vec{r}_B$ for each two arbitrary points $A$ on the stellar surface and $B$ in the disk $R$-$z$ plane,
where $\vec{r}_A=\vec{r}$ and $\vec{r}_B=(R,0,z)$, we obtain the line-of-sight vector $\vec{d}$ (see Fig.~\ref{irrad}), 
connecting points $A$ and $B$. 
The (squared) magnitude $d$ of the line-of-sight vector $\vec{d}$ is (see Sect~\ref{smakec}, cf.~\citet{smak})
\begin{align}\label{dist2}
d^2=R^2+z^2+r^2-2r\left(R\,\text{sin}\,\theta\,\text{cos}\,\phi+z\,\text{cos}\,\theta\right).
\end{align}
Using Eq.~\eqref{dist2} we express the cosine of the deviation angle $\alpha$ 
between the vector of effective gravity $\vec{g}_{\text{eff}}$ and the stellar surface line-of-sight vector $\vec{d}$
(cf.~e.g,~\citet{smak,giletka}),
\begin{align}\label{cosgefd}
\text{cos}\,\alpha=\frac{\left(\frac{1}{r^2}-\frac{\Omega^2r}{GM_\star}\right)
         \left(R\,\text{cos}\,\phi-r\,\text{sin}\,\theta\right)\text{sin}\,\theta+
         \frac{\text{cos}\,\theta}{r^2}\left(z-r\,\text{cos}\,\theta\right))}
         {\left[\left(\frac{1}{r^2}-\frac{\Omega^2r}{GM_\star}\right)^2\text{sin}^2\theta+
         \frac{\text{cos}^2\theta}{r^4}\right]^{1/2}d}\geq 0,
\end{align}
where the inequality condition $\text{cos}\,\alpha\geq 0$ excludes the line-of-sight vectors that connect each point in the disk $R$-$z$ plane with the ``invisible'' 
points on the stellar surface. 
Following the (rather simplifying) ``flaring disk'' linear interpolation approximation, $\text{tan}\,\gamma=(z/H)\,\Delta H/\Delta R$ (where $H=H(R,0)$ 
is the disk vertical scale height \eqref{kepscaleheight} where $\Delta H$ and $\Delta R$ are 
the differences of these quantities within the radially neighboring computational cells), we obtain the normal vector $\vec{\nu}$ 
(see Fig.~\ref{irrad}) to each disk surface element $\text{d}S$ in the 
arbitrary disk $R$-$z$ plane point $B\left(R,0,z\right)$. 
It may be written in the component form as $\vec{\nu}=\left[z-z\left(1+\Delta a/a\right)\left(1+\Delta R/R\right)^{3/2},\,0,\,\Delta R\right]$,
where $a$ is the speed of sound and $\Delta a$ is its difference within the radially neighboring computational cells.
Calculation of the $\text{sin}\,\beta$ angle (Fig.~\ref{irrad}) between this interpolated "disk surface" plane at 
arbitrary point $B$ in the disk $R$-$z$ plane and the line-of-sight vector $\vec{d}$ gives 
the following relation,
\begin{align}\label{sinbeta}
\text{sin}\,\beta=\frac{z\left[\left(1+\frac{\Delta a}{a}\right)\left(1+\frac{\Delta R}{R}\right)^{3/2}-1\right]\left(R-r\,\text{sin}\,\theta\,\text{cos}\,\phi\right)+
\Delta R\left(r\,\text{cos}\,\theta-z\right)}
{\left\{z^2\left[\left(1+\frac{\Delta a}{a}\right)\left(1+\frac{\Delta R}{R}\right)^{3/2}-1\right]^2+(\Delta R)^2\right\}^{1/2}d}\geq 0,
\end{align}
where in the isothermal case the term in bracket containing the sound speed $a$ is equal to 1. The inequality condition $\text{sin}\,\beta\geq 0$ excludes the line-of-sight vectors 
that may direct below the local "disk surface" plane.
This condition is however yet supplemented by the stellar surface ``visible domain'' condition, $\text{cos}\,\theta\geq(z/H)\,H(R_\text{eq})/r$,
where $r$ is the spherical radial coordinate of the stellar surface point at the vertical level that corresponds to the inner boundary disk vertical scale height $H(R_\text{eq})$.
This condition excludes the stellar irradiation that comes from the star's equatorial belt that is below the vertical (flaring) level $z/H$.

Following Eqs.~\eqref{effcoseps}-\eqref{sinbeta} together with the supplementary conditions, 
we integrate the (frequency integrated) stellar irradiative flux impinging each point $B$ in the disk radial-vertical plane
over the whole ``visible'' domain of the stellar disk. We get the following relation (cf.~Eq.~\eqref{CenterFlux2}),
\begin{align}\label{centerflux}
\mathcal{F}_{\text{irr}}=\frac{1}{\pi}\!\!\iint_{\theta,\phi}\!\! F_{\!\star}(\Omega,\theta)\,\text{d}S_{\!\!\star}\,
\frac{\left(1-u+u\,\text{cos}\,\alpha\right)\,\text{cos}\,\alpha\,\text{sin}\,\beta}{\left(1-u/3\right)\,d^2},
\end{align}
where the radiative flux $ F_{\!\star}(\Omega,\theta)$ that emerges from the stellar surface is equal to the magnitude of the radiative flux vector
$\vec{F}_{\text{rad}}(\Omega,\theta)$ given in Eq.~\eqref{omegacgamacflux} and $u$ is the coefficient of stellar linear limb darkening
(see the description in Sect.~\ref{smakec}). 
% The explicit numerical forms of the equations introduced in this section are
% schematically given in Appendix~\ref{dvojdimradous}.
\subsection{First step towards the calculation of the disk thermal structure}\label{templajznik}
\begin{figure}[t]
\begin{center}
\includegraphics[width=11.5cm]{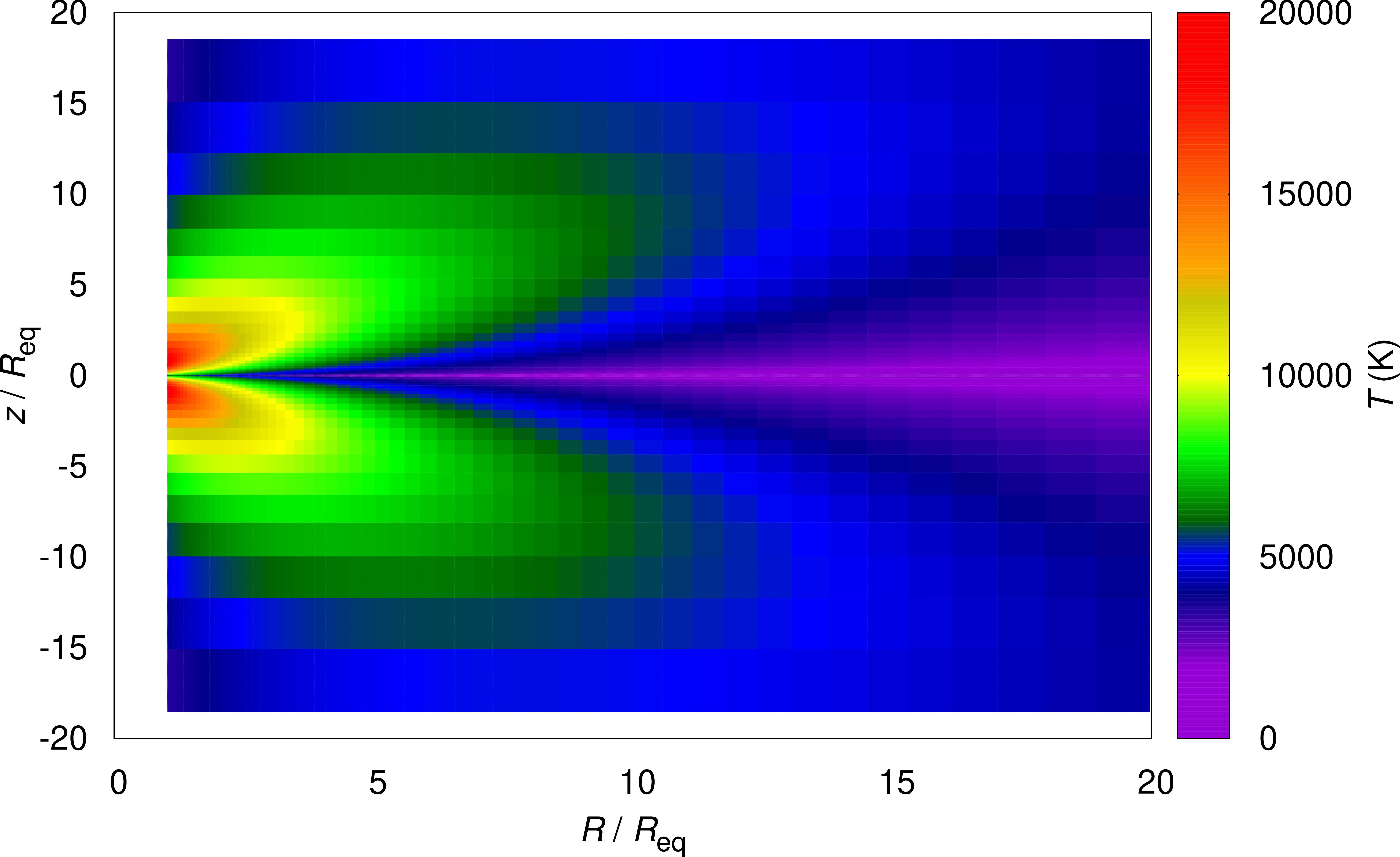}
\caption{The color graph maps the distribution of the temperature in the $R$-$z$ plane of the outflowing disk of the B0-type of star with $T_{\mathrm{eff}}=30\,000\,\text{K}$
(the parameters are given in Sect.~\ref{timemod}) calculated on the orthogonal two-dimensional 
logarithmic grid using the method described in Sect.~\ref{templajznik} and in Appendix~\ref{dvojdimradous} up to 20 stellar radii.
The selected stellar mass loss rate is $\dot{M}=10^{-11}\,M_{\odot}\,\text{yr}^{-1}$. The calculated optical depth throughout the disk is demonstrated in Fig.~\ref{tauoptik}.
The temperature near the disk midplane is lower than in the upper layers due to the shielding of the optically thick gas.}
\label{tempicekdisk11}
\end{center}
\end{figure}
\begin{figure}[t]
\begin{center}
\includegraphics[width=11.5cm]{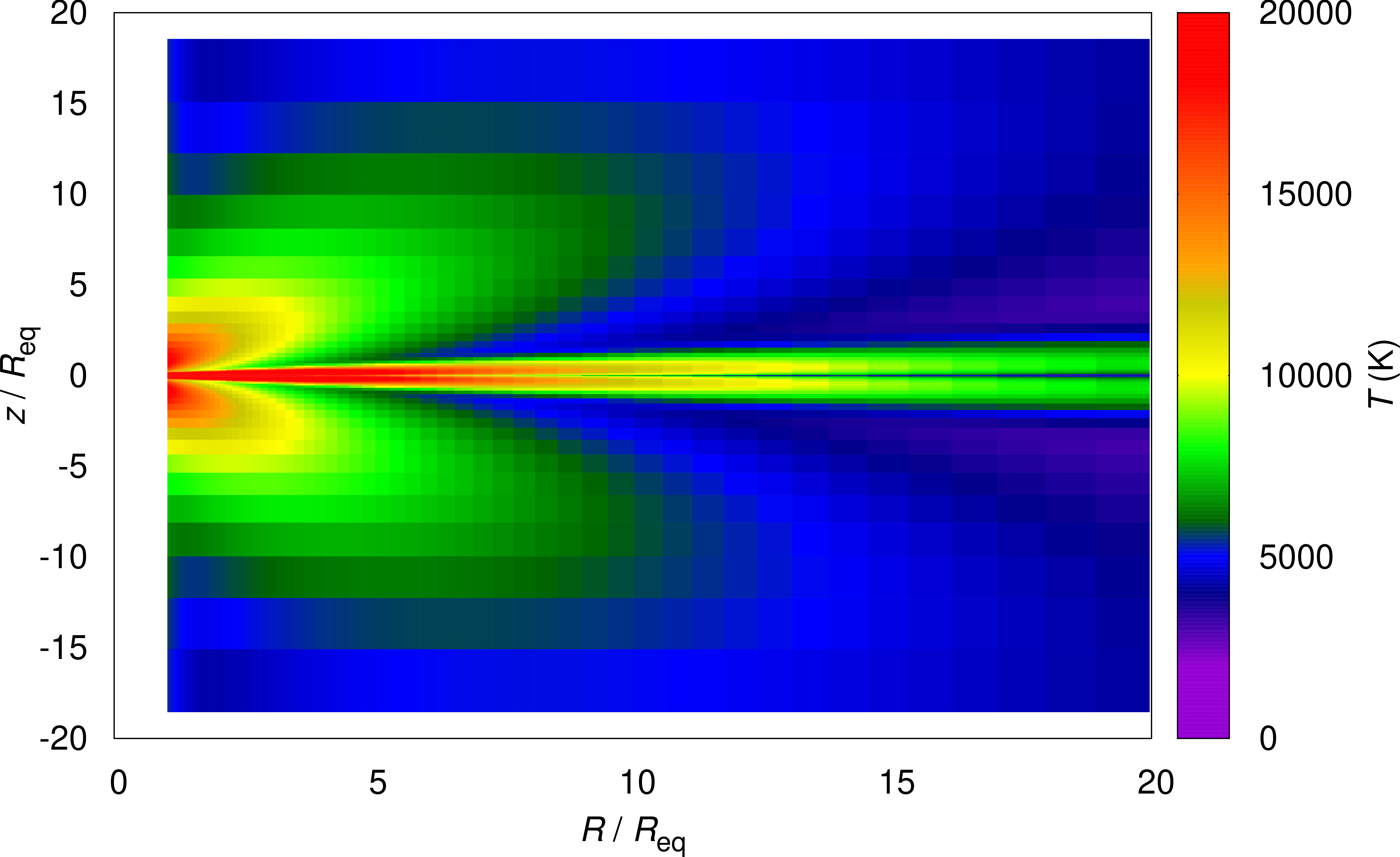}
\caption{As in Fig.~\ref{tempicekdisk11}, however
for the selected stellar mass loss rate $\dot{M}=10^{-7}\,M_{\odot}\,\text{yr}^{-1}$.
There is the significant strip of increased temperature near the disk midplane caused by the viscous heat flux.}
\label{tempicekdisk7}
\end{center}
\end{figure}

Following the basic properties of the irradiation from the central star described in Sect.~\ref{vikvous} and employing Eq.~\eqref{presgradour2}, 
we assume the mean opacity for Thomson electron scattering in ionized hydrogen, $\kappa_\text{es}=0.04\,\text{m}^2\text{kg}^{-1}$, as the opacity $\kappa$.
we calculate the limit in the disk where the optical depth may be regarded 
as $\tau=2/3$ (denoting it as $\tau_{2/3}$). For a simplicity, we integrate for the purpose of calculation of the optical depth the stellar radiation along 
a ray that emerges from the stellar pole, which is however able to penetrate into the geometrically deepest layers of the disk. We difference the optical depth $\tau$ 
using the ray tracing method of either long- or short-characteristics
\citep[see, e.g,][see also Figs.~\ref{lajzlik}a, \ref{lajzlik}b and the description of the schema of corresponding short-characteristics grid interpolation given in 
Appendix \ref{dvojdimradous}]{hausna}. 
We use the piecewise linear interpolation of $\chi=\kappa\rho$, giving 
\begin{align}\label{tauchel}
\Delta\tau_i=(\chi_i+\chi_{i+1})|s_{i+1}-s_i|/2, 
\end{align}
where $s$ is the ray trace coordinate, $(\Delta s)^2=(\Delta R)^2+(\Delta z)^2$,
calculating self-consistently the hydrodynamic (according to Sect.~\ref{hydrostatous}) and thermal properties of the gas. We also currently develop the corresponding short-characteristics solver,
since we cannot due to the specific logarithmic or flaring grid geometry directly use the common schemes given, e.g., in \citet{norda,hulas}, etc. 

In various steps of our calculations of the thermal structure we include either merely the 
stellar irradiation or we employ also the contribution of the viscous disk heating while comparing the range and distribution of both these effects for different parameters of the 
selected mass loss rate. 
We calculate (self-consistently with the density structure) the temperature $T_{2/3}$ at the optical depth limit $\tau_{2/3}$ simply via the 
(frequency integrated) Eddington approximation \citep[see, e.g.,][for the details]{Mihalas1}
$\sigma T^4_{2/3}=\frac{3}{4}(\tau_{2/3}+\frac{2}{3})(\mathcal{F}_{\text{irr}}+{F}_{\text{vis}})$,
where $\mathcal{F}_{\text{irr}}$ is given by Eq.~\eqref{centerflux}, ${F}_{\text{vis}}$ is given by Eqs.~\eqref{kocour} and \eqref{kocourek} and $\sigma$ is the Stefan-Boltzmann constant.
In the optically thick domain where $\tau>\tau_{2/3}$ (cf.~Fig.~\ref{tauoptik}) is the disk thermal and density structure, affected by the irradiation, calculated 
via the disk vertical thermal and 
hydrostatic equilibrium equations using the diffusion approximation
\eqref{presour}-\eqref{radacek}, while the viscous heating is determined by Eqs.~\eqref{kocour} and \eqref{kocourek}. 
The computational convergence is provided by repeating the calculation until the disk midplane condition $F_{\text{vis}}(z=0)=0$ is satisfied \citep[since there is no source 
of viscous heating ``below'' the disk midplane, cf.][]{Lee}. In the optically thin domain ($\tau<\tau_{2/3}$) we extrapolate the thermal structure in the similar way.
We demonstrate the comparison of particular current models of the limited range of the inner disk region 
with various selected parameters of mass loss rate $\dot{M}$ in Figs.~\ref{tempicekdisk11} and \ref{tempicekdisk7}. 

Comparing the basic features of the calculations with the published models, we see that the temperature at the disk inner boundary $T(R_\text{eq})\approx 20\,000\,\text{K}$. This corresponds 
to 66\% of the stellar effective temperature $T_{\text{eff}}\approx 30\,000\,\text{K}$, what agrees with, e.g., the calculations of 
\citet{Karfiol08}, who found approximately $T(R_\text{eq})/T_{\text{eff}}\approx 0.6$ within the 
nearly isothermal solution of Keplerian viscous disk with the stellar mass loss rate $\dot{M}=5\cdot 10^{-11}\,M_\odot\,\text{yr}^{-1}$.
The reason of this temperature drop at the base of the disk is the self-shadowing of the star itself due to the stellar oblate geometry and equatorial darkening together with 
the self-shadowing of the optically thick gas in the inner disk region. Unlike \citet{Karfiol08}, we have not obtained the rapid temperature increase back to 
to the value of approximately 60\% of $T_{\text{eff}}$
at the radius $R\approx 5R_{\text{eq}}$,
which results from taking into account the optically thin radiative equilibrium temperature in significantly rarefied gas in larger distance.
\citet{Sigi1} derive the disk base temperature $T(R_\text{eq})=13\,500\,\text{K}$ in a circumstellar viscous disk of Be star $\gamma$ Cas with $T_{\text{eff}}\approx 25\,000\,\text{K}$.
Several different density models of \citet{Sigi1} with $\rho(R_{\text{eq}})=10^{-9}\,\text{kg}\,\text{m}^{-3}$, 
$\rho(R_{\text{eq}})=5\cdot 10^{-9}\,\text{kg}\,\text{m}^{-3}$ and $\rho(R_{\text{eq}})=5\cdot 10^{-8}\,\text{kg}\,\text{m}^{-3}$ 
(the mass loss rate $\dot{M}=10^{-11}\,M_\odot\,\text{yr}^{-1}$ in Fig.~\ref{tempicekdisk11}
approximately corresponds to $\rho(R_{\text{eq}})=10^{-8}\,\text{kg}\,\text{m}^{-3}$) show in the two-dimensional temperature distributions of the $\gamma$ Cas disk 
the development of a cool region near and at the equatorial plane with a fairly strong vertical temperature gradients. The temperature in those models vertically ranges 
approximately from 10\,000 K in the disk midplane to the maximum of 15\,000 K in the 
highest density model and approximately from 8\,000 K in the disk midplane to the maximum of 15\,000 K in the lowest density model, within the radial domain up to 5 stellar radii. 
In our model, presented in Fig.~\ref{tempicekdisk11}, the disk vertical temperature profile is similar 
(9\,000\,K-15\,000\,K), however, only in the very short radial region up to 1-2 stellar radii. This is 
again likely caused by the fact that we did not take into account the optically thin radiative equilibrium temperature in larger distance.

The currently performed calculations are based on a diffusion approximation within the optically thick domain of the disk, where we consider the optical depth $\tau\geq 2/3$.
Since this approximation cannot be used in the optically thin domains and the non-local character of the radiative
transfer must be taken into account, the calculations have to be therefore regarded merely as a first step towards the more advanced calculations that involve the NLTE radiative transfer
algorithm for a two-dimensional (possibly a three-dimensional) hydrodynamics simulations. In this point we note that the use of any of well-known and widely used
astrophysical radiative codes (e.g.,~the code TLUSTY, which has the implemented routines for calculation of the radiative transfer in accretion disks, etc.) 
does not seem to be appropriate for the purpose of the self-consistent calculations of the disks, due to the specific disk geometry and the varying (in space and time) profiles
of hydrodynamic quantities (the specialized radiative codes such as H-DUST or BEDISK are not available to us).

Within the future work we prepare the calculations of the disk thermal structure that may be based on some of the following approaches: one 
possible way might be to employ the thermal balance of electrons \citep[the calculations follow the principles of the nebular approximation introduced e.g., in]
[cf.~also \citealt{kubes1,kubes2}]{ferlan}. Another way is to include the full NLTE calculations,
where the radiative transfer in the given radiative field from the central star may be calculated, e.g.,~via the escape probability approximation
(similarly to the principles of the BEDISK code developed by \citet{Sigut,Sigi1}, etc.). We may also compute the radiative heating \citep[following, e.g.,][]{brulda} 
using the the short-characteristics
formal solver \citep{Mihalas1,olsic,kunacek1,kunacek2} via a self-consistent computational method that combines the radiative transfer and hydrodynamics by starting with an existing 
hydrodynamics code and adding the necessary radiation processes \citep[e.g.,][etc.]{melasa,turban,belous,milous}. There are also methods of a multi-dimensional radiative transfer designed 
specifically for use with parallel hydrodynamics codes \citep{richkonik}, called hybrid characteristics, 
which can neither be classified as long, nor as short characteristics, applying the underlying principles of both (on the adaptive mesh).

The question of whether to maintain the full self-consistent calculations of the density and thermal 
structure, or to include the pre-calculated disk density as a fixed foundation for the calculation of the thermal structure, 
will be solved according to the tested computational feasibility of various principles of calculations
(cf.~Sect.~\ref{hydrostatous}).
In any case, the further detailed modeling will finally have to be an evolving stepwise compromise between maximally realistic approach and computational efficiency.
In this point, it is also absolutely necessary for very large problems (that involve many spatial and/or angular grid points) to
parallelize the calculation for a large number of processors using some of the corresponding procedures (see Sect.~\ref{paralelous}).

\section{Flaring coordinate system}\label{flarecoords}
\subsection{Basic transformations and differential operators}\label{genflarevecs}
\noindent Circumstellar disks are the typical cases, which may require the introduction of 
a unique coordinate system that will best fit their geometry (see Sect.~\ref{hydrostatous}). 
The disks are basically axisymmetric and in vertical 
hydrostatic equilibrium, they however rapidly flare with radius. The inclusion of the inclination angle of disk vertical layers causes major 
difficulties in numerical schemes and leads to fundamental restrictions, when described in standard cylindrical or spherical geometry. To avoid the 
geometrical discrepancy, we started developing the unique coordinate system for grid modeling and we call it flaring coordinate system. 

Figure~\ref{conus2} illustrates this system in radial-vertical ($R$-$z$) plane. Since the radial and azimuthal coordinates are identical to
standard cylindrical coordinates, we denote them $R$ and $\phi$. The third coordinate $\xi$ is defined as tangent of spherically 
symmetric coordinate surface, $\xi=z/R=\varw/R_{\text{eq}}$ (the meaning of the quantity $\varw$ is illustrated in Fig.~\ref{conus2}). 
Free parameters that adjust the scale and proportions of the grid (besides the selected stellar equatorial radius $R_{\text{eq}}$) are  
maximum radial distance $R=R_{\text{max}}$ and tangent of selected maximum flaring angle, denoted as $\text{TG}$. Maximum vertical 
grid extent $\varw=W_{\text{max}}(R_{\text{eq}})$ is then equal to $\text{TG}\cdot R_{\text{eq}}$.
\begin{figure}[t]
\begin{center}
\includegraphics[width=13cm,angle=0]{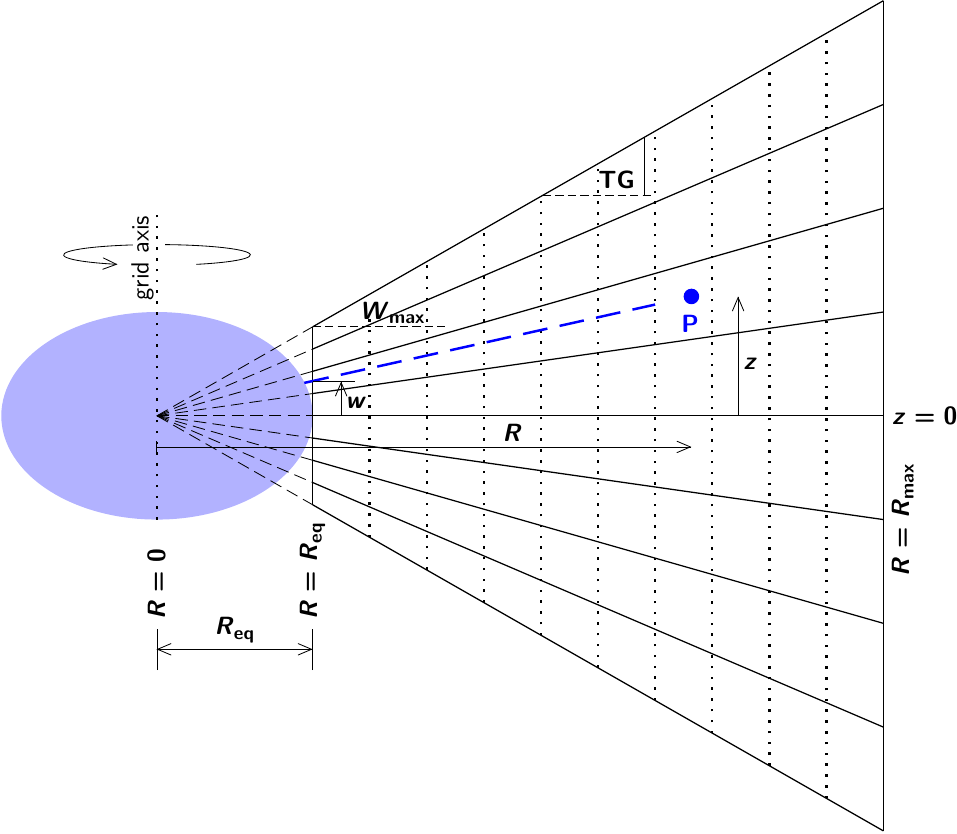}
\caption{\small{Schema of the flaring (cylindrical wedge) coordinate system in radial-vertical ($R$-$z$) plane. Coordinates 
of arbitrary general point $P$ are $R,\,\phi,\,\xi$, intersection of coordinate surface $\xi=z/R$ with the $R$-$z$ plane is highlighted by the 
blue dashed line.}}
\label{conus2}
\end{center}
\end{figure} 
The axis of symmetry intersects the center of star and is perpendicular to its equatorial plane ($\varw=0$).
Transformation equations from the flaring into Cartesian coordinates are
\begin{align}\label{flare1}
x=R\,\text{cos}\,\phi,\quad y=R\,\text{sin}\,\phi,\quad z=R\xi.
\end{align} 
Analogously to Eqs.~\eqref{trafocylunit1} and \eqref{trafosphereunit1}, transformation equations of 
the unit basis vectors from Cartesian to the flaring coordinates are 
\begin{align}\label{flare3}
\mathbf{\hat{\boldsymbol{R}}}=\mathbf{\hat{x}}\,\text{cos}\,\phi+\mathbf{\hat{y}}\,\text{sin}\,\phi,\quad
\mathbf{\hat{\boldsymbol{\phi}}}=-\mathbf{\hat{x}}\,\text{sin}\,\phi+\mathbf{\hat{y}}\,\text{cos}\,\phi,\quad
\mathbf{\hat{\boldsymbol{\xi}}}=
\frac{\mathbf{\hat{z}}-\left(\mathbf{\hat{x}}\,\text{cos}\,\phi+\mathbf{\hat{y}}\,\text{sin}\,\phi\right)\xi}{\sqrt{1+\xi^2}},
\end{align}
while the inverse transformation of the unit basis vectors is
\begin{align}\label{flare4} 
\mathbf{\hat{x}}=\mathbf{\hat{\boldsymbol{R}}}\,\text{cos}\,\phi-\mathbf{\hat{\boldsymbol{\phi}}}\,\text{sin}\,\phi,\quad
\mathbf{\hat{y}}=\mathbf{\hat{\boldsymbol{R}}}\,\text{sin}\,\phi+\mathbf{\hat{\boldsymbol{\phi}}}\,\text{cos}\,\phi,\quad
\mathbf{\hat{z}}=\mathbf{\hat{\boldsymbol{R}}}\xi+\mathbf{\hat{\boldsymbol{\xi}}}\sqrt{1+\xi^2}.
\end{align}
In flaring coordinate system all the basis vectors are non-constant, their directions vary in dependence on azimuthal and flaring angle. 
Angular and time derivatives of the unit basis vectors are
\begin{align}
\label{flare5}
&\frac{\partial\mathbf{\hat{\boldsymbol{R}}}}{\partial{\phi}}=\mathbf{\hat{\boldsymbol{\phi}}},\quad &&\frac{\partial\mathbf{\hat{\boldsymbol\phi}}}{\partial\phi}=-\mathbf{\hat{\boldsymbol{R}}},\quad
&&\frac{\partial\mathbf{\hat{\boldsymbol\xi}}}{\partial\phi}=-\mathbf{\hat{\boldsymbol{\phi}}}\frac{\xi}{\sqrt{1+\xi^2}},\nonumber\\
&\frac{\partial\mathbf{\hat{\boldsymbol{R}}}}{\partial{\xi}}={0},\quad &&\frac{\partial\mathbf{\hat{\boldsymbol{\phi}}}}{\partial{\xi}}={0},\quad
&&\frac{\partial\mathbf{\hat{\boldsymbol{\xi}}}}{\partial{\xi}}=-\mathbf{\hat{\boldsymbol{R}}}\frac{1}{\sqrt{1+\xi^2}}-\mathbf{\hat{\boldsymbol{\xi}}}\frac{\xi}{1+\xi^2},\\
&\frac{\partial\mathbf{\hat{\boldsymbol{R}}}}{\partial t}
=\mathbf{\hat{\boldsymbol{\phi}}}\dot{\phi},\quad
&&\frac{\partial\mathbf{\hat{\boldsymbol{\phi}}}}{\partial t}
=-\mathbf{\hat{\boldsymbol{R}}}\dot{\phi},\quad
&&\frac{\partial\mathbf{\hat{\boldsymbol{\xi}}}}{\partial t}
=-\mathbf{\hat{\boldsymbol{R}}}\frac{\dot{\xi}}{\sqrt{1+\xi^2}}
-\mathbf{\hat{\boldsymbol{\phi}}}\frac{\xi\dot{\phi}}{\sqrt{1+\xi^2}}-\mathbf{\hat{\boldsymbol{\xi}}}\frac{\xi\dot{\xi}}{1+\xi^2}.\nonumber
\end{align}
Covariant and contravariant metric tensors $g_{ij}$ and $g^{ij}$ of the system for coordinates $R,\,\phi,\,\xi$, respectively, 
and the transformation Jacobians $J=\sqrt{|\text{det}\,g_{ij}|},\,J^{-1}=\sqrt{|\text{det}\,g^{ij}|}$ from Cartesian to the flaring coordinate system and vice versa, are
\begin{align}
g_{ij}=
\begin{pmatrix}
1+\xi^2 & 0 & 
R\xi \\
0 & R^2 & 0 \\
R\xi & 0 & R^2
\end{pmatrix},\quad
g^{ij}=
\begin{pmatrix}
\label{trans2b} \displaystyle{1} & \displaystyle{0} & \displaystyle{-\frac{\xi}{R}}\\
\displaystyle{0} & \displaystyle{\frac{1}{R^2}} & \displaystyle{0} \\
\displaystyle{-\frac{\xi}{R}} & \displaystyle{0} & \displaystyle{\frac{1+\xi^2}{R^2}}
\end{pmatrix},\quad J=R^2,\quad J^{-1}=\frac{1}{R^2}. 
\end{align}

Gradient of a scalar function $f=f\,(R,\,\phi,\,\xi)$ in this coordinate system (cf.~Eq.~\eqref{gradcylvector}) is
\begin{align}\label{flare12}
\vec{\nabla}f=\mathbf{\hat{\boldsymbol{R}}}\frac{\partial f}{\partial R}+
\mathbf{\hat{\boldsymbol{\phi}}}\frac{1}{R}\frac{\partial f}{\partial\phi}+
\mathbf{\hat{\boldsymbol{\xi}}}\frac{\sqrt{1+\xi^2}}{R}\frac{\partial f}{\partial\xi}.
\end{align}
The gradient of an arbitrary vector field $\vec{A}(R,\,\phi,\,\xi)$ is defined as action of a gradient operator on a vector, where the vector 
components are scalar functions. Using matrix formalism, we obtain
\begin{align}\label{flare13}
\vec{\nabla}\vec{A}=\,\,\,\,\,\,\bordermatrix{~&\mathbf{\hat{\boldsymbol{R}}} & \mathbf{\hat{\boldsymbol{\phi}}} & \mathbf{\hat{\boldsymbol{\xi}}} \cr
           \mathbf{\hat{\boldsymbol{R}}} & \displaystyle{\frac{\partial A_R}{\partial R}} & 
           \displaystyle{\frac{\partial A_{\phi}}{\partial R}} & \displaystyle{\frac{\partial A_\xi}{\partial R}} \cr
           \mathbf{\hat{\boldsymbol{\phi}}} & \displaystyle{\frac{1}{R}\frac{\partial A_R}{\partial\phi}-\frac{A_{\phi}}{R}} & 
           \displaystyle{\frac{1}{R}\frac{\partial A_\phi}{\partial\phi}+\frac{A_R}{R}}-\frac{A_\xi}{R}\frac{\xi}{\sqrt{1+\xi^2}} & 
           \displaystyle{\frac{1}{R}\frac{\partial A_\xi}{\partial\phi}} \cr
           \mathbf{\hat{\boldsymbol{\xi}}} & \displaystyle{\frac{\sqrt{1+\xi^2}}{R}\frac{\partial A_R}{\partial\xi}-\frac{A_\xi}{R}} & 
           \displaystyle{\frac{\sqrt{1+\xi^2}}{R}\frac{\partial A_\phi}{\partial\xi}} & 
           \displaystyle{\frac{\sqrt{1+\xi^2}}{R}\frac{\partial A_\xi}{\partial\xi}-\frac{A_\xi}{R}\frac{\xi}{\sqrt{1+\xi^2}}} \cr}.
\end{align}
Divergence of a common vector field $\vec{A}\,(R,\,\phi,\,\xi)$ is defined as a dot product 
of gradient vector (see Eq.~\eqref{flare12}) and a common vector 
$A_R\mathbf{\hat{\boldsymbol{R}}}+A_{\phi}\mathbf{\hat{\boldsymbol{\phi}}}+A_\xi\mathbf{\hat{\boldsymbol{\xi}}}$, 
where, unlike the orthogonal coordinate systems, the dot products of different basis vectors may not generally be zero. 
In particular, the nonzero dot product there is
$\displaystyle
\mathbf{\hat{\boldsymbol{R}}}\cdot\mathbf{\hat{\boldsymbol{\xi}}}=-{\xi}\left({1+\xi^2}\right)^{-1/2}.
$ 
By executing this we obtain the divergence of a vector 
\begin{align}\label{flare14b}
\vec{\nabla}\cdot\vec{A}=
\frac{1}{R}\frac{\partial}{\partial R}\left(R A_R\right)+
\frac{1}{R}\frac{\partial A_{\phi}}{\partial\phi}+
\frac{\sqrt{1+\xi^2}}{R}\frac{\partial A_\xi}{\partial\xi}-\frac{\xi}{R\sqrt{1+\xi^2}}
\frac{\partial}{\partial R}\left(R A_\xi\right)-\frac{\xi}{R}\frac{\partial A_R}{\partial\xi}.
\end{align}
Rotation of a vector field $\vec{A}\,(R,\,\phi,\,\xi)$ is defined as a cross product of 
gradient vector (see Eq.~\eqref{flare12}) and a common vector 
$A_R\mathbf{\hat{\boldsymbol{R}}}+A_{\phi}\mathbf{\hat{\boldsymbol{\phi}}}+A_\xi\mathbf{\hat{\boldsymbol{\xi}}}$, 
where however the nonzero cross products of different basis vectors are
\begin{align}\label{flare17} 
\mathbf{\hat{\boldsymbol{R}}}\times\mathbf{\hat{\boldsymbol{\phi}}}=
\mathbf{\hat{\boldsymbol{R}}}\xi+\mathbf{\hat{\boldsymbol{\xi}}}\sqrt{1+\xi^2},\,\,\,\,\,\,\,\,\,\,\,\,\,\,\,\,\,\,\,\,
\mathbf{\hat{\boldsymbol{\phi}}}\times\mathbf{\hat{\boldsymbol{\xi}}}=
\mathbf{\hat{\boldsymbol{R}}}\sqrt{1+\xi^2}+\mathbf{\hat{\boldsymbol{\xi}}}\xi,\,\,\,\,\,\,\,\,\,\,\,\,\,\,\,\,\,\,\,\,
\mathbf{\hat{\boldsymbol{\xi}}}\times\mathbf{\hat{\boldsymbol{R}}}=
\frac{\mathbf{\hat{\boldsymbol{\phi}}}}{\sqrt{1+\xi^2}}.
\end{align}
By applying this, one obtains the vector of rotation in the form
\begin{align}\label{flare18}
\vec\nabla\times\vec{A}=&\,\mathbf{\hat{\boldsymbol{R}}}
\left\{\frac{\xi}{R}\left[\frac{\partial}{\partial R}\left(R A_\phi\right)-\frac{\partial A_R}{\partial\phi}\right]+
\frac{\sqrt{1+\xi^2}}{R}\left(\frac{\partial A_\xi}{\partial\phi}-\sqrt{1+\xi^2}\frac{\partial A_\phi}{\partial\xi}\right)\right\}+\nonumber\\
+&\,\mathbf{\hat{\boldsymbol{\phi}}}
\left\{\frac{1}{R}\left[\frac{\partial A_R}{\partial\xi}-
\frac{1}{\sqrt{1+\xi^2}}\frac{\partial}{\partial R}\left(R A_\xi\right)\right]\right\}+\nonumber\\
+&\,\mathbf{\hat{\boldsymbol{\xi}}}
\left\{\frac{\sqrt{1+\xi^2}}{R}\left[\frac{\partial}{\partial R}\left(R A_\phi\right)-\frac{\partial A_R}{\partial\phi}\right]+
\frac{\xi}{R}\left(\frac{\partial A_\xi}{\partial\phi}-\sqrt{1+\xi^2}\frac{\partial A_\phi}{\partial\xi}\right)\right\}.
\end{align}\\
Since the Laplacian is defined as a dot product of two identical gradient vectors, by applying this on Eq.~\eqref{flare12}, the 
Laplacian operator in the flaring coordinates becomes
\begin{align}\label{flare19} 
\Delta=\vec{\nabla}\cdot\vec{\nabla}=\frac{1}{R}\frac{\partial}{\partial R}\left(R\frac{\partial}{\partial R}\right)+
\frac{1}{R^2}\frac{\partial^2}{\partial\phi^2}+
\frac{\sqrt{1+\xi^2}}{R^2}\frac{\partial}{\partial\xi}\left(\sqrt{1+\xi^2}\frac{\partial}{\partial\xi}\right)
-\frac{2\xi}{R}\frac{\partial^2}{\partial R\,\partial\xi}.
\end{align}
\subsection{Volumes and surfaces}\label{flarevolsurfs}
\noindent Volumes and surfaces of computational grid cells directly enter the calculations in the method of finite volumes 
(see Sect.~\ref{vanleerix}). The Jacobian of the coordinate transformation is $R^2$ (see Eq.~\eqref{trans2b}).
The volume of a space, bounded by coordinate surfaces (i.e.,~the surfaces with the coordinates $R_2,\,R_1,\,\phi_1,\,\phi_2,\,\varw_1\,\varw_2$ being constant) is 
\begin{align}\label{flare20}
V=\int_{R_1}^{R_2}\!\!\!\!\!R^2\,\text{d}R\!\!\!\int_{\phi_1}^{\phi_2}\!\!\!\!\!\text{d}\phi\!\!\!
\int_{\xi_1}^{\xi_2}\!\!\!\!\!\text{d}\xi=\frac{R_2^3-R_1^3}{3}\Big(\phi_2-\phi_1\Big)\frac{w_2-w_1}{R_{\text{eq}}}.
\end{align} 
Square roots of subdeterminants in covariant metric tensor in Eq.~\eqref{trans2b} that correspond to coordinate surfaces are 
\begin{align}\label{flare21}
J'_{R}=R^2,\quad J'_{\phi}=R,\quad J'_{\xi}=R\sqrt{1+\xi^2},
\end{align} 
where the subscripts refer to the constant surface coordinate. Areas of particular grid cell surfaces are calculated from the integrals
\begin{align}
&S_{R}=R^2\!\!\!
\int_{\phi_1}^{\phi_2}\!\!\!\!\!\text{d}\phi\!\!\!\int_{\xi_1}^{\xi_2}\!\!\!\!\!\text{d}\xi=
R^2\,\big(\phi_2-\phi_1\big)\,\frac{w_2-w_1}{R_{\text{eq}}},\\\label{trans72}
&S_{\phi}=\int_{\xi_1}^{\xi_2}\!\!\!\!\!\text{d}\xi\!\!\!\int_{R_1}^{R_2}\!\!\!\!\!R\,\text{d}R=
\frac{R_2^2-R_1^2}{2}\,\frac{w_2-w_1}{R_{\text{eq}}},\\
&S_{\xi}=\sqrt{1+\xi^2}\!\!\!\int_{R_1}^{R_2}\!\!\!\!\!R\,
\text{d}R\!\!\!\int_{\phi_1}^{\phi_2}\!\!\!\!\!\text{d}\phi=\frac{R_2^2-R_1^2}{2}
\,\big(\phi_2-\phi_1\big)\,\frac{\sqrt{R_{\text{eq}}^2+w^2}}{R_{\text{eq}}}.\label{trans62}
\end{align}
\subsection{Velocity and acceleration}\label{flarevelacc}
\noindent Referring to basic vectorial transformations that are described in 
Eqs.~\eqref{flare1}-\eqref{flare4}, the position vector in the flaring coordinate system 
(analogously to the description in Appendix~\ref{velacccylinder} and in Appendix~\ref{velaccsphere}) takes the form
\begin{align}\label{flare22}
\vec{r}=x\mathbf{\hat{x}}+y\mathbf{\hat{y}}+z\mathbf{\hat{z}}=\mathbf{\hat{\boldsymbol{R}}}R\left(1+\xi^2\right)+
\mathbf{\hat{\boldsymbol{\xi}}}R\xi\sqrt{1+\xi^2}.
\end{align}
The components of the velocity vector $\vec{V}$ and the acceleration vectors $\vec{a}$ are defined as 
\begin{align}\label{flare23}
\vec{V}=V_R\mathbf{\hat{\boldsymbol{R}}}+
v_\phi\mathbf{\hat{\boldsymbol{\phi}}}+v_\xi\mathbf{\hat{\boldsymbol{\xi}}},\quad\quad\vec{a}=\frac{\text{d}\vec{V}}{\text{d}t}=a_R\mathbf{\hat{\boldsymbol{R}}}+a_\phi\mathbf{\hat{\boldsymbol{\phi}}}+
a_\xi\mathbf{\hat{\boldsymbol{\xi}}}.  
\end{align}
The velocity vector we obtain by differentiation of Eq.~\eqref{flare22} with use of Eq.~\eqref{flare5},
\begin{align}\label{flare24}
\vec{V}=\frac{\text{d}\vec{r}}{\text{d}t}=
\mathbf{\hat{\boldsymbol{R}}}\left[\dot{R}\left(1+\xi^2\right)+R\xi\dot{\xi}\right]+\mathbf{\hat{\boldsymbol{\phi}}}\,R\dot{\phi}+
\mathbf{\hat{\boldsymbol{\xi}}}\sqrt{1+\xi^2}\left(\dot{R}\xi+R\dot{\xi}\right).
\end{align}
Differentiation of Eq.~\eqref{flare24} according to time gives the components of acceleration vector in this coordinate system in the form
\begin{align}
a_R&=\left(1+\xi^2\right)\ddot{R}-R\dot{\phi}^2+2\dot{R}\xi\dot{\xi}+R\xi\ddot{\xi}&&
=\frac{\text{d}V_R}{\text{d}t}-R\dot{\phi}^2-\dot{R}\xi\dot{\xi}-R\dot{\xi}^2,\label{wedgerac1}\\
a_\phi&=R\ddot{\phi}+2\dot{R}\dot{\phi}&&
=\frac{\text{d}V_\phi}{\text{d}t}+\dot{R}\dot{\phi},\label{wedgerac2}\\
a_\xi&=\sqrt{1+\xi^2}\left(\ddot{R}\xi+2\dot{R}\dot{\xi}+R\ddot{\xi}\right)&&
=\frac{\text{d}V_\xi}{\text{d}t}-\frac{\xi\dot{\xi}}{\sqrt{1+\xi^2}}\left(\dot{R}\xi+R\dot{\xi}\right).
\label{wedgerac}\end{align}
From Eqs.~\eqref{flare24} and \eqref{wedgerac1}-\eqref{wedgerac} we derive the main components of the velocity vector,
\begin{align}\label{flare27}
\dot{R}=V_R-\frac{\xi V_\xi}{\sqrt{1+\xi^2}},\quad
\dot{\phi}=\frac{V_\phi}{R},\quad
\dot{\xi}=\frac{-\xi V_R+\sqrt{1+\xi^2}V_\xi}{R}.
\end{align}
Using Eqs.~\eqref{flare27} and following the identity $\text{d}\vec{V}/\text{d} t=\partial\vec{V}/\partial t+\vec{V}\cdot\vec{\nabla}\vec{V}$, 
we write the components of
acceleration vector in flaring coordinate system in terms of velocity vector components in the form
\begin{align}\label{acwedgrterm}
a_R&=\frac{\partial V_R}{\partial t}+\underbrace{V_R\frac{\partial V_R}{\partial R}+
\frac{V_\phi}{R}\frac{\partial V_R}{\partial\phi}
+V_\xi\frac{\sqrt{1+\xi^2}}{R}\frac{\partial V_R}{\partial\xi}}_{\left(\vec{V}\cdot\vec{\nabla}\right)V_R}-
\frac{V_\phi^2+V_\xi^2}{R}+\frac{\xi V_R V_\xi}{R\sqrt{1+\xi^2}},\\
\label{acwedgphiterm}
a_\phi&=\frac{\partial V_\phi}{\partial t}+\underbrace{V_R\frac{\partial V_\phi}{\partial R}+
\frac{V_\phi}{R}\frac{\partial V_\phi}{\partial\phi}
+V_\xi\frac{\sqrt{1+\xi^2}}{R}\frac{\partial V_\phi}{\partial\xi}}_{\left(\vec{V}\cdot\vec{\nabla}\right)V_\phi}+
\frac{V_R V_\phi}{R}-\frac{\xi V_\phi V_\xi}{R\sqrt{1+\xi^2}},\\
\label{acwedgthetaterm}
a_\xi&=\frac{\partial V_\xi}{\partial t}+\underbrace{V_R\frac{\partial V_\xi}{\partial R}+
\frac{V_\phi}{R}\frac{\partial V_\xi}{\partial\phi}
+V_\xi\frac{\sqrt{1+\xi^2}}{R}\frac{\partial V_\xi}{\partial\xi}}_{\left(\vec{V}\cdot\vec{\nabla}\right)V_\xi}+
\frac{\xi^2V_R V_\xi}{R\left(1+\xi^2\right)}-\frac{\xi V_\xi^2}{R\sqrt{1+\xi^2}}.
\end{align}
The meaning of the right-hand side terms in Eqs.~\eqref{acwedgrterm}-\eqref{acwedgthetaterm} is analogous to cylindrical and spherical case 
(cf.~Eqs.~\eqref{trafounit8}-\eqref{trafounit10} and Eqs.~\eqref{trafosphereunit10}-\eqref{trafosphereunit12}).
\chapter{Summary}\label{sumarix}
In order to have own efficient numerical modeling tool, flexible for any desired hydrodynamic problem, 
we have developed the two-dimensional (magneto-) hydrodynamic computational code (and partly the radiative extension), which is 
currently adapted mainly for calculations in cylindrical $R$-$z$ plane of circumstellar disks (we 
programmed also $R$-$\phi$ plane and three-dimensional routines, we however do not demonstrate these variants here).
The accuracy and the computational efficiency of the code has been properly tested on 
essential physical problems (Riemann-Sod shock tube, Kelvin-Helmholtz instability, etc., see Sects.~\ref{rymashock}, \ref{Keshock}) as well as on simple astrophysical simulations
(e.g., the spherically symmetric CAK wind, cf.~Sect.~\ref{astrochecks}).

Within the one-dimensional hydrodynamic approach
we have computed axisymmetric, vertically integrated, time-dependent models of decretion (outflowing) disks of
critically (or subcritically) rotating stars \citep{kurfek}.
Our code involves the full Navier-Stokes shear viscosity terms described in Sect.~\ref{largemodelix} as well as the artificial viscosity terms described, e.g., in Appendix~\ref{adjustik}.
We parameterize the disk temperature profile using the power law relation (see Eq.~\eqref{temperature}). 
Various temperature profiles give significantly different slopes of integrated density $\Sigma$ decrease through major part of the disk.
The radial dependence of the disk equatorial density $\rho_0$ may in various regions remarkably differ
from the analytically derived profile of the density ($\rho_0\sim R^{-7/2}$), frequently used in models focused on calculation of the disk thermal structure \citep[e.g.,][]{Sigi1,giletka}.
Since the radial profile of the $\alpha$ viscosity parameter
is basically uncertain, we parameterize it via an independent power law radial dependence (Eq.~\eqref{alpvis}).

Our time-dependent one-dimensional models \citep[see also][]{kurfek} 
confirm the basic results obtained in the preliminary stationary models (see Sect.~\ref{statcalc}),
in respect of the sonic point distance and of the distance of the region that may be regarded as the disk ``outer edge`` (i.e., the approximative radius where the
rotational velocity begins to rapidly drop) on parameterized temperature and viscosity profiles.
This ``disk outer edge'' radius is 
strongly affected by the temperature and viscosity profiles, however,
it does not exceed the distance where we expect the disk equatorial density may fall to the average density of the interstellar medium.
The sonic point is located at larger radii in the models with steeper temperature decline, 
while its radius very weakly depends on the viscosity profile. 
The analytical approximations in Sect.~\ref{radthin} provide an instructive and adequate picture of the disk structure behavior.

We examine the total angular momentum 
contained in the disk as well as the total mass of the disk, both these characteristics increase if the temperature and viscosity profiles decrease.
The disk mass loss rate is determined by the angular momentum loss needed to
keep the star at (or near to) the critical rotation. The larger is the extension of the
disk, the larger has to be the specific angular momentum of the disk material (for a fixed $\dot{M}$) while the mass loss rate is lower (for fixed $\dot{J}$, cf.~Eq.~\eqref{sonicestL}). 
Close to the sonic point the profile of the disk angular momentum
loss rate flattens (see e.g.,~Figs.~\ref{B0pt0}, \ref{B0pt0n04}, etc.) and becomes radially independent \citep[e.g.,][]{Krticka,kurfek},
we can thus evaluate the disk mass loss
rate from the maximum angular momentum loss rate at the sonic point radius 
that can be roughly regarded as an
effective outer disk radius \citep[][see also Eq.~\eqref{rsrout} and Sect.~\ref{magnetousci}]{Krticka}.
The unphysical decline of the disk rotational velocity and the angular momentum loss at
large radii, which is obviously present in the isothermal models with constant viscosity
parameters, can be avoided in the models with decreasing temperature and
viscosity parameters \citep{kurfek}. 

Within the time-dependent modeling we recognize the wave that transforms the initial
state of the calculated quantities to their final stationary state. We assume that during the disk developing phase
a similar wave occurs and its amplitude and velocity depends namely on the distribution of the density in the stellar neighborhood.
We regard the velocity of the wave as the velocity of the spreading of the disk.
The models of the disk behavior in case that the central star rotates subcritically (see Sect.~\ref{subcritic}) indicate that
if the stellar equatorial rotational velocity exceeds the limit of approximately $97\%$ of the critical velocity,
the inner disk pulsates up to some $2$-$3$ stellar equatorial radii, while 
for lower stellar equatorial rotational velocity the disk characteristics are
unstable and gradually decline; the lower $V_{\phi}(R_{\text{eq}})$, the faster is their decrease \citep{kurfek}.
The models with ``zero'' reference disk density (Sect.~\ref{zerodens}) show the bumps in flow characteristics near the disk outer boundary (cf. Fig.~\ref{zerovcak}), which may indicate 
similarity with the stellar winds bow shocks, in any case this may lead to a promising future investigation of the physics of the interactions between the disk 
and interstellar medium.

In Sect.~\ref{magnetousci} we provide the one-dimensional study of the magnetorotational instabilities (MRI) in stellar outflowing disks. 
The magnetohydrodynamic simulations of the disk MRI can be performed either as the local problem (within a box with a dimension of a few scale heights, see, e.g.,
\citet{Hawba,skala}, cf.~Sect.~\ref{nonlinbalbus}) or as the global problem \citep[e.g.,][]{kralik,Penna}.
Since the global MHD simulations (which have to be in principle at least two-dimensional) due to enormous computational cost fail to describe the global profiles of the MRI characteristics
at the distance of the order of hundreds stellar radii, we have to
employ the semianalytic approach to estimate the global radial MRI behavior. 
Our future MHD simulations thus certainly do not alter the basic results of the calculation (see also \citet{Krticka_2014}):
MRI disappears approximately in the sonic point region, i.e., at the radius of several hundreds of stellar radii.
The time-dependent simulations however show that the disks may
extend to the infinity even in the case of vanishing MRI \citep[e.g.,][]{kurfek}. We suppose that the outward transport of the disk material is driven
by the gas high radial velocity that determines the disk behavior in the distant supersonic regions \citep{Krticka_2014}.

Within the disk $R$-$z$ plane two-dimensional calculations of the self-consistent time-dependent 
hydrodynamic density-velocity structure (Sect.~\ref{hydrostatous}) we perform the first step toward the more evolved 
models of the disk thermal structure (Sect.~\ref{teplous}), including 
the effects of impinging irradiation of the rotationally oblate star (see Sects.~\ref{irrad} and \ref{vikvous}) and the disk internal viscous heating. 
The fundamental technical difficulty is however in this point given by the enormous unproportionality of the disk vertical scale height even within a
radial distance of a few stellar radii. We overcome this problem by using either the standard orthogonal cylindrical grid that is logarithmically scaled
in both computational directions, or we propose a specific nonorthogonal “flaring” grid (see Sect.~\ref{flarecoords} for mathematical background). 
For this reason we are still not able to calculate the two-dimensional global disk thermal structure (nor the MHD structure) up to thousands stellar radii,
the only feasible way is to model a local problem within a box with a dimension of a few scale heights \citep{Balbus}.
The advantages and viability of these
grid systems (and other issues) will be yet the subject of the proper future testing (cf.~the description in Sect.~\ref{hydrostatous}).

\clearpage

\begin{appendix}
\addtocontents{toc}{\def\protect\cftchappresnum{\appendixname{} }}
\renewcommand\chaptername{Appendix}

\chapter{Hydrodynamic equations in curvilinear coordinate systems}\label{appendix1}
\section{Transformation of vectors}\label{vectortrans}
\subsection{Differential operators in cylindrical polar coordinates}\label{diffcylinder}
\noindent Cylindrical polar coordinate system is defined by radial, azimuthal and vertical coordinates $R,\,\phi,\,z$, respectively. 
Transformation equations from cylindrical into Cartesian coordinates are \citep[see the formalisms introduced, e.g.,~in][etc.]{Arfken}
\begin{align}\label{trafocyl1}
x=R\,\text{cos}\,\phi,\quad
y=R\,\text{sin}\,\phi,\quad
z=z.
\end{align} 
We denote corresponding unit vectors of the cylindrical basis as $\mathbf{\hat{\boldsymbol{R}}},\,\mathbf{\hat{\boldsymbol{\phi}}},\,\mathbf{\hat{\boldsymbol{z}}}.$ 
Transformation equations of 
the basis vectors from Cartesian to cylindrical coordinates are
\begin{align}\label{trafocylunit1}
\mathbf{\hat{\boldsymbol{R}}}=\mathbf{\hat{\boldsymbol{x}}}\,\text{cos}\,\phi+\mathbf{\hat{\boldsymbol{y}}}\,\text{sin}\,\phi,\quad
\mathbf{\hat{\boldsymbol{\phi}}}=-\mathbf{\hat{\boldsymbol{x}}}\,\text{sin}\,\phi+\mathbf{\hat{\boldsymbol{y}}}\,\text{cos}\,\phi,\quad
\mathbf{\hat{\boldsymbol{z}}}=\mathbf{\hat{\boldsymbol{z}}},
%\phantom{\hspace{24.2cm}}
\end{align}
where $\mathbf{\hat{\boldsymbol{x}}},\,\mathbf{\hat{\boldsymbol{y}}},\,\mathbf{\hat{\boldsymbol{z}}}$ are the
unit vectors of the standard Cartesian basis (see Fig.~\ref{cylda}).
\begin{figure}[t]
\begin{center}
\includegraphics[width=3.3cm]{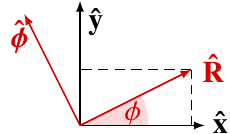}
\caption{Transformation schema of the basis vectors between Cartesian and cylindrical polar coordinate systems 
(see Eqs.~\eqref{trafocylunit1} and \eqref{trafocylunit2}).}
\label{cylda}
\end{center}
\end{figure}
The equations of inverse unit vectors transformation from cylindrical to Cartesian coordinate basis are
\begin{align}\label{trafocylunit2}
\mathbf{\hat{\boldsymbol{x}}}=\mathbf{\hat{\boldsymbol{R}}}\,\text{cos}\,\phi-\mathbf{\hat{\boldsymbol{\phi}}}\,\text{sin}\,\phi,\quad
\mathbf{\hat{\boldsymbol{y}}}=\mathbf{\hat{\boldsymbol{R}}}\,\text{sin}\,\phi+\mathbf{\hat{\boldsymbol{\phi}}}\,\text{cos}\,\phi,\quad
\mathbf{\hat{\boldsymbol{z}}}=\mathbf{\mathbf{\hat{\boldsymbol{z}}}}.
\end{align}
In Cartesian coordinate system all the basis vectors are constant (their length and direction does not vary), 
$\hat{\boldsymbol{z}}$ is the only constant basis vector in cylindrical polar system. 
Nonzero spatial derivatives and the time derivatives of the unit basis vectors are
\begin{align}\label{trafocylunit2a}
\frac{\partial\mathbf{\hat{\boldsymbol{R}}}}{\partial\phi}=\mathbf{\hat{\boldsymbol{\phi}}},\quad
\frac{\partial\mathbf{\hat{\boldsymbol{\phi}}}}{\partial\phi}=-\mathbf{\hat{\boldsymbol{R}}},\quad
\frac{\partial\mathbf{\hat{\boldsymbol{R}}}}{\partial t}=\dot{\phi}\,\mathbf{\hat{\boldsymbol{\phi}}},\quad
\frac{\partial\mathbf{\hat{\boldsymbol{\phi}}}}{\partial t}=-\dot{\phi}\,\mathbf{\hat{\boldsymbol{R}}},\quad
\frac{\partial\mathbf{\hat{\boldsymbol{z}}}}{\partial t}=0.
\end{align}
Covariant and contravariant metric tensors $g_{ij}$ and $g^{ij}$ of the cylindrical system for coordinates $R,\,\phi,\,z$, respectively, 
and the transformation Jacobians from Cartesian to cylindrical coordinate system, and vice versa, are
\begin{align}
g_{ij}=
\begin{pmatrix}\label{cylmetriccovartensorek}
1 & 0 & 0\\
0 & R^2 & 0 \\
0 & 0 & 1
\end{pmatrix},\quad
g^{ij}=
\begin{pmatrix}
1 & 0 & 0\\
0 & \displaystyle{\frac{1}{R^2}} & 0\\
0 & 0 & 1
\end{pmatrix},\quad J=\sqrt{\left|\text{det}\,g_{ij}\right|}=R,\quad J^{-1}=\sqrt{\left|\text{det}\,g^{ij}\right|}=\frac{1}{R}. 
\end{align}
\noindent Since the expressions for gradient, divergence, rotation and Laplacian 
in cylindrical coordinate system are standard and well known \citep[see, e.g.,][]{kvasek,Arfken}, 
we introduce here only basic principles needed for understanding 
the derivation of operators described in Sect.~\ref{flarecoords} and the formalisms used in Sects.~\ref{Massconserve}-\ref{energyconserve} and
Sect.~\ref{sheainst}. 
Using Eq.~\eqref{trafocylunit2} to expand the Cartesian gradient of a scalar function $f$, defined as $\vec{\nabla}f=\mathbf{\hat{x}}\,\partial f/\partial x+\mathbf{\hat{y}}\,\partial f/\partial y+
\mathbf{\hat{z}}\,\partial f/\partial z$, we obtain the gradient of a scalar function $f(R,\,\phi,\,z)$ in cylindrical coordinates,
\begin{align}\label{gradcylvector} 
\vec{\nabla}f=\mathbf{\hat{\boldsymbol{R}}}\frac{\partial f}{\partial R}+
\mathbf{\hat{\boldsymbol{\phi}}}\frac{1}{R}\frac{\partial f}{\partial\phi}+\mathbf{\hat{z}}\frac{\partial f}{\partial z}.
\end{align}
Using the matrix formalism, the gradient of a vector field $\vec{A}(R,\,\phi,\,z)$ (obtained by acting of the gradient operator $\vec{\nabla}$ on
a vector) is the 2nd order tensor
\begin{align}\label{gradcylvectorvector}
\vec{\nabla}\vec{A}=\,\,\,\,\,\,\bordermatrix{~&\mathbf{\hat{\boldsymbol{R}}} & \mathbf{\hat{\boldsymbol{\phi}}} & \mathbf{\hat{z}} \cr
           \mathbf{\hat{\boldsymbol{R}}} & \displaystyle{\frac{\partial A_R}{\partial R}} & 
           \displaystyle{\frac{\partial A_{\phi}}{\partial R}} & \displaystyle{\frac{\partial A_z}{\partial R}} \cr
           \mathbf{\hat{\boldsymbol{\phi}}} & \displaystyle{\frac{1}{R}\frac{\partial A_R}{\partial\phi}-\frac{A_{\phi}}{R}} & 
           \displaystyle{\frac{1}{R}\frac{\partial A_\phi}{\partial\phi}+\frac{A_R}{R}} & 
           \displaystyle{\frac{1}{R}\frac{\partial A_z}{\partial\phi}} \cr
           \mathbf{\hat{z}} & \displaystyle{\frac{\partial A_R}{\partial z}} & \displaystyle{\frac{\partial A_{\phi}}{\partial z}} & 
           \displaystyle{\frac{\partial A_z}{\partial z}} \cr}.
\end{align}
Divergence of a vector field $\vec{A}\,(R,\,\phi,\,z)$ is defined as a dot product 
of gradient vector $\vec{\nabla}$ and a common vector 
$A_R\mathbf{\hat{\boldsymbol{R}}}+A_{\phi}\mathbf{\hat{\boldsymbol{\phi}}}+A_z\mathbf{\hat{\boldsymbol{z}}}$. In cylindrical coordinates
the divergence of a vector $\vec{A}\,(R,\,\phi,\,z)$ is
\begin{align}\label{divcylvector} 
\vec{\nabla}\cdot\vec{A}=
\frac{1}{R}\frac{\partial}{\partial R}\left(R A_R\right)+\frac{1}{R}\frac{\partial A_{\phi}}{\partial\phi}+
\frac{\partial A_z}{\partial z}.
\end{align}
\noindent Rotation of a vector field $\vec{A}\,(R,\,\phi,\,z)$ is defined as a cross product of 
gradient vector $\vec{\nabla}$ and a common vector 
$A_R\mathbf{\hat{\boldsymbol{R}}}+A_{\phi}\mathbf{\hat{\boldsymbol{\phi}}}+A_z\mathbf{\hat{\boldsymbol{z}}}$, 
the vector rotation in cylindrical coordinates takes the form
\begin{align}\label{rotcylvector}
\vec\nabla\times\vec{A}=
\left(\frac{1}{R}\frac{\partial A_z}{\partial\phi}-\frac{\partial A_{\phi}}{\partial z}\right)\mathbf{\hat{\boldsymbol{R}}}+
\left(\frac{\partial A_R}{\partial z}-\frac{\partial A_z}{\partial R}\right)\mathbf{\hat{\boldsymbol{\phi}}}+
\frac{1}{R}\left[\frac{\partial}{\partial R}\left(R A_{\phi}\right)-\frac{\partial A_R}{\partial\phi}\right]\mathbf{\hat{z}}.
\end{align}
The Laplacian operator in cylindrical coordinates is
\begin{align}\label{laplacecyl} 
\Delta=\vec{\nabla}\cdot\vec{\nabla}
=\frac{1}{R}\frac{\partial}{\partial R}\left(R\frac{\partial}{\partial R}\right)+
\frac{1}{R^2}\frac{\partial^2}{\partial\phi^2}+\frac{\partial^2}{\partial z^2}.
\end{align}
In orthogonal coordinate systems are the identical vector transformations \citep[e.g.,][]{kvasek,Arfken} produced using the 
covariant derivatives (denoted as $\nabla_j$), 
involving Lam\'e coefficients $h$ defined as 
$h_ih_j=g_{ij}\delta_{ij},\,h^ih^j=g^{ij}\delta^{ij}$, and the so-called Christoffel symbols $\Gamma$ defined as
$\Gamma_{jk}^{l}=1/2\,g^{lm}\left(\partial g_{km}/\partial x_j+\partial g_{jm}/\partial x_k-
\partial g_{jk}/\partial x_m\right)$ where $l,\,m$ are free indices.
The only nonzero Christoffel symbols in cylindrical coordinates are $\Gamma_{\phi\phi}^R=-R,$ 
$\Gamma_{\phi R}^{\phi}=\Gamma_{R\phi}^{\phi}=1/R$, and, e.g.,~the tensorial equation \eqref{gradcylvectorvector} is
$
\vec\nabla\vec{A}=\nabla_jA_k=
1/(h_jh_k)\left[\partial/\partial x_j(h_kA_k)-\Gamma_{jk}^lh_lA_l\right]\mathbf{\hat{x}}_j\mathbf{\hat{x}}_k.
$
\subsection{Velocity and acceleration in cylindrical polar coordinates}\label{velacccylinder}
According to Eq,~\eqref{trafocylunit2a}, the position and velocity vectors in the cylindrical coordinate system are
\begin{align}\label{trafounit3}
\vec{r}=x\mathbf{\hat{\boldsymbol{x}}}+y\mathbf{\hat{\boldsymbol{y}}}+z\mathbf{\hat{\boldsymbol{z}}}=R\mathbf{\hat{\boldsymbol{R}}}+z\mathbf{\hat{\boldsymbol{z}}},\quad\quad
\vec{V}=\frac{\text{d}\vec{r}}{\text{d}t}=\frac{\text{d}\left(R\mathbf{\hat{\boldsymbol{R}}}+z\mathbf{\hat{z}}\right)}{\text{d}t}=
\mathbf{\hat{\boldsymbol{R}}}\dot{R}+\mathbf{\hat{\boldsymbol{\phi}}}{R}\dot{\phi}+\mathbf{\hat{z}}\dot{z}.
\end{align}
Since the velocity and acceleration vectors are simultaneously defined as
\begin{align}\label{trafounit5}
\vec{V}=V_R\mathbf{\hat{\boldsymbol{R}}}+
V_\phi\mathbf{\hat{\boldsymbol{\phi}}}+V_z\mathbf{\hat{z}},\quad\quad
\vec{a}=\frac{\text{d}\vec{V}}{\text{d}t}=a_R\mathbf{\hat{\boldsymbol{R}}}+a_\phi\mathbf{\hat{\boldsymbol{\phi}}}+a_z\mathbf{\hat{z}},
\end{align}
differentiation of Eq.~\eqref{trafounit3} with respect to time gives the components of the acceleration vector,
\begin{align}\label{trafounit7}
a_R=\ddot{R}-R\dot{\phi}^2
=\frac{\text{d}V_R}{\text{d}t}-R\dot{\phi}^2,\quad\quad
a_\phi=R\ddot{\phi}+2\dot{R}\dot{\phi}=\frac{\text{d}V_\phi}{\text{d}t}+\dot{R}\dot{\phi},\quad\quad
a_z=\ddot{z}=\frac{\text{d}V_z}{\text{d}t}.
\end{align}
Noting that $\text{d}/\text{d}t=\partial/\partial t+\vec{V}\cdot\vec{\nabla}$, we write the acceleration vector expressed in terms of the 
velocity vector components in generally used form
\begin{align}\label{trafounit8}
a_R&=\frac{\partial V_R}{\partial t}+\underbrace{V_R\frac{\partial V_R}{\partial R}+
\frac{V_\phi}{R}\frac{\partial V_R}{\partial\phi}
+V_z\frac{\partial V_R}{\partial z}}_{\left(\vec{V}\cdot\vec{\nabla}\right)V_R}-\frac{V_\phi^2}{R},\\
\label{trafounit9}
a_\phi&=\frac{\partial V_\phi}{\partial t}+\underbrace{V_R\frac{\partial V_\phi}{\partial R}+
\frac{V_\phi}{R}\frac{\partial V_\phi}{\partial\phi}
+V_z\frac{\partial V_\phi}{\partial z}}_{\left(\vec{V}\cdot\vec{\nabla}\right)V_\phi}+\frac{V_R V_\phi}{R},\\
\label{trafounit10}
a_z&=\frac{\partial V_z}{\partial t}+\underbrace{V_R\frac{\partial V_z}{\partial R}+
\frac{V_\phi}{R}\frac{\partial V_z}{\partial\phi}
+V_z\frac{\partial V_z}{\partial z}}_{\left(\vec{V}\cdot\vec{\nabla}\right)V_z},
\end{align}
where the braced terms are components of the advection terms and the last terms on the right-hand side express acceleration due to noninertial forces -  
the magnitude of centrifugal acceleration (cf. Eq.~\eqref{rotframe2}) induced by azimuthally directed rotation in Eq.~\eqref{trafounit8} and half
of the Coriolis acceleration in Eq.~\eqref{trafounit9} (the second half became included in advection part, cf. Eq.~\eqref{trafounit7}).

We can compare the expressions derived in this section with the derivation of the formulas for the noninertial (fictitious) forces, given, e.g., in
\citet{Landau1} in a rotating reference frame with a fixed rotation axis: setting the quantities in rotating frame as primed 
and the quantities in inertial frame without a prime, for the velocity vector we obtain $\vec{V}=\vec{V}\,'+\vec{\Omega}\times\vec{R}\,'$,
where $\Omega$ is the angular velocity of the rotating frame (cf.~Eq.~\eqref{trafounit3}, assuming the fixed rotation axis is the 
$z$-axis). Substituting the velocity $\vec{V}\,'$ into the Lagrangian of the inertial frame motion
$L=mV^2/2$ and differentiating it, we obtain the expression for the acceleration term in rotating frame,
\begin{align}\label{rotframe2}
\frac{\text{d}\vec{V}\,'}{\text{d}t}=\frac{\text{d}\vec{V}}{\text{d}t}-\vec{\Omega}\times\left(\vec{\Omega}\times\vec{R}\,'\right)-
\frac{\text{d}\vec{\Omega}}{\text{d}t}\times\vec{R}\,'-2\,\vec{\Omega}\times\vec{V}\,'.
\end{align}
The second term on the right-hand side of Eq.~\eqref{rotframe2} represents the centrifugal acceleration that in case of stationary axisymmetric rotation can be 
written as $\Omega^2R'=V_{\phi}^2/R'=V_{\phi}^2/R$ with $R'\equiv R$ being the radial distance. 
The third term is the so-called Euler acceleration that in uniformly and stationary rotating frame vanishes and it explicitly corresponds to
the first four terms on the right-hand side of Eq.~\eqref{trafounit9}. The last term 
is the Coriolis acceleration that in case of stationary axisymmetric rotation is perpendicular to $\vec{V}\,'$
and can be written as $-2\,\Omega V'=-2\,V_{\phi}V'/R$ where $R$ is the radial distance of considered fluid parcel.
\subsection{Differential operators in spherical polar coordinates}\label{diffsphere}
Spherical polar coordinate system is defined by radial, spherical and azimuthal coordinates $r,\,\theta,\,\phi$, respectively. 
Transformation equations from spherical into Cartesian coordinates (cf. Fig.~\ref{sphereda}) are 
\begin{align}\label{trafosphere1}
x=r\,\text{sin}\,\theta\,\text{cos}\,\phi,\quad
y=r\,\text{sin}\,\theta\,\text{sin}\,\phi,\quad
z=r\,\text{cos}\,\theta.
\end{align} 
We denote corresponding unit vectors of the spherical basis as $\hat{\boldsymbol{r}},\,\hat{\boldsymbol{\theta}},\,\hat{\boldsymbol{\phi}}.$ 
Transformation equations of 
the basis vectors from Cartesian to spherical coordinates and the equations of inverse basis vectors transformation from spherical to Cartesian coordinates are (see Fig.~\ref{sphereda})
\begin{align}\label{trafosphereunit1}
&\hat{\boldsymbol{r}}=\hat{\boldsymbol{x}}\,\text{sin}\,\theta\,\text{cos}\,\phi+
\hat{\boldsymbol{y}}\,\text{sin}\,\theta\,\text{sin}\,\phi+\hat{\boldsymbol{z}}\,\text{cos}\,\theta,&&\hat{\boldsymbol{x}}=\hat{\boldsymbol{r}}\,\text{sin}\,\theta\,\text{cos}\,\phi+\hat{\boldsymbol{\theta}}\,\text{cos}\,
\theta\,\text{cos}\,\phi-\hat{\boldsymbol{\phi}}\,\text{sin}\,\phi,\nonumber\\
&\hat{\boldsymbol{\theta}}=\hat{\boldsymbol{x}}\,\text{cos}\,\theta\,\text{cos}\,\phi+
\hat{\boldsymbol{y}}\,\text{cos}\,\theta\,\text{sin}\,\phi-\hat{\boldsymbol{z}}\,\text{sin}\,\theta,&&\hat{\boldsymbol{y}}=\hat{\boldsymbol{r}}\,\text{sin}\,\theta\,\text{sin}\,\phi+
\hat{\boldsymbol{\theta}}\,\text{cos}\,\theta\,\text{sin}\,\phi+\hat{\boldsymbol{\phi}}\,
\text{cos}\,\phi,\\
&\hat{\boldsymbol{\phi}}=-\hat{\boldsymbol{x}}\,\text{sin}\,\phi+\hat{\boldsymbol{y}}\,\text{cos}\,\phi,&&\hat{\boldsymbol{z}}=\hat{\boldsymbol{r}}\,\text{cos}\,\theta-\hat{\boldsymbol{\theta}}\,\text{sin}\,\theta\nonumber.
\end{align}
In spherical coordinate system all the basis vectors are non-constant, their directions vary in dependence on spherical and azimuthal angle.
\begin{figure}[t]
\begin{center}
\includegraphics[width=6cm]{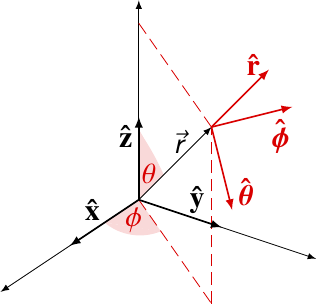}
\caption{Schema of the basis vectors configuration in Cartesian and spherical polar coordinate systems 
(see Eqs.~\eqref{trafosphere1} and \eqref{trafosphereunit1}).}
\label{sphereda}
\end{center}
\end{figure}
The angular and time derivatives of the unit basis vectors are \citep[e.g.,][]{Arfken}
\begin{align}\label{trafosphereunit3}
&\frac{\partial\mathbf{\hat{\boldsymbol{r}}}}{\partial{\theta}}=\mathbf{\hat{\boldsymbol{\theta}}},&&\frac{\partial\mathbf{\hat{\boldsymbol\theta}}}{\partial\theta}=-\mathbf{\hat{\boldsymbol{r}}},&&
\frac{\partial\mathbf{\hat{\boldsymbol\phi}}}{\partial\theta}=0,\nonumber\\
&\frac{\partial\mathbf{\hat{\boldsymbol{r}}}}{\partial{\phi}}=\mathbf{\hat{\boldsymbol{\phi}}}\,\text{sin}\,\theta,\quad
&&\frac{\partial\mathbf{\hat{\boldsymbol{\theta}}}}{\partial{\phi}}=\mathbf{\hat{\boldsymbol{\phi}}}\,\text{cos}\,\theta,\quad
&&\frac{\partial\mathbf{\hat{\boldsymbol{\phi}}}}{\partial{\phi}}=
-\mathbf{\hat{\boldsymbol{r}}}\,\text{sin}\,\theta-\mathbf{\hat{\boldsymbol{\theta}}}\,\text{cos}\,\theta,\\
&\frac{\partial\mathbf{\hat{\boldsymbol{r}}}}{\partial t}=\mathbf{\hat{\boldsymbol\theta}}\dot{\theta}+
\mathbf{\hat{\boldsymbol\phi}}\dot{\phi}\,\text{sin}\,\theta,
&&\frac{\partial\mathbf{\hat{\boldsymbol\theta}}}{\partial t}=-\mathbf{\hat{\boldsymbol{r}}}\dot{\theta}+
\mathbf{\hat{\boldsymbol\phi}}\dot{\phi}\,\text{cos}\,\theta,
&&\frac{\partial\mathbf{\hat{\boldsymbol\phi}}}{\partial t}=
-\mathbf{\hat{\boldsymbol{r}}}\dot{\phi}\,\text{sin}\,\theta-\mathbf{\hat{\boldsymbol\theta}}\dot{\phi}\,\text{cos}\,\theta.\nonumber
\end{align}
Analogously to Eq.~\eqref{cylmetriccovartensorek}, the covariant and contravariant metric tensors $g_{ij}$ and $g^{ij}$ of the spherical system for coordinates $R,\,\theta,\,\phi$, respectively, and the 
transformation Jacobians from Cartesian to spherical coordinate system, and vice versa, are
\begin{align}
g_{ij}=
\begin{pmatrix}\label{sphermetriccovartensorek}
1 & 0 & 0\\
0 & r^2 & 0 \\
0 & 0 & r^2\,\text{sin}^2\theta
\end{pmatrix},\quad
g^{ij}=
\begin{pmatrix}
1 & 0 & 0\\
0 & \displaystyle{\frac{1}{r^2}} & 0 \\
0 & 0 & \displaystyle\frac{1}{r^2\,\text{sin}^2\theta}
\end{pmatrix},\quad J=r^2\,\text{sin}\,\theta,\quad J^{-1}=\frac{1}{r^2\,\text{sin}\,\theta}. 
\end{align}
Following the same principles as in Appendix~\ref{diffcylinder} (see Eqs.~\eqref{gradcylvector}-\eqref{laplacecyl}), 
gradient of a scalar function $f$ in spherical coordinates is 
\begin{align}\label{gradspherevector} 
\vec{\nabla}f=\mathbf{\hat{\boldsymbol{r}}}\frac{\partial f}{\partial r}+
\mathbf{\hat{\boldsymbol{\theta}}}\frac{1}{r}\frac{\partial f}{\partial\theta}+
\mathbf{\hat{\boldsymbol{\phi}}}\frac{1}{r\,\text{sin}\,\theta}\frac{\partial f}{\partial\phi}.
\end{align}
Tensor of a gradient of a common vector field $\vec{A}(r,\,\theta,\,\phi)$ in matrix formalism has the form
\begin{align}\label{gradspherevectorvector}
\vec{\nabla}\vec{A}=\,\,\,\,\,\bordermatrix{~&\mathbf{\hat{\boldsymbol{r}}} & \mathbf{\hat{\boldsymbol{\theta}}} & \mathbf{\hat{\boldsymbol{\phi}}} \cr
           \mathbf{\hat{\boldsymbol{r}}} & \displaystyle{\frac{\partial A_r}{\partial r}} & 
           \displaystyle{\frac{\partial A_{\theta}}{\partial r}} & \displaystyle{\frac{\partial A_\phi}{\partial r}} \cr
           \mathbf{\hat{\boldsymbol{\theta}}} & \displaystyle{\frac{1}{r}\frac{\partial A_r}{\partial\theta}-\frac{A_{\theta}}{r}} & 
           \displaystyle{\frac{1}{r}\frac{\partial A_\theta}{\partial\theta}+\frac{A_r}{r}} & 
           \displaystyle{\frac{1}{r}\frac{\partial A_\phi}{\partial\theta}} \cr
           \mathbf{\hat{\boldsymbol{\phi}}} & \displaystyle{\frac{1}{r\,\text{sin}\,\theta}\frac{\partial A_r}{\partial\phi}-\frac{A_\phi}{r}} & 
           \displaystyle{\frac{1}{r\,\text{sin}\,\theta}\frac{\partial A_\theta}{\partial\phi}-\frac{A_\phi}{r}\,\text{cot}\,\theta} & 
           \displaystyle{\frac{1}{r\,\text{sin}\,\theta}\frac{\partial A_\phi}{\partial\phi}+\frac{A_r}{r}+
           \frac{A_\theta}{r}\,\text{cot}\,\theta} \cr}.
\end{align}
In spherical coordinates
the divergence of a vector $\vec{A}\,(r,\,\theta,\,\phi)$ is
\begin{align}\label{divspherevector} 
\vec{\nabla}\cdot\vec{A}=
\frac{1}{r^2}\frac{\partial}{\partial r}\left(r^2A_r\right)+
\frac{1}{r\,\text{sin}\,\theta}\frac{\partial}{\partial\theta}\left(\text{sin}\,\theta\,A_{\theta}\right)+
\frac{1}{r\,\text{sin}\,\theta}\frac{\partial A_{\phi}}{\partial\phi}.
\end{align}
Vector rotation in spherical coordinates takes the form
\begin{align}\label{rotspherevector}
\vec\nabla\times\vec{A}=\frac{1}{r\,\text{sin}\,\theta}
\left[\frac{\partial}{\partial\theta}\left(\text{sin}\,\theta\,A_\phi\right)-
\frac{\partial A_\theta}{\partial\phi}\right]\mathbf{\hat{\boldsymbol{r}}}+
\frac{1}{r}\left[\frac{1}{\text{sin}\,\theta}\frac{\partial A_r}{\partial\phi}-
\frac{\partial}{\partial r}\left(rA_\phi\right)\right]\mathbf{\hat{\boldsymbol{\theta}}}
+\,\frac{1}{r}\left[\frac{\partial}{\partial r}\left(rA_\theta\right)-\frac{\partial A_r}{\partial\theta}\right]\mathbf{\hat{\boldsymbol{\phi}}}
\end{align}
and for the Laplacian operator in spherical coordinates one obtains
\begin{align}\label{laplacesphere} 
\Delta=\vec{\nabla}\cdot\vec{\nabla}=\frac{1}{r^2}\frac{\partial}{\partial r}\left(r^2\frac{\partial}{\partial r}\right)+
\frac{1}{r^2\,\text{sin}\,\theta}\frac{\partial}{\partial\theta}\left(\text{sin}\,\theta\frac{\partial}{\partial\theta}\right)+
\frac{1}{r^2\,\text{sin}^2\theta}\frac{\partial^2}{\partial\phi^2}.
\end{align}
The nonzero spherical Christoffel symbols are $\Gamma_{\theta\theta}^r=-r$, $\Gamma_{r\theta}^{\theta}=
\Gamma_{r\phi}^{\phi}=1/r$, $\Gamma_{\phi\phi}^r=-r\,\text{sin}^2\theta,$
$\Gamma_{\phi\phi}^\theta=-\text{sin}\,\theta\,\text{cos}\,\theta,\,\Gamma_{\theta\phi}^\phi=\text{cot}\,\theta$
(where, due to the orthogonality of the system, are the symbols that are symmetrical in lower indices identical, 
see the formalism introduced in Sect.~\ref{diffcylinder}).
\subsection{Velocity and acceleration in spherical polar coordinates}\label{velaccsphere}
The position and velocity vectors in the spherical coordinate system according to Eq,~\eqref{trafosphereunit3} are
\begin{align}\label{trafosphereunit5}
\vec{r}=x\mathbf{\hat{\boldsymbol{x}}}+y\mathbf{\hat{\boldsymbol{y}}}+z\mathbf{\hat{\boldsymbol{z}}}=r\mathbf{\hat{\boldsymbol{r}}},\quad\quad
\vec{\varv}=\frac{\text{d}\vec{r}}{\text{d}t}=\frac{\text{d}\left(r\mathbf{\hat{\boldsymbol{r}}}\right)}{\text{d}t}=
\mathbf{\hat{\boldsymbol{r}}}\dot{r}+{r}\left(\mathbf{\hat{\boldsymbol{\theta}}}\dot{\theta}+
\mathbf{\hat{\boldsymbol{\phi}}}\dot{\phi}\,\text{sin}\,\theta\right).
\end{align}
Since the velocity and acceleration vectors are simultaneously defined as
\begin{align}\label{trafosphereunit7}
\vec{\varv}=\varv_r\mathbf{\hat{\boldsymbol{r}}}+
\varv_\theta\mathbf{\hat{\boldsymbol{\theta}}}+\varv_\phi\mathbf{\hat{\boldsymbol{\phi}}},\quad\quad
\vec{a}=\frac{\text{d}\vec{\varv}}{\text{d}t}=a_r\mathbf{\hat{\boldsymbol{r}}}+a_\theta\mathbf{\hat{\boldsymbol{\theta}}}+
a_\phi\mathbf{\hat{\boldsymbol{\phi}}}, 
\end{align}
differentiation of Eq.~\eqref{trafosphereunit5} with respect to time gives the components of the acceleration vector,
\begin{align}\label{trafosphereunit9}
a_r&=\ddot{r}-r\dot{\theta}^2-r\dot{\phi}^2\text{sin}^2\theta&&
=\frac{\text{d}\varv_r}{\text{d}t}-r\dot{\theta}^2-r\dot{\phi}^2\text{sin}^2\theta,\\
a_\theta&=r\ddot{\theta}+2\dot{r}\dot{\theta}-r\dot{\phi}^2\text{sin}\,\theta\,\text{cos}\,\theta&&\label{trafosphereunit9ab}
=\frac{\text{d}\varv_\theta}{\text{d}t}+\dot{r}\dot{\theta}-r\dot{\phi}^2\text{sin}\,\theta\,\text{cos}\,\theta,\\
a_\phi&=r\ddot{\phi}\,\text{sin}\,\theta+2\dot{r}\dot{\phi}\,\text{sin}\,\theta+2r\dot{\theta}\dot{\phi}\,\text{cos}\,\theta&&\label{trafosphereunit9ac}
=\frac{\text{d}\varv_\phi}{\text{d}t}+\dot{r}\dot{\phi}\,\text{sin}\,\theta+r\dot{\theta}\dot{\phi}\,\text{cos}\,\theta.
\end{align}
Noting again that $\text{d}/\text{d}t=\partial/\partial t+\vec{\varv}\cdot\vec{\nabla}$, we write the acceleration vector expressed in terms of the 
velocity vector components in generally used form
\begin{align}\label{trafosphereunit10}
a_r&=\frac{\partial\varv_r}{\partial t}+\underbrace{\varv_r\frac{\partial\varv_r}{\partial r}+
\frac{\varv_\theta}{r}\frac{\partial\varv_r}{\partial\theta}
+\frac{\varv_\phi}{r\,\text{sin}\,\theta}\frac{\partial\varv_r}{\partial\phi}}_{\left(\vec{\varv}\cdot\vec{\nabla}\right)\varv_r}-
\frac{\varv_\theta^2+\varv_\phi^2}{r},\\
\label{trafosphereunit11}
a_\theta&=\frac{\partial\varv_\theta}{\partial t}+\underbrace{\varv_r\frac{\partial\varv_\theta}{\partial r}+
\frac{\varv_\theta}{r}\frac{\partial\varv_\theta}{\partial\theta}
+\frac{\varv_\phi}{r\,\text{sin}\,\theta}\frac{\partial\varv_\theta}{\partial\phi}}_{\left(\vec{\varv}\cdot\vec{\nabla}\right)\varv_\theta}+
\frac{\varv_r\,\varv_\theta}{r}-\frac{\varv^2_\phi\,\text{cot}\,\theta}{r},\\
\label{trafosphereunit12}
a_\phi&=\frac{\partial\varv_\phi}{\partial t}+\underbrace{\varv_r\frac{\partial\varv_\phi}{\partial r}+
\frac{\varv_\theta}{r}\frac{\partial\varv_\phi}{\partial\theta}
+\frac{\varv_\phi}{r\,\text{sin}\,\theta}\frac{\partial\varv_\phi}{\partial\phi}}_{\left(\vec{\varv}\cdot\vec{\nabla}\right)\varv_\phi}+
\frac{\varv_r\,\varv_\phi}{r}+\frac{\varv_\theta\,\varv_\phi\,\text{cot}\,\theta}{r},
\end{align}
where, similarly to the cylindrical case, the braced terms are components of the advection terms and the terms on the right-hand sides express acceleration due to noninertial forces,
i.e., the magnitude of centrifugal acceleration (cf. Eq.~\eqref{rotframe2}) induced by spherically and azimuthally directed rotation in Eq.~\eqref{trafosphereunit10}, half
of the Coriolis acceleration and spherically dependent magnitudes of centrifugal acceleration in Eqs.~\eqref{trafosphereunit11} and \eqref{trafosphereunit12} 
(cf. Eqs.~\eqref{trafosphereunit9}-\eqref{trafosphereunit9ac}).
\section{Continuity and momentum equation}\label{mainappend}
\subsection{Equation of continuity}\label{contappend}
The general form of continuity equation is introduced in Eqs.~\eqref{masscylinder001} and Eq.~\eqref{masscylinder002} in Sect.~\ref{Massconserve}. 
To transform the continuity equation into \textit{cylindrical} coordinate system 
we follow equation of a vector divergence \eqref{divcylvector}, the explicit cylindrical form of continuity equation becomes
\begin{align}\label{conticylap}
\frac{\partial\rho}{\partial t}+
\frac{1}{R}\frac{\partial}{\partial R}\left(R\,\rho V_R\right)+\frac{1}{R}\frac{\partial\,(\rho V_{\phi})}{\partial\phi}+
\frac{\partial\left(\rho V_z\right)}{\partial z}=0.
\end{align}
In quite similar way, following Eq.~\eqref{divspherevector} we express the continuity equation in \textit{spherical} polar coordinates, 
its explicit form is 
\begin{align}\label{contispherelap} 
\frac{\partial\rho}{\partial t}+
\frac{1}{r^2}\frac{\partial}{\partial r}\left(r^2\rho\,\varv_R\right)+
\frac{1}{r\,\text{sin}\,\theta}\frac{\partial}{\partial\theta}\left(\text{sin}\,\theta\,\rho\,\varv_{\theta}\right)+
\frac{1}{r\,\text{sin}\,\theta}\frac{\partial\,(\rho\,\varv_{\phi})}{\partial\phi}=0.
\end{align}
In axisymmetric or spherically symmetric case the terms with angular derivatives vanish, the cylindrical expression \eqref{conticylap} is thus 
reduced to the axisymmetric cylindrical continuity equation 
\begin{align}\label{masscylinder1}
\frac{\partial\rho}{\partial t}+
\frac{1}{R}\frac{\partial}{\partial R}\left(R\,\rho V_R\right)+\frac{\partial}{\partial z}\left(\rho V_z\right)=0.
\end{align}
Omitting the vertical motion leads to radial cylindrical continuity equation, while
the spherically symmetric continuity equation (see Eq.~\eqref{contispherelap}) becomes 
\begin{align}\label{masscylinder002}
\frac{\partial\rho}{\partial t}+
\frac{1}{r^2}\frac{\partial}{\partial r}\left(r^2\rho\,\varv_R\right)=0. 
\end{align}
We introduce all these expressions explicitly, since the various forms of the continuity equation are frequently referred to in the text.
% In axisymmetric case the second term in spatial derivatives in Eq.~\eqref{conticylap} drops, while the spherically symmetric case obviously leads to 
% vanishing of the second and third term in spatial derivatives in Eq.~\eqref{contispherelap}.
\subsection{Momentum equation}\label{momappend}
In this section we provide the explicit cylindrical and spherical form of equation of motion (momentum equation) introduced in 
Sect.~\ref{momentumconserve} as well as in Eqs.~\eqref{radmomconserve} and \eqref{phimcon} in Sect.~\ref{largemodelix}.
We however omit here  
the viscosity terms in the right-hand side of the equation, which are described in detail in Appendix~\ref{stressappend}.
Following the acceleration terms expressed in Eqs.~\eqref{trafounit8}-\eqref{trafounit10} 
and the expression of the gradient \eqref{gradcylvector}, the explicit form of the radial component of the vectorial momentum equation 
in \textit{cylindrical} coordinate system is
\begin{align}\label{cylmomrad}
\frac{\partial V_R}{\partial t}+V_R\frac{\partial V_R}{\partial R}+
\frac{V_\phi}{R}\frac{\partial V_R}{\partial\phi}
+V_z\frac{\partial V_R}{\partial z}=\frac{V_\phi^2}{R}-\frac{1}{\rho}\frac{\partial P}{\partial R}+{F_R},%-\frac{\partial\Phi}{\partial R},
\end{align}
where the terms on the right-hand side are noninertial (centrifugal) acceleration term, scalar pressure gradient term 
(acceleration induced by a pressure force) divided by density and the term $\vec{F}$ that represents the radial component of 
vector of global acceleration 
induced by external forces. Most important item in this $\vec{F}$ term is the gravitational acceleration
which is a negative gradient of the gravitational potential $\Phi$ \eqref{gravcylexpli}. The azimuthal component of 
the momentum equation is
\begin{align}
\label{cylmomphi}
\frac{\partial V_\phi}{\partial t}+V_R\frac{\partial V_\phi}{\partial R}+
\frac{V_\phi}{R}\frac{\partial V_\phi}{\partial\phi}
+V_z\frac{\partial V_\phi}{\partial z}=-\frac{V_R V_\phi}{R}-\frac{1}{\rho R}\frac{\partial P}{\partial\phi}+{F_\phi},
\end{align}
where the meaning of the right-hand side terms is similar (cf.~Eqs.~\eqref{trafounit8}-\eqref{trafounit10}). 
The vertical component of the momentum equation is
\begin{align}
\label{cylmomvert}
\frac{\partial V_z}{\partial t}+V_R\frac{\partial V_z}{\partial R}+
\frac{V_\phi}{R}\frac{\partial V_z}{\partial\phi}
+V_z\frac{\partial V_z}{\partial z}=-\frac{1}{\rho}\frac{\partial P}{\partial z}+{F_z}.
\end{align}
Regarding the gravitational acceleration in the term $\vec{F}$ (external forces), the most frequent source of external gravity is a spherically symmetric body 
(since the nonspherical deviation of gravitational force induced by rotationally oblate star's 
mass distribution is proved to be negligible, cf. the note in Sect.~\ref{teplous}), the gravitational potential $\Phi=-GM_{\star}/r$, where $M_{\star}$ is the stellar mass
and $r$ denotes the spherical polar radial distance (in cylindrical coordinates it is $r^2=R^2+z^2$). Including this expression, the nonzero components of the gravitational acceleration
term $-\vec{\nabla}\Phi$ in cylindrical coordinates (see also Sect.~\ref{thindisk}) are 
\begin{align}
\label{gravcylexpli}
g_R=-\frac{GM_{\star}R}{\left(R^2+z^2\right)^{3/2}},\quad\quad g_z=-\frac{GM_{\star}z}{\left(R^2+z^2\right)^{3/2}}.
\end{align}

In \textit{spherical} coordinate system, using the acceleration terms expressed in Eqs.~\eqref{trafosphereunit10}-\eqref{trafosphereunit12} 
and the spherical form of the gradient \eqref{gradspherevector}, we write the three components of the momentum equation 
in the following general form. The radial component is
\begin{align}\label{spheremomrad}
\frac{\partial\varv_r}{\partial t}+\varv_r\frac{\partial\varv_r}{\partial r}+
\frac{\varv_\theta}{r}\frac{\partial\varv_r}{\partial\theta}
+\frac{\varv_\phi}{r\,\text{sin}\,\theta}\frac{\partial\varv_r}{\partial\phi}
=\frac{\varv_\theta^2+\varv_\phi^2}{r}-\frac{1}{\rho}\frac{\partial P}{\partial r}-\frac{\partial\Phi}{\partial r},
\end{align}
the spherical component is
\begin{align}\label{spheremomrotat}
\frac{\partial\varv_\theta}{\partial t}+\varv_r\frac{\partial\varv_\theta}{\partial r}+
\frac{\varv_\theta}{r}\frac{\partial\varv_\theta}{\partial\theta}
+\frac{\varv_\phi}{r\,\text{sin}\,\theta}\frac{\partial\varv_\theta}{\partial\phi}=
\frac{\varv^2_\phi\,\text{cot}\,\theta}{r}-\frac{\varv_r\,\varv_\theta}{r}-\frac{1}{\rho r}\frac{\partial P}{\partial\theta}-\frac{1}{r}\frac{\partial\Phi}{\partial\theta},
\end{align}
and the azimuthal component is
\begin{align}\label{spheremomphi}
\frac{\partial\varv_\phi}{\partial t}+\varv_r\frac{\partial\varv_\phi}{\partial r}+
\frac{\varv_\theta}{r}\frac{\partial\varv_\phi}{\partial\theta}
+\frac{\varv_\phi}{r\,\text{sin}\,\theta}\frac{\partial\varv_\phi}{\partial\phi}=
-\frac{\varv_\theta\,\varv_\phi\,\text{cot}\,\theta}{r}-\frac{\varv_r\,\varv_\phi}{r}-
\frac{1}{\rho\,r\,\text{sin}\,\theta}\frac{\partial P}{\partial\theta}-\frac{1}{r\,\text{sin}\,\theta}\frac{\partial\Phi}{\partial\theta}.
\end{align}
The meaning of the right-hand side terms is described in Eqs.~\eqref{trafosphereunit10}-\eqref{trafosphereunit12}. 
Analogously to Eq.~\eqref{gravcylexpli}, the external gravity induced by a spherically symmetric body is
\begin{align}\label{spheregravka}
g_r=-\frac{GM_{\star}}{r^2}.
\end{align}
\section{Stress tensor}\label{stressappend}
The general properties of the (Cauchy) stress tensor \citep[e.g.,][]{Landau1,Mihalas2} are described in Sect.~\ref{stresstens}. The two of its equivalent forms are 
(see Eqs.~\eqref{stresik1} and \eqref{stresik2}, including the detailed description)
\begin{eqnarray}\label{stresikapp1}
T_{ij}=-P\,\delta_{ij}+\eta\left(\nabla_iV_j+\nabla_jV_i\right)+(\zeta-\frac{2}{3}\eta)\,\nabla_kV^k\delta_{ij},
\end{eqnarray}
or, involving the formalism of the Cauchy strain tensor $E_{ij}$ \citep{Mihalas2},
\begin{eqnarray}\label{stresikapp2}
T_{ij}=-P\,\delta_{ij}+2\eta\,E_{ij}+(\zeta-\frac{2}{3}\eta)\,\nabla_kV^k\delta_{ij},\quad\text{with}\quad E_{ij}=\frac{1}{2}\left(\nabla_iV_j+\nabla_jV_i\right), 
\end{eqnarray}
where the symbol $\nabla_i$ represents the covariant derivative, whose general explicit form is introduced in Sect.~\ref{diffcylinder}.
In Cartesian coordinates all the Lam\'e coefficients $h=1$ and all the Christoffel symbols vanish
(see Appendix~\ref{diffcylinder}). The components of stress and the strain tensor \eqref{stresikapp2} 
thus take the elementary form
\begin{align}\label{strainkartus}
T_{ij}=-P\delta_{ij}+\eta\left(\frac{\partial V_j}{\partial x_i}+\frac{\partial V_i}{\partial x_j}\right)
+(\zeta-\frac{2}{3}\eta)\nabla_kV^k\delta_{ij},\quad\quad
E_{ij}=\frac{1}{2}\left(\frac{\partial V_j}{\partial x_i}+\frac{\partial V_i}{\partial x_j}\right). 
\end{align}
\subsection{Stress tensor in cylindrical polar coordinates}\label{stresscylappend}
Following Eq.~\eqref{stresikapp2} we write all independent components of 
symmetric stress tensor in cylindrical coordinates,
\begin{align}
T_{RR}&=-P+2\eta\left(\frac{\partial V_R}{\partial R}\right)+(\zeta-\frac{2}{3}\eta)\vec{\nabla}\cdot\vec{V},
&&T_{R\phi}=\eta\left[\frac{1}{R}\frac{\partial V_R}{\partial \phi}+R\frac{\partial}{\partial R}\left(\frac{V_\phi}{R}\right)\right],\nonumber\\
T_{\phi\phi}&=-P+2\eta\left(\frac{1}{R}\frac{\partial V_\phi}{\partial \phi}+\frac{V_R}{R}\right)+(\zeta-\frac{2}{3}\eta)\vec{\nabla}\cdot\vec{V},
&&T_{Rz}=\eta\left(\frac{\partial V_z}{\partial R}+\frac{\partial V_R}{\partial z}\right),\label{stresikvalecexplirphi}\\
T_{zz}&=-P+2\eta\left(\frac{\partial V_z}{\partial z}\right)+(\zeta-\frac{2}{3}\eta)\vec{\nabla}\cdot\vec{V},
&&T_{\phi z}=\eta\left(\frac{1}{R}\frac{\partial V_z}{\partial \phi}+\frac{\partial V_\phi}{\partial z}\right)\nonumber,
\end{align}
where $P$ is the diagonal component of stress tensor (scalar pressure), 
while all the nondiagonal terms form the viscous stress tensor that is yet subdivided to bulk viscosity terms 
(the terms containing the velocity divergence) and the shear stress terms.
Including the stress tensor, the general expression for $i$-th component of momentum equation (Eqs.~\eqref{cylmomrad}-\eqref{cylmomvert}) takes the form 
\begin{eqnarray}\label{genmomstreform}
\rho a_i=\nabla_j T_{ij}-\rho\nabla_i\Phi,
\end{eqnarray}
where the left-hand side represents the (density multiplied) acceleration terms from Eqs.~\eqref{trafounit8}-\eqref{trafounit10}, 
the first term on the right-hand side is the 
covariant derivative of stress tensor and the last term on the right-hand side is the $i$-th component of the density of the gravitational force. The explicit form
of covariant derivative of a tensor \citep[see, e.g.,][see also the formalism introduced in Sect.~\ref{diffcylinder}]{Arfken} is
\begin{align}\label{covarderitensor}
\nabla_j T_{ij}=
\frac{1}{h_jh_ih_j}\left[\frac{\partial}{\partial x_j}(h_ih_jT_{ij})-\Gamma_{ij}^k\,h_kh_jT_{kj}\right].%\mathbf{\hat{x}}_j\mathbf{\hat{x}}_i\mathbf{\hat{x}}_j.
\end{align}
Following Eqs.~\eqref{genmomstreform} and \eqref{covarderitensor} and including stress tensor 
components from Eq.~\eqref{stresikvalecexplirphi}, 
the radial component of momentum equation in cylindrical coordinates can be written as
\begin{align}
\rho a_R&=F_R+\nabla_R T_{RR}+\nabla_\phi T_{R\phi}+\nabla_z T_{Rz}      
                    =F_R+\partial_R T_{RR}+\frac{1}{R}\partial_\phi T_{R\phi}+\partial_z T_{Rz}+\frac{1}{R}T_{RR}-\frac{1}{R}T_{\phi\phi}\nonumber\\
                    &=F_R-\frac{\partial P}{\partial R}+\frac{\partial}{\partial R}\left[2\eta\left(\frac{\partial V_R}{\partial R}\right)+(\zeta-
                    \frac{2}{3}\eta)(\vec{\nabla}\cdot\vec{V})\right]\nonumber\\
                    &+\frac{1}{R}\frac{\partial}{\partial \phi}\left\{\eta\left[\frac{1}{R}\frac{\partial V_R}{\partial\phi}
                    +R\frac{\partial}{\partial R}\left(\frac{V_\phi}{R}\right)\right]\right\}
                    +\frac{\partial}{\partial z}\left[\eta\left(\frac{\partial V_z}{\partial R}+\frac{\partial V_R}{\partial z}\right)\right]
                    +\frac{2\eta}{R}\left[R\frac{\partial}{\partial R}\left(\frac{V_R}{R}\right)-\frac{1}{R}\frac{\partial V_\phi}{\partial \phi}\right],
                    \label{appendrmomcylinder1}
                    \end{align}
		    where $F_R$ is the density of the radial gravitational force component $\rho g_R$ (see Eq.~\eqref{gravcylexpli}). Analogously 
		    the azimuthal component of momentum equation is
                    \begin{align}
\rho a_\phi&=F_\phi+\nabla_R T_{\phi R}+\nabla_\phi T_{\phi\phi}+\nabla_z T_{\phi z}     
                    =F_\phi+\partial_R T_{\phi R}+\frac{1}{R}\partial_\phi T_{\phi\phi}+\partial_z T_{\phi z}+\frac{2}{R}T_{\phi R}\nonumber\\
                    &=F_\phi-\frac{1}{R}\frac{\partial P}{\partial\phi}+\frac{\partial}{\partial R}\left\{\eta\left[\frac{1}{R}\frac{\partial V_R}{\partial \phi}
                    +R\frac{\partial}{\partial R}\left(\frac{V_\phi}{R}\right)\right]\right\}
                    +\frac{1}{R}\frac{\partial}{\partial\phi}\left[2\eta\left(\frac{1}{R}\frac{\partial V_\phi}{\partial\phi}
                    +\frac{V_R}{R}\right)+(\zeta-\frac{2}{3}\eta)(\vec{\nabla}\cdot\vec{V})\right]\nonumber\\
                    &+\frac{\partial}{\partial z}\left[\eta\left(\frac{1}{R}\frac{\partial V_z}{\partial\phi}+\frac{\partial V_\phi}{\partial z}\right)\right]
                    +\frac{2\eta}{R}\left[\frac{1}{R}\frac{\partial V_R}{\partial \phi}
                    +R\frac{\partial}{\partial R}\left(\frac{V_\phi}{R}\right)\right],\label{appendphimomcylinder1}
                    \end{align}
                    and the vertical component of momentum equation is
\begin{align}
\rho a_z&=F_z+\nabla_R T_{zR}+\nabla_\phi T_{z\phi}+\nabla_z T_{zz}\nonumber     
                    =F_z+\partial_R T_{zR}+\frac{1}{R}\partial_\phi T_{z\phi}+\partial_z T_{zz}+\frac{1}{R}T_{Rz}\nonumber\\
                    &=F_z-\frac{\partial P}{\partial z}+\frac{\partial}{\partial R}\left[\eta\left(\frac{\partial V_z}{\partial R}
                    +\frac{\partial V_R}{\partial z}\right)\right]
                    +\frac{1}{R}\frac{\partial}{\partial\phi}\left[\eta\left(\frac{1}{R}\frac{\partial V_z}{\partial\phi}
                    +\frac{\partial V_\phi}{\partial z}\right)\right]\nonumber\\
                    &+\frac{\partial}{\partial z}\left[2\eta\left(\frac{\partial V_z}{\partial z}\right)
                    +(\zeta-\frac{2}{3}\eta)(\vec{\nabla}\cdot\vec{V})\right]
                    +\frac{\eta}{R}\left(\frac{\partial V_z}{\partial R}
                    +\frac{\partial V_R}{\partial z}\right).
                    \label{appendzmomcylinder1}
\end{align}
\subsection{Stress tensor in spherical polar coordinates}\label{stresssphereappend}
Similarly to Appendix~\ref{stresscylappend}, following Eq.~\eqref{stresikapp2} and employing
Christoffel symbols formalism introduced in Appendix~\ref{diffsphere}, we explicitly write all independent components of symmetric stress tensor in spherical coordinates,
\begin{align}\label{stresikkulaexpli}
T_{rr}&=-P+2\eta\left(\frac{\partial\varv_r}{\partial r}\right)+(\zeta-\frac{2}{3}\eta)\vec{\nabla}\cdot\vec{\varv},
&&T_{r\theta}=\eta\left[\frac{1}{r}\frac{\partial\varv_r}{\partial\theta}+r\frac{\partial}{\partial r}\left(\frac{\varv_\theta}{r}\right)\right],\nonumber\\
T_{\theta\theta}&=-P+2\eta\left(\frac{1}{r}\frac{\partial\varv_\theta}{\partial\theta}+\frac{\varv_r}{r}\right)+(\zeta-\frac{2}{3}\eta)\vec{\nabla}\cdot\vec{\varv},
&&T_{r\phi}=\eta\left[\frac{1}{r\,\text{sin}\,\theta}\frac{\partial\varv_r}{\partial\phi}+r\frac{\partial}{\partial r}\left(\frac{\varv_\phi}{r}\right)\right],\\
T_{\phi\phi}&=-P+2\eta\left(\frac{1}{r\,\text{sin}\,\theta}\frac{\partial\varv_\phi}{\partial\phi}+\frac{\varv_r+\varv_\theta\,\text{cot}\,\theta}{r}\right)+
(\zeta-\frac{2}{3}\eta)\vec{\nabla}\cdot\vec{\varv},
&&T_{\theta\phi}=\eta\left[\frac{1}{r\,\text{sin}\,\theta}\frac{\partial\varv_\theta}{\partial\phi}
+\frac{\text{sin}\,\theta}{r}\frac{\partial}{\partial\theta}\left(\frac{\varv_\phi}{\text{sin}\,\theta}\right)\right].\nonumber
\end{align}
The meaning of particular terms and quantities is identical to Eq.~\eqref{stresikvalecexplirphi} in Appendix~\ref{stresscylappend}.
Following Eqs.~\eqref{genmomstreform} and \eqref{covarderitensor} and including spherical stress tensor 
components introduced in Eq.~\eqref{stresikkulaexpli}, we write 
the radial component of momentum equation in spherical coordinates in the explicit form
\begin{align}
\rho a_r&=F_r+\nabla_r T_{rr}+\nabla_\theta T_{r\theta}+\nabla_\phi T_{r\phi}\nonumber\\           
                    &=F_r+\frac{1}{r^2}\partial_r\left(r^2\,T_{rr}\right)
                    +\frac{1}{r\,\text{sin}\,\theta}\partial_\theta\left(\text{sin}\,\theta\,T_{r\theta}\right)
                    +\frac{1}{r\,\text{sin}\,\theta}\partial_\phi T_{r\phi}-\frac{1}{r}\left(T_{\theta\theta}+T_{\phi\phi}\right)\nonumber\\
                    &=F_r-\frac{\partial P}{\partial r}+\frac{\partial}{\partial r}\left[2\eta\left(\frac{\partial\varv_r}{\partial r}\right)+(\zeta-
                    \frac{2}{3}\eta)(\vec{\nabla}\cdot\vec{\varv})\right]\nonumber\\
                    &+\frac{1}{r}\frac{\partial}{\partial\theta}\left\{\eta\left[\frac{1}{r}\frac{\partial\varv_r}{\partial\theta}
                    +r\frac{\partial}{\partial r}\left(\frac{\varv_\theta}{r}\right)\right]\right\}
                    +\frac{1}{r\,\text{sin}\,\theta}\frac{\partial}{\partial\phi}\left\{\eta\left[\frac{1}{r\,\text{sin}\,\theta}
                    \frac{\partial\varv_r}{\partial\phi}+r\frac{\partial}{\partial r}\left(\frac{\varv_\phi}{r}\right)\right]\right\}\nonumber\\
                    &+\frac{\eta}{r}\left[4r\frac{\partial}{\partial r}\left(\frac{\varv_r}{r}\right)
                    -\frac{2}{r\,\text{sin}\,\theta}\frac{\partial}{\partial\theta}\left(\varv_\theta\,\text{sin}\,\theta\right)
                    -\frac{2}{r\,\text{sin}\,\theta}\frac{\partial\varv_\phi}{\partial\phi}
                    +r\,\text{cot}\,\theta\frac{\partial}{\partial r}\left(\frac{\varv_\theta}{r}\right)
                    +\frac{\text{cot}\,\theta}{r}\frac{\partial\varv_r}{\partial\theta}\right]\label{appendrmomkula1},
                    \end{align}
                    where $F_r$ is mainly the density of the radial gravitational force $\rho g_r$ (see Eq.~\ref{spheregravka}). 
                    The spherical component of momentum equation can be written as
\begin{align}
\rho a_\theta&=F_\theta+\nabla_r T_{\theta r}+\nabla_\theta T_{\theta\theta}+\nabla_\phi T_{\theta\phi}\nonumber\\         
                    &=F_\theta+\frac{1}{r^2}\partial_r\left(r^2\,T_{\theta r}\right)
                    +\frac{1}{r\,\text{sin}\,\theta}\partial_\theta\left(\text{sin}\,\theta\,T_{\theta\theta}\right)
                    +\frac{1}{r\,\text{sin}\,\theta}\partial_\phi T_{\theta\phi}+\frac{1}{r}\left(T_{\theta r}-\text{cot}\,\theta\,T_{\phi\phi}\right)\nonumber\\
                    &=F_\theta-\frac{1}{r}\frac{\partial P}{\partial\theta}+\frac{\partial}{\partial r}
                    \left\{\eta\left[\frac{1}{r}\frac{\partial\varv_r}{\partial\theta}
                    +r\frac{\partial}{\partial r}\left(\frac{\varv_\theta}{r}\right)\right]\right\}
                    +\frac{1}{r}\frac{\partial}{\partial\theta}\left[2\eta\left(\frac{1}{r}\frac{\partial\varv_\theta}{\partial\theta}+\frac{\varv_r}{r}\right)+(\zeta-
                    \frac{2}{3}\eta)(\vec{\nabla}\cdot\vec{\varv})\right]\nonumber\\
                    &+\frac{1}{r\,\text{sin}\,\theta}\frac{\partial}{\partial\phi}\left\{\eta\left[\frac{1}{r\,\text{sin}\,\theta}
                    \frac{\partial\varv_\theta}{\partial\phi}+\frac{\text{sin}\,\theta}{r}\frac{\partial}{\partial\theta}
                    \left(\frac{\varv_\phi}{\text{sin}\,\theta}\right)\right]\right\}\nonumber\\
                    &+\frac{\eta}{r}\left\{\frac{2\,\text{cot}\,\theta}{r}\left[\text{sin}\,\theta\frac{\partial}{\partial\theta}
                    \left(\frac{\varv_\theta}{\text{sin}\,\theta}\right)-\frac{1}{\text{sin}\,\theta}\frac{\partial\varv_\phi}{\partial\phi}\right]
                    +3r\frac{\partial}{\partial r}\left(\frac{\varv_\theta}{r}\right)
                    +\frac{3}{r}\frac{\partial\varv_r}{\partial\theta}\right\}\label{appendphimomkula1}.
                    \end{align}
                    The azimuthal component of momentum equation can be written as
\begin{align}
\rho a_\phi&=F_\phi+\nabla_r T_{\phi r}+\nabla_\theta T_{\phi\theta}+\nabla_\phi T_{\phi\phi}\nonumber\\           
                    &=F_\phi+\frac{1}{r^2}\partial_r\left(r^2\,T_{\phi r}\right)
                    +\frac{1}{r\,\text{sin}\,\theta}\partial_\theta\left(\text{sin}\,\theta\,T_{\phi\theta}\right)
                    +\frac{1}{r\,\text{sin}\,\theta}\partial_\phi T_{\phi\phi}+\frac{1}{r}\left(T_{\phi r}+\text{cot}\,\theta\,T_{\theta\phi}\right)\nonumber\\
                    &=F_\phi-\frac{1}{r\,\text{sin}\,\theta}\frac{\partial P}{\partial\phi}+\frac{\partial}{\partial r}
                    \left\{\eta\left[\frac{1}{r\,\text{sin}\,\theta}\frac{\partial\varv_r}{\partial\phi}
                    +r\frac{\partial}{\partial r}\left(\frac{\varv_\phi}{r}\right)\right]\right\}
                    +\frac{1}{r}\frac{\partial}{\partial\theta}\left\{\eta\left[\frac{1}{r\,\text{sin}\,\theta}
                    \frac{\partial\varv_\theta}{\partial\phi}+\frac{\text{sin}\,\theta}{r}\frac{\partial}{\partial\theta}
                    \left(\frac{\varv_\phi}{\text{sin}\,\theta}\right)\right]\right\}\nonumber\\
                    &+\frac{1}{r\,\text{sin}\,\theta}\frac{\partial}{\partial\phi}\left[2\eta\left(\frac{1}{r\,\text{sin}\,\theta}\frac{\partial\varv_\phi}{\partial\phi}
                    +\frac{\varv_r}{r}+\frac{\varv_\theta\,\text{cot}\,\theta}{r}\right)+(\zeta-
                    \frac{2}{3}\eta)(\vec{\nabla}\cdot\vec{\varv})\right]\nonumber\\
                    &+\frac{\eta}{r}\left\{\frac{2\,\text{cot}\,\theta}{r}\left[\text{sin}\,\theta\frac{\partial}{\partial\theta}
                    \left(\frac{\varv_\phi}{\text{sin}\,\theta}\right)+\frac{1}{\text{sin}\,\theta}\frac{\partial\varv_\theta}{\partial\phi}\right]
                    +3r\frac{\partial}{\partial r}\left(\frac{\varv_\phi}{r}\right)
                    +\frac{3}{r\,\text{sin}\,\theta}\frac{\partial\varv_r}{\partial\phi}\right\}\label{appendzmomkula1}.
\end{align}
\section{Energy equation}\label{enappend}
Following the formalism of the stress tensor introduced in Appendix~\ref{stressappend}, 
we perform in this section the explicit development of the term $T^{ij}\nabla_jV_i$
in the internal energy equation (see Eq.~\eqref{internalenergy} and its detailed description)
\begin{align}\label{internalenergyappend}
\rho\frac{\text{d}\epsilon}{\text{d}t}=T^{ij}\nabla_jV_i-\nabla_jq^j.
\end{align}
\noindent Using Eq.~\eqref{stresikapp2}, we expand the first 
term on right-hand side of Eq.~\eqref{internalenergyappend} into expression 
\begin{align}\label{stresstensordetailappend}
T^{ij}\nabla_jV_i\!=\!\left[\!-P\delta^{ij}\!+2\eta E^{ij}\!+\left(\zeta\!-\!\frac{2}{3}\eta\right)\!\nabla_kV^k\delta^{ij}\!\right]\!\nabla_jV_i\!
=\!-P\,\nabla_iV^i\!+\!2\eta E^{ij}\nabla_jV_i\!+\left(\zeta\!-\!\frac{2}{3}\eta\right)\left(\nabla_iV^i\right)^2\!,
\end{align}
where in the right-hand side term we contract the indices using Kronecker deltas.
Using the definition of the symmetric strain tensor given in Eq.~\eqref{stresikapp2}, we expand 
the second term on the right-hand side of Eq.~\eqref{stresstensordetailappend}, obtaining
\begin{align}
E^{ij}\nabla_jV_i=\frac{1}{2}\left(\nabla^jV^i+\nabla^iV^j\right)\nabla_jV_i=\frac{1}{2}\left(\nabla^jV^i\nabla_jV_i+
\nabla^iV^j\nabla_jV_i\right).
\end{align} 
The symmetricity of the strain tensor $E^{ij}=E^{ji}$ in indices $i,\,j$ as well as the orthogonality of the coordinates, implying $E^{ij}E_{ij}=E_{ij}E^{ij}$ 
(cf. Eq.~\eqref{cylmetriccovartensorek} and Eq.~\eqref{sphermetriccovartensorek}), gives the identity
\begin{align}
E^{ij}E_{ij}&=\frac{1}{4}\left(\nabla^jV^i+\nabla^iV^j\right)\left(\nabla_jV_i+\nabla_iV_j\right)
=\frac{1}{4}\left(\nabla^jV^i\nabla_jV_i+\nabla^jV^i\nabla_iV_j+\nabla^iV^j\nabla_jV_i+\nabla^iV^j\nabla_iV_j\right)\nonumber\\
&=\frac{1}{2}\left(\nabla^jV^i\nabla_jV_i+\nabla^iV^j\nabla_jV_i\right)=E_{ij}E^{ij}.
\end{align}
Equation \eqref{stresstensordetailappend} can be written as 
\begin{align}\label{picapendour}
T^{ij}\nabla_jV_i=-P\,\nabla_iV^i+2\eta E_{ij}E^{ij}+\left(\zeta-\frac{2}{3}\eta\right)\left(\nabla_iV^i\right)^2,
\end{align}
where the first term on the right-hand side represents the reversible work done by the expanding matter, while the 
last two terms on the right-hand side, which represent the irreversible conversion of mechanical energy into heat, are called the dissipation function,
\begin{align}\label{Phiappend}
\Psi=2\eta E_{ij}E^{ij}+\left(\zeta-\frac{2}{3}\eta\right)\left(\vec\nabla\cdot\vec{V}\right)^2.
\end{align}
The quadratic terms in Eq.~\eqref{Phiappend} show that the term $\Psi$ is always positive
(this function is usually denoted as $\Phi$, see, e.g.,~\citet{Mihalas2}, however, we call it $\Psi$ in order not to be confused 
with gravitational potential).
Using Eq.~\eqref{strainkartus}, the explicit form of the dissipation function in \textit{Cartesian} coordinates becomes
\begin{align}\label{cartisip}
\Psi=2&\eta\left[\left(\frac{\partial V_x}{\partial x}\right)^2+\left(\frac{\partial V_y}{\partial y}\right)^2
                    +\left(\frac{\partial V_z}{\partial z}\right)^2\right.\nonumber\\
                    +&\left.\frac{1}{2}\left(\frac{\partial V_x}{\partial y}+\frac{\partial V_y}{\partial x}\right)^2
                    +\frac{1}{2}\left(\frac{\partial V_x}{\partial z}+\frac{\partial V_z}{\partial x}\right)^2                                                                                               
                    +\frac{1}{2}\left(\frac{\partial V_y}{\partial z}+\frac{\partial V_z}{\partial y}\right)^2\right]
                    +\left(\zeta-\frac{2}{3}\eta\right)\left(\vec\nabla\cdot\vec{V}\right)^2.
\end{align}
In \textit{cylindrical} polar coordinates the dissipation function can be written as
\begin{align}\label{cylisip}
\Psi=2&\eta\left\{\left(\frac{\partial V_R}{\partial R}\right)^2
                    +\left(\frac{1}{R}\frac{\partial V_\phi}{\partial \phi}+\frac{V_R}{R}\right)^2
                    +\left(\frac{\partial V_z}{\partial z}\right)^2\right.\nonumber\\
                    +&\left.\frac{1}{2}\left[\frac{1}{R}\frac{\partial V_R}{\partial \phi}+R\frac{\partial}{\partial R}\left(\frac{V_\phi}{R}\right)\right]^2
                    +\frac{1}{2}\left(\frac{\partial V_z}{\partial R}+\frac{\partial V_R}{\partial z}\right)^2                                                                                               
                    +\frac{1}{2}\left(\frac{1}{R}\frac{\partial V_z}{\partial \phi}+\frac{\partial V_\phi}{\partial z}\right)^2\right\}\\
                    +&\left(\zeta-\frac{2}{3}\eta\right)\left[\frac{1}{R}\frac{\partial}{\partial R}\left(RV_R\right)
                    +\frac{1}{R}\frac{\partial V_{\phi}}{\partial\phi}+\frac{\partial V_z}{\partial z}\right]^2.\nonumber
\end{align}
In \textit{spherical} polar coordinates the full explicit form of the dissipation function is
\begin{align}\label{kulisip}
\Psi=2&\eta\left\{\left(\frac{\partial\varv_r}{\partial r}\right)^2
                    +\left(\frac{1}{r}\frac{\partial\varv_\theta}{\partial\theta}+\frac{\varv_r}{r}\right)^2
                    +\left(\frac{1}{r\,\text{sin}\,\theta}\frac{\partial\varv_\phi}{\partial\phi}+\frac{\varv_r}{r}+
                    \frac{\varv_\theta\,\text{cot}\,\theta}{r}\right)^2\right.\nonumber\\
                    +&\left.\frac{1}{2}\left[\frac{1}{r}\frac{\partial\varv_r}{\partial\theta}+r\frac{\partial}{\partial r}\left(\frac{\varv_\theta}{r}\right)\right]^2
                    +\frac{1}{2}\left[\frac{1}{r\,\text{sin}\,\theta}\frac{\partial\varv_r}{\partial\phi}+
                    r\frac{\partial}{\partial r}\left(\frac{\varv_\phi}{r}\right)\right]^2\right.\\                                                                                               
                    +&\left.\frac{1}{2}\left[\frac{1}{r\,\text{sin}\,\theta}\frac{\partial\varv_\theta}{\partial\phi}
                    +\frac{\text{sin}\,\theta}{r}\frac{\partial}{\partial\theta}\left(\frac{\varv_\phi}{\text{sin}\,\theta}\right)\right]^2\right\}
                    +\left(\zeta-\frac{2}{3}\eta\right)\left[\frac{1}{r^2}\frac{\partial}{\partial r}\left(r^2\varv_r\right)
                    +\frac{1}{r\,\text{sin}\,\theta}\frac{\partial}{\partial\theta}\left(\text{sin}\,\theta\,\varv_\theta\right)
                    +\frac{1}{r\,\text{sin}\,\theta}\frac{\partial\varv_\phi}{\partial\phi}\right]^2\!\!\nonumber.
\end{align}

\chapter{Details of the computational codes}\label{appendix2}
\begin{figure}[h!]
\begin{center}
\includegraphics[width=13.2cm]{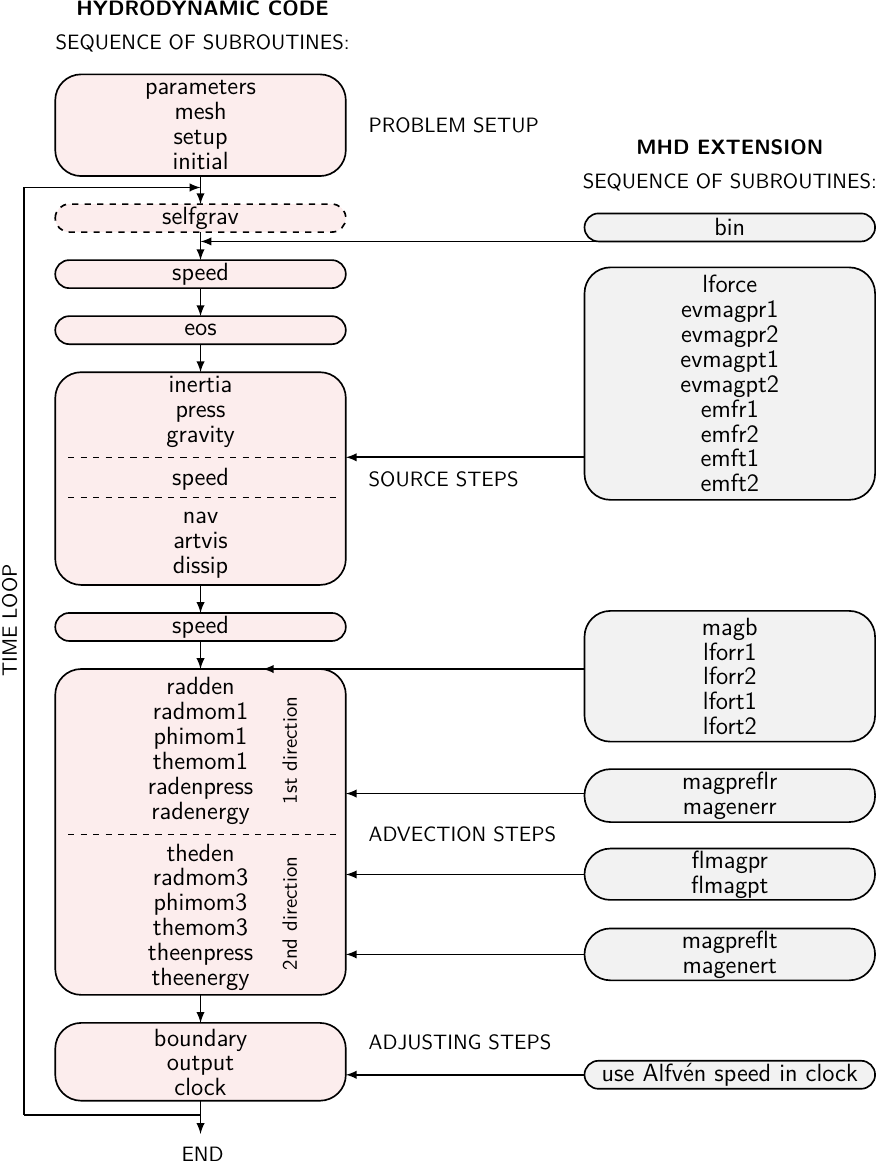}
\caption{Schematic flow chart of the time loop of the sequence of the basic subroutines of hydrodynamic time-dependent Eulerian code
with separated magnetohydrodynamic part. The position of magnetohydrodynamic subroutines in the hydrodynamic time loop is indicated by the arrows.
The meaning and content of particular subroutines is described in Appendix~\ref{dvojdimhydrous} \citep[cf.][]{Normi,Stony1a,Stony1b}.}
\label{hydroloop}
\end{center}
\end{figure}
\begin{figure}[t]
\begin{center}
\includegraphics[width=6.5cm]{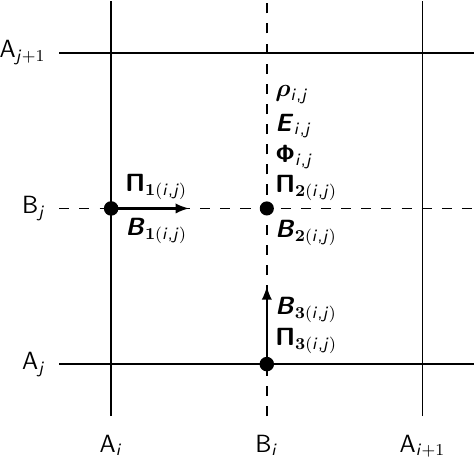}
\caption{Schema of the positions of particular scalar or vector quantities within one A-grid cell of the (magneto)hydrodynamic code.
The scalar quantities, i.e., density $\rho$, energy $E$, gravitational potential $\Phi$ as well as the momentum component 
$\Pi_2=\rho V_2$ and the magnetic induction component $B_2$, which are perpendicular to both 
coordinate directions of the longitudinal (radial)-vertical two-dimensional grid, are located at the BB-mesh points. The momentum component 
$\Pi_1=\rho V_1$ and the magnetic induction component $B_1$ streaming along 
the coordinate direction $i$ are located at AB-mesh points, the momentum component $\Pi_3=\rho V_3$ and the magnetic induction component $B_3$ streaming along 
the coordinate direction $j$ are located at BA-mesh points (where the first capital letter denotes the grid in longitudinal (radial) $i$-direction and the second capital letter
denotes the grid in latitudinal (azimuthal, vertical or spherical) $j$-direction, according to selected coordinate system. In the axisymmetric 
models of stellar outflowing disks the quantities $\Pi_1$, $\Pi_3$ correspond to $\Pi_R$, $\Pi_z$ while $\Pi_2$ corresponds to the angular momentum $J$. Adapted from \citet{Stony1a}.}
\label{gridak}
\end{center}
\end{figure}
\section{Two-dimensional hydrodynamic code}\label{dvojdimhydrous}
The structure of our 2D hydrodynamic time-dependent code is based on and combines the schemes introduced in \citet{Normi,Stony1a,felda}. 
The general properties of the code are given in Sect.~\ref{timedependix}. In this section we describe the basic computational subroutines of the 
hydrodynamic code. The sequence of the subroutines corresponds to the order in which these subroutines are called by the main program. The schema of the  
sequenced code structure (including the subroutines of the magnetohydrodynamic part of the code - see Appendix~\ref{dvojdimmagnetohydrous})
is shown in Fig.~\ref{hydroloop}.
\subsection{Problem setup}\label{setous}
\subsubsection{Subroutine \textit{parameters} - definition of the problem parameters}
This subroutine contains the selection of basic problem parameters, namely the type of the grid (e.g., equidistant, logarithmic, etc.) and 
the geometry of the problem (Cartesian, cylindrical, spherical, flaring), size of the computational domain given by its limiting points
$x_{\text{min}}$ (at the mesh points $i,j=3$) and $x_{\text{max}}$ (at the points $i,j=ni+2,nj+2$) and number $ni,nj$ of the grid points in all the coordinate 
directions (see Fig.~\ref{stagis}), initial and terminal time of computation, number of calculated timesteps and the fraction of the timesteps plotted to the 
output file. We set also the type of the boundary values (i.e., reflecting, outflowing, periodic, etc.) and the values of all physical
constants involved in the particular problem.
\subsubsection{Subroutine \textit{mesh} - parameters of the computational grid}
In this subroutine all geometrical grid characteristics of the staggered mesh (see Fig.~\ref{stagis}) are defined. 
The coordinate intervals of the grid cell edges in one-dimensional general scheme are 
\begin{align}\label{delfinci}
\Delta x^{\text{A}}=x^{\text{B}}_{i}-x^{\text{B}}_{i-1},\quad\quad\Delta x^{\text{B}}=x^{\text{A}}_{i+1}-x^{\text{A}}_i.  
\end{align}
This applies in the both specified computational coordinate directions $i,\,j$, 
while in the third remaining coordinate direction we set the unit interval of the grid cell edge. 
According to the geometric nature of the problem we basically use two types of grid cells spatial scaling, the equidistant grid and the 
grid that is logarithmic in one or even in both directions. The grid points in the equidistant and logarithmic mode, respectively, are defined (in 
one-dimensional general scheme) as (see also Fig.~\ref{stagis})
\begin{align}\label{cylmesh}
x_i\,(\text{equ})=x_{\text{min}}+\left(x_{\text{max}}-x_{\text{min}}\right)\frac{i-3}{ni},\quad\quad 
x_i\,(\text{log})=x_{\text{min}}\left(\frac{x_{\text{max}}}{x_{\text{min}}}\right)^\frac{i-3}{ni},
\end{align}
The basic curvilinear coordinate systems variables are described in Appendix~\ref{diffcylinder} 
and in Appendix \ref{diffsphere}, the special flaring coordinate system is described in Sect.~\ref{flarecoords}. For the calculations using the finite volume method 
we need to specify the so-called control surfaces $S^{\!\text{A}}_{i,j},\,S^{\text{B}}_{i,j}$ (for both meshes A, B and for all three coordinate directions) 
and the control volumes $\Omega^{\text{A}}_{i,j},\,\Omega^{\!\text{B}}_{i,j}$ of the grid cells (see Sect.~\ref{vanleerix}).
In Cartesian coordinate system the control surfaces (where the subscript denotes the variable referring to corresponding coordinate surface, cf.~the
description of the flaring coordinate system in Sect.~\ref{flarevolsurfs}) and the control volumes are $S_x=\Delta y\,\Delta z$, $S_y=\Delta z\,\Delta x$,
$S_z=\Delta x\,\Delta y$ and $\Omega_x=\Delta x\,\Delta y\,\Delta z$ (where the $\Delta$ intervals correspond to Eq.~\eqref{delfinci}).
Since Cartesian, cylindrical as well as spherical systems are orthogonal, the expressions for the surfaces and volumes are identical for the both grids A and B.
In the cylindrical coordinate system, preferably used for the disk modeling, these geometrical quantities (see also Appendix~\ref{diffcylinder}) take
the general geometrical form 
\begin{align}\label{cyl20}
S\!_R=R\,\Delta\phi\,\Delta z,\quad S\!_\phi=\Delta z\,\Delta R,\quad
S\!_z=\frac{R_2^2-R_1^2}{2}\,\Delta\phi,\quad\Omega=\frac{R_2^2-R_1^2}{2}\,\Delta\phi\,\Delta z,
\end{align}
where $\Delta R=\left(R_2-R_1\right),\,\Delta\phi=\left(\phi_2-\phi_1\right),\,\Delta z=\left(z_2-z_1\right)$.
In the spherical coordinate system (used dominantly for the stellar physics modeling) the corresponding quantities (see also Appendix~\ref{diffsphere}) are
\begin{align}\label{kul20} 
S\!_r=r^2\left|\Delta\,\text{cos}\,\theta\right|\,\Delta\phi,\quad
S\!_\theta=\frac{r_2^2-r_1^2}{2}\,\text{sin}\,\theta\,\Delta\phi,\quad
S\!_\phi=\frac{r_2^2-r_1^2}{2}\,\Delta\theta,\quad
\Omega=\frac{r_2^3-r_1^3}{3}\left|\Delta\,\text{cos}\,\theta\right|\,\Delta\phi.
\end{align}
The meaning of the points with subscript 2 in Eqs.~\eqref{cyl20} and \eqref{kul20} is $i,j$ on the A-mesh while it is $i+1,j+1$ on the B-mesh and 
the meaning of the points with subscript 1 is $i-1,j-1$ on the A-mesh while it is $i,j$ on the B-mesh. This applies in general also for the interval $\phi_2-\phi_1$, 
in case of the two-dimensional $R-z$ plane disk modeling we however set $\phi_2-\phi_1=1$.
\subsubsection{Subroutine \textit{setup}}
The subroutine defines the initial functions of the variable quantities of the external or internal scalar or vector fields which do not change during the time evolution
of the system, e.g. the radially dependent sound speed $a(R)$, the radially dependent viscosity parameter $\alpha(R)$ (see Sects.~\ref{eqraddisk}, \ref{thindisk}),
radial and vertical components of the gravitational acceleration $g_R$, $g_z$ induced by an external object (central star in case of the circumstellar disk), etc.
\subsubsection{Subroutine \textit{initial} - definition of the initial profiles of (magneto)hydrodynamic quantities}
In this subroutine we set the analytic functions that characterize the approximated initial profiles of the hydrodynamic conservative quantities defined 
in Sect.~\ref{operous}, i.e., of the density $\rho$, the momentum (angular momentum) components $\rho\vec{V}$ and $R\times\rho\vec{V}$ and the total energy $E$,
as well as the initial profiles of the components of magnetic induction vector $\vec{B}$, when employing the MHD part of the code.
The initial function is here defined within the whole range of the grid, i.e., from $i,j=1$ to $i,j=ni+4,nj+4$ (see Fig.~\ref{stagis} for details).
For example, in case of two-dimensional circumstellar outflowing disk we set the initial density profile, using the 
assumption of the vertically integrated density $\Sigma(R)\sim R^{-2}$ \citep[][see also Sect.~\ref{thindisk}]{okac91} and combining 
Eqs.~\eqref{fullvertigo}, \eqref{Sigmasegva} and the expression for the disk vertical scale height $H=a/\Omega$ (see Sect.~\ref{vertikalekdisk}), as 
\begin{align}\label{ininumerden}
\rho(R,z)=\Sigma\big(R_\text{eq}\big)\left(\frac{R_\text{eq}}{R}\right)^2\!\sqrt{\frac{GM_\star}{2\pi a^2R^3}}\,\,
\text{exp}\left[\frac{GM_\star}{a^2R}\left(\frac{R}{\sqrt{R^2+z^2}}-1\right)\right].
\end{align}
The inner boundary surface density is estimated via Eq.~\eqref{angularmomentumfluxix0} in the following way: we select the value of mass loss rate $\dot{M}$ and 
set the initial inner boundary radial flow velocity $V_R\big(R_\text{eq}\big)=1$. The resulting profiles of all hydrodynamic quantities do 
not depend on the absolute value of $\Sigma\big(R_\text{eq}\big)$, thus we can reversely re-establish this value using the proper converged value of $V_R\big(R_\text{eq}\big)$.
In case of one-dimensional disk models we use the simple initial surface density relation
\begin{align}\label{ininumersurfden}
\Sigma(R)=\Sigma\big(R_\text{eq}\big)\left(\frac{R_\text{eq}}{R}\right)^2.
\end{align}
The initial profile of the outflowing disk angular momentum $J$ takes in two-dimensional models the explicit form
\begin{align}\label{ininumerangmom}
J(R,z)=\rho(R,z)\sqrt{g_RR^3},
\end{align}
where the radial component $g_R$ of the external gravity is given by Eq.~\eqref{gravcylexpli}. In one-dimensional calculations this term
becomes only radially dependent and the density $\rho(R,z)$ becomes the surface density $\Sigma(R)$. The initial radial and vertical components of momentum 
(cf.~Eq.~\eqref{sourcicek1})
\begin{align}\label{ininumerradm}
\Pi_R=\rho(R,z)V_R=0,\quad\quad\quad \Pi_z=\rho(R,z)V_z=0
\end{align}
in the whole computational domain, 
(regardless of the necessary initial approximation of $V_R\big(R_\text{eq}\big)$
in Eq.~\eqref{angularmomentumfluxix0}), 
since any arbitrary initial profile of radial velocity $V_R$ does not affect the calculations
and according to assumed vertical hydrostatic equilibrium.
The further important initial functions of particular quantities we describe in detail within the sections that refer to the specific models.
\subsection{Source steps}\label{koropticek}
The source steps add the operator-splitted (Eq.~\eqref{oprourix}) source functions, i.e. the force terms from right-hand sides of hydrodynamic equations (see Sect.~\eqref{operous}).
The computation (as well as in the following advection steps block) is performed merely within the grid computational domain (marked in Fig.~\ref{stagis}),
i.e.~from $i,j=4$ to $i,j=ni+2,nj+2$ for the for the A-grid positioned quantities (vector components advected in the corresponding direction) 
and from $i,j=3$ to $i,j=ni+2,nj+2$ for the B-grid positioned quantities (scalars and vector components advection directed otherwise).
\subsubsection{Subroutine \textit{selfgrav} - the Poisson equation solver}
This subroutine solves the Poisson equation $\Delta\Phi=4\pi G\rho$ via the tridiagonal matrix in the second spatial derivatives of the gravitational 
potential using the LAPACK (linear algebra) package \citep{anders}. Since the influence of the self-gravity in case of the
stellar outflowing disks is negligible, we do not employ the subroutine in the disks modeling, nor we describe it in detail.
\subsubsection{Subroutine \textit{speed} - the components of the velocity of hydrodynamic flow}
This subroutine is in fact not the part of the source steps, it however must be initialized right after the computation enters the main part (time loop) of the program and 
it is yet updated between two parts of source block and between source and advection part of the code \citep[][see, Fig.~\ref{hydroloop} for the chart of the code]{Normi,richter}.
Each velocity component is calculated from the corresponding momentum component 
at the corresponding grid point (see Figs.~\ref{stagis} and \ref{gridak}).%; 
% in the following relations we indicate the type of the mesh, i.e., A or B, only in the terms where we need to prevent the possible ambiguity. In all other terms the mesh type corresponds 
% to the particular quantity).
We write
\begin{align}\label{velonumrad}
V^\text{AB}_{\!R\,(i,j)}=\frac{2\Pi_{R\,(i,j)}^\text{AB}}{\rho_{i-1,j}^\text{BB}+\rho_{i,j}^\text{BB}},\quad\quad 
V^\text{BA}_{\!z\,(i,j)}=\frac{2\Pi_{z\,(i,j)}^\text{BA}}{\rho_{i,j-1}^\text{BB}+\rho_{i,j}^\text{BB}},\quad\quad
V^\text{BB}_{\!\phi\,(i,j)}=\frac{J_{i,j}^\text{BB}}{\rho_{i,j}^\text{BB}\,R_{i}^\text{B}},
\end{align}
where the first mesh-type superscript letter 
refers to the radial grid direction while the second refers to the vertical grid direction.
The flow velocities 
at the other staggered mesh points are consequently linearly interpolated.
\subsubsection{Subroutine \textit{eos} - equation of state}
The subroutine contains a simple algorithm that calculates the pressure and consequently the sound speed from Eqs.~\eqref{statix5} or \eqref{statix6},
and \eqref{gengenenustateeq}. We do not need this subroutine in the current disk calculations 
(nor the energy equation itself - see the description in Sect.~\ref{operous}) since the temperature is 
there parameterized via the power law dependence (see Eq.~\eqref{temperature}). The subroutine does not update the conservative 
quantities - it is also not the true source step.
\subsubsection{Subroutine \textit{inertia} - contribution of inertial force}
The contribution of the inertial term, i.e.,~the first right-hand side term in radial cylindrical momentum equation~\eqref{cylmomrad} is explicitly 
programmed as the approximation of two adjacent B-mesh cells flow in the form 
\begin{align}\label{inertiumrad}
\Pi^{\text{AB},\,n+a/m_1}_{R\,(i,j)}=\Pi^{\text{AB},\,n}_{R\,(i,j)}+\Delta t\,\left(\rho_{i-1}^{\text{BB},\,n}+\rho_{i}^{\text{BB},\,n}\right)\,\,
\left(\frac{J_{i-1,j}^{\text{BB},\,n}}{R^\text{B}_{i-1}\rho_{i-1,j}^{\text{BB},\,n}}+\frac{J_{i,j}^{\text{BB},\,n}}{R^\text{B}_{i}\rho_{i,j}^{\text{BB},\,n}}\right)^2\,\Big(8R^\text{A}_{i}\Big)^{-1},
\end{align}
for the radial component of momentum where the meaning of $a$ is given in remarks to Eq.~\eqref{multikorektorix} and $m_1$ 
is the number of source steps (see Eq.~\eqref{oprourix}) that update the radial momentum (the density source update is not needed, while
$m_2$ refers to angular momentum, $m_3$ to vertical momentum and $m_4$ to energy).
We do not add the analogous term in azimuthal cylindrical momentum equation \eqref{cylmomphi} as a source term because it is 
implicitly involved in the advection part of the angular momentum $J$ cylindrical flow (cf.~Eqs.~\eqref{gennumerang},
\eqref{sourcicek3} and \eqref{phimcon} for the one-dimensional flow, see \citet{Normi}).
\subsubsection{Subroutine \textit{press} - contribution of pressure force}
The pressure force contribution in the directions $R,\,z$ is calculated from the corresponding pressure gradients using the schema and the notation of 
Eq.~\eqref{oprourix}. We explicitly write
\begin{align}\label{pressonumrad}
\Pi^{\text{AB},\,n+2a/m_1}_{R\,(i,j)}=\Pi^{\text{AB},\,n+a/m_1}_{R\,(i,j)}-\Delta t\,\frac{P_{i,j}^{\text{BB},\,n}-P_{i-1,j}^{\text{BB},\,n}}{R_{i}^\text{B}-R_{i-1}^\text{B}}.
\end{align}
We regard the pressure $P$ in disk calculations as the isothermal pressure given by Eq.~\eqref{statix6} (see Sect.~\ref{statak}).
The relation for the vertical component is quite analogous, however, the disk vertical hydrostatic balance makes it unnecessary, since the 
derivative of the vertical momentum component and the vertical gravity force component $\rho g_z$ in each timestep cancel.
\subsubsection{Subroutine \textit{gravity} - contribution of gravitational force}
The subroutine updates (3rd source update of $\Pi_R$) the momenta by adding the gravitational force components contribution, using the similar schema as Eq.~\eqref{pressonumrad}
where we replace the pressure gradient with the components of the gravity term $\rho\vec{g}$ given in Eqs.~\eqref{gravcylexpli} and \eqref{spheregravka}. 
\subsubsection{Subroutine \textit{nav} - including the Navier-Stokes shear viscosity}
This is the fundamental step that involves the shear stress tensor terms (see Eq.~\eqref{appendphimomcylinder1} for the disk) in the models.
In the axisymmetric models of the disk the vertically and azimuthally dependent derivatives as well as the vertical 
velocity (momentum) components (due to the assumed vertical hydrostatic equilibrium) are omitted.
The explicitly programmed axisymmetric shear viscosity relation (Eq.~\eqref{phimcon} with use of Eq.~\eqref{angfl}) thus becomes
($b$ is analogous to $a$ in Eq.~\eqref{pressonumrad}, however for $J$)
\begin{align}\label{shearixnumrad}
J^{\,\text{BB},\,n+b/m_2}_{i,j}&=J^{\,\text{BB},\,n}_{i,j}-\Delta t\,\Big\{\,\Big[\Big(\alpha_{i}^\text{B}+\alpha_{i+1}^\text{B}\Big)
\,\Big(a^{2\,\,\text{BB}}_{i,j}+a^{2\,\,\text{BB}}_{i+1,j}\Big)\,\Big(R^{\text{A}}_{i+1}\Big)^2
\Big(\rho_{i,j}^{\text{BB},\,n}+\rho_{i+1,j}^{\text{BB},\,n}\Big)\nonumber\\
&-\,\Big(\alpha_{i-1}^\text{B}+\alpha_{i}^\text{B}\Big)\,\Big(a^{2\,\,\text{BB}}_{i-1,j}+a^{2\,\,\text{BB}}_{i,j}\Big)
\,\Big(R^{\text{A}}_{i}\Big)^2\Big(\rho_{i-1,j}^{\text{BB},\,n}+\rho_{i,j}^{\text{BB},\,n}\Big)\Big]\,
\Big[8R^\text{B}_{i}\Big(R^\text{A}_{i+1}-R^\text{A}_{i}\Big)\Big]^{-1}\nonumber\\
&+\,\Big[\Big(\alpha_{i}^\text{B}+\alpha_{i+1}^\text{B}\Big)\,\Big(a^{2\,\,\text{BB}}_{i,j}+a^{2\,\,\text{BB}}_{i+1,j}\Big)
\,\Big(R^{\text{A}}_{i+1}\Big)^3\Big(\rho_{i,j}^{\text{BB},\,n}+\rho_{i+1,j}^{\text{BB},\,n}\Big)\,
\big(V^{\text{AB},\,n}_{\phi\,(i+1,j)}\big)^{-1}\nonumber\\
&-\,\Big(\alpha_{i-1}^\text{B}+\alpha_{i}^\text{B}\Big)\,\Big(a^{2\,\,\text{BB}}_{i-1,j}+a^{2\,\,\text{BB}}_{i,j}\Big)\,
\Big(R^{\text{A}}_{i}\Big)^3\Big(\rho_{i-1,j}^{\text{BB},\,n}+\rho_{i,j}^{\text{BB},\,n}\Big)\,\big(V^{\text{AB},\,n}_{\phi\,(i,j)}\big)^{-1}\Big]\nonumber\\
&\times\,\Big(V^{\text{AB},\,n}_{\phi\,(i+1,j)}-V^{\text{AB},\,n}_{\phi\,(i,j)}\Big)\,\Big[8R^\text{B}_{i}\Big(R^\text{A}_{i+1}-R^\text{A}_{i}\Big)^2\Big]^{-1}\nonumber\\
&+\,\alpha_{i}^\text{B}\,a^{2\,\,\text{BB}}_{i,j}\,\Big(R^\text{B}_{i}\Big)^2\rho_{i,j}^\text{BB}\,\Big(V^{\text{BB},\,n}_{\phi\,(i-1,j)}-2V^{\text{BB},\,n}_{\phi\,(i,j)}+
V^{\text{BB},\,n}_{\phi\,(i+1,j)}\Big)\,
\Big[V^{\text{BB},\,n}_{\phi\,(i,j)}\,\Big(R^{\text{A}}_{i+1}-R^{\text{A}}_{i}\Big)^2\Big]^{-1}\,\Big\},
\end{align}
where the $\alpha$ viscosity parameter (scalar quantity) varies merely in the radial direction while the squared sound speed $a^2$ may vary in the both directions $R,\,z$. 
\subsubsection{Subroutine \textit{artvis} - artificial viscosity}
The subroutine solves the equation \eqref{artivis}, the physical effects of the artificial viscosity and its role is described in Sect.~\ref{operous}. 
The scalar quantity $Q_k$ contributes to the corresponding momentum $k$-component advection in the corresponding direction,
its contribution to the radial advection of the radial component of momentum for example is
\begin{align}\label{artiviskorektorix}
\Pi_{R\,(i,j)}^{\text{AB},\,n+5a/m_1}=\Pi_{R\,(i,j)}^{\text{AB},\,n+4a/m_1}-\Delta t\frac{Q^{\text{BB},\,n}_{R\,(i,j)}-Q^{\text{BB},\,n}_{R\,(i-1,j)}}
{R^{\text{B}}_{i}-R^{\text{B}}_{i-1}},
\end{align}
the mesh position of the velocity vectors in Eq.~\eqref{artivis} algorithm  must therefore lead to the resulting BB-mesh position of the scalar artificial viscosity $Q$.
In the energy equation its contribution is
\begin{align}\label{artivisenergokorektorix}
E^{\text{BB},\,n+d/m_4}_{i,j}=E^{\text{BB},\,n}_{i,j}-\Delta t\,Q^{\text{BB},\,n}_{R\,(i,j)}\frac{V^{\text{AB},\,n}_{R\,(i+1,j)}-V^{\text{AB},\,n}_{R\,(i,j)}}
{R^{\text{A}}_{i+1}-R^{\text{A}}_{i}}-\Delta t\,Q^{\text{BB},\,n}_{z\,(i,j)}\frac{V^{\text{BA},\,n}_{z\,(i,j+1)}-V^{\text{BA},\,n}_{z\,(i,j)}}
{z^{\text{A}}_{j+1}-z^{\text{A}}_{j}}, 
\end{align}
where $d$ is analogous to $a$ in Eq.~\eqref{pressonumrad}, however for the energy $E$.
\subsubsection{Subroutine \textit{dissip} - including the dissipation function into energy equation}
The inclusion of the dissipation function $\Psi$ (Eq.~\eqref{cylisip}) into the energy equation is currently not used in the outflowing disk models due to the temperature 
parameterization via the power law (see Eq.~\eqref{temperature}). Its efficiency is however properly tested in the introductory two-dimensional models 
(see Sect.~\ref{testingleerix}).
\subsection{Advection steps \textmd{(transport steps)}}\label{advecticek}
The subroutines included in the block of advection steps use the algorithm described in Eqs.~\eqref{vanlicek}-\eqref{multikorektorix} in Sect.~\ref{vanleerix}.
The sequence of the subroutines is grouped into two coordinate directional streams that enter the computation respectively. 
\subsubsection{Subroutines \textit{radden, theden} - radial and vertical density flow}
Substituting the density $\rho$ (as well as any other scalar quantity) for the general variable $f$ into the referred equations \eqref{vanlicek}-\eqref{multikorektorix}, 
we obtain the advection step for the density of the matter.
In the algorithm for the radial density flow (subroutine radden) we substitute for the general surface $S^{\!\text{A}}$ 
into Eq.~\eqref{multikorektorix} the surface $S_{\!R}^{\text{AB}}$ from Eq.~\eqref{cyl20},
in the algorithm for the vertical density flow in the subroutine theden (where the abbreviation refers to spherical theta as a generalized third direction)
we substitute $S_{\!z}^{\text{BA}}$ from the same equation \eqref{cyl20} while in both subroutines we use the volume element $\Omega^{\text{BB}}$ from Eq.~\eqref{cyl20}.
\subsubsection{Subroutines \textit{radmom, phimom, themom (1,3)} - radial and vertical momentum advection}
The six subroutines contain the algorithms for radial and vertical momentum advection of all the momentum components $R,\,\phi,\,z$ 
where the both coordinate directions of advection are denoted as 1 (radial) and 3 (vertical). The equations \eqref{vanlicek}-\eqref{multikorektorix}
analogously apply for the advection of vector quantities, there is however necessary to involve the corresponding mesh points for the particular terms of
particular components advection, regarding the two possible (positive or negative) orientations of the flow. For example, the equation \eqref{multikorektorix} for the 
\textit{radial} advection of the 
\textit{radial} momentum component becomes the exact form (where $I$ is the interface interpolant described in Eq.~\eqref{predator} 
and $S$ is the control surface described in Eq.~\eqref{multikorektorix}),
\begin{align}\label{multimomentradkorektorix}
\Pi_{R\,(i,j)}^{\text{AB},\,n+1}=\Pi_{R\,(i,j)}^{\text{AB},\,n}-\frac{\Delta t}{\Omega^{\text{AB}}_{i,j}}
\left(I_{i,j}^{\,\text{BB},\,n+a}V^{\text{BB},\,n+a}_{R\,(i,j)}S^{\text{BB}}_{\!R\,(i,j)}-
I_{i-1,j}^{\,\text{BB},\,n+a}V^{\text{BB},\,n+a}_{R\,(i-1,j)}S^{\text{BB}}_{\!R\,(i-1,j)}\right),
\end{align}
while for example the equation \eqref{multikorektorix} for the 
\textit{radial} advection of the 
\textit{vertical} momentum component becomes the exact form (where $c$ is analogous to $a$ in Eqs.~\eqref{pressonumrad} and \eqref{multimomentradkorektorix}, for $\Pi_{z}$)
\begin{align}\label{multimomentvertkorektorix}
\Pi_{z\,(i,j)}^{\text{BA},\,n+1}=\Pi_{z\,(i,j)}^{\text{BA},\,n}-\frac{\Delta t}{\Omega^{\text{BA}}_{i,j}}
\left(I_{i,j}^{\,\text{AA},\,n+c}V^{\text{AA},\,n+c}_{R\,(i,j)}S^{\text{AA}}_{\!z\,(i,j)}-
I_{i-1,j}^{\,\text{AA},\,n+c}V^{\text{AA},\,n+c}_{R\,(i-1,j)}S^{\text{AA}}_{\!z\,(i-1,j)}\right),
\end{align}
and analogously for the radial and vertical advection of the remaining momentum components (noting that the 
$\phi$-momentum is advected as a scalar quantity).
\subsubsection{Subroutines \textit{radenpress, theenpress, radenergy, theenergy} - advection of the enthalpy flow}
These subroutines provide the calculation of the energy equation \eqref{gengenenumerenergy}. Subroutine radenpress calculates 
the radial advection of the pressure flow $\mathcal{P}\vec{V}$ (where we currently include only the scalar pressure $P$ within the pressure tensor, however, it is 
potentially easy to employ also the nondiagonal viscous terms), subroutine theenpress calculates its vertical advection. The subroutines radenergy and theenergy
calculate the radial and vertical advection of the energy flow (it is likely possible to combine the corresponding component subroutines into one 
subroutine for the enthalpy $E+\mathcal{P}$ directional advection (cf.~Eq.~\eqref{gengenenumerenergy}). The structure of the subroutines is identical with
the structure of the subroutines radden, theden, since we advect the scalar quantities (they are however not currently used for the outflowing disk calculations).
\subsection{Adjusting steps}\label{adjustik}
The subroutines provide the final adjustment of the computational algorithm within each timestep. %We include there also the 
%the artificial viscosity contribution \eqref{artivis} since it does not result from explicitly expressed physical force effects (it is thus not the true source term).
\subsubsection{Subroutine \textit{boundary} - implementation of the boundary conditions}
Boundary conditions are specified for all conservative hydrodynamic quantities (scalars and vector components) $\rho,\,\Pi_R,\,\Pi_z,\,J$ and $E$, 
in the magnetohydrodynamic code extension (Appendix~\ref{dvojdimmagnetohydrous}) the components of magnetic field induction vector $\vec{B}$ are added.
In the code we may currently employ 4 types of boundary conditions in general (that may also differ for particular quantities). Using the notation for the grid 
points indices from Fig.~\ref{stagis} and denoting $f$ the arbitrary (component of) conservative quantity, in the arbitrary coordinate direction $i$ 
(with $ni$ grid points within the computational domain, see Fig.~\ref{stagis}) we set
the following types of the boundary conditions (b. c.):
\begin{align}\label{borderixa}
&f^{\text{B}}_{2,j}=f^{\text{B}}_{2,j},&&f^{\text{A}}_{3,j}=f^{\text{A}}_{3,j},&&
f^{\text{A or B}}_{ni+3,j}=f^{\text{A or B}}_{ni+3,j}&& &&\text{fixed b. c.},\\
&f^{\text{B}}_{2,j}=f^{\text{B}}_{3,j},&&f^{\text{A}}_{3,j}=-f^{\text{A}}_{4,j},&&\label{borderixb}
f^{\text{A or B}}_{ni+3,j}=\mp f^{\text{A or B}}_{ni+2,j}&& &&\text{reflecting (solid wall) b. c.},\\
&f^{\text{B}}_{2,j}=f^{\text{B}}_{ni+2,j},&&f^{\text{A}}_{3,j}=f^{\text{A}}_{ni+3,j},&&\label{borderixc}
f^{\text{A or B}}_{ni+3,j}=f^{\text{A or B}}_{3,j}&& &&\text{periodic b. c.},\\
&f^{\text{B}}_{2,j}=f^{\text{B}}_{3,j},&&f^{\text{A}}_{3,j}=f^{\text{A}}_{4,j},&&\label{borderixd}
f^{\text{A or B}}_{ni+3,j}=f^{\text{A or B}}_{ni+2,j}&& &&\text{outflow (free) b. c.},
\end{align}
where the conditions $f^\text{A}$ refer to the vector components advected in the corresponding direction while the conditions $f^\text{B}$ refer
to the scalar advection and the vector components advection directed otherwise (while the mesh type that corresponds to the second direction is arbitrary). Since the computations in the source and advection steps 
run only within the computational domain (marked in Fig.~\ref{stagis}), the fixed boundary conditions in each timestep re-read the values given by the initial function.
The boundary conditions for the ghost zones $f^{\text{B}}_{1,j}$, $f^{\text{B}}_{ni+4,j}$ 
(see the description in Fig.~\ref{stagis}) are specified similarly: in the fixed or outflow (free or extrapolated) case the 
grid point indices on the left-hand sides of Eqs.~\eqref{borderixa} or \eqref{borderixd} are shifted to the edge values for the given grid type,
in the reflecting case all the grid point indices in Eqs.~\eqref{borderixb} are shifted symmetrically while 
in the periodic case all the grid point indices in Eqs.~\eqref{borderixc} are shifted simultaneously.

In case of the stellar decretion disk models (see Sect.~\ref{timemod}) we fix the inner (left) boundary condition in the radial direction for the density $\rho$ or $\Sigma$ (which is
in fact irrelevant because the density values do not affect the calculations) as well as for the angular momentum $J$ (we assume the constant 
Keplerian rotation velocity at the constant stellar equatorial radius) and the vertical momentum component $\Pi_z$ (zero vertical momentum flow due to the assumed 
vertical hydrostatic equilibrium)
while we set free inner (left) boundary condition in the radial direction for the radial momentum flow $\Pi_R$.
We set the outer (right) boundary conditions in the radial direction as free for all the quantities.
The lower and upper (left and right) boundary conditions in the vertical direction in two-dimensional models we set as outflow (or alternatively periodic, which does not significantly
affect the current results).
\subsubsection{Subroutine \textit{output} - printing calculated data to output file}
The subroutine controls the sending of the selected data (in dependence of the spatial coordinates) 
calculated in the selected timesteps for plotting into the data file (or files). 
The number of the plotted timesteps is given in the problem parameters subroutine, usually it is up to several hundreds.
The subroutine also controls some additional functions, e.g., checks the Courant stability theorem violation 
(see the following subroutine clock for the description)
or continuously displays important operational data (e.g., the number of plotted timestep)
or the estimated total central processing unit time of the calculation (CPU time), etc.
\subsubsection{Subroutine \textit{clock} - control of the computational stability of the code}
From the Fourier decomposition of the numerical perturbations known as the 
von Neumann stability analysis \citep{kraken,Hirsch2} results that an arbitrary time-dependent computational code is numerically stable if 
the so-called \textit{Courant-Friedrichs-Lewy} number (or shortly the \textit{Courant} number denoted as cfl) satisfies the inequality
\begin{align}\label{kurna}
\text{cfl}=\frac{V\Delta t}{\Delta x}\leq 1\qquad\quad(\text{Courant stability condition}),
\end{align}
where $V$ is the velocity of the hydrodynamic flow and $\Delta x$ corresponds to \eqref{delfinci}.
The subroutine primarily adjusts the time interval $\Delta t$ via the (directionally splitted) algorithm \citep{Stony1a}
\begin{align}\label{kurna2}
\Delta t_{\,R}=\frac{R_{i+1}-R_{i}}{\text{max}\big(|V_{R\,(i,j)}|,\,\text{tiny}\big)},\quad\quad\quad
\Delta t_{\,z}=\frac{z_{j+1}-z_{j}}{\text{max}\big(|V_{z\,(i,j)}|,\,\text{tiny}\big)}
\end{align}
where the number $\text{\textit{tiny}}$ denotes the very small value (of the order of $10^{-15}$) that  
prevents computational singularity. The time interval $\Delta t$ has yet to be modified taking into account the sound speed $a$ (plus the Alfv\'en speed $V_{\!A}$
in magnetohydrodynamic simulations) leading to the relation \citep{Stony1a,Stony1b}
\begin{align}\label{kurna3}
\Delta t_{\,a}=\frac{\text{min}\left(R_{i+1}-R_{i},\,z_{j+1}-z_{j}\right)}{\text{max}\left(\!\!\sqrt{a^2_{i,j}+V^2_{\!A\,R\,(i,j)}},\,
\sqrt{a^2_{i,j}+V^2_{\!A\,\phi\,(i,j)}},\,\sqrt{a^2_{i,j}+V^2_{\!A\,z\,(i,j)}},\,\text{tiny}\right)},
\end{align}
where the Alfv\'en speed $V_{\!A\,R}=B_R/\!\!\!\sqrt{\mu\rho}$, $V_{\!A\,\phi}=B_\phi/\!\!\!\sqrt{\mu\rho}$, $V_{\!A\,z}=B_z/\!\!\!\sqrt{\mu\rho}$ (see Sect.~\ref{sheainst}).
The contribution of the numerical viscosity to the time interval modification we denote as $\Delta t_{\,\text{vis}}$ whose resulting value 
minimizes its directional splitting explicitly written as
\begin{align}\label{kurna4}
\Delta t_{\,\text{vis}}=\text{min}\left[\frac{R_{i+1}-R_{i}}{\text{max}\left(4\text{C}_2|V_{R\,(i+1,j)}-V_{R\,(i,j)}|,\,\text{tiny}\right)},\,
\frac{z_{j+1}-z_{j}}{\text{max}\left(4\text{C}_2|V_{z\,(i,j+1)}-V_{z\,(i,j)}|,\,\text{tiny}\right)}\right], 
\end{align}
where the factor $4$ in the denominator arises from the change of the mathematical nature of the momentum equation to a diffusion equation due to the inclusion of the 
artificial viscosity. The stable explicit diffusion schemes apply the constraint $\Delta t\leq (\Delta x)^2/4\nu$ where $\nu$ is the coefficient of kinematic viscosity
described in Eq.~\eqref{kincl} \citep[see,][for details]{Stony1a}. The timestep is calculated using the relation \citep{Roache,Normi,Stony1a} 
\begin{align}\label{kurna5}
\Delta t=\text{cfl}\,\left[\big(\Delta t_{\,R}\big)^{-2}+
\big(\Delta t_{\,z}\big)^{-2}+\big(\Delta t_{\,a}\big)^{-2}+\big(\Delta t_{\,\text{vis}}\big)^{-2}\right]^{-1/2},
\end{align}
the algorithm thus checks the Courant stability theorem by controlling the timestep according to varying physical conditions.
\section{Two-dimensional magnetohydrodynamic (MHD) extension of the hydrodynamic code}\label{dvojdimmagnetohydrous}
In this section we briefly describe the MHD code structure. Although we do not currently use the MHD extension of the code in the disk models introduced 
within this thesis,
we will however employ it within the subsequent calculations of the magnetorotational disk instabilities or other problems connected with 
the stellar magnetic fields. Compared to hydrodynamic equations,
the MHD numerical solution is associated with two fundamental difficulties \citep{Stony1b}. The first one is the permanent divergence constraint resulting from 
Maxwell's equations ($\vec{\nabla}\cdot\vec{B}=0$) whose numerical solution cannot be calculated separately like in the hydrodynamic equation of state and must 
be directly incorporated into the MHD equations algorithm using the so-called constrained transport (CT) method \citep{evohol}: employing the second 
Maxwell equation \eqref{mhd7a} written in the integral form (Faraday's law of induction) and using the Stokes' theorem, we write
\begin{align}\label{kurna6}
\frac{\partial\Phi_S}{\partial t}=-\int_S(\vec{\nabla}\times\vec{\mathcal{E}})\cdot\vec{n}\,\text{d}S=-\int_{\partial S}\vec{\mathcal{E}}\cdot\text{d}\vec{\ell},
\end{align}
where $\Phi_S$ is the magnetic flux through a surface $S$ bounded by a closed curve $\partial S$ and 
$\vec{\mathcal{E}}$ is the electromotive force (EMF) \citep{evohol,Stony1b} defined with use of Eq.~\eqref{mhd1b} as $\vec{\mathcal{E}}=-\vec{V}\times\vec{B}$.
We construct the finite-difference approximation of (directionally splitted) Eq.~\eqref{kurna6} for the two surfaces demonstrated in 
Fig.~\ref{emfik}, 
where the two oppositely directed components of the EMF with the same indices on the top and bottom edges cancel, 
in the following way \citep{Stony1b} (where $\Delta 2$ is the vertical cell's edge length):
\begin{align}\label{kurna7}
\frac{\Phi^{n+1}_{1\,(i,j)}-\Phi^{n}_{1\,(i,j)}}{\Delta t}=\big[\mathcal{E}_{2\,(i,j+1)}-\mathcal{E}_{2\,(i,j)}\big]\,\Delta 2,\quad\quad
\frac{\Phi^{n+1}_{3\,(i,j)}-\Phi^{n}_{3\,(i,j)}}{\Delta t}=\big[\mathcal{E}_{2\,(i,j)}-\mathcal{E}_{2\,(i+1,j)}\big]\,\Delta 2.%\label{kurna8}
\end{align}
We note that Eq.~\eqref{kurna7} is not the part of the code, merely the demonstration of the method.
Assuming that the EMF components are constant along one grid cell's edges as well as the magnetic induction components are constant within one grid cell, we 
may in Cartesian system write (cf.~Eqs.~\eqref{cyl20} and \eqref{kul20} for curvilinear coordinate systems) $\Phi_1=B_1\Delta 2\,\Delta 3$ and 
$\Phi_3=B_3\Delta 1\,\Delta 2$. 
We use this sequence of indices in order to achieve 
the correspondence
with the cylindrical system $R,\,\phi,\,z$ where the first and the third components are poloidal while the second component is toroidal.
When we express the total flux through whole grid cell's volume, the contributions of the EMFs 
in combined Eqs.~\eqref{kurna7} along each of the grid cell's edge cancel (oppositely stream in neighboring surfaces).
The divergence-free constraint is thus guaranteed (applying the Gauss's theorem) by the expression $\sum_S\Phi\cdot S=0$
(see \citet{evohol,Stony1b} for detailed explanation).

The second fundamental difficulty \citep{Stony1b} results from the existence of the \textit{incompressible} (since the property of the waves is 
the absence of any fluctuations in fluid density and pressure) transverse Alfv\'en waves \citep[][cf.~also Eq.~\eqref{mhd35}]{Bittyk}
that may cause a large dispersion error when applying the same 
hydrodynamic evolution schemes (Appendix~\ref{advecticek}). There is no analog to these waves in hydrodynamic equations. 
To eliminate this, \citet{Stony1b} proposed the method of characteristics
(MOC) that is applied to solve the predictor (half-time) step of the field components (see subroutines \textit{emfr}, \textit{emft} for
basic numerical schemes of the evolution 
\begin{figure}[t]
\begin{center}
\includegraphics[width=7cm]{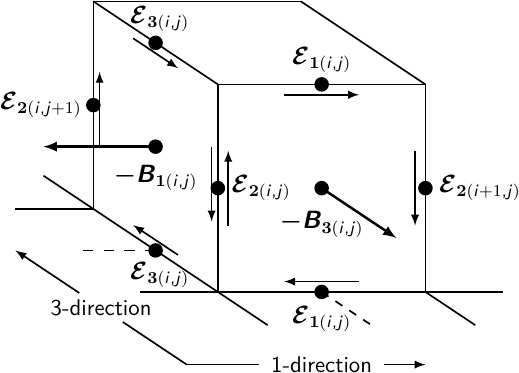}
\caption{The illustration of the loops of electromotive force (EMF) denoted as $\mathcal{E}$ (cf.~Eq.~\eqref{kurna7}). 
The arrows show the direction of integration of EMF loops around the closed
contours of the schematic cube (see Eq.~\eqref{kurna7}) in order to evolve the 
components of the magnetic field (which are shown here as negatively oriented).
Adapted from \citet{Stony1b}.}
\label{emfik}
\end{center}
\end{figure}
of the poloidal field components and subroutines \textit{evmagpr}, \textit{evmagpt} for the evolution of the toroidal field component):
the determining simplification of this method is that only the (noncompressible) Alfv\'en wave characteristics are used for the computation of the predictor
step for the evolution of momenta and magnetic induction components. 
\begin{figure}[t]
\begin{center}
\includegraphics[width=7cm]{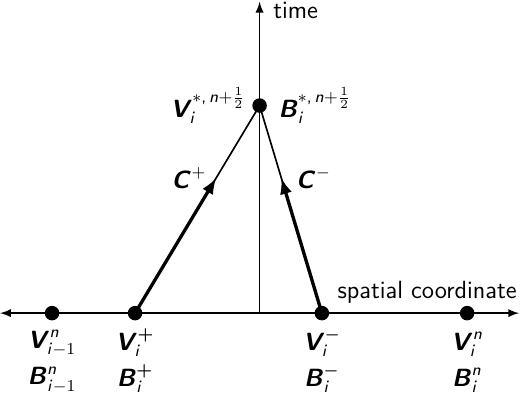}
\caption{The one-dimensional schematic graph of the spacetime positions of the backward characteristics ${C}^-$ and the forward characteristics ${C}^+$
(see Appendix~\ref{dvojdimmagnetohydrous}). We use these characteristics for the calculation of half-timestep evolution of the components (marked with * )
of the velocity vector $\vec{V}$ 
and magnetic field vector $\vec{B}$.
By using the characteristic speed $(V_i\mp V_{A,i})$ (see Eq.~\eqref{charakticek}) and employing the analog of Eq.~\eqref{predator} we interpolate the quantities 
$V_i^-,\,V_i^+\,B_i^-,\,B_i^+$ on the footpoints of both characteristics $C^-,\,C^+$, the values $V_i^{*,\,n+1/2}$ and $B_i^{*,\,n+1/2}$
are then calculated combining Eqs.~\eqref{charakticek1} and \eqref{charakticek1a}.
Adapted from \citet{Stony1b}.}
\label{charcik}
\end{center}
\end{figure}
This is allowed by the fact that other (compressive) linear MHD wave modes 
are thermodynamically coupled with hydrodynamics, the hydrodynamic algorithms thus evolve these compressive MHD waves accurately \citep[similarly 
to compressive sound waves, see,][]{Stony1b}.

We consider the one-dimensional (i.e.,~only $R$-dependent or only $z$-dependent) incompressible flow 
($\vec{\nabla}\cdot\vec{V}=0$), which implies $\partial V_R/\partial R\equiv 0$ or $\partial V_z/\partial z\equiv 0$, while the restrictions $\partial B_R/\partial R\equiv 0$ and
$\partial B_z/\partial z\equiv 0$ result from the divergence-free constraint. 
The basic MHD Eqs.~\eqref{mhd7} and \eqref{mhd8}, respectively (see also Eqs.~\eqref{mhd13}, \eqref{mhd15}, \eqref{mhd21} and \eqref{mhd23}), simplify to symmetric pair of equations,
\begin{align}\label{simplicek}
\frac{\partial\vec{V}}{\partial t}=\frac{B_R}{\mu_0\rho}\frac{\partial\vec{B}}{\partial R}-V_R\frac{\partial\vec{V}}{\partial R},\quad\quad\quad\quad\quad
\frac{\partial\vec{B}}{\partial t}=B_R\frac{\partial\vec{V}}{\partial R}-V_R\frac{\partial\vec{B}}{\partial R}.
\end{align}
By multiplying the second equation by $(\mu_0\rho)^{-1/2}$ and then summing and subtracting them, we obtain for each 
evolved grid component the characteristic form (where the Lam\'e coefficients $h_z\,(\text{or}\,\,h_3)$ 
for the cylindrical and spherical coordinate systems are defined in Appendix~\ref{diffcylinder}),
\begin{align}\label{charakticek}
\frac{\partial V_R}{\partial t}+\left(V_z\mp\frac{B_z}{\sqrt{\mu_0\rho}}\right)\frac{1}{h_z}\frac{\partial V_R}{\partial z}\pm
\frac{1}{\sqrt{\mu_0\rho}}\left[\frac{\partial B_R}{\partial t}+\left(V_z\mp
\frac{B_z}{\sqrt{\mu_0\rho}}\right)\frac{1}{h_z}\frac{\partial B_R}{\partial z}\right]=0,\\
\frac{\partial V_z}{\partial t}+\left(V_R\mp\frac{B_R}{\sqrt{\mu_0\rho}}\right)\frac{\partial V_z}{\partial R}\pm
\frac{1}{\sqrt{\mu_0\rho}}\left[\frac{\partial B_z}{\partial t}+\left(V_R\mp\label{charakticeka}
\frac{B_R}{\sqrt{\mu_0\rho}}\right)\frac{\partial B_z}{\partial R}\right]=0,
\end{align}
where the terms $(V_z\mp V_{A,z})$ and $(V_R\mp V_{A,R})$ in brackets are the characteristic speeds used for 
the computation of the upwind values (in this point we may employ the analog of Eq.~\eqref{predator}) 
$V_i^-\!,\,B_i^-\,\text{and}\,V_i^+\!,\,B_i^+$ on the footpoints of both characteristics $C^-\,\text{and}\,\,C^+$ (depicted in Fig.~\ref{charcik}).
The two equations \eqref{charakticek} and \eqref{charakticeka} may be expressed in the Lagrangean form
(where, due to the incompressibility, the equations describe only the propagation of information in a moving fluid via Alfv\'en waves and do not admit any compressive wave modes \citep{Stony1b}),
\begin{align}\label{charakterobecnicek}
\frac{\text{d}{V}_R}{\text{d}t}\pm\frac{1}{\sqrt{\mu_0\rho}}\frac{\text{d}{B}_R}{\text{d}t}=0,\quad\quad\quad
\frac{\text{d}{V}_z}{\text{d}t}\pm\frac{1}{\sqrt{\mu_0\rho}}\frac{\text{d}{B}_z}{\text{d}t}=0.
\end{align}
The half-time step values $V_i^{*,\,n+1/2}$ and $B_i^{*,\,n+1/2}$ (the superscript * hence denotes the quantities calculated using MOC) 
result as the direct solution of the system of differenced characteristic equations along $C^-,\,C^+$, respectively, 
\begin{align}\label{charakticek1}
\big(V_i^{*,\,n+1/2}-V_i^{-,\,n}\big)+\big(B_i^{*,\,n+1/2}-B_i^{-,\,n}\big)\big(\mu_0\rho^{-,\,n}_i\big)^{-1/2}=0,\\
\big(V_i^{*,\,n+1/2}-V_i^{+,\,n}\big)-\big(B_i^{*,\,n+1/2}-B_i^{+,\,n}\big)\big(\mu_0\rho^{+,\,n}_i\big)^{-1/2}=0,\label{charakticek1a}
\end{align}
where the interpolated densities are $\rho^-=\rho_i$ and $\rho^+=\rho_{i-1}$.
Once the predictor step calculation is complete for all quantities at every cell corner, we calculate the EMF \eqref{mhd1b} via the reduced relation \eqref{magbik1}
in the subroutine \textit{magb} (in the following description of the subroutines as well as the flow chart of the code in Fig.~\ref{emfik} 
we apply mainly the schema introduced in \citet{Stony1b}).
\subsection{MHD source steps extension}\label{magorissourcis}
%The structure of the subroutines as well as the flow chart of the code applies the schema introduced in \citet{Stony1b}.
\subsubsection{Subroutine \textit{bin} - positions of the magnetic induction components on the grid}
Similarly to setting the hydrodynamic velocity grid components in subroutine speed we set the location of the magnetic field components 
for all directions of advection by interpolation of the three basic components $B_R^{\text{AB}}$, $B_\phi^{\text{BB}}$, $B_z^{\text{BA}}$ given in the 
subroutine initial. 
The magnetic field components are located at the same points as the corresponding velocity components, 
the subscripts $R,\,\phi,\,z$ do not necessarily denote cylindrical coordinates and may refer to generalized directions $1,\,2,\,3$.
\subsubsection{Subroutine \textit{lforce} - adding the Lorentz force terms}
This subroutine updates the radial and vertical momenta by adding the part of the Lorentz force terms from the momentum equation Eqs.~\eqref{mhd7} 
explicitly expressed in Eqs.~\eqref{mhd13} and \eqref{mhd15} (omitting the $\phi$-derivatives). Since the terms containing $B_Z\,(\partial B_R/\partial z)$ in Eq.~\eqref{mhd13}
and $B_R\,(\partial B_z/\partial R)$ in Eq.~\eqref{mhd15} describe the propagation of the Alfv\'en waves (within Eqs.~\eqref{charakticek} and \eqref{charakticeka}),
we split the updating into two parts. In this subroutine we add merely the gradients of magnetic pressure, while we add the Alfv\'en wave source 
terms within the 
subroutines \textit{lforr}, \textit{lfort}, after the MOC and CT evolution is complete.
The explicit form of the update of these momentum components is
\begin{align}\label{kramous1}
\Pi^{\text{AB},\,n+4a/m_1}_{R\,(i,j)}\!&=\!\Pi^{\text{AB},\,n+3a/m_1}_{R\,(i,j)}\!\!-\!\frac{\Delta t}{\mu_0}\!
\left[\!B_{z\,(i,j)}^{\text{AB},\,n}\!\frac{h_z^{\text{B}}B_{z\,(i,j)}^{\text{BB},\,n}\!-\!h^{-}_zB_{z\,(i-1,j)}^{\text{BB},\,n}}{h_z^{\text{A}}(R^{\text{B}}_{i}-R^{\text{B}}_{i-1})}\!+\!
B_{\phi\,(i,j)}^{\text{AB},\,n}\!\frac{h_\phi^{\text{B}}B_{\phi\,(i,j)}^{\text{BB},\,n}\!-\!h^{-}_\phi B_{\phi\,(i-1,j)}^{\text{BB},\,n}}{h_\phi^{\text{A}}(R^{\text{B}}_{i}-R^{\text{B}}_{i-1})}\!\right]\!,\\
\Pi^{\text{BA},\,n+4c/m_3}_{z\,(i,j)}\!&=\!\Pi^{\text{BA},\,n+3c/m_3}_{z\,(i,j)}\!\!-\!\frac{\Delta t}{\mu_0}\!\left[\!B_{R\,(i,j)}^{\text{BA},\,n}\!
\frac{B_{R\,(i,j)}^{\text{BB},\,n}\!-\!B_{R\,(i,j-1)}^{\text{BB},\,n}}{h_z^{\text{B}}(z^{\text{B}}_{j}-z^{\text{B}}_{j-1})}\!+\!
B_{\phi\,(i,j)}^{\text{BA},\,n}\!\frac{h_\phi^{\text{B}}B_{\phi\,(i,j)}^{\text{BB},\,n}\!-\!h^-_\phi B_{\phi\,(i,j-1)}^{\text{BB},\,n}}{h_\phi^{\text{A}}h_z^{\text{B}}(z^{\text{B}}_{j}-z^{\text{B}}_{j-1})}\!\right],
\label{kramous1a}
\end{align}
where the minus superscript at the Lam\'e coefficients $h_\phi,\,h_z\,(\text{or}\,\,h_\phi,\,h_\theta\,\,\text{in spherical system})$ 
denotes the $i-1$ or $j-1$ grid location, according to the grid position of the coupled hydrodynamic quantity. For example the $j$-independent coefficient 
$h_\phi^-$ in cylindrical coordinates is $R_{i-1}$ or $R_{i}$ while the spherical $j$-dependent coefficient $h_\phi^-$ is $r\,\text{sin}\,\theta_{\,i-1,j}$ 
or $r\,\text{sin}\,\theta_{\,i,j-1}$. Coefficient $h_z^-$ is $1$ in cylindrical coordinates and the spherical $j$-independent coefficient $h_z^-$ is $r_{i-1}$ or $r_{i}$.
\subsubsection{Subroutines \textit{emfr}1, \textit{emfr}2, \textit{emft}1, \textit{emft}2 - partial updating of the EMF calculations}
The subroutines solve the coupled equations \eqref{charakticek1} splitted into two parts which correspond to the two 
characteristics $C^-\,\text{and}\,\,C^+$. Using the characteristic speed $(V_i\mp V_{A,i})$ (employing e.g., the analog of Eq.~\eqref{predator}) we compute the 
upwind interpolations of the velocity and magnetic field components, 
$V_i^-,\,V_i^+\,B_i^-,\,B_i^+$ (see Eqs.~\eqref{charakticek}). Since the solution for the radial and vertical component are of the same form,
we perform each partial update (for each position on the nodes of the computational grids A, B) within the single relation: 
\begin{align}\label{emfr1}
V^{-,\,n+1/2}_{R,\,z}=\frac{\sqrt{\rho^{-,\,n}}V^{-,\,n}_{R,\,z}+B^{-,\,n}_{R,\,z}\big{/}\!\!\!\sqrt{\mu_0}}{\sqrt{\rho^{-,\,n}}+\sqrt{\rho^{+,\,n}}},\quad\quad\quad
B^{-,\,n+1/2}_{R,\,z}=\frac{\sqrt{\rho^{+,\,n}}B^{-,\,n}_{R,\,z}+\sqrt{\mu_0\rho^{-,\,n}\rho^{+,\,n}}V^{-,\,n}_{R,\,z}}{\sqrt{\rho^{-,\,n}}+\sqrt{\rho^{+,\,n}}},\\
\label{emfr2}
V^{+,\,n+1/2}_{R,\,z}=\frac{\sqrt{\rho^{+,\,n}}V^{+,\,n}_{R,\,z}-B^{+,\,n}_{R,\,z}\big{/}\!\!\!\sqrt{\mu_0}}{\sqrt{\rho^{-,\,n}}+\sqrt{\rho^{+,\,n}}},\quad\quad\quad
B^{+,\,n+1/2}_{R,\,z}=\frac{\sqrt{\rho^{-,\,n}}B^{+,\,n}_{R,\,z}-\sqrt{\mu_0\rho^{-,\,n}\rho^{+,\,n}}V^{+,\,n}_{R,\,z}}{\sqrt{\rho^{-,\,n}}+\sqrt{\rho^{+,\,n}}}.
\end{align}
To obtain the complete solution of Eqs.~\eqref{charakticek1} we simply sum the backward 
and forward updates for the corresponding components of the velocity and magnetic induction (obtaining the right-hand side terms in Eq.~\eqref{magbik1}).
\subsubsection{Subroutine \textit{magb} - updating of the poloidal components of magnetic induction}
We may explicitly write the calculation of the electromotive force from Eq.~\eqref{mhd1b} in the form
\begin{align}\label{magbik1}
\mathcal{E}^{\,n+1/2}_{\phi\,(i,j)}=V^{*,\,n+1/2}_{R\,(i,j)}B^{*,\,n+1/2}_{z\,(i,j)}-V^{*,\,n+1/2}_{z\,(i,j)}B^{*,\,n+1/2}_{R\,(i,j)},
\end{align}
for each A, B node separately. We can 
calculate the first (radial) and the third (vertical) components of the magnetic induction (using Eqs.~\eqref{mhd1b}, \eqref{mhd8} 
and Eqs.~\eqref{rotcylvector},
\eqref{rotspherevector}) as 
\begin{align}\label{magbik2}
B^{\text{AB},\,n+1}_{R\,(i,j)}&=B^{\text{AB},\,n}_{R\,(i,j)}+\frac{\Delta t}{h_\phi^{\text{B}}h_z^{\text{A}}}
\frac{h^+_\phi\mathcal{E}^{\text{AA},\,n+1/2}_{\phi\,(i,j+1)}-h_\phi^{\text{A}}\mathcal{E}^{\text{AA},\,n+1/2}_{\phi\,(i,j)}}{z_{j+1}^\text{A}-z_{j}^\text{A}},\\
B^{\text{BA},\,n+1}_{z\,(i,j)}&=B^{\text{BA},\,n}_{z\,(i,j)}-\frac{\Delta t}{h_\phi^{\text{B}}}\label{magbik3}
\frac{h^+_\phi\mathcal{E}^{\text{AA},\,n+1/2}_{\phi\,(i+1,j)}-h_\phi^{\text{A}}\mathcal{E}^{\text{AA},\,n+1/2}_{\phi\,(i,j)}}{R_{i+1}^\text{A}-R_{i}^\text{A}},
\end{align}
where the superscript $+$ at the Lam\'e coefficients $h_\phi\,(\text{or}\,\,h_2\,\,\text{in general})$ 
denotes the $i+1$ or $j+1$ grid point location, according to the grid position of the coupled quantity.
\subsubsection{Subroutines \textit{lforr}1, \textit{lforr}2, \textit{lfort}1, \textit{lfort}2 - updating the Alfv\'en wave part of the Lorentz force}
The subroutines complete the evolution of the poloidal Lorentz force terms by including 
the Alfv\'en wave source terms that were not updated (due to the fact that the CT method used for the calculation of the EMFs is not an 
operator splitted schema - see \citet{Stony1b})  within the EMF calculation in the subroutines \textit{emfr} 1,2 and \textit{emft} 1,2, i.e., 
the terms containing expressions $B_Z\,(\partial B_R/\partial z)$ 
in Eq.~\eqref{mhd13}
and $B_R\,(\partial B_z/\partial R)$ in Eq.~\eqref{mhd15}. 
Since the momentum equation is operator splitted, we can effectively express the Lorentz force term in the \textit{Lagrangean}, i.e., in the \textit{comoving} frame
(where the explicitly velocity dependent terms in the characteristic equations \eqref{charakticek1} and \eqref{charakticek1a} are dropped).
We thus evaluate the new radially and vertically updated half-time step values of magnetic induction components $B_R^{\,**,\,n+1/2}$ and $B_z^{\,**,\,n+1/2}$\!, 
respectively, using the relations analogous to Eqs.~\eqref{simplicek}-\eqref{charakticek1a} (i.e., for the Alfv\'en waves propagating in the z-direction and the R-direction).
The superscript$^{**}$ hence denotes the 
quantities evaluated using this Lagrangian upwind interpolation.
The total derivatives of the radial and vertical components are 
\begin{align}
\frac{\text{d}}{\text{d}t}=\frac{\partial}{\partial t}\mp\frac{B_z}{\!\!\sqrt{\mu_0\rho}}\frac{1}{h_z}\frac{\partial}{\partial z},\quad\quad\quad
\frac{\text{d}}{\text{d}t}=\frac{\partial}{\partial t}\mp\frac{B_R}{\!\!\sqrt{\mu_0\rho}}\frac{\partial}{\partial R}.
\end{align}
The Alfv\'en wave part of the Lorentz force source term (the Lagrangian updating is consistently added after the first hydrodynamic velocity update - see Fig.~\ref{hydroloop})
takes the explicit form
\begin{align}\label{alfoucek1}
\Pi^{\text{AB},\,n+6a/m_1}_{R\,(i,j)}&=\Pi^{\text{AB},\,n+5a/m_1}_{R\,(i,j)}+\frac{\Delta t}
{\mu_0}B_{z\,(i,j)}^{\text{AB},\,n}\frac{B_{R\,(i,j+1)}^{\text{AA}\,**,\,n+1/2}-B_{R\,(i,j)}^{\text{AA}\,**,\,n+1/2}}{h_z^{\text{B}}(z_{j+1}^\text{A}-z_{j}^\text{A})},\\
\Pi^{\text{BA},\,n+6c/m_3}_{z\,(i,j)}&=\Pi^{\text{BA},\,n+5c/m_3}_{z\,(i,j)}-\frac{\Delta t}
{\mu_0}B_{R\,(i,j)}^{\text{BA},\,n}\frac{h^+_zB^{\text{AA}\,**,\,n+1/2}_{z\,(i+1,j)}-h_z^{\text{A}}B^{\text{AA}\,**,\,n+1/2}_{z\,(i,j)}}{h_z^{\text{B}}(R_{i+1}^\text{A}-R_{i}^\text{A})}.
\end{align}
\subsubsection{Subroutines \textit{evmagr}1, \textit{evmagr}2, \textit{evmagt}1, \textit{evmagt}2 - evolving of the toroidal source terms}
Following the formalism described in Eqs.~\eqref{simplicek}, we can write the toroidal components of basic MHD equations \eqref{mhd7} and \eqref{mhd8}
(see also Eqs.~\eqref{mhd18} and \eqref{mhd26}) in the general form (where the components $R,\,\phi,\,z$ may denote the components $1,2,3$ in arbitrary 
orthogonal coordinate system)
\begin{align}\label{torevik1}
\frac{\partial{V_\phi}}{\partial t}&=-\frac{V_R}{h_\phi}\frac{\partial}{\partial R}(h_\phi V_\phi)-\frac{V_z}{h_\phi h_z}\frac{\partial}{\partial z}(h_\phi V_\phi)
+\frac{B_R}{\mu_0\rho}\frac{1}{h_\phi}\frac{\partial}{\partial R}(h_\phi B_\phi)+\frac{B_z}{\mu_0\rho}\frac{1}{h_\phi h_z}\frac{\partial}{\partial z}(h_\phi B_\phi),\\
\frac{\partial{B_\phi}}{\partial t}&=-\frac{1}{h_z}\frac{\partial}{\partial R}(h_z V_RB_\phi)-\frac{1}{h_z}\frac{\partial}{\partial z}(V_zB_\phi)\label{torevik1a}
+B_Rh_\phi\frac{\partial}{\partial R}\left(\frac{V_\phi}{h_\phi}\right)+B_z\frac{h_\phi}{h_z}\frac{\partial}{\partial z}\left(\frac{V_\phi}{h_\phi}\right),
\end{align}
where in each of these equations the first two terms are obviously the advection terms while the last two terms are the source terms
that are responsible for torsional Alfv\'en wave motion \citep{Stony1b}. We may expand the evolution of the toroidal source terms
(where $\xi$ and $\eta$ substitute the quantities $V_\phi/h_\phi$ and $h_\phi B_\phi$) in a similar manner as the evolution of the 
poloidal field source terms (see Eqs.~\eqref{charakticek} and ~\eqref{charakticeka}), we thus obtain
\begin{align}\label{torevik2}
\frac{\partial{\xi}}{\partial t}=\left(\frac{1}{h_\phi}\right)^2\left(\frac{B_R}{\mu_0\rho}\frac{\partial\eta}{\partial R}+
\frac{B_z}{\mu_0\rho}\frac{1}{h_z}\frac{\partial\eta}{\partial z}\right),
\quad\quad\quad\quad
\left(\frac{1}{h_\phi}\right)^2\frac{\partial{\eta}}{\partial t}=B_R\frac{\partial\xi}{\partial R}+\frac{B_z}{h_z}\frac{\partial\xi}{\partial z}.
\end{align}
Adding and subtracting these coupled equations we obtain the compact characteristic equation (cf.~Eqs.~\eqref{charakterobecnicek})
that can be differenced along the characteristics $C^-$ and $C^+$,
\begin{align}\label{torevik3}
\frac{\text{d}{\xi}}{\partial t}\pm\left(\frac{1}{h_\phi}\right)^2\!\!\!\!\frac{1}{\sqrt{\mu_0\rho}}\,\frac{\text{d}\eta}{\partial t}=0.
\end{align}
Since we do not need to calculate EMFs in this case (in two-dimensional calculations the toroidal field satisfies the divergence free
constraint at all times due to the symmetry, we therefore apply here only the MOC schema and not the CT method) we 
calculate the toroidal field evolution in the \textit{Lagrangian} frame without the 
explicitly velocity dependent terms in the characteristic equations \eqref{charakticek1} and \eqref{charakticek1a} (see the more detailed explanation of the difference
between Eulerian and Lagrangian upwind interpolation within the previous description of the subroutines \textit{lforr}1-\textit{lfort}2). 
This leads to the set of coupled equations basically similar to Eqs.~\eqref{charakticek1} and \eqref{charakticek1a},
\begin{align}\label{charakticektoro1}
\big(\xi_i^{\,*,\,n+1/2}-\xi_i^-\big)+\big(\eta_i^{\,*,\,n+1/2}-\eta_i^-\big)\big/\big(h_\phi^2\!\!\sqrt{\mu_0\rho}\big)^-_i=0,\\
\big(\xi_i^{\,*,\,n+1/2}-\xi_i^+\big)-\big(\eta_i^{\,*,\,n+1/2}-\eta_i^+\big)\big/\big(h_\phi^2\!\!\sqrt{\mu_0\rho}\big)^+_i=0.\label{charakticektoro1a}
\end{align}
These partially updated values are used in the first (e.g.,~radial) direction evolution of equations \eqref{torevik1} and \eqref{torevik1a}
by adding the source terms,
\begin{align}\label{kramoustoro1}
J^{\text{BB},\,n+2b/m_2}_{i,j}&=J^{\text{BB},\,n+b/m_2}_{i,j}+\frac{\Delta t}{\mu_0}\,B_{R\,(i,j)}^{\text{BB},\,n}\,
\frac{\eta_{i+1,j}^{\text{AB}\,*,\,n+1/2}-\eta_{i,j}^{\text{AB}\,*,\,n+1/2}}{R_{i+1}^\text{A}-R_{i}^\text{A}},\\
B^{\text{BB},\,n+b^\prime\!/2}_{\phi\,(i,j)}&=B^{\text{BB},\,n}_{\phi\,(i,j)}+\frac{\Delta t}{\mu_0}\,B_{R\,(i,j)}^{\text{BB},\,n}\,h_\phi^{\text{B}}\,
\frac{\xi_{i+1,j}^{\text{AB}\,*,\,n+1/2}-\xi_{i,j}^{\text{AB}\,*,\,n+1/2}}
{R_{i+1}^\text{A}-R_{i}^\text{A}},
\label{kramoustoro1a}
\end{align}
where the equation \eqref{kramoustoro1} is the azimuthal analog of Eqs.~\eqref{kramous1}, \eqref{kramous1a} and $b^\prime$ in Eq.~\eqref{kramoustoro1a} is the analog of 
$b$ in Eq.~\eqref{kramoustoro1}, however for $B_\phi$.
Repeating this process in the second direction we complete the source terms' adding evolution of equations \eqref{torevik1} and \eqref{torevik1a}, obtaining
\begin{align}\label{kramoustoro2}
J^{\text{BB},\,n+3b/m_2}_{i,j}&=J^{\text{BB},\,n+2b/m_2}_{i,j}+\frac{\Delta t}{\mu_0}\,B_{z\,(i,j)}^{\text{BB},\,n}\,
\frac{\eta_{i,j+1}^{\text{BA}\,*,\,n+1/2}-\eta_{i,j}^{\text{BA}\,*,\,n+1/2}}{h_z^\text{B}(z_{j+1}^\text{A}-z_{j}^\text{A})},\\
B^{\text{BB},\,n+b^\prime}_{\phi\,(i,j)}&=
B^{\text{BB},\,n+b^\prime\!/2}_{\phi\,(i,j)}+\frac{\Delta t}{\mu_0}\,B_{z\,(i,j)}^{\text{BB},\,n}\,h_\phi^{\text{B}}\,
\frac{\xi_{i,j+1}^{\text{BA}\,*,\,n+1/2}\!-\!\xi_{i,j}^{\text{BA}\,*,\,n+1/2}}{h_z^\text{B}(z_{j+1}^\text{A}-z_{j}^\text{A})}.
\label{kramoustoro2a}
\end{align}
We note that for three-dimensional calculations the toroidal symmetry for the divergence free constraint does not apply and the full MOC-CT formalism has to be used
\citep{Stony1b}.
\subsection{MHD advection steps extension}\label{magorisadvectis}
\subsubsection{Subroutines \textit{flmagpr}, \textit{flmagpt} - toroidal field advection}
Once the source terms MOC update \eqref{kramoustoro1}-\eqref{kramoustoro2a} is finished we advect the quantities $J$ and $B_\phi$. 
The advection of $J$ is described in Appendix~\ref{dvojdimhydrous} (subroutines \textit{phimom1} and \textit{phimom3}). 
In order to minimize the diffusion of the toroidal magnetic field relative to the mass, as required by flux-freezing \citep{Stony1b}, we advect the quantity $B_\phi/\rho$
instead of the field itself. The predictor step first evolves the density $\rho$ and the quantity $B_\phi/\rho$ separately, forming the new half-time step quantity 
$\mathcal{F}^{\,n+1/2}_{i,j}=\rho^{n+1/2}_{i,j}(B_\phi/\rho)^{\,n+1/2}_{i,j}V^{\,n+1/2}_{i,j}$.
The advection step is performed via the directionally splitted algorithm (cf.~Eqs.~\eqref{multimomentradkorektorix}, \eqref{multimomentvertkorektorix}),
\begin{align}\label{flumagis1}
B_{\phi\,(i,j)}^{\text{BB},\,n+1}&=B_{\phi\,(i,j)}^{\text{BB},\,n}+\frac{\Delta t}{\Omega^{\text{BB}}_{i,j}}
\bigg[\mathcal{F}_{i+1,j}^{\,\text{AB},\,n+\frac{1}{2}}V^{\text{AB},\,n+\frac{1}{2}}_{R\,(i+1,j)}S^{\text{AB}}_{\!R\,(i+1,j)}-
\mathcal{F}_{i,j}^{\,\text{AB},\,n+\frac{1}{2}}V^{\text{AB},\,n+\frac{1}{2}}_{R\,(i,j)}S^{\text{AB}}_{\!R\,(i,j)}\bigg],\\
B_{\phi\,(i,j)}^{\text{BB},\,n+1}&=B_{\phi\,(i,j)}^{\text{BB},\,n}+\frac{\Delta t}{\Omega^{\text{BB}}_{i,j}}
\bigg[\mathcal{F}_{i,j+1}^{\,\text{BA},\,n+\frac{1}{2}}V^{\text{BA},\,n+\frac{1}{2}}_{z\,(i,j+1)}S^{\text{BA}}_{\!z\,(i,j+1)}-
\mathcal{F}_{i,j}^{\,\text{BA},\,n+\frac{1}{2}}V^{\text{BA},\,n+\frac{1}{2}}_{z\,(i,j)}S^{\text{BA}}_{\!z\,(i,j)}\bigg].
\label{flumagis2}
\end{align}
\subsubsection{Subroutines \textit{magpreflr}, \textit{magpreflt} - magnetic pressure term advection}
Using the advection schema described in Eqs.~\eqref{multimomentradkorektorix}, \eqref{multimomentvertkorektorix} we update the total energy by advection of the magnetic pressure flux term
$\displaystyle -B^2/(2\mu_0)\vec{V}=(B_R^2+B_\phi^2+B_z^2)/(2\mu_0)\vec{V}$ (second term in square bracket in energy equation \eqref{mhd10}) in $R$ and $z$ direction.
\subsubsection{Subroutines \textit{magenerr}, \textit{magenert} - adding the divergence of magnetic tension force}
The subroutines update the total energy by adding the directionally splitted magnetic tension force flux term
$\displaystyle (\vec{B}\cdot\vec{V})\vec{B}/\mu_0$ (first term in square bracket in energy equation \eqref{mhd10}) using the finite volumes method 
(cf.~Eqs.~\eqref{multimomentradkorektorix} and \eqref{multimomentvertkorektorix}).

\section{Radiative extension of the hydrodynamic code (in progress)}\label{dvojdimradous}
\begin{figure}[t]
\begin{center}
\includegraphics[width=7.25cm]{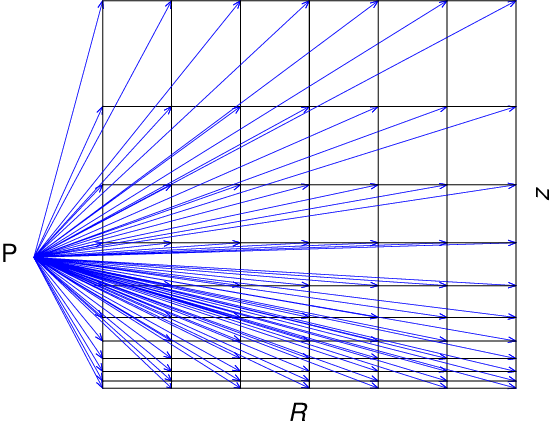}\quad
\includegraphics[width=7.25cm]{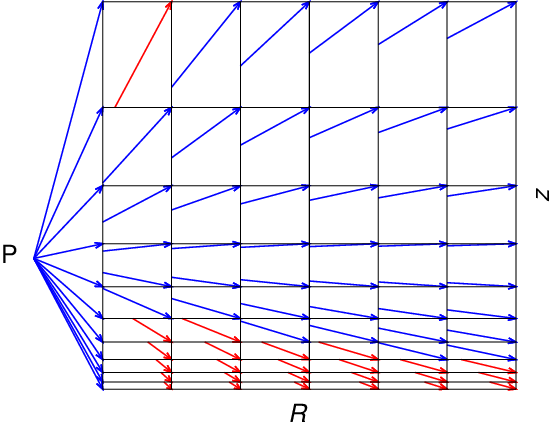}\vspace{0.1cm}
\small{Fig.~\ref{lajzlik}a}\quad\quad\quad\quad\quad\quad\quad\quad\quad\quad\quad\quad\quad\quad\quad\quad\quad\quad\small{Fig.~\ref{lajzlik}b}
\caption{Comparison of the long (a) and short (b) characteristics
method for the same schematic disk $R$-$z$ plane, represented by the two-dimensional logarithmic grid and the source of radiation denoted by point $P$. 
For the long characteristics method, the closer 
to the source we get, the more rays pass through one cell, resulting in a large number of redundant
calculations. The short characteristics method avoids this by the interpolation of the quantities from
cells that have been dealt with previously, so only the short ray sections that pass
from cell to cell need to be computed. The footpoints of the ray sections that are located on the vertical / horizontal grid cells interface lines are denoted in Eqs.~\eqref{svetlousek3} 
and \eqref{svetlousek4} as $z^\prime$ or $R^\prime$, respectively.
The rays that pass through the footpoints $z^\prime$ or $R^\prime$ are distinguished by the blue or the red color. 
The idea is adapted from \citet{richkonik}.}
\label{lajzlik}
\end{center}
\end{figure}
\begin{figure}[t]
\begin{center}
\includegraphics[width=11cm]{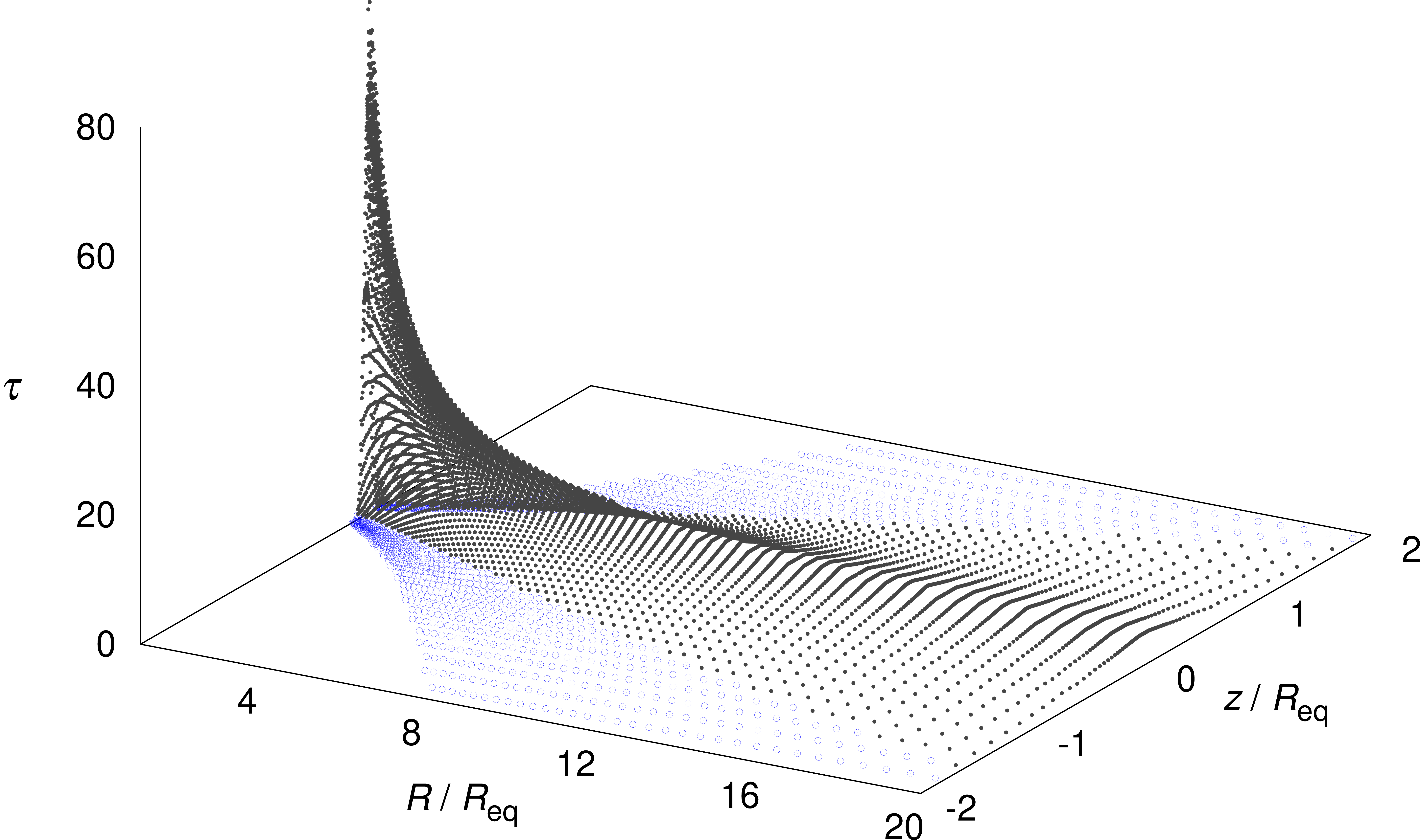}
\caption{The map of the disk $R$-$z$
plane optical depth $\tau$ calculated via the method described in Sect.~\ref{dvojdimradous} in the disk. 
The optical depth $\tau=2/3$ marks the limit that separates disk regions regarded as optically thick (black colored) 
and optically thin (blue colored). 
The irregularities in the optically thick peak near the very inner disk boundary are enhanced due to the extreme radial as well as vertical logarithmic narrowing of the grid cells
in the midplane region.
They however do not affect the calculations of the disk thermal structure in these optically thick central regions, 
since the optical depth in the black domain we currently approximate using Eq.~\eqref{presgradour2}.
Further improvement of the numerical coordination between the grid configuration and the short-characteristics algorithm is desired
within the future work.}
\label{tauoptik}
\end{center}
\end{figure}
We computed the optical depths along the rays described in Sect.~\ref{templajznik} that determine the disk thermal equilibrium calculation using the long- or short-characteristics 
method. For the future work we prefer the following
short-characteristics ray tracing computational schema (see Fig.~\ref{lajzlik}b) which avoids the large number of rays that intersect the grid cells near the radiative 
source (cf.~Fig.~\ref{lajzlik}a).
Within the orthogonal $R$-$z$ grid configuration we have to distinguish two fundamental variants: in the first one the source of radiation vertically 
coincides with an arbitrary grid cell interface $z_j$ level
(in our particular problem this means $z_j=R_\star$ for some $j$) while in the second one the same source is located somewhere between the particular 
grid cell interfaces, i.e., $z_j<R_\star<z_{j+1}$ for some $j$.
There is a small probability of such coincidence in case of the single source of radiation, however, we anticipate more advanced calculations with large number of 
radiation sources set on the stellar surface and therefore with low chance to check their position ``manually''.

We denote $\delta$ the angle of a diagonal of a computational grid cell, in the first variant we distinguish the two different cases 
(referring to the ``upper'' half-space of the star-disk system), $z_j<R_\star$ and $z_j\geq R_\star$, leading to
\begin{align}\label{svetlousek1}
\left(\text{tan}\,\delta\right)_{\,i,j}=\frac{z_{j}-z_{j+1}}{R_{i+1}-R_{i}}\quad\quad\text{and}\quad\quad\left(\text{tan}\,\delta\right)_{\,i,j}=\frac{z_{j}-z_{j-1}}{R_{i+1}-R_{i}},
\end{align}
respectively.
The second variant however leads to three different cases, $z_j<R_\star\wedge z_{j+1}<R_\star$, $z_j<R_\star\wedge z_{j+1}>R_\star$ and $z_j>R_\star$, implying  
\begin{align}\label{svetlousek1a}
|\text{tan}\,\delta|_{\,i,j}=\frac{z_{j+1}-z_{j}}{R_{i+1}-R_{i}}.
\end{align}
% for the first and the latter case, 
% while for the radiative source level cells with $z_j<R_\star<z_{j+1}$ we do not need to specify the $\delta$ angle.
Denoting $z^\prime$ and $R^\prime$ the intersections of the ray with vertical or horizontal grid cell interfaces, we may distinguish the following principal cases
with respect to the grid cell position ``below'' or ``above`` the source of radiation, regardless of the variant of the radiative source position. The two of them,
\begin{align}\label{svetlousek2}
|\text{tan}\,\delta|_{\,i,j}<\frac{|z_{j}-R_\star|}{R_{i}},\quad\quad|\text{tan}\,\delta|_{\,i,j-1}<\frac{|z_{j}-R_\star|}{R_{i}},
\end{align}
for the ``below'' and ``above`` position, respectively,
lead to employing the $R^\prime$ intersection while we have to employ the $z^\prime$ intersection in all other cases. For the vertical grid cell intersections $z^\prime$ we may write
\begin{align}\label{svetlousek3}
z^\prime_{i,j}=z_j+\frac{(R_\star-z_{j})(R_{i}-R_{i-1})}{R_{i}},
\end{align}
regardless of the cell position, where in the first variant we obtain a single particular $z^\prime$ for each horizontal strip
of grid cells (cf.~Fig~\ref{lajzlik}b) while in the second one for the radiative source level cells with $z_j<R_\star<z_{j+1}$ 
we have to implement two points $z^\prime_1$ and $z^\prime_2$. Regarding the horizontal grid cell intersection $R^\prime$, the relation however depends on the cell position.
Distinguishing for the first variant the principal cases $z_j<R_\star$ and $z_j\geq R_\star$, respectively,
we may write
\begin{align}\label{svetlousek4}
R^\prime_{i,j}=\frac{R_{i}\,(R_\star-z_{j+1})}{R_\star-z_{j}},\quad\quad R^\prime_{i,j}=\frac{R_{i}\,(z_{j-1}-R_\star)}{z_{j}-R_\star}.
\end{align}
For the second variant the relation \eqref{svetlousek4} for the two cases $z_j<R_\star\wedge z_{j+1}<R_\star$ and $z_j>R_\star$
is identical
(we note that for the radiative source level cells with $z_j<R_\star<z_{j+1}$ we never need to specify the point $R^\prime$).

The optical depth $\tau$ is calculated using Eq.~\eqref{tauchel} where 
the ray trace intervals $\Delta s$ are calculated as the length of the segments depicted in Fig.~\ref{lajzlik}b. 
The graph that demonstrates the calculation of the disk optical depth $\tau=2/3$ in the disk $R$-$z$ plane (denoting the optical depth limit $\tau=2/3$ that separates  
the optically thick and optically thin regions) is depicted in Fig.~\ref{tauoptik}. 
The thermal structure is calculated via Eqs.~\eqref{kocour}-\eqref{radacek} that are
differenced on corresponding grid (see Fig.~\ref{stagis}).
\section{Parallelization of calculations}\label{paralelous}
To accelerate and even to achieve the feasibility of very large (one- or even multi-dimensional) computational problems it is necessary to
parallelize the calculations for a large number of processors (processes). The principle of the code parallelization is to split the one overall spatial area (cf.~Fig.~\ref{stagis})
into number of separate domains (ranks). These ranks are, according to the nature of the problem, calculated either fully separately, or, if there is mutual relation between the 
ranks (e.g., in hydrodynamic calculations are the resulting values in the neighboring domains spatially coupled due to the link to given boundary conditions), the developed procedures enable the 
communication on the corresponding boundaries of the layers (without any noticeable deceleration of the computational speed). 

We use the standard library
MPI (Message-Passing Interface) version 3.1, developed by a group of researchers from academia and industry to function on a wide variety of parallel computers
(the official source of the library is on the website \url{http://www.mpi-forum.org/}). It is programmed to move
the data from the address space of one process to
that of another process through cooperative operations on each process (so-called point-to-point communication between two processors). 

Since this is a very wide and specialized discipline, we do not describe here the techniques of the parallel programming. The main goal of using the method is the 
significant (however expected) acceleration of the computational time 
in case of either separated spatial domains or the domains where the mutual boundary ``send and receive'' communication is necessary. In many cases 
the computational process on a single processor is even not feasible (the file even cannot be compiled).
The MPI library is designed for various computer programming languages such as Fortran, C, C++ and Java, there however may be substantial computational differences 
(there is for example a different grid cells sequence in 2D parallelization using MPI for Fortran, where the computation ``goes'' within each rank at first in ``vertical'' direction, 
while the MPI for C language starts to calculate horizontally).

We have currently programmed most of the described one- and two-dimensional programs in the parallel mode and
we perform the parallelized computations via the simultaneous execution of separate portions of a program by employing (up to) 200 processors within a computer cluster. 
We mainly use the computational cluster METACENTRUM which is virtual organization that operates and manages distributed computing infrastructure 
of co-operative academic centers within the Czech Republic. The computing and storage facilities are owned
by parties and projects contributing to the Czech National Grid Infrastructure, provided under the program “Projects of Large Infrastructure
for Research, Development, and Innovations” (LM2010005).
METACENTRUM computational cluster sites are located at: 
Masaryk University in Brno (Center CERIT-SC and Faculty of Science), 
Institute of Organic Chemistry and Biochemistry (UOCHB) in Prague, University of West Bohemia in Pilsen (Faculty of Applied Sciences), 
University of South Bohemia (Faculty of Science's cluster), Academy of Sciences (FZU), 
Departement of Telecommunication Engineering FEL ČVUT, and CESNET z.s.p.o..
The official web site is \url{http://metavo.metacentrum.cz/}.

\chapter{Brief description of some of the referred astrophysical processes}\label{briefmention}
\section{Stellar winds}\label{stelwindis}
\subsection{Parker coronal (solar) wind}\label{parkeris}
Following the isotropic spherical continuity and momentum equations~\eqref{contispherelap} and \eqref{spheremomrad}, we obtain the (stationary) relation, known as
Parker's equation or Parker's solar wind model \citep{parek,parkan}, for the 
spherically symmetric wind (see the notation introduced in Appendix~\ref{appendix1}),
\begin{align}\label{parkousek}
\frac{1}{\varv_r}\left(\varv_r^2-a^2\right)\frac{\text{d}\varv_r}{\text{d}r}=\frac{2a^2}{r}-\frac{GM_{\star}}{r^2}+\frac{\text{d}a^2}{\text{d}r},
\end{align}
whose integration gives five solutions where only one of them is however physically relevant in case of a stationary wind:
the plasma leaves the hot base layers (solar corona) with a small velocity that increases towards the sonic point
(which can be regarded as a kind of nozzle). After passing through it, 
the plasma propagates with asymptotically supersonic velocity.
In the isothermal case we obtain a simple analytic relation for the sonic (critical) point distance (where $\varv_r=a$), 
\begin{align}\label{parkousek2}
r_\text{s}=\frac{GM_{\star}}{2a^2}.
\end{align}
The coronal wind is produced by the thermal expansion of stellar (solar) corona and comes from the various coronal regions which determine the 
wind velocity profile (slow vs.~fast wind, etc.). 
\subsection{Spherically symmetric line-driven stellar winds}\label{lajncakis}
The fundamental work that describes the hot stars' stellar winds accelerated by scattering in an ensemble of stellar atmospheric spectral lines 
(especially from metal ions) is given by \citet{CAK1}. 
Equations of the wind dynamics follow from isotropic spherical continuity and momentum equations~\eqref{contispherelap} and \eqref{spheremomrad}, 
where the radiative acceleration has to be taken into account. We may
express the total radiative acceleration $g_{\text{rad}}$ as the sum of radiative acceleration induced by continuum electron scattering opacity 
and by the spectral line absorption, $g_{\text{rad}}=g_{\text{rad,es}}+g_{\text{rad},L}$.
In isotropic case the radiative transfer equation integrated over all frequencies (in the diffusion approximation) gives the equation of radiative equilibrium,
\begin{align}\label{CAK3}
\frac{\text{d}P_{\!\!\text{rad}}}{\text{d}r}=-\frac{\rho\kappa F_{\!\!\text{rad}}}{c},
\end{align}
where $P_{\!\!\text{rad}}$ is the radiation pressure, $\kappa$ is the frequency averaged opacity, $F_{\!\!\text{rad}}$ is the total 
(frequency integrated) radiative flux and $c$ denotes the speed of light.
Comparing the radiative and hydrostatic equilibrium, Eq.~\eqref{CAK3} gives the 
radiative acceleration driven by the electron scattering opacity, $g_{\text{rad,es}}=\kappa_{\text{es}}F_{\!\!\text{rad}}/c$.
We may express the radiative acceleration $g_{\text{rad},L}$ produced by the absorption of radiation in spectral lines as
\begin{align}\label{CAK4}
g_{\text{rad},L}=g_{\text{rad,es}}M(t),
\end{align}
where $M(t)$ is the so-called line force multiplier, which expresses the effect of all spectral lines \citep{CAK1}.
In a static medium the optical depth $\tau$ is defined as $\displaystyle\text{d}\tau=-\kappa\rho\,\text{d}r$.
In a moving medium where the basic relations follow the Sobolev approximation \citep{sobol1}, the optical depth (denoted in case of the moving medium  
as $t$) is \citep{CAK1}
\begin{align}\label{CAK6}
t=\kappa_{\text{es}}\rho\,\varv_{\text{th}}\left|\frac{\text{d}r}{\text{d}\varv}\right|.
\end{align}
From the Maxwellian velocity distribution we have the thermal velocity on the star's surface $\varv_{\text{th}}=(2\mathcal{R}T_{\text{eff}})^{1/2}$, 
where $\mathcal{R}$ is the specific gas constant. It is defined as $\mathcal{R}=k/(\mu m_u)$, where $k$ is the Boltzmann constant, $\mu$ is the mean molecular weight
and $m_u$ is the atomic mass unit (see Sect.~\ref{statak}).
At a given $T_{\text{eff}}$, \citet{CAK1} introduce the relation for the line force multiplier in case of the point source of radiation in the parametric form
\begin{align}\label{CAK7}
M(t)=kt^{-\alpha},
\end{align}
where the parameter $\alpha$ represents basically the ratio of the radiative force produced by the absorption in optically thick lines to the overall radiative force and the 
parameter $k$ scales the magnitude of the radiative force \citep[][other frequently used parameter is $\bar{Q}$, introduced by \citet{galose} that 
scales the magnitude of the radiative force in units of the radiative force produced by electron scattering, we however do not adopt it here]{CAK1,maeder}. 
These parameters vary with temperature, we may however for simple approximations use the 
mean values \citep{CAK1} $k=0.035$ and $\alpha=0.62$.

The stationary equation of motion \eqref{spheremomrad} is now modified into the form 
\begin{align}\label{CAK8}
\varv\frac{\text{d}\varv}{\text{d}r}=-\frac{1}{\rho}\frac{\text{d}P_{\text{g}}}{\text{d}r}-\frac{GM_{\!\star}}{r^2}+
\frac{\kappa_{\text{es}}F_{\!\!\text{rad}}}{c}\left[1+M(t)\right],
\end{align}
where we denote the $P_\text{g}$ in order to distinguish the gas and radiation pressure and
the two terms in the square bracket distinguish $g_{\text{rad,es}}$ and $g_{\text{rad},L}$. Including the mass conservation 
equation \eqref{contispherelap} in the spherically symmetric form $\dot{M}=4\pi r^2\rho\varv$ into the relation for the moving medium optical depth
\eqref{CAK6}, we obtain
\begin{align}\label{CAK9}
t=\kappa_{\text{es}}\frac{\dot{M}}{4\pi r^2\varv}\varv_{\text{th}}\left|\frac{\text{d}r}{\text{d}\varv}\right|.
\end{align}
Using the above relations and neglecting the gas pressure (which in fact represents here only the minor contribution in supersonic part), we write 
the equation of motion \eqref{CAK8} as (cf.~\citet{CAK1})
\begin{align}\label{CAK10}
\underbrace{r^2\varv\frac{\text{d}\varv}{\text{d}r}}_{\varw'}=-GM_{\!\star}(1-\Gamma)+
\underbrace{\Gamma GM_{\!\star}k\left
(\frac{4\pi}{\dot{M}\varv_{\text{th}}\kappa_{\text{es}}}\right)^{\alpha}}_{C}\underbrace{
\left(r^2\varv\left|\frac{\text{d}\varv}{\text{d}r}\right|\right)^{\alpha}}_{(\varw')^\alpha},
\end{align}
where $\Gamma=\kappa_{\text{es}}L_\star/4\pi GM_{\!\star}c$ ($L_\star$ is the stellar luminosity)
is the Eddington factor (cf.~Eq.~\eqref{fomisgamma}). 

Since the star is however not the point source of radiation (at least to some considerable distance), 
we include into the calculation of the force-multiplier $M(t)$ (see Eqs.~\eqref{CAK4}, \eqref{CAK7} and \eqref{CAK8}) 
also the finite spatial angle of impinging stellar irradiation (\textit{finite disk correction factor} - FDCF), 
that is expressed as \citep{CAK1,Lamercas} 
\begin{align}\label{CAK39}
M_{\text{corr}}(t)=\frac{(1+\sigma)^{\alpha+1}-(1+\sigma\mu_{\star}^2)^{\alpha+1}}
{(1-\mu_{\star}^2)(\alpha+1)\sigma(1+\sigma)^{\alpha}}\,M(t),
\end{align}
where from the Sobolev approximation relations \citep[Sobolev optical depth, see, e.g.,][]{sobol1,Lamercas} follows the factor $\sigma$ while the factor $\mu_{\star}$
is the projection cosine of stellar radius (limiting angle of the stellar irradiation cone). We write them
\begin{align}\label{CAK40}
\sigma=\frac{\text{d}\,\text{ln}\,\varv}{\text{d}\,\text{ln}\,r}-1\quad\text{and}
\quad\mu_{\star}=\sqrt{1-\left(\frac{R_{\star}}{r}\right)^2}.
\end{align}
Using the substituted underbraced quantities, 
the differentiation of Eq.~\eqref{CAK10} in $\varw^\prime$ gives \citep[e.g.,][]{Lamercas}
\begin{align}\label{CAK32}
1-\alpha\text{C}(\varw^\prime)^{\alpha-1}=0,\quad\quad\text{yielding}\quad\quad C(\varw')^\alpha=\frac{\varw'}{\alpha}.
\end{align}
Inserting the quantity $C(\varw')^\alpha$ from Eq.~\eqref{CAK32} into Eq.~\eqref{CAK10} we get $\varw'=GM_{\!\star}(1-\Gamma)\,\alpha/(1-\alpha)$
and after rearrangements we obtain 
the constant mass loss rate written in the form \citep[e.g.,][]{Lamercas}
\begin{align}\label{CAK30}
\dot{M}=\frac{4\pi GM_{\!\star}}{\kappa_{\text{es}}\varv_{\text{th}}}\,\alpha\,(1-\alpha)^{(1-\alpha)/\alpha}
\,(\Gamma\,k)^{1/\alpha}\,(1-\Gamma)^{(\alpha-1)/\alpha}.
\end{align}
Inserting Eq.~\eqref{CAK32} into \eqref{CAK10}, this equation takes the quite simple form
\begin{align}\label{CAK29}
r^2\varv\frac{\text{d}\varv}{\text{d}r}=\frac{\alpha}{1-\alpha}\,GM_{\!\star}(1-\Gamma).
\end{align} 
Integrating the equation \eqref{CAK29} from the stellar surface radius $R_{\star}$ to some arbitrary distance $r$,
(where we consider the wind velocity at the stellar surface as negligible) 
we obtain the analytic approximation for the wind velocity profile 
\begin{align}\label{CAK41}
\varv(r)=\varv_{\infty}\left({1-\frac{R_{\star}}{r}}\right)^{\beta}\quad\text{with}\quad
\beta=1/2\,\,\,\text{and}\,\,\,\varv_{\infty}^2=\frac{\alpha}{1-\alpha}\varv_{\text{esc}}^2,
\end{align}
where the (squared) stellar escape velocity $\varv_{\text{esc}}^2=2GM_{\!\star}(1-\Gamma)/R_{\star}$.
Taking into account the finite spatial angle of impinging stellar irradiation (cf.~Eq.~\eqref{CAK39}),
the factor $\beta$ usually takes the value $\beta\approx 0.8-1.0$ \citep[see, e.g.,][]{CAK1,Lamercas}.
\subsection{Line-driven stellar winds from rotating, gravity-darkened stars}
\label{vonZeipel}
The spherically symmetric stellar wind model introduced in Sect.~\ref{lajncakis} assumes fixed stellar surface brightness.
Considering stellar rotation, the standard CAK formalism involves the effect of centrifugal force 
(where from the conservation of angular momentum must apply 
$r\varv_{\phi}=R_{\star}\varv(R_\text{eq})$) and equation~\eqref{CAK8} becomes \citep{maeder}
\begin{align}\label{fomis}
\varv\frac{\text{d}\varv}{\text{d}r}+\frac{GM_{\!\star}}{r^2}+\frac{1}{\rho}\frac{\text{d}P_{\text{g}}}{\text{d}r}-\frac{\varv^2(R_\text{eq})R^2(R_\text{eq})}
{r^3}-g_{\text{rad}}=0.
\end{align}
Assuming purely radial wind driving, Eq.~\eqref{fomis} would lead to the formation of dense equatorial disk-like flow 
with ${\Sigma}_{\text{disk}}\sim\varv(R_\text{eq})$, and relatively fast and low-density polar wind (cf.~the WCD paradigm in Sect.~\ref{wicom}).

Noting that the vector $\vec{g}_{\text{tot}}$ of the total stellar gravitational acceleration is the vector sum 
\begin{align}\label{geff}
\vec{g}_{\text{tot}}=\vec{g}_{\text{eff}}+\vec{g}_{\text{rad}}=\vec{g}_{\text{grav}}+\vec{g}_{\text{rot}}+\vec{g}_{\text{rad}},
\end{align} 
with $\vec{g}_{\text{eff}}$ being the vector sum of the gravitational and centrifugal acceleration (Eq.~\eqref{fomis1}), 
we introduce the so-called $\Omega\Gamma$ limit which represents the superposed rotational ($\Omega$ or break-up) limit and the Eddington (radiative or $\Gamma$) limit. These limits express the 
maximum of the centrifugal or the radiative force above which the upper stellar layers are no longer gravitationally bound:
to express the local Eddington factor $\Gamma(\Omega,\theta)$, we set $\vec{g}_{\text{tot}}=\vec{0}$ locally \eqref{geff}, which implies $\vec{g}_{\text{rad}}=-\vec{g}_{\text{eff}}$.
From Eq.~\eqref{CAK3} (where $\vec{g}_{\text{rad}}=\kappa\vec{F}_{\text{rad}}/c$) we express the $\Omega\Gamma$ limiting radiative flux as
\begin{align}\label{limitingomegacgamacflux}
\vec{F}_{\text{lim}}(\Omega,\theta)=-\frac{c}{\kappa(\Omega,\theta)}\vec{g}_{\text{eff}}(\Omega,\theta).
\end{align}
We define the Eddington $\Gamma(\Omega,\theta)$ limit at the stellar surface of a rotating star as a ratio of the actual to the limiting local stellar radiative flux. Eqs.~\eqref{omegacgamacflux} and \eqref{limitingomegacgamacflux} give
\begin{align}\label{fomisgamma}
\Gamma(\Omega,\theta)=\frac{{F}_{\text{rad}}(\Omega,\theta)}{{F}_{\text{lim}}(\Omega,\theta)}=
\frac{\kappa(\Omega,\theta)\,L_\star}{4\pi c GM_{\!\star}\left(1-\frac{\Omega^2}{2\pi G\overline{\rho}_M}\right)}=-\frac{{g}_{\text{rad}}(\Omega,\theta)}{{g}_{\text{eff}}(\Omega,\theta)}.
\end{align} 
Using Eq.~\eqref{fomisgamma}, the rotationally and latitudinally dependent total gravity can be written in a simple form as $\vec{g}_{\text{tot}}(\Omega,\theta)=\vec{g}_{\text{eff}}(\Omega,\theta)[1-\Gamma(\Omega,\theta)]$.
Rewriting the stationary CAK global mass loss rate $\dot{M}$ equation~\eqref{CAK30} into the form
\begin{align}\label{kamni}
\dot{M}=\frac{4\pi}{\kappa_{\text{es}}\varv_{\text{th}}}\left(\frac{k\,\alpha\,\kappa_{\text{es}} L_\star}{4\pi c}\right)^{1/\alpha}
\left(\frac{1-\alpha}{\alpha}\right)^{(1-\alpha)/\alpha}\left[GM_{\!\star}(1-\Gamma)\right]^{(\alpha-1)/\alpha},
\end{align}
and inserting the expressions for $\vec{g}_{\text{tot}}(\Omega,\theta)$  from Eq.~\eqref{geff}, for $\vec{F}_{\text{rad}}(\Omega,\theta)$ from Eq.~\eqref{omegacgamacflux}
and for ${T}_{\text{eff}}(\Omega,\theta)$ from Eq.~\eqref{teploukis}
into Eq.~\eqref{kamni}, 
we obtain the equation of proportionality for the stellar mass flux at a given colatitude $\theta$ per unit stellar surface $\Delta S$ \citep{maeder},
\begin{align}\label{lmln}
\frac{\Delta\dot{M}(\Omega,\theta)}{\Delta S}\sim \frac{(\kappa_{\text{es}}\,k\alpha)^{1/\alpha}}{\kappa_{\text{es}}}\left(\frac{1-\alpha}{\alpha}\right)^{(1-\alpha)/\alpha}
\!\!\!\!\!\!\sigma^{1/8}\left[\frac{L}{4\pi GM_{\!\star}
\left(1-\frac{\Omega^2}{2\pi G\overline{\rho}_M}\right)}\right]^{(1/\alpha-1/8)}
\!\!\!\!\!\!\frac{{g}_{\text{eff}}^{7/8}(\theta)}{\big[1-\Gamma(\Omega,\theta)\big]^{(1/\alpha-1)}}.
\end{align}
The asphericity of the stellar winds may be thus enhanced by the ${g}_{\text{eff}}$ effect and by the ${\kappa}$ effect (see Sect.~\ref{wicom}, Sect.~\ref{bistab} and Fig.~\ref{figtwo} for the description).
        
We obtain the total mass loss rate of the rotating, gravitationally darkened star by integration of Eq.~\eqref{lmln} over the whole stellar surface, calculating the surface averaged 
effective gravity as $\overline{{g}_{\text{eff}}}=\iint\vec{g}_{\text{eff}}\cdot\text{d}\vec{S}=4\pi GM_{\!\star}
\left(1-\frac{\Omega^2}{2\pi G\overline{\rho}_M}\right)\,\big[S(\Omega)\big]^{-1}$. The total surface of the rotationally oblate star $S(\Omega)$ 
is calculated by integration of Eq.~\eqref{effsurarea}, with use of Eqs.~\eqref{effpotroot} and \eqref{effcoseps}.
The ratio of the total (latitudinally averaged) mass loss rate of a rotating star and to that of a non-rotating star (Eq.~\eqref{kamni}),
with the same position on HR diagram, can be written as \citep{maeder}
\begin{align}\label{mnoucek}
\frac{\dot{M}(\Omega)}{\dot{M}(0)}=\frac{\big(1-\Gamma\big)^{(1/\alpha-1)}}{\left[1-\frac{\Omega^2}
{2\pi G\overline{\rho}_M}\right]^{(1/\alpha-7/8)}
\big[1-\Gamma(\Omega)\big]^{(1/\alpha-1)}},
\end{align}
where the Eddington factor $\Gamma$ in the numerator refers to non-rotating star, while $\Gamma(\Omega)$ in the denominator refers to the rotating star.
If $\Omega=1$, this ratio is equal to 1. 
Table~\ref{maratabulak} gives the comparison of various numerically calculated ratios ${\dot{M}(\Omega)}/\dot{M}(0)$.

\begin{table}[t]
\begin{center}
\small\begin{tabular}{@{}rlr@{.}lr@{.}lr@{.}lr@{.}l@{}}
\hline
\rule{0mm}{4.5mm}${M_{\!\star}}_{\text{ini}}$ & $\Gamma$ & \multicolumn{2}{l}{$\frac{\dot{M}(\Omega)}{\dot{M}(0)}$} & 
\multicolumn{2}{l}{$\frac{\dot{M}(\Omega)}{\dot{M}(0)}$} & \multicolumn{2}{l}{$\frac{\dot{M}(\Omega)}{\dot{M}(0)}$} &
\multicolumn{2}{l@{}}{$\frac{\dot{M}(\Omega)}{\dot{M}(0)}$}\\
& & \multicolumn{2}{l}{$\alpha=0.52$} & \multicolumn{2}{l}{$\alpha=0.24$} & \multicolumn{2}{l}{$\alpha=0.17$} & \multicolumn{2}{l@{}}{$\alpha=0.15$}\\
\hline
\rule{0mm}{2.5mm}120 & 0.903 & \multicolumn{2}{l}{crit} & \multicolumn{2}{l}{crit} & \multicolumn{2}{l}{crit} & \multicolumn{2}{l@{}}{crit}\\
                  %85 & 0.691 & \multicolumn{2}{l}{crit}  & \multicolumn{2}{l}{crit} & \multicolumn{2}{l}{crit} & \multicolumn{2}{l@{}}{crit}\\
                  60 & 0.527 & 4&00 & 101&8 & \multicolumn{2}{l}{1196} & \multicolumn{2}{l@{}}{3731}\\
                  40 & 0.356 & 2&26 & 14&4 & 58&5 & 112&1\\
                  25 & 0.214 & 1&86 & 7&43 & 21&3 & 34&5\\
                  %20 & 0.156 & 1&77 & 6&21 & 16&1 & 25&0\\
                  15 & 0.097 & 1&69 & 5&33 & 12&8 & 19&1\\
                  %12 & 0.063 & 1&66 & 4&95 & 11&4 & 16&7\\
                   9 & 0.034 & 1&63 & 4&67 & 10&4 & 15&0\\
\hline
\end{tabular}
\caption{Example of the numerical calculations \citep{salda,lampa} of the critically rotating star's mass loss rate to the equally massive non-rotating star's mass loss rate ratio, 
analytically expressed in Eq.~\eqref{mnoucek} (metallicity $Z=0.02$), for various CAK $\alpha$ parameters. 
The label ``crit'' means that due to $\Gamma$ and $\Omega$ the surface layers are unbound even before reaching $\varv_{\text{crit}}$. Adapted from \citet{maeder}.}
\label{maratabulak}
\end{center}
\end{table}
Including the \textit{nonradial} component of $g_{\text{rad},L}$, which is antiparallel with the vector of the effective gravity $\vec{g}_{\text{eff}}$ (see Fig.~\ref{epsideviatik}), 
leads in case of rotating stars to inhibition of the WCD paradigm, described in Sect.~\ref{wicom} \citep{ovoce1}.
The models that include the nonradial forces and gravity-darkening \citep[see, e.g.,][see also Figs.~\ref{owo2}]{ovorot} show 
the lower equatorial wind mass flux $\dot{M}(\Omega,\theta)$, due to the latitudinally reduced $\vec{g}_{\text{eff}}(\Omega,\theta)$ (see, e.g.,~Eq.~\eqref{lmln}).
The equatorial wind density is thus reduced relative to that at higher latitudes.
There is a further reason why the radiative driving is ill-suited to form a dense equatorial disk-like outflow:
in optically thin stellar winds (or within the optically thin spherical envelopes, e.g., in planetary nebulae, see Sects.~\ref{Beforbphen}, \ref{Lumbluvars} and \ref{postaks}) a multiple photon scattering 
maintains a strong outward radiative force. In an optically thick disk the radiative acceleration $g_\text{rad,L}$ is produced by single scattering,
for a disk of a transverse optical thickness $\tau>1$ the radiative acceleration is 
$g_\text{rad,thick}\approx g_\text{rad,thin}/\tau.$
Unless the Eddington factor $\Gamma\gg 1$, the radiation acceleration $g_\text{rad}$ in the disk is unable even to balance the $g_\text{grav}$
\citep{ovoce6,ovorot}.

\section{Magnetically confined stellar wind density enhancements}\label{confidenti}
We attach a brief description of the processes connected with the disk-like 
density enhancements possibly driven by stellar $\vec B$ field (see~Sect.~\ref{postaks}).
Stellar magnetic field is generally characterized by its dominant component, i.e., by the dipole field with the magnetic axis
arbitrarily tilted to the axis of rotation \citep{Towny2}. In case of the magnetic field dominance
$\displaystyle\eta_*>1\text{ where }\eta_*=B^2/(\mu_0\rho\varv^2)$ is the ratio of magnetic and kinetic energy densities. Such field may gives
rise to a thin outflowing disk in the magnetic equatorial plane \citep{dula1}.

\subsection{Equations of magnetically confined matter in the field of oblique rotators}\label{eqconfirot}
Force equilibrium for a mass element $\delta m$ located in the position $\vec{r}$ in stellar proximity is given by the balance of the components of
the gravitational and centrifugal forces tangential to the local magnetic field line \citep{prus04},
\begin{align}\label{prusik1}
(\vec{F}_G+\vec{F}_C)\cdot\vec{B}=\vec{0}. 
\end{align}
We may write the expressions for centrifugal force in cylindrical frame and in spherical frame, respectively, in the form
(where the boldface-typed quantities with ``hat`` are unit vectors, see the notation introduced in Appendix~\ref{appendix1})
\begin{align}\label{prusik2}
\vec{F}_C=\delta m\,\Omega^2R\mathbf{\hat{\boldsymbol{R}}},\quad\quad\quad\quad\vec{F}_C=\delta m\,\Omega^2r
\left[\mathbf{\hat{\boldsymbol{r}}}-\mathbf{\hat{\boldsymbol{\Omega}}}\left(\mathbf{\hat{\boldsymbol{r}}}\cdot\mathbf{\hat{\boldsymbol{\Omega}}}\right)\right].
\end{align}
Considering the dipole stellar magnetic field, we denote the angle between magnetic and rotation axes as $\psi$ and the azimuthal angle of 
the magnetic moment vector $\vec{m}$ as $\phi$. 
Magnetic dipole field with a magnetic dipole moment $\vec{m}=m\mathbf{\hat{\boldsymbol{m}}}$ in spherical coordinates is \citep[e.g.,][]{Landau2}
\begin{align}\label{prusik3}
\vec{B}(\vec{r})=\frac{\mu_0m}{4\pi r^3}
\left[3\left(\mathbf{\hat{\boldsymbol{m}}}\cdot\mathbf{\hat{\boldsymbol{r}}}\right)\mathbf{\hat{\boldsymbol{r}}}-\mathbf{\hat{\boldsymbol{m}}}\right]. 
\end{align}
Inserting Eqs.~\eqref{prusik2} (spherical equation) and \eqref{prusik3} together with the expression for the gravitational force 
$\displaystyle \vec{F}_G=-\mathbf{\hat{\boldsymbol{r}}}\,GM_{\!\star}\delta m/r^2$ in Eq.~\eqref{prusik1} consequently yields \citep{prus04},
\begin{align}\label{prusik4}
\left[\mathbf{\hat{\boldsymbol{r}}}\left(1-\frac{GM_{\!\star}}{\Omega^2r^3}\right)
-\mathbf{\hat{\boldsymbol{\Omega}}}\left(\mathbf{\hat{\boldsymbol{r}}}\cdot\mathbf{\hat{\boldsymbol{\Omega}}}\right)\right]\cdot
\bigg[3\left(\mathbf{\hat{\boldsymbol{m}}}\cdot\mathbf{\hat{\boldsymbol{r}}}\right)\mathbf{\hat{\boldsymbol{r}}}-\mathbf{\hat{\boldsymbol{m}}}\bigg]=0,
\end{align}
where $\displaystyle (GM_{\!\star}/\Omega^2)^{1/3}=R_{\text{co}}$ denotes the \textit{corotation radius}.
We distinguish three possible configurations of distribution of magnetically confined circumstellar matter in oblique rotators \citep{prus04}:
\begin{itemize}
\item Aligned rotator where $\psi=0$ $(\mathbf{\hat{\boldsymbol{m}}}=\mathbf{\hat{\boldsymbol{\Omega}}})$ gives the following equilibrium condition
for the confined matter,
\begin{align}\label{prusik5}
\left\{2\left[1-\left(\frac{R_{\text{co}}}{r}\right)^3\right]
-3\left(\mathbf{\hat{\boldsymbol{r}}}\cdot\mathbf{\hat{\boldsymbol{\Omega}}}\right)^2+1\right\}
\big(\mathbf{\hat{\boldsymbol{r}}}\cdot\mathbf{\hat{\boldsymbol{\Omega}}}\big)=0.
\end{align}
Equation \eqref{prusik5} implies two solutions:
\begin{enumerate}
\item$\displaystyle\text{cos}\,\theta\equiv\mathbf{\hat{\boldsymbol{r}}}\cdot\mathbf{\hat{\boldsymbol{\Omega}}}=0$:
accumulation of matter in (coinciding magnetic and rotational) equatorial plane of the star.
\item$\displaystyle\mathbf{\hat{\boldsymbol{r}}}\cdot\mathbf{\hat{\boldsymbol{\Omega}}}\neq 0,\,\,
\text{cos}^2\theta=1-\frac{2}{3}\left(\frac{R_{\text{co}}}{r}\right)^3$:
matter accumulation in chimney-shaped surfaces above and below the equatorial plane whose axes coincide with the 
stellar rotational axis. This solution only exists for 
$r\geq(2/3)^{1/3}R_{\text{co}}$, the stability tests however show that the ``chimney'' solution is 
unstable while the equatorial solution is stable for $r>R_{\text{co}}$ \citep{prus04}.
\end{enumerate}
\item Perpendicular rotator, $\psi=\pi/2$ $(\mathbf{\hat{\boldsymbol{m}}}\cdot\mathbf{\hat{\boldsymbol{\Omega}}}=0)$ where
Eq.~\eqref{prusik4} gives the following equilibrium,
\begin{align}\label{prusik6}
\left\{2\left[1-\left(\frac{R_{\text{co}}}{r}\right)^3\right]
-3\left(\mathbf{\hat{\boldsymbol{r}}}\cdot\mathbf{\hat{\boldsymbol{\Omega}}}\right)^2\right\}
\big(\mathbf{\hat{\boldsymbol{r}}}\cdot\mathbf{\hat{\boldsymbol{m}}}\big)=0.
\end{align}
Equation \eqref{prusik6} again implies two solutions:
\begin{enumerate}
\item$\displaystyle\text{cos}\,\theta\equiv\mathbf{\hat{\boldsymbol{r}}}\cdot\mathbf{\hat{\boldsymbol{m}}}=0$:
the solution corresponds to the equatorial plane, however now with respect to the magnetic axis.
\item$\displaystyle\mathbf{\hat{\boldsymbol{r}}}\cdot\mathbf{\hat{\boldsymbol{m}}}\neq 0,\,\,
\text{cos}^2\theta=\frac{2}{3}\left[1-\left(\frac{R_{\text{co}}}{r}\right)^3\right]$:
chimney-shaped structure of confined matter that is axisymmetric with respect to the rotation axis of the star. 
The solution only exists for $r>R_{\text{co}}$. The stability analysis in this case shows that both solutions are
unstable near the star (up to $1$-$2$ $R_{\text{co}}$) while they are stable further out. \citep{prus04}.
\end{enumerate}
\item Oblique rotators where the equilibrium condition
from Eq.~\eqref{prusik4} gives
\begin{align}\label{prusik7}
\left\{\left[2\left(1-\left(\frac{R_{\text{co}}}{r}\right)^3\right)
-3\left(\mathbf{\hat{\boldsymbol{r}}}\cdot\mathbf{\hat{\boldsymbol{\Omega}}}\right)^2\right]\mathbf{\hat{\boldsymbol{m}}}
+\text{cos}\,\psi\,\mathbf{\hat{\boldsymbol{\Omega}}}\right\}
\cdot\mathbf{\hat{\boldsymbol{r}}}=0.
\end{align}
This indicates more complicated structure of the equilibrium regions: in case of small $\psi$ there forms a disk-like and a chimney-
shaped structure (with a disk plane in the magnetic equatorial plane and with the the chimney axis tilted with respect to the rotation axis)
while for large $\psi$ the chimney shapes become more curved and tilted and the disk becomes somewhat warped (see \citep{prus04} for details).
\end{itemize}
The study of \citet{prus04} presents a formulation of a strong magnetic field limit based on the condition of the balance of forces that are tangential to the field lines and 
maps out the complex surfaces on which the circumstellar material can accumulate \citep{Towny2}. This approach (supported by other prior similar studies based on minimization of
effective potential, e.g.,~\citet{nakone}) was used as a foundation for the rigidly rotating magnetosphere model (see Sect.~\ref{rindzak}).
\subsection{Rigidly rotating magnetosphere (RRM) model}\label{rindzak}
The model of a magnetosphere that is at rest in a corotating reference frame (cf.~Sect.~\ref{macom}) whose
basic principles follow the considerations of \citet{Towny2}: if the magnetic field line 
potential $\Psi(s)$ exhibits an extremum along the field line (where $s$ is the coordinate direction along the field line), 
so that $\text{d}\Psi/\text{d}s\equiv\Psi^\prime=0$ at some point, then the plasma parcel remains at rest.
Whether the parcel can remain at such an equilibrium point at rest over significant 
timescales depends however on the nature of the extremum. At a local maximum where 
$\text{d}^2\Psi/\text{d}s^2\equiv\Psi''<0$ the equilibrium is unstable: small displacements away from the extremal point perpetually grow. 
On the other hand, at a local minimum with $\Psi''>0$, the equilibrium is stable:
any small displacement along the local magnetic field line produces a restoring force directed toward the equilibrium point. Such minima 
represent the locations for circumstellar matter to accumulate, it forms a magnetosphere that is at rest in a corotating reference frame \citep{Towny2}.

Comparing the potentials that arise from Eq.~\ref{prusik1}:
within the Roche limit (where the most of the stellar mass is assumed to be concentrated centrally with a spherically symmetric distribution) 
the effective potential $\Psi$  
is in spherical coordinates given by Eq.~\eqref{fomis1}.
Using the dimensionless coordinate 
$\xi=r/R_{\text{co}}$, Eq. \eqref{fomis1} becomes
\begin{align}\label{rigid29}
\Psi({\xi})=\frac{GM_{\!\star}}{R_{\text{co}}}\left(-\frac{1}{\xi}-\frac{1}{2}\xi^2\,\text{sin}^2\,\theta\right).
\end{align}
We introduce the dimensionless potential $\Xi$, independent of the angular (rotational) velocity $\Omega$:  
\begin{align}\label{rigid30}
\Xi({\xi})=\frac{R_{\text{co}}}{GM_\star}\Psi({\xi})=-\frac{1}{\xi}-\frac{1}{2}\xi^2\,\text{sin}^2\,\theta.
\end{align}
In Eq.~\eqref{rigid30} we can identify two regimes: if $r$ is much smaller than 
the corotation radius $R_{\text{co}}$ ($\xi\ll 1$), the potential $\Xi$ is spherically symmetric
and increases outwards. Conversely, if $r$ greatly exceeds $R_{\text{co}}$ 
($\xi\,\text{sin}\,\theta\gg 1$), the potential $\Xi$ exhibits the cylindrical symmetry about the same axis and decreases outwards.
In the latter regime the field line potential exhibits the minimum near which circumstellar plasma accumulates.

In a corotating spherical frame with aligned axes Eq.~\eqref{prusik3} becomes
\begin{align}\label{rigid1corot}
\vec{B}(\vec{r})=\frac{\mu_0}{4\pi}\frac{m}{r^3}\left(2\,\text{cos}\,\theta\,\mathbf{\hat{\boldsymbol{r}}}+\text{sin}\,\theta\,\mathbf{\hat{\boldsymbol{\theta}}}\right).
\end{align}
We obtain the projection $B_s$ of the field measured along
the field line as $B_s=(B_r^2+B_\theta^2)^{1/2}=B$,
\begin{align}\label{rigid1param1}
B=\frac{\mu_0}{4\pi}\frac{m}{r^3}\sqrt{1+3\,\text{cos}^2\,\theta}.
\end{align}
By integrating the spherical field-line identity $\text{d}r/B_r=r\,\text{d}\theta/B_{\theta}$, where from Eq.~\eqref{rigid1corot} follows
$B_r/B_{\theta}=2\,\text{cos}\,\theta/\text{sin}\,\theta$, we obtain the parametric equation
\begin{align}\label{rigid1param}
\xi=\gamma\,\text{sin}^2\theta,
\end{align}
where the parameter $\gamma$ specifies the maximum radius $R_{\text{peak}}$ ($\theta=\pi/2$) 
of the field line in units of the corotation radius $\gamma=R_{\text{peak}}/R_{\text{co}}$.
Inserting Eq.~\eqref{rigid1param} into Eq.~\eqref{rigid30}, we obtain
the dimensionless potential of a dipole field line
in case of \textit{aligned} dipole configuration $(\mathbf{\hat{\boldsymbol{m}}}\cdot\mathbf{\hat{\boldsymbol{\Omega}}}=1)$,
\begin{align}\label{rigid32}
\Xi(\theta)=-\frac{1}{\gamma\,\text{sin}^2\theta}-\frac{1}{2}\gamma^2\,\text{sin}^6\theta,
\end{align}
By integrating the field-line identity $\text{d}s/B=r\,\text{d}\theta/B_{\theta}$, from Eqs. \eqref{rigid1corot},
\eqref{rigid1param} and \eqref{rigid1param1} we obtain 
\begin{align}\label{rigid1param2}
\frac{\text{d}s}{\text{d}\theta}=R_{\text{co}}\gamma\,\text{sin}\,\theta\,\sqrt{1+3\,\text{cos}^2\,\theta}.
\end{align}
For $\theta=\pi/2$ we use the identity $\xi=\gamma$. Differentiation
$\Xi^{\prime\prime}=(\theta^\prime)^2(\text{d}^2\Xi/\text{d}\theta^2)+\theta^{\prime\prime}(\text{d}\Xi/\text{d}\theta)$ gives
\begin{align}\label{rigid35}
\frac{\text{d}^2\Xi}{\text{d}s^2}=\Xi^{\prime\prime}=\frac{1}{R_{\text{co}}^2}\left(-\frac{2}{\xi^3}+3\right).
\end{align}
Since $\Xi^{\prime\prime}$ must be positive in order to constitute an accumulation surface,
the inner truncation radius is thus given by $\xi_\text{in}\approx 0.87$ at which $\Xi''$ changes from positive $(\xi>\xi_\text{in})$
to negative $(\xi<\xi_\text{in})$ values. 
Throughout the region in the equatorial plane between this truncation radius $\xi_\text{in}$ and the corotation radius ($\xi=1$)
magnetic tension supports material against the net inward pull caused by gravity that exceeds here the centrifugal force. Beyond this region,
when ($\xi>1$) the centrifugal force surpasses gravity and the effect of magnetic tension holds the material down against 
the net outward pull \citep{Towny2}.
In Keplerian disks the gravitational and centrifugal force is in exact balance, this is not required in a RRM
inasmuch the magnetic tension can absorb any net resultant force perpendicular to field lines \citep{prus04}.

Hydrostatic stratification along the field line is governed by the equation of hydrostatic balance
$\displaystyle\text{d}P/\text{d}s=-\rho\,\text{d}\Psi/\text{d}s$
where the gas pressure is given by Eq.~\eqref{statix2}. For simplicity we assume the constant temperature $T$, 
by integrating the hydrostatic equilibrium condition we obtain the density distribution along the field line,
\begin{align}\label{rigid40}
\rho(s)=\rho_m\,\text{exp}\left[-\mu m_{\text{u}}\frac{\Psi(s)-\Psi_m}{kT}\right],
\end{align}
where the subscript $m$ denotes the value at the potential minimum where $s=s_m$.
Taylor expansion of the effective potential $\Psi(s)$ about this minimum gives
\begin{align}\label{rigid41}
\Psi(s)=\Psi_m+\frac{1}{2}\Psi''(s-s_m)^2+\,\ldots\,,
\end{align}
where we have used the fact that by the definition $\Psi'_m=0$. 
In the neighborhood of the minimum the density distribution 
Eq.~\eqref{rigid40} may be thus well approximated by
\begin{align}\label{rigid42}
\rho(s)\approx\rho_m\,\text{exp}\left[-\mu m_{\text{u}}\frac{\Psi''(s-s_m)^2}{2kT}\right]
\approx\rho_m\,\text{exp}\left[-\frac{(s-s_m)^2}{h_m^2}\right].
\end{align}
The density scale height $h_m$ of the RRM (using Eq.~\eqref{rigid30} and Eq.~\eqref{rigid35} that gives $\Xi''=3/R_{\text{co}}^2$ for
$\xi\gg 1$) therefore is
\begin{align}\label{rigid43}
h_m=\sqrt{\frac{2kT}{\mu m_{\text{u}}}\frac{1}{\Psi''}}=\sqrt{\frac{2kT}{\mu m_{\text{u}}}\frac{R_{\text{co}}}{GM_{\!\star}}}\sqrt{\frac{1}{\Xi''}},\quad\quad
h_m=\sqrt{\frac{2kT}{3\mu m_{\text{u}}GM_{\!\star}}}\,R_{\text{co}}^{3/2}\,\,\text{for}\,\,r\gg R_{\text{co}}.
\end{align}
Similar vertical stratification formally applies for Keplerian disks ($H\sim R^{3/2}$, see Sect.~\ref{vertikalekdisk}), 
however for the RRM this remains constant even far from the origin (it does not produce the flaring disk).
By integrating Eq.~\eqref{rigid42} over the Gaussian hydrostatic stratification, we obtain the relation for the local surface density $\sigma_m$ 
\begin{align}\label{rigid45}
\sigma_m=\int_{-\infty}^{\infty}\rho(s)\,\text{d}s
\approx\mu_m\,\rho_m\int_{-\infty}^{\infty}\,\text{exp}\left[-\frac{(s-s_m)^2}{h_m^2}\right]\,\text{d}s,\quad\text{i.e.}
\quad\sigma_m\approx\mu_m\,\rho_m\!\!\sqrt{\pi}h_m,
\end{align}
(cf.~Eq.~\eqref{Sigmasegva} for Keplerian disk) where $\mu_m$ denotes the projection cosine to the surface normal.

Model of the global distribution of the surface density that is proportional to the accumulation rate of material
loaded from the star's radiatively driven wind has been proposed by \citet{Towny2}: 
for a dipole flux-tube bundle intersecting the stellar surface at $r=R_{\star}$ with 
a projection cosine $\mu_{\star}$ and having a cross-sectional area $\text{d}A_{\star}$, the rate of mass increase is
\begin{align}\label{rigid46}
\dot{m}=\frac{2\mu_{\star}\dot{M}}{4\pi R_{\star}^2}\,\text{d}A_{\star},
\end{align}
where the factor 2 takes into account the mass injection at two distinct footpoints. 
Considering the simple case with a single minimum at field line coordinate $s_m$ where the flux-tube area is $\text{d}A_m$
and the projection cosine to the accumulation surface normal is $\mu_m$, the corresponding rate of increase of the surface density 
(where $\dot{\sigma}_m\,\text{d}A_m=\dot{m}\mu_m$) can be written as
\begin{align}\label{rigid47}
\dot{\sigma}_m=\mu_m\frac{2\mu_{\star}\dot{M}}{4\pi R_{\star}^2}\,\frac{\text{d}A_{\star}}{\text{d}A_m}.
\end{align} 
Due to the divergence free constraint $\vec{\nabla}\cdot\vec{B}=0$ we have the identity $\text{d}A_{\star}B_{\star}=\text{d}A_mB_m$,
whose substitution into Eq.~\eqref{rigid47} gives
\begin{align}\label{rigid48}
\dot{\sigma}_m=\mu_m\frac{2\mu_{\star}\dot{M}}{4\pi R_{\star}^2}\,\frac{B_m}{B_{\star}}.
\end{align}
For a dipole field thus the material feeding rate of the disk obviously declines with radius, 
according to $\dot{\sigma}_m\sim B\sim r^{-3}$ (cf.~Eq.~\eqref{rigid1corot},
see \citet{Towny2} for further details).

\clearpage

\end{appendix}
\newpage

\HlavickaLiteratura

\addcontentsline{toc}{chapter}{References}
\renewcommand{\bibname}{References}

\bibliographystyle{aa} % style apj.bst
\bibliography{bibliography}

\newpage

\HlavickaPublikace

\addcontentsline{toc}{chapter}{List of publications}
% \pdfbookmark{List of publications}{List of publications}
% \vloz{Appendix}
{\chapter*{List of publications}}
\section*{Refereed papers}
\begin{itemize}
 \item Kurf\"urst, P., Feldmeier, A. \& Krti\v cka, J., \textit{Modeling of extended decretion disks of critically 
      rotating stars}, {2014}, \href{http://adsabs.harvard.edu/abs/2014A\%26A...569A..23K}{A\&A, 569, 23}
 \item Krti\v cka, J., Kurf\" urst, P. \& Krti\v ckov\'a, I., 
      \textit{Magnetorotational instability in the decretion disks of critically rotating stars and the outer structure of Be and Be/X-ray disks},
      2014, \href{http://adsabs.harvard.edu/abs/2015A\%26A...573A..20K}{A\&A, 573, A20}
\end{itemize}
\section*{Conference proceedings}
\begin{itemize}
 \item Kurf\"urst, P., \textit{Hydrodynamics of Decretion Disks} in From Interacting
        Binaries to Exoplanets: Essential Modeling Tools, ed. M. T. Richards, \&
        I. Hubeny, 2012, Cambridge University Press, \href{http://adsabs.harvard.edu/abs/2012IAUS..282..257K}{IAUS, 282, 257}
 \item Kurf\"urst, P., \& Krti\v cka,
        J., \textit{Hydrodynamics of Outer Parts of Viscous Decretion Disks}, in Circumstellar Dynamics at High Resolution, ed. A. Carciofi,
        \& Th. Rivinius, 2012, \href{http://adsabs.harvard.edu/abs/2012ASPC..464..223K}{ASPC, 464, 223}
 \item Kurf\"urst, P., Feldmeier, A. \& Krti\v cka, J., \textit{Time-dependent modeling of extended thin decretion disks},
      in Massive Stars: From $\alpha$ to $\Omega$, 2013, \href{http://adsabs.harvard.edu/abs/2013arXiv1309.3505K}{arXiv:1309.3505} [astro-ph.SR]
 \item Kurf\"urst, P., \& Krti\v cka, J., \textit{Time-dependent modeling of extended decretion disks}, in Bright Emissaries, ed. C. Jones, \& A. Sigut, 2014, ASPC, in press
\end{itemize}
\section*{Textbook}
\begin{itemize}
\item Kurf\"urst, P., \textit{Po\v cetn\'i praktikum} (\textit{Computing practice}), 2015, study material of fundamental
mathematics for physicists, bio- and astrophysicists and set of examples for practicing: \url{http://www.physics.muni.cz/~petrk/skripta.pdf}
\end{itemize}

\newpage

%%%%%%%%%%%%%%%%%%%%%%%%%%%%%%%%%%%%%%%%%%%%%%%%
%%%%%%%%%%% PRAZDNA STRANA NA ZAVER %%%%%%%%%%%%
%%%%%%%%%%%%%%%%%%%%%%%%%%%%%%%%%%%%%%%%%%%%%%%%

\newpage
\thispagestyle{empty}
\fancyhf{}
\newpage
\mbox{}

\end{document}